%% file: 00-Main.tex
\documentclass[12pt, a4paper, twoside, openright]{book}
\usepackage{msor-thesis}
\usepackage{palatino} 
\usepackage{url} 
\usepackage[normalem]{ulem}
\usepackage[T1]{fontenc}
\usepackage{amssymb}
\usepackage{amsmath}
\usepackage{graphicx}\graphicspath{ {figures/} }
\usepackage{amsfonts}
\usepackage[usenames,dvipsnames]{xcolor}
\usepackage[pagebackref=true]{hyperref}   
\hypersetup{colorlinks=true, 
citecolor=MidnightBlue, linkcolor=MidnightBlue, urlcolor=Cyan}
\usepackage{tikz}
\usepackage{bm}
\usepackage{amsthm, physics}
\usepackage{mathrsfs}
\usepackage{multicol}
\usepackage{float, array, tabu}
\usepackage{tensor}
\usepackage{tocloft}
\usepackage{enumerate}
\usepackage{bigints}
\usepackage{appendix}
\usepackage{setspace}
\usepackage{cancel}
\usepackage[parfill]{parskip}
\usepackage{physics, tensor, float, subcaption}
\usepackage{doi}

\definecolor{applegreen}{rgb}{0.55, 0.71, 0.0}

\newtheorem{theorem}{Theorem}
\newcommand{\e}{\mathrm{e}}
\newcommand{\V}{\mathcal{V}}
\def\O{{\mathcal{O}}}
\DeclareMathOperator\diag{diag}
\DeclareMathOperator\sign{sign}
\def\tr{{\mathrm{tr}}}
\def\diag{{\mathrm{diag}}}
\def\cof{{\mathrm{cof}}}
\def\pdet{{\mathrm{pdet}}}
\def\d{{\mathrm{d}}}
\def\Kerr{{\scriptscriptstyle{\mathrm{Kerr}}}}
\def\eos{{\scriptscriptstyle{\mathrm{eos}}}}
\newcommand*{\Chi}{\mbox{\Large$\chi$}}
\def\Chi{\mathcal{X}}
\def\emph{\textit}
\renewcommand{\emph}{\textit}
\renewcommand{\underline}{\textit}

\begin{document}

\frontmatter


\title{Excising Curvature Singularities from General Relativity}
\author{Alex Simpson}

\subject{Mathematics}
\abstract{\addcontentsline{toc}{chapter}{Abstract}
This thesis operates within the framework of general relativity without curvature singularities. The motivation for this framework is explored, and several conclusions are drawn with a look towards future research. There are many ways to excise curvature singularities from general relativity; a full list of desirable constraints on candidate geometries is presented. Several specific candidate spacetimes in both spherical symmetry and axisymmetry are rigorously analysed, typically modelling (charged or uncharged) regular black holes or traversable wormholes. Broadly, these are members of the family of black-bounce spacetimes, and the family of black holes with asymptotically Minkowski cores. Related thin-shell traversable wormhole constructions are also explored \emph{via} the Darmois--Israel formalism, as well as a brief look at the viability of thin-shell Dyson mega-spheres. The eye of the storm geometry is analysed, and discovered to be very close to an idealised candidate geometry within this framework. It is found to contain highly desirable features, and is not precluded by currently available measurements. For all spacetimes discussed, particular focus is placed on the extraction of (potential) astrophysical observables in principle falsifiable/verifiable by the observational and experimental communities. An examination of the spin one and spin zero quasinormal modes on a background regular black hole with asymptotically Minkowski core is performed by employing the relativistic Cowling approximation. A cogent effort is made to streamline the discourse between theory and experiment, and to begin filling the epistemological gap, which will enable the various communities involved to optimise the advancement of physics \emph{via} the newly available observational technologies (such as LIGO/Virgo, and the upcoming LISA). Furthermore, three somewhat general theorems are presented, and two new geometries are introduced for the first time to the literature.
}

\ack{\addcontentsline{toc}{chapter}{Acknowledgements}
First, Matt. You have been an incredible mentor, and a great friend. The amount of time you make for your students is truly noteworthy, especially for a researcher of your caliber and prominence. Massive thanks for all the conversations, and all the time spent putting up with my at times overly earnest and dogged behaviour. I will never forget the teachings that you have passed on.

To my wonderful and loving partner, Harri. You have been absolute granite for me through the entire duration of this PhD (and beyond). You have given me outrageous levels of dedication and support; I simply could not have done what I've achieved without you. Your kindness and capacity for love is a true beacon in what can be an at times challenging world, and I am honoured and privileged to bathe in it. Thank you.

To my parents, Liana and Glenn. Your support from the day I was born has enabled simply incredible opportunities for me and my future. Every day I am grateful for the sacrifices you have both made, putting my education and optionality above all else. Thanks also for all the hospitality, whether it was tolerating me as a flatmate, or feeding me Sunday dinners, your constant assistance through this process has been invaluable. Much love!

To my grandfather, Ian. Thanks for all the chats. I have found in you and fantastic companion to explore the scientific world with, and your insights never fail to amaze and surprise me. I'm looking forward to many more conversations; do give me a call if you have any questions on the content in the thesis! Love you grandpa.

To my grandmother Pat, who sadly passed away during my PhD while the COVID lockdown was happening in 2020. Thank you for your warmth and your love in life. I am grateful to have shared many spirited conversations with you, and your imparted wisdom has helped to guide me through this journey. I hope you have found peace.

To my other grandmother, Flora. Your strength of character has always prevailed. The difficulty of the life decisions you have made which have allowed me the opportunity to be where I am can not be overstated. You are an incredible and inspirational force, and I am ever grateful for your love and support.
\clearpage

I would like to thank Thomas Berry. You have become a close friend, and I thoroughly enjoyed our time working on GR together. Specific thanks go towards your contributions on the various publications we have co-authored; more generally thanks for the many many productive conversations, insights, and banter. Looking forward to the boundless opportunities in the future for us to work together!

Extending this, I would like to thank all of my co-authors in my career to-date. More specifically, those who co-authored papers from which content has been drawn for this thesis are: Matt Visser, Thomas Berry, Francisco Lobo, Edgardo Franzin, Stefano Liberati, Jacopo Mazza, Manuel Rodrigues, Marcos de Silva, and Prado Mart\'{i}n--Moruno. Thank you all for your contributions.

To Ratu, hah. You're the man, Ratu. You inspired me to learn GR in the first instance, and ever since I have been grateful for our friendship. The conversations we have are stimulating, fun, challenging, and help me to retain my zest and passion for life. Cheers.

To the flat at Ellice Street. Gorilla grip Gabe, Vita, Kieran, sweet James, Ratu, and Tahlia. I could not have imagined a more supportive and exciting living environment to be a part of while I embarked on this little black hole campaign. Every single one of you was a fantastic housemate, and exceptional friend. Thanks guys.

To the friend group in general, you're all the best. I am super lucky to be surrounded by such a caring, supportive, and outright zany troupe. To Max, thanks for the vend. To Mez, Bek, Maylis, DT, Josh, Henry, Nikolai, Theo, Boots, Dickie, Marcus, Chloe, Scotty, Gus, Katie; you're all living legends. If there's anyone I've missed, I'm sorry, rip into me about it. Much love.

\begin{center}
It's been a ride.
\end{center}
}

\phd


\maketitle

{\addcontentsline{toc}{chapter}{\textbf{Contents}}}
\tableofcontents

{\addcontentsline{toc}{chapter}{\textbf{List of figures}}}
\newpage
\listoffigures


\mainmatter



\include{01-Introduction}


\include{02-Black-bounce/1-SVog}

%
%
%
%

\include{02-Black-bounce/4-SVCBB}


\include{02-Black-bounce/3-SVNBB}
\include{03-AMC/1-AMCog}


\include{03-AMC/2-AMCQNM}


\include{03-AMC/3-AMCEOS}


\include{02-Black-bounce/2-SVthinshell}


\include{04-Dyson}


\include{05-Conclusion}

%
%
\bibliographystyle{acm}
\addcontentsline{toc}{chapter}{Bibliography}
\bibliography{10-myrefs.bib}

\end{document}

%% file: 01-Introduction.tex
\chapter{Introduction}\label{C:intro}

%
%
%
%
%
\vspace*{8pt}
Scientific progress is traditionally marked by experimental or observational falsification/verification of hypotheses.
%
%
General relativity (GR) and quantum mechanics (QM) are the two most successful physical theories in human history according to this metric. Given the constraints provided by available data, both have well-understood domains of validity. The major role of the theoretical physicist is to probe the domains which remain unspoken for. Very broadly speaking, there are two main areas for which humanity does not currently possess satisfactory fundamental theories:

\begin{itemize}
    \item The large-scale structure of spacetime. These theories fall under the field of cosmology, and typical approaches are to assume that standard GR holds up to some distance scale before introducing dark matter and dark energy terms to account for several observed phenomena (almost always a classical formulation). Progress in this field is challenging due to the difficult nature of obtaining reliable data on such large distance scales. This thesis will only touch \textit{very} briefly on problems belonging to this domain.
    \item Quantum gravity. The intersection of QM (or more maturely, quantum electrodynamics and quantum chromodynamics) with GR has been the ``ultimate'' question for theoretical physicists for decades now. A satisfactory, phenomenologically verifiable theory of quantum gravity is still elusive. Many different approaches have been taken, all with their specific set of motivations. This thesis will concern itself primarily with problems occupying this domain, with a focus on excising classical curvature singularities from standard GR.
\end{itemize}
%
%
\clearpage

The environment in which a ``complete'' theory of quantum gravity is required can be cogently summarised by the following diagram:\footnote{Please see~\href{https://en.wikipedia.org/wiki/Planck_units}{PlanckUnits-Wikipedia}, and~\href{https://en.wikipedia.org/wiki/Compton_wavelength}{ComptonWavelength-Wikipedia}, for further information on the Planck mass and Compton wavelength respectively.}
\vspace*{7pt}

\begin{figure}[htb!]
\begin{center}
    \includegraphics[scale=0.68]{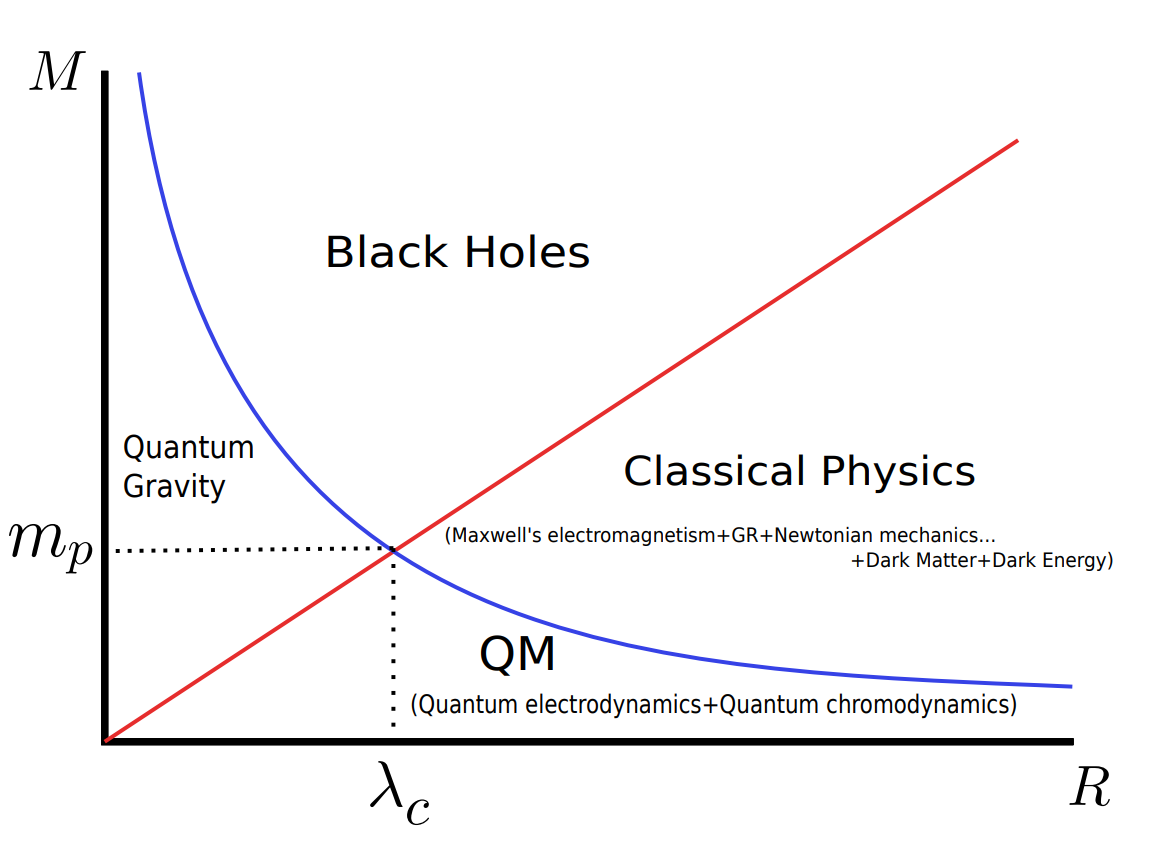}
    \caption[Graphical representation of the different physical regimes, expressed with respect to mass $M$ and distance scale $R$]{Graphical representation of the different physical regimes, expressed with respect to mass $M$ and distance scale $R$. Here, $m_p$ is the Planck mass, and $\lambda_c$ the Compton wavelength.}\label{IntroQG}
\end{center}
\end{figure}
%
\vspace*{7pt}

The three dots in the ``Classical Physics'' regime are employed to indicate that dark matter and dark energy formulations are only required on extremely large distance scales; usually of order the kiloparsec scale for dark matter (\textit{e.g.} intra-galactic separation scale), and the megaparsec scale for dark energy (\textit{e.g.} inter-galactic separation scale).

In the absence of a better alternative, it is common to extrapolate classical physics into the black hole region, and to then compare predicted phenomenology with what little empirical data is available. Many theoreticians agree that this extrapolation is fairly reasonable in the region between an event (outer) and Cauchy (inner) horizon.\footnote{Though even this is up for debate; ``near-horizon'' physics contains many peculiarities~\cite{GVP2:Visser:1996}, and indeed the Mazur--Mottola gravastars predict no horizon formation at all~\cite{GCS:Mazur:2002, GVCS:Mazur:2004, DEACS:Mazur:2004, SG:Visser:2004}. See \S~\ref{BHMint} for further detail on these objects. See Appendix~\ref{B:GRbasics} for definition of various types of horizon.} Going any further into the interior is effectively guesswork, and when doing so, extrapolating GR predicts the existence of curvature singularities. These almost always occur in precisely the region where quantum gravity is expected to take over; this is discussed further in \S~\ref{Intro:knob}. It is worth noting that typical problems on the QM distance scale are ``low-mass''; the mass can often be approximated to zero. As such, one usually either ignores gravity entirely, or at worst embeds the problem in a (quasi)-Newtonian framework. There is no straightforward way to extrapolate QM into the high-gravity regimes.

Attempts to resolve the mystery of quantum gravity range far and wide. There was a period for many years when observational and experimental capacity in physics increased only very slowly. In parallel, there is the inherent difficulty of working with black holes --- constructing experiments or discovering effective observational techniques which are able to probe behind a horizon is, by nature, \textit{very hard}. The ability to falsify/verify theories was hence highly constrained, and in the absence of an immediate path to progress \textit{\`{a} la} the traditional scientific method, creativity amongst theoreticians proliferated. Various approaches emerged --- these range from completely new frameworks and ideas, such as string theory~\cite{ST1:Witten:1987, ST2:Witten:2012, ST:Polcinski:2005}, to more minimally modified theories built on the shoulders of the robust ``GR$+$QM'' duo, such as loop quantum gravity~\cite{QG:Rovelli:2004, CLQG:Rovelli:2015, ITLQGTC:Ashtekar:2007}, or causal dynamical simplicial manifolds~\cite{QGVCDT:Ambjorn:2014, CDTATSFATOQG:Ambjorn:2015}. Failure to convert any of these approaches into a phenomenologically verifiable theory led to many critiques bemoaning the stagnation of theoretical physics; \textit{e.g.} Lee Smolin's ``The Trouble With Physics''~\cite{TTWP:Smolin:2007}, or Peter Woit's ``Not Even Wrong"~\cite{NEW:Woit:2011}. However, we have now entered a new age of instrumentalism and observational scope. The advent of collaborations such as the Event Horizon Telescope (EHT)~\cite{FM87EHTR5:Akiyama:2019, FM87EHTR4:Akiyama:2019, FM87EHTR3:Akiyama:2019, FM87EHTR2:Akiyama:2019, FM87EHTR1:Akiyama:2019, FM87EHTR:Akiyama:2019}, LIGO/Virgo~\cite{BBHMITFALIGOOR:Abbott:2016, OOGWFABBHM:Abbott:2016, OOA50SMBBH:Abbott:2017, AIOTBBHM:Abbott:2016, POTBBHM:Abbott:2016, GW23SM:Abbott:2020, GWTC1:Abbott:2019}, and the James Webb Space Telescope~\cite{JamesWebb1, JamesWebb2, JamesWebb3, JamesWebb4, JamesWebb5, JamesWebb6}, enable incredible new opportunities.\footnote{See~\href{https://en.wikipedia.org/wiki/LIGO}{LIGO-Wikipedia}, \href{https://en.wikipedia.org/wiki/Event_Horizon_Telescope}{EHT-Wikipedia}, and~\href{https://en.wikipedia.org/wiki/James_Webb_Space_Telescope}{JamesWebb-Wikipedia} for further details.} Arguably even more impressive and promising is the upcoming LISA project~\cite{LISA:Barausse:2020}, fully funded and scheduled for launch in $2037$.\footnote{See~\href{https://en.wikipedia.org/wiki/Laser_Interferometer_Space_Antenna}{LISA-Wikipedia} for further details.} Consequently, one can advocate for a grounding of the discourse; a political shift closer to the aforementioned tried and true traditional scientific method. Theoreticians have an obligation to appeal as directly as possible to observational or experimental capacity when making their predictions. A compelling line of argument is that the most effective way of doing this is to turn one theoretical knob at a time...

%
\clearpage

\section{Turning one knob at a time}\label{Intro:knob}
\enlargethispage{30pt}

In a ``controlled experiment'' a scientist adjusts only one independent variable for each iteration of said experiment. This concept of ``iterative minimal adjustment'' is fundamental to the traditional scientific method. Abstracting this notion, one can think of the individual physical theories which make up the body of all theoretical physics as the experiments, and the axiomata from which each physical theory is derived as the independent variables. Applying the analogy of ``iterative minimal adjustment'' when attempting to progress theoretical physics, one chooses to begin with the full list of axioms from which an already-well-understood theory is derived, and make one well-motivated modification to the chosen list at a time before examining the consequences regarding predicted phenomena. This philosophy on ``how to do theoretical physics'' has the distinct advantage over other approaches in that it is extremely straightforward to coherently record progress in the literature; one thus hopes to avoid expending time and resources on reformulations of known results. The fact each iteration is only ``minimally'' adjusted also tends to lead to far more tractable and digestible discourse when compared with constructing a brand new theory from scratch. Exploring this analogy in the context of attempting to unify quantum gravity is certainly worthwhile. Given the magnificent success of \textit{both} GR and QM, ``iterative minimal adjustment'' ought to look like starting from either unadulterated QM, or unadulterated GR, and making only \textit{one} modification to the respective list of axioms. Which direction one approaches from should be a decision guided by one's own cumulative expertise and available resources. For the purposes of this thesis, the framework will always be that of standard GR, and the singular modification to the axioms shall be to forbid classical curvature singularities (discussed further below).

To keep the discourse somewhat self-contained, a formulation of standard GR, with a compact list of the specific axioms, is provided in Appendix~\ref{A:GR}. Also included is Appendix~\ref{B:GRbasics}, where standard GR definitions used throughout this thesis are presented and discussed. Unless explicitly stated otherwise, the metric signature $(-,+,+,+)$ is adopted for all candidate spacetimes and associated discussions (specifically, the only exception to this is in Chapter~\ref{C:SVNBB}, where a signature of $(+,-,-,-)$ is used instead). Unless stated otherwise, geometrodynamic units (where $c=G_{N}=1$) are used, and Greek indices $\alpha,\beta,\mu,\nu,\ldots$ index four-dimensional spacetime, whilst Latin indices $i,j,k,\ldots$ index three-dimensional space.

%
Black holes in GR have a curious history in the literature. The unique solution to the vacuum Einstein equations in static spherical symmetry was first discovered by Karl Schwarzschild in 1916~\cite{Uber:Schwarzschild:1916},\footnote{In 1925, Birkhoff~\cite{RAMP:Birkhoff:1923} realised that ``static'' was not needed as an input assumption;  instead it is a consequence of applying the vacuum Einstein equations in spherical symmetry.} though it was some $50$ years before the community realised the solution could be extrapolated inwards and describe a black hole region with a classical point-like curvature singularity at its core.\footnote{As defined in Appendix~\ref{B:GRbasics}, a curvature or gravitational singularity is mathematically characterised by a coordinate location where one of the nonzero orthonormal Riemann tensor components is infinite.} It wasn't until 1963 that Roy Kerr discovered the unique solution to the vacuum Einstein equations in stationary axisymmetry~\cite{GFOASMAAEOASM:Kerr:1963}; this time the rotation of the spacetime induces a \textit{ring} singularity in the deep core. The Kerr solution is one of the great triumphs of GR --- given astrophysical sources rotate, it is the most appropriate geometry for the majority of astrophysical contexts.
\enlargethispage{30pt}

The presence of infinities in \textit{any} physical theory has always prompted theorists to be concerned, and to more closely examine the underlying fundamentals. Perturbed by the presence of these infinities in the exact solutions, Roger Penrose explored the implications of trapped surfaces forming under gravitational collapse, ultimately resulting in his famous ``Singularity Theorem''~\cite{GC:Penrose:1969}. This is usually understood\footnote{In fact, the actual statement of the theorem does not point to a complete inevitability of curvature singularities, but depends on a number of conditions. An alternative to a curvature singularity allowed by the theorem is null incompleteness~\cite{RPK}.} as the statement that given certain assumptions in classical GR, a curvature singularity will always be the final state of gravitational collapse. For both the Schwarzschild and Kerr solutions, as well as for the majority of generalised geometries subject to Penrose's singularity theorem, the curvature singularities occur at a distance scale that only a mature theory of quantum gravity can adequately describe. At these distance scales, there is certainly a breakdown in the predictability of the theory.

Consequently, despite the fact that the exterior of black hole regions is pathology-free, the deep core seems to be riddled with problems~\cite{TTNODCO:Cardoso:2019}. The presence of curvature singularities is but one of these; also one finds generically that the (maximally extended) Kerr family of solutions harbours closed timelike curves, and features Cauchy horizons~\cite{SOS:Penrose:1978, APFODIGR:Reall:2018}. As dictated theoretically by the weak cosmic censorship conjecture~\cite{GC:Penrose:1969, GCACC:Wald:1999}, spacetime singularities are cloaked by horizons and are therefore inaccessible to distant observers \textit{via} standard observational methods. In fact, there are still many subtle and interesting issues going on in black hole physics.  Deep issues of principle still remain, despite decades of work on the subject, and in many cases it is worthwhile to carefully reanalyse and reassess work from several decades ago~\cite{BHIGR:Visser:2008, SDH:Barcelo:2009}. See also recent phenomenological discussions such as~\cite{POOH:Visser:2014, PAOBHBGR:Carballo-Rubio:2018, OTVORBH:Carballo-Rubio:2018, GCBH:Carballo-Rubio:2020, OTPBATCOBH:Carballo-Rubio:2020, IOASBHR:Doran:2008}.

%
Appealing to the strategy of ``turning one knob at a time'', and beginning with standard GR, it follows that one of the most accessible axiomatic changes to enforce is to forbid classical curvature singularities. Mathematically, this means enforcing that every nonzero component of the Riemann curvature tensor is globally finite with respect to an orthonormal basis. It is precisely within this ``standard GR $+$ no curvature singularities'' framework that most of the research in this thesis is performed.

\section{Nonsingular ``black hole mimickers''}\label{BHMint}

%
%

Depending on geometric constraints (\textit{e.g.} spherical symmetry \textit{versus} axisymmetry), there are many ``tricks'' one can play in order to deviate from Schwarzschild/Kerr and eliminate the classical curvature singularities at their cores. This allows one to enter the realm of ``black hole mimickers''; classes of nonsingular object which are amenable to the extraction of astrophysical observables falsifiable/verifiable by the observational community. Observational results in the ensuing decades will determine whether any one of these classes of object should replace classical black holes due to modelling astrophysical reality with higher accuracy. The majority of ``black hole mimickers'' fall under one of the following classes of object: regular black holes, traversable wormholes, gravastars, or Dyson mega-spheres. Through the lens of ``standard GR $+$ no curvature singularities'', this thesis concerns itself with thorough analysis of various regular black hole and traversable wormhole geometries, as well as a brief analysis of thin-shell Dyson mega-spheres. It is worthwhile to provide historical context and literature review for each of these classes of object before seguing into the specific analyses.


%
%
%
\textbf{Regular black holes:} A subset of the nonsingular geometries of interest are the so-called ``regular black holes'' (RBHs). By regular, one means in the sense of James Bardeen~\cite{Tbilisi:Bardeen:1968}, who presented what is widely regarded as the first RBH model in GR at the GR5 conference held in Tbilisi, 1968.\footnote{Sadly, James Bardeen recently passed away on June $20^{\text{th}}$, 2022. He will be forever remembered for his contributions to GR, and to theoretical physics in general. For details on his life, please see~\href{https://en.wikipedia.org/wiki/James_M._Bardeen}{J.Bardeen-Wikipedia}.} As stated above, regularity is achieved \textit{via} enforcing global finiteness on orthonormal curvature tensor components and Riemann curvature invariants. By finding a suitable source term the Bardeen RBH~\cite{Tbilisi:Bardeen:1968} was reinterpreted as an exact solution of the Einstein equations in reference~\cite{RBHIGRCTNE:Ayon-Beato:1998}; it remains the most comprehensively studied RBH in the literature. After Bardeen, the Hayward~\cite{FAEONBH:Hayward:2006} and Frolov~\cite{ILPAABHMWACAH:Frolov:2014} RBHs are likely the next most prominent, largely due to their high tractability. Beyond these three geometries there are many candidate RBHs in the literature --- in fact the ``Simpson--Visser'' spacetime analysed in Chapter~\ref{C:SVog} has become one of the most popular recent RBH models in GR (promoted, by finding a suitable source term, to an exact solution of the Einstein equations in reference~\cite{FSFSVS:Bronnikov:2022}; see \S~\ref{SimpViss} for details). Generally, in both spherical symmetry and axisymmetry, RBHs have a well-established lineage both in the historical and recent literature~\cite{SORRBH:Abdujabbarov:2016, OTVORBH:Carballo-Rubio:2018, OTPBATCOBH:Carballo-Rubio:2020, GCBH:Carballo-Rubio:2020, HOD:Carballo-Rubio:2021, GCBHILG:Carballo-Rubio:2022, Tbilisi:Bardeen:1968, TBMAANMM:Ayon-Beato:2000, FAEONBH:Hayward:2006, RPBH:Bronnikov:2006, RRBH:Bambi:2013, MTKSPORBHFTS:Bambi:2014, GLBRW:Jusufi:2018, RRBHICMG:Jusufi:2020, QMQO:Jusufi:2021, KBHWPH:Herdeiro:2016, ILPAABHMWACAH:Frolov:2014, ANRBH:Ghosh:2015, RRBHATWEC:Neves:2014, RRBHS:Abdujabbarov:2014, TSOGKBHWEM:Tinchev:2015, CORBHIGR:Fan:2016, GRRBHIGRCTNE:Toshmatov:2017, NOTCOTGRCRBHIGRCNE:Toshmatov:2017, BTTW:Simpson:2019, VSBATW:Moruno:2019, NBB:Lobo:2021, RBHWAMC:Simpson:2019, TLQG:Brahma:2021, ANFORBHM:Franzin:2021, CBBS:Franzin:2021}.

%
RBH spacetimes often contain unusual/intriguing underlying physics. It was recently shown that the spacetime structure of regular spherically symmetrical black holes generically entails the violation of the strong energy condition (SEC)~\cite{RBHAEC:Zaslavskii:2010}. It has been shown that the SEC is violated in any static region within the event horizon in such a way that the Tolman mass becomes negative~\cite{RBHAEC:Zaslavskii:2010}. In the nonstatic case, there is a constraint of another kind which, for a perfect fluid, entails the violation of the dominant energy condition (DEC)~\cite{RBHAEC:Zaslavskii:2010}.

%
Furthermore, a general procedure for constructing exact RBH solutions has been presented, in the presence of electric or magnetic charges in GR coupled to nonlinear electrodynamics (NLED)~\cite{RMBHAMFNE:Bronnikov:2001, CORBHIGR:Fan:2016, COCORBHIGR:Bronnikov:2017}. One obtains a two-parameter family of spherically symmetric black hole solutions, where the singularity at the spacetime centre can be eliminated by moving to a certain region in the parameter space --- consequently the black hole solutions become regular everywhere. The global properties of the solutions were then studied and the first law of thermodynamics readily derived. The procedure has also been generalised to include a cosmological constant, and RBH solutions that are asymptotic to an anti-de Sitter spacetime can be constructed. See reference~\cite{RBHAEC:Zaslavskii:2010} for details.

The study of RBHs has also been generalised to modified theories of gravity and their relation with the energy conditions~\cite{ECIMG:Capozziello:2014, GECIETOG:Capozziello:2015}. For instance, a class of RBH solutions has been obtained in four-dimensional $f(R)$ gravity, where $R$ is the curvature scalar, coupled to a nonlinear electromagnetic source~\cite{RBHIf(R)GCTNE:Junior:2016}. Using the metric formalism and assuming static and spherically symmetric spacetimes, the resulting $f(R)$ and NLED functions are characterised by a one-parameter family of solutions which are generalisations of known RBHs in GR coupled to NLED~\cite{RMBHAMFNE:Bronnikov:2001, SI(2R):Gibbons:1996, NRBHSFNE:Ayon-Beato:1999, RBHIGRCTNE:Ayon-Beato:1998, RECVSWDSCINECTGR:Dymnikova:2004, GSBNE:Arellano:2007, ESOf(R):Hollenstein:2008, EDO(2+1):Balart:2009, NCFENESBH:Garcia:2012, Pf(R)BHINE:Olmo:2011, RBHWANES:Balart:2014, NBHINGCTEE:Guerrero:2020}. The related RBHs of GR can then be recovered when the free parameter vanishes, and where consequently the Einstein--Hilbert action is recovered, \textit{i.e.}, $f(R) \propto R$. The regularity of the solutions has been further analysed, and it was shown that there are particular solutions that violate \textit{only} the SEC, which is consistent with the results attained in~\cite{RBHAEC:Zaslavskii:2010}.

This analysis was then generalised by leaving both the function $f(R)$ and the NLED Lagrangian unspecified in the model, and regular solutions were constructed through an appropriate choice of the mass function~\cite{GFRBHOGRTf(R)G:Fabris:2016}. It was shown that these solutions have two horizons, namely, an event horizon and a Cauchy horizon. All energy conditions are satisfied throughout the spacetime, except the SEC, which is violated near the Cauchy horizon. Regular solutions of GR coupled with NLED were also found by considering general mass functions and then imposing the constraint that the weak energy condition (WEC) and the DEC are simultaneously satisfied~\cite{UDAWECFBNCORBH:Junior:2018}. Further solutions of RBHs have been found by considering both magnetic and electric sources~\cite{BRBHWAES:Rodrigues:2018}, or by adding rotation~\cite{RRBH:Bambi:2013, RRBHATWEC:Neves:2014, RRBHS:Abdujabbarov:2014, GRRBHSWC:Azreg:2014, RRECBHASINEMCTG:Dymnikova:2015, ORRBH:Fayos:2017}, or indeed by considering alternative modified theories of gravity~\cite{RBHIQG:Berej:2006, RBHIf(T):Houndjo:2015, RMBHIf(G):Rodrigues:2019, RBHIf(G):Rodrigues:2018, RBHIRG:Junior:2020, RORSBHC:Cano:2021}.


%
%
%
\textbf{Wormholes:} One can tentatively trace back the origin of wormhole physics all the way to Flamm's work in 1916~\cite{BZEG:Flamm:1916}, and then to the ``Einstein--Rosen bridge'' wormhole-type solutions considered by Einstein and Rosen in 1935~\cite{TPPITGTOR:Einstein:1935}. However, the field lay dormant for approximately two decades until 1955, when John Wheeler became interested in topological issues in GR~\cite{G:Wheeler:1955}. He considered multiply-connected spacetimes, where two widely separated regions were connected by a tunnel-like gravitational-electromagnetic entity, which he denoted as a ``geon''. These were hypothetical solutions to the coupled Einstein--Maxwell field equations. Subsequently, isolated pieces of work do appear, such as Homer Ellis' drainhole concept~\cite{EFTAD:Ellis:1973, TEFD:Ellis:1979}, Bronnikov's tunnel-like solutions~\cite{STTASC:Bronnikov:1973}, and Clement's five-dimensional axisymmetric regular multiwormhole solutions~\cite{ARMSIFGR:Clement:1984}, until the full-fledged renaissance of wormhole physics in 1988, through the seminal paper by Morris and Thorne~\cite{WISATUFIT:Morris:1988}. The compact characterisation of a traversable wormhole geometry was one of the key results of the work done by Morris and Thorne~\cite{WISATUFIT:Morris:1988}, and can be best summarised as follows --- A ``traversable wormhole'' is a horizon-free geometry with a centralised throat hypersurface connecting two asymptotically Minkowski regions of spacetime and satisfying the ``flare-out'' condition for the area function: $A''(r_{\text{throat}})>0$.
\enlargethispage{20pt}

In fact, the modern incarnation of Lorentzian wormholes (and specifically traversable wormholes) now has over 30 years of history. Early work dates from the late 1980s~\cite{WISATUFIT:Morris:1988, WTMATWEC:Morris:1988, TWFSMSS:Visser:1989, TWSSE:Visser:1989, WBUAC:Visser:1990, QMSOMSW:Visser:1990, WAIT:Visser:1989}.
Lorentzian wormholes became considerably more mainstream in the 1990s~\cite{QWILS:Visser:1990, LWIHOGT:Hochberg:1990, PEIWATM:Frolov:1990, QW:Visser:1991, WWATC:Visser:1991, FWTTM:Visser:1993, VVD:Visser:1994, NWAGL:Cramer:1995, EWATWEC:Kar:1994, ELW:Kar:1996, TSW:Poisson:1995, LW:Visser:1995, GSOTGSTWT:Hochberg:1997, GWT:Hochberg:1997, TWTRR:Visser:1997, RTW:Teo:1998, NECIDW:Hochberg:1998, DWASAEC:Hochberg:1998, GDWAVOFNEC:Hochberg:1999, TWVTSEC:Hochberg:1999, TWFMCCSF:Barcelo:1999}, including work on energy condition violations~\cite{GVP1:Visser:1996, GVP2:Visser:1996, GVP3:Visser:1996, GVP4:Visser:1997, GVP:Visser:1997, ECATCI:Barcelo:2000, CASEC:Moruno:2017}, with significant work continuing into the decades 2000--2009~\cite{BS:Barcelo:2000, barcelo2000scalar, R=0:Dadhich:2002, TFTEC:Barcelo:2002, TWWASECV:Dadhich:2003, QECVITW:Dadhich:2004, FLOWDS:Lobo:2004, EWGWNE:Arellano:2006, GCOBW:Lobo:2007, ESIGR:Lobo:2007, CSTW:Bohmer:2007, WGWCM:Bohmer:2008, GCOWGICWG:Lobo:2008, WGf(R):Lobo:2009} and 2010--2019~\cite{WGSBANCC:Garcia:2010, NCCWWMSTNEC:Garcia:2011, WGIMTGATEC:Bohmer:2012, GSSDTTWISGR:Garcia:2012, LSAOGTS:Garcia:2015, MWWEM:Harko:2013, ETWSTWEC:Lobo:2015, TWSTWECITLG:Dehgani:2015, ANATTWAA:Bouhmadi:2018, EMRATW:Boonserm:2018, BTTW:Simpson:2019, VSBATW:Moruno:2019, WWDAEC:Lobo:2017, WIGHMGOTMNECE:Lemos:2018}.
For the purposes of this thesis, particular focus is placed on the thin-shell formalism~\cite{UDGDSAU:Sen:1924, FVDMIDEG:Lanczos:1924, MDSMXXV:Darmois:1927, TRDLGEDL:Lichnerowicz:1955, SHATSIGR:Israel:1966}, first applied to Lorentzian wormholes in~\cite{TWFSMSS:Visser:1989, TWSSE:Visser:1989}, and subsequently further developed in that and other closely related settings by many other authors~\cite{LSOCTW:Eiroa:2004, LSAOTWWACC:Crawford:2003, ASAABH:Frauendiener:1990, SOASAABH:Brady:1991, RS:Goncalves:2002, MWWACC:Lemos:2003, ECTWADS:Lobo:2005, SSOATSSATW:Lobo:2004, PSTWIAASB:Lemos:2004, WSBAPE:Sushkov:2005, PETW:Lobo:2005, SAODTS:Crawford:2005, SOPW:Lobo:2005, CTW:Eiroa:2004, TWIDG:Eiroa:2005, TWIETWAGT:Eiroa:2006, TSWIHDET:Chakraborty:2006, TWITHST:Chakraborty:2007, TWAWGCS:Bejarano:2007, SOCGTW:Eiroa:2007, PSTW:Lemos:2008, TWIBG:Eiroa:2008, TWWAGCG:Eiroa:2009, SOTWWSS:Eiroa:2008, SGAOTWWCS:Eiroa:2010, SOTWSBNMIEG:Amirabi:2010, TWIdGR:Dias:2010, SOBTW:Gao:2011, PSATC:DeBenedictis:2008, JATIGRUCA:Lake:1996, SOTSSTAW:Ishak:2002}.\footnote{Much of the technical machinery for the thin-shell formalism was in fact pioneered by Werner Israel~\cite{SHATSIGR:Israel:1966}. Sadly, Werner Israel recently passed away on May $18^{\text{th}}$, 2022. He will be forever remembered for his contributions to gravitational physics. Please see~\href{https://en.wikipedia.org/wiki/Werner_Israel}{W.Israel-Wikipedia} for further details.} The notation will largely follow that of Hawking and Ellis~\cite{TLSSOS:Ellis:1973}.
Even more specifically, in Part~\ref{PartIII} the primary focus is on the technical machinery built up regarding spherically-symmetric thin-shell spacetimes in references~\cite{TWFSMSS:Visser:1989, LW:Visser:1995, GSSDTTWISGR:Garcia:2012, LSAOGTS:Garcia:2015, ANATTWAA:Bouhmadi:2018}, and for the bulk spacetimes (away from the thin-shell), attention is restricted either to the recently developed black-bounce spacetimes of references~\cite{BTTW:Simpson:2019, VSBATW:Moruno:2019}, or to the Schwarzschild case (for related work, the reader is referred to~\cite{RBHABU:Bronnikov:2007, GCOBBH:Bronnikov:2003}).

\textbf{Gravastars:} The ``gravastar'' (\textit{gra}vitational \textit{va}cuum \textit{star}) was developed by Mazur and Mottola~\cite{GCS:Mazur:2002, DEACS:Mazur:2004, DEACS:Mazur:2005, GVCS:Mazur:2004, SG:Visser:2004, GMHAP:Cattoen:2005, SDES:Lobo:2006, GSBNE:Arellano:2007} in the early 2000s. It remains one of the more curious black hole mimickers, and represents a serious challenge to the standard conception of a black hole. Hypothetically, gravastars would form due to an effective phase transition at or near the location where one would expect event horizon formation. The interior of what would have classically been a black hole region is replaced by a suitably chosen segment of de Sitter space. These constructions possess many layers; as such the thin-shell formalism is typically employed for analyses. While Lorentzian wormholes are in general very different from Mazur--Mottola gravastars, it is worth pointing out that in the thin-shell approximation there are very many technical similarities --- quite often a thin-shell wormhole calculation can be modified to provide a thin-shell gravastar calculation at the cost of flipping a few strategic minus signs~\cite{GTG:Garcia:2012, NSAOTG:Garcia:2018}.

%
%
%
\textbf{Dyson mega-spheres:} In 1960 Freeman Dyson~\cite{SFASSOIR:Dyson:1960} mooted the idea that an arbitrarily advanced civilization, 
(at least a  Kardashev type II civilization~\cite{TOIBEC:Kardashev:1964})\footnote{See~\href{https://en.wikipedia.org/wiki/Kardashev_scale}{Kardashev-Wikipedia} for information on the Kardashev scale.}, 
might seek to control and utilise the energy output of an entire star by building a spherical mega-structure to completely enclose the star, trap all its  radiant emissions, and use the energy flux to do ``useful'' work.
Dyson's idea has led to observational searches~\cite{IRAS:Carrigan:2009}, extensive technical discussions~\cite{ADSAABH:Hsiao:2021}, and more radical proposals such as reverse-Dyson configurations~\cite{LUABS:Opatrny:2017} (where one harvests the CMB and dumps waste heat into a central black hole), and ``hairy'' Dyson spheres~\cite{GHODSVCAHB:Kaloper:2011} (with Gallileon ``hair''). Somewhat strangely, relativistic analyses of these objects appear few and far between (the vast majority of related work has been performed in (quasi)-Newtonian gravity). Recognising that there are scenarios where the Dyson mega-structure of interest would require a fully relativistic treatment, it is worth attempting to extract associated (potential) astrophysical observables \textit{via} standard GR analysis. The astronomers may then have more clues to check for these objects on an ongoing basis as part of the search for advanced civilizations.

%
\section{Two useful theorems}

%
Below are two useful theorems which in the correct context can shorten various calculations and make for more efficient analysis. Theorem~\ref{Theorem:Kretsch} enables one to conclude as to the curvature-regularity of static spacetimes \textit{via} examination of the global finiteness of the Kretschmann scalar. Theorem~\ref{Theorem:PGLT2} allows for one to conclude as to the multiplicative separability of the massive or massless minimally coupled Klein--Gordon equation (scalar wave equation) on the background spacetime when invoking the relativistic Cowling approximation~\cite{Cowling}: where one permits field fluctuations whilst keeping the candidate geometry fixed. This is a crucial ingredient in being able to perform standard spin zero quasi-normal modes analyses for candidate spacetimes. Both of these theorems will be used where appropriate throughout this thesis, either for exposition, or in place of unnecessarily lengthy calculations.

\subsection*{Regularity of  static spacetimes}\label{NBB:appendix}

In reference~\cite{BHCAED:Bronnikov:2013}, Bronnikov and Rubin showed that for a spherically symmetric and static spacetime, finiteness of the Kretschmann scalar is enough to forbid a curvature singularity. This observation allows one to state the following somewhat more general theorem that does not appeal to spherical symmetry. This theorem was first presented in reference~\cite{NBB:Lobo:2021}.
\begin{theorem}\label{Theorem:Kretsch}
    In the strictly static region of any static spacetime, the Kretschmann scalar is positive semi-definite, being a sum of squares of the nonzero components $R^{\hat{\mu}\hat{\nu}}_{\ \ \hat{\alpha}\hat{\beta}}$. Then if this scalar is finite, all the orthonormal components of the Riemann tensor must be finite.
\end{theorem}
{\it Proof:}
First, for any arbitrary spacetime in terms of any orthonormal basis, the Kretschmann scalar is defined as
\begin{equation}
K = R_{\mu\nu\alpha\beta}R^{\mu\nu\alpha\beta} = R_{\hat{\mu}\hat{\nu}\hat{\alpha}\hat{\beta}}R^{\hat{\mu}\hat{\nu}\hat{\alpha}\hat{\beta}} \ .
\end{equation}
Now, assuming that one can distinguish space from time, split the indices into space and time: $\hat{\mu}=(\hat{t},\hat{i})$, so that
\begin{equation}
K = R_{\hat{i}\hat{j}\hat{k}\hat{l}}R^{\hat{i}\hat{j}\hat{k}\hat{l}}
+ 4R_{\hat{t}\hat{i}\hat{j}\hat{k}}R^{\hat{t}\hat{i}\hat{j}\hat{k}}
+ 4R_{\hat{t}\hat{i}\hat{t}\hat{j}}R^{\hat{t}\hat{i}\hat{t}\hat{j}} + 4R_{\hat{t}\hat{t}\hat{t}\hat{i}}R^{\hat{t}\hat{t}\hat{t}\hat{i}} + R_{\hat{t}\hat{t}\hat{t}\hat{t}}R^{\hat{t}\hat{t}\hat{t}\hat{t}} \ .
\end{equation}
But the last two terms vanish in view of the symmetries of the Riemann tensor, and so
\begin{equation}
K = R_{\hat{i}\hat{j}\hat{k}\hat{l}}R^{\hat{i}\hat{j}\hat{k}\hat{l}}
+ 4R_{\hat{t}\hat{i}\hat{j}\hat{k}}R^{\hat{t}\hat{i}\hat{j}\hat{k}}
+ 4R_{\hat{t}\hat{i}\hat{t}\hat{j}}R^{\hat{t}\hat{i}\hat{t}\hat{j}} \ .
\end{equation}
%

But since, in the strictly static region where the $t$ coordinate is timelike, one has  $g_{\hat{\mu}\hat{\nu}}=\eta_{\hat{\mu}\hat{\nu}}= {\rm diag} \{-1,1,1,1\}$, this reduces to
\begin{equation}
K = R_{\hat{i}\hat{j}\hat{k}\hat{l}}R_{\hat{i}\hat{j}\hat{k}\hat{l}}
- 4R_{\hat{t}\hat{i}\hat{j}\hat{k}}R_{\hat{t}\hat{i}\hat{j}\hat{k}}
+ 4R_{\hat{t}\hat{i}\hat{t}\hat{j}}R_{\hat{t}\hat{i}\hat{t}\hat{j}} \ .
\end{equation}
Furthermore, in the strictly static region where the $t$ coordinate is timelike, the four-metric is block-diagonalisable: $g_{\mu\nu}=\left(-N^2\right)\oplus (g_{ij})$, where $N$ is the ``lapse function''~\cite{LMCELTDLR:Painleve:1921, LGDLMDNEDLMD:Painleve:1921, ALDSEIDEG:Gullstrand:1922, ANFOTKS:Doran:2000, TRMOBH:Hamilton:2008, PCFTKS:Natario:2009, RCSFSAOSS:Martel:2001, WPCF:Faraoni:2020} (note the shift vector from the ADM decomposition~\cite{TDOGR:Arnowitt:2008, G:Misner:2000} is automatically zero in this context). More to the point, the extrinsic curvature of the constant-$t$ spacial slices is then zero, and hence by the Gauss--Codazzi--Mainardi~\cite{DGOCAS:Carmo:2016, GR:Wald:2010, G:Misner:2000} equations one has $R_{\hat{t}\hat{i}\hat{j}\hat{k}}=0$.

Thence as long as the spacetime is static one can make the split: spacetime $\rightarrow$ space$+$time in such a manner that
\begin{equation}
K = R_{\hat{i}\hat{j}\hat{k}\hat{l}}R_{\hat{i}\hat{j}\hat{k}\hat{l}}
+ 4R_{\hat{t}\hat{i}\hat{t}\hat{j}}R_{\hat{t}\hat{i}\hat{t}\hat{j}}\geq 0 \ .
\end{equation}
Consequently in any static spacetime if the Kretschmann scalar is globally finite, then all the orthonormal components $R_{\hat{\mu}\hat{\nu}\hat{\alpha}\hat{\beta}}$ of the Riemann tensor must be globally finite. \textit{QED.}

{\it Corollary:} One can determine the regularity of a static spacetime simply by checking if the Kretschmann scalar is globally finite.

It is worth noting that similar comments can be made about the Weyl tensor:
\begin{equation}
C_{\mu\nu\alpha\beta} C^{\mu\nu\alpha\beta} =
C_{\hat{i}\hat{j}\hat{k}\hat{l}}C_{\hat{i}\hat{j}\hat{k}\hat{l}}
- 4C_{\hat{t}\hat{i}\hat{j}\hat{k}}C_{\hat{t}\hat{i}\hat{j}\hat{k}}
+ 4C_{\hat{t}\hat{i}\hat{t}\hat{j}}C_{\hat{t}\hat{i}\hat{t}\hat{j}} \ .
\end{equation}
But the static condition implies that \textit{both} the four-metric and the Ricci tensor are block-diagonalisable. Thence both  $g_{\mu\nu}=\left(-N^2\right)\oplus (g_{ij})$ and $R_{\mu\nu} = R_{tt}\oplus R_{ij}$. This now implies that in static spacetimes $C_{\hat{t}\hat{i}\hat{j}\hat{k}} = R_{\hat{t}\hat{i}\hat{j}\hat{k}} = 0$. So as long as the spacetime is static one can split spacetime $\rightarrow$ space$+$time in such a manner that
\begin{equation}
C_{\mu\nu\alpha\beta} C^{\mu\nu\alpha\beta} =
C_{\hat{i}\hat{j}\hat{k}\hat{l}}C_{\hat{i}\hat{j}\hat{k}\hat{l}}
+ 4C_{\hat{t}\hat{i}\hat{t}\hat{j}}C_{\hat{t}\hat{i}\hat{t}\hat{j}}\geq 0 \ .
\end{equation}
Consequently in any static spacetime if the Weyl scalar $C_{\mu\nu\alpha\beta} C^{\mu\nu\alpha\beta}$ is globally finite, then all the orthonormal components $C_{\hat{\mu}\hat{\nu}\hat{\alpha}\hat{\beta}}$ of the Weyl tensor must be globally finite.

%
%
\subsection*{Separability of the Klein--Gordon equation}
\vspace*{7pt}

In order to analyse the behaviour of test fields propagating in a background spacetime it is standard to invoke the relativistic Cowling approximation --- where one permits the scalar/vector field of interest to oscillate in the presence of a fixed background geometry. For spin zero scalar fields, one is specifically interested in the behaviour of the massive or massless minimally coupled Klein--Gordon equation (scalar wave equation; possibly with a mass term):
\begin{equation}
{1\over\sqrt{-g}} \partial_\alpha \left[ \sqrt{-g}\, g^{\alpha\beta} \, \partial_\beta \, \Phi(t,r,\theta,\phi) \right] = \mu^2 \, \Phi(t,r,\theta,\phi) \ .
\end{equation}
Specifically, if the Klein--Gordon equation separates \textit{via} multiplication on the background spacetime, then one has a separable scalar wave form when performing calculations for spin zero test fields --- allowing one to utilise the ``standard'' techniques when analysing the associated ringdown and quasi-normal modes. This is discussed further in Chapter~\ref{C:AMCQNM}.
\clearpage

A precursor for separability of the Klein--Gordon equation is the existence of a nontrivial Killing tensor $K_{\mu\nu}$. However, by itself this is not enough to guarantee separability. An explicit check needs to be carried out. There are two ways to proceed --- either \textit{via} direct calculation/brute force, or indirectly by studying the commutativity properties of certain differential operators. The latter option is often more expedient; achieved using Theorem~\ref{Theorem:PGLT2} below.

In reference~\cite{PGLT2:Baines:2021}, a refinement was made to Proposition~$1.3$ from reference~\cite{TCTATPAIPOKS:Giorgi:2021}. It should also be noted that this result is implicitly present in older work by Benenti and Francaviglia~\cite{ROCSSATATGR:Benenti:1979}. The result is best summarised as follows:
%
%
\begin{theorem}\label{Theorem:PGLT2}
    Let $(\mathcal{M},g_{\mu\nu})$ be a Lorentzian manifold possessing a nontrivial Killing tensor $K_{\mu\nu}$. Then upon definition of the Carter operator as: $\mathcal{K}\Phi=\nabla_{\mu}\left(K^{\mu\nu}\nabla_{\nu}\Phi\right)$, and the D'Alembertian scalar wave operator as: $\Box\Phi=\nabla_{\mu}\left(g^{\mu\nu}\nabla_{\nu}\Phi\right)$, there is the following result:
    \begin{equation}
        [\mathcal{K},\Box]\Phi = \frac{2}{3}\left(\nabla_{\mu}[R,K]^{\mu}{}_{\nu}\right)\nabla^{\nu}\Phi \ .
    \end{equation}
    A sufficient condition for this operator commutator to vanish is therefore the commutativity of the Ricci and Killing tensors \textit{via} matrix multiplication; $R^{\mu}{}_{\alpha}K^{\alpha}{}_{\nu}=K^{\mu}{}_{\alpha}R^{\alpha}{}_{\nu}\Longrightarrow[R,K]^{\mu}{}_{\nu}=0$. Hence for any candidate spacetime with a nontrivial Killing tensor, commutativity of the Ricci and Killing tensors \textit{via} matrix multiplication is sufficient to conclude that the massive or massless minimally coupled Klein--Gordon equation is separable on the candidate geometry.
\end{theorem}


{\it Proof:}
In the recent reference~\cite[page 9, Proposition 1.3]{TCTATPAIPOKS:Giorgi:2021} that author demonstrated explicitly that
\begin{eqnarray}\label{Prop1.3}
[\mathcal{K},\Box]\Phi &=& \left\{ \left(\nabla_\alpha R - {4\over3} \nabla_\beta R^\beta{}_\alpha \right)   K^\alpha{}_\nu \right.
\nonumber\\
&&
\left.
+ {2\over3}\left( R^{\beta\alpha} \nabla_\beta K_{\alpha\nu} - R^\alpha{}_\nu\nabla_\beta K^\beta{}_\alpha 
- \left(\nabla_\alpha R^\beta{}_\nu \right) K^\alpha{}_\beta \right) \right\} \nabla^\nu \Phi \ . \nonumber \\
&&
\end{eqnarray}
Note that if $\left[\mathcal{K},\Box\right]\,\Phi=0$, then the matrix representations of the operators $\mathcal{K},\Box$ are simultaneously diagonalisable; earlier work by Carter~\cite{HASSSOEE:Carter:1968, GSOTKFOGF:Carter:1968} demonstrates why this implies a separable Klein--Gordon equation on the background geometry.

From Eq.~(\ref{Prop1.3}), now use the (twice contracted) Bianchi identity, in the opposite direction from what one might expect, to temporarily make things more complicated:
\begin{equation}
\nabla_\alpha R = 2\nabla_\beta R^\beta{}_\alpha \ .
\end{equation}
Then Eq.~(\ref{Prop1.3}) becomes
\begin{eqnarray}
[\mathcal{K},\Box]\Phi &=& \Bigg\{ \left( + {2\over3} \nabla_\beta R^\beta{}_\alpha \right) K^\alpha{}_\nu \nonumber \\
&& \nonumber \\
&& \qquad
+ {2\over3}\left( R^\beta{}_\alpha \nabla_\beta K^\alpha{}_\nu - R^\alpha{}_\nu\nabla_\beta K^\beta{}_\alpha 
- \left(\nabla_\alpha R^\beta{}_\nu \right) K^\alpha{}_\beta \right) \Bigg\} \nabla^\nu \Phi \ . \nonumber \\
&&
\end{eqnarray}
That is,
\begin{eqnarray}
[\mathcal{K},\Box]\Phi &=& {2\over3} 
\Big\{ 
R^\beta{}_\alpha \nabla_\beta K^\alpha{}_{\nu} - R^\alpha{}_\nu\nabla_\beta K^\beta{}_\alpha \nonumber \\
&& \nonumber \\
&& \qquad \qquad
- (\nabla_\alpha R^\beta{}_\nu ) K^\alpha{}_\beta  +  (\nabla_\beta R^\beta{}_\alpha)    K^\alpha{}_\nu 
\Big\} \nabla^\nu \Phi \ .
\end{eqnarray}
Relabelling some indices:
\begin{eqnarray}
[\mathcal{K},\Box]\Phi &=& {2\over3} 
\Big\{ 
R^\beta{}_\alpha \nabla_\beta K^\alpha{}_{\nu} - R^\alpha{}_\nu\nabla_\beta K^\beta{}_\alpha \nonumber \\
&& \nonumber \\
&& \qquad \qquad
- \left(\nabla_\beta R^\alpha{}_\nu \right) K^\beta{}_\alpha  +  \left(\nabla_\beta R^\beta{}_\alpha\right)    K^\alpha{}_\nu 
\Big\} \nabla^\nu \Phi \ .
\end{eqnarray}
That is,
\begin{equation}
[\mathcal{K},\Box]\Phi = {2\over3} 
\nabla_\beta \left( R^\beta{}_\alpha  K^\alpha{}_{\nu} - K^\beta{}_\alpha  R^\alpha{}_{\nu} \right)  \nabla^\nu \Phi \ .
\end{equation}
Finally, rewrite this as:
\begin{equation}
[\mathcal{K},\Box]\Phi = {2\over3} 
\left( \nabla_\beta [R,K]^\beta{}_\nu   \right)  \nabla^\nu \Phi \ .
\end{equation}
\textit{QED.}

%
%
%
%
%
Having motivated the research into the framework of ``GR $+$ no curvature singularities'', and with thorough understanding of the framework now in-hand, together with well-established histories of the various objects involved and two newly established theorems to assist the analyses, one is well-armed to begin exploration into families of nonsingular candidate spacetimes. Part~\ref{PartI} of this thesis summarises and analyses the family of ``black-bounce'' spacetimes, first introduced to the literature in reference~\cite{BTTW:Simpson:2019}. In Part~\ref{PartII}, the family of regular black holes with asymptotically Minkowski cores is thoroughly explored, with the ringdown analysis of the quasi-normal modes being performed in Chapter~\ref{C:AMCQNM}, and the highly desirable ``eye of the storm'' geometry from reference~\cite{EOS:Simpson:2021} being examined in Chapter~\ref{C:AMCEOS}. Finally, Part~\ref{PartIII} deep-dives into various related thin-shell constructions, before concluding remarks and discourse surrounding potential for future research are presented in the Conclusions in Chapter~\ref{C:con}.

An effort is made to keep the discussion at least \textit{somewhat} self-contained, consequently suitably modified/updated work and figures from many publications are represented, including some extremely limited content from the current author's own MSc thesis~\cite{Masters:Simpson:2019}, and Mr Thomas Berry's MSc thesis~\cite{Berrythesis}.\footnote{Please see the declaration in Appendix~\ref{C:Publications} for specifics.} A reader new to the field of GR would do well to accompany the reading of this thesis with one of the standard textbooks of GR; the author recommends a combination of Misner, Thorne, and Wheeler's ``Gravitation''~\cite{G:Misner:2000}, together with Hawking and Ellis' ``The Large Scale Structure of Spacetime''~\cite{TLSSOS:Ellis:1973}. For certain passages, Visser's ``Lorentzian Wormholes''~\cite{LW:Visser:1995} provides the most closely-related summary in terms of both content and communication style. Experienced readers should have no trouble.

%

%% file: 02-Black-bounce/1-SVog.tex
\part{The family of black-bounce spacetimes}\label{PartI}

\chapter{Beyond Simpson--Visser spacetime}\label{C:SVog}

One family of candidate geometries which model alternatives to classical black holes is the family of ``black-bounce'' spacetimes~\cite{BTTW:Simpson:2019, NBB:Lobo:2021, VSBATW:Moruno:2019, DTBTW:Lobo:2020, ANFORBHM:Franzin:2021, CBBS:Franzin:2021}. These stem from the original static spherically symmetric, so-called ``Simpson--Visser'' (SV) spacetime~\cite{BTTW:Simpson:2019}. All of these ``black hole mimickers'' are globally free from curvature singularities, pass all weak-field observational tests of standard GR, and smoothly interpolate between either regular black holes or traversable wormholes. The original SV spacetime is the black-bounce candidate geometry which effectively acts as the analog to Schwarzschild; key results for this geometry are presented in \S~\ref{SimpViss}. The Vaidya black-bounce extension to SV spacetime~\cite{VSBATW:Moruno:2019}, where one allows for very carefully controlled dynamics in spherical symmetry, is then explored in \S~\ref{SVVaidya}. Given the community is already largely familiar with both SV spacetime and the Vaidya extension, the results displayed here are only cursory.\footnote{Explicit and detailed analyses can be found either in the author's MSc Thesis~\cite{Masters:Simpson:2019}, or in the original publications~\cite{BTTW:Simpson:2019, VSBATW:Moruno:2019}. The results presented herein are largely to keep the discourse somewhat self-contained.} In \S~\ref{SVbbRN}, the more recently constructed black-bounce Reissner--Nordstr\"{o}m spacetime~\cite{CBBS:Franzin:2021} is thoroughly explored --- this is the black-bounce analog to standard Reissner--Nordstr\"{o}m spacetime.
\enlargethispage{10pt}

All candidate spacetimes in Chapters~\ref{C:SVog} and~\ref{C:SVNBB} are spherically symmetric samples from the black-bounce family. A spherically symmetric thin-shell traversable wormhole variant of SV spacetime is rigorously analysed in Part~\ref{PartIII}, Chapter~\ref{C:SVthinshell}. The more astrophysically relevant regime of stationary axisymmetry is explored in Chapter~\ref{C:SVCBB}, \textit{via} analysis of the black-bounce Kerr, and black-bounce Kerr--Newman spacetimes. All members of the family of black-bounce spacetimes are amenable to highly tractable analysis and straightforward extraction of (potential) astrophysical observables.

%
\section{``Simpson--Visser'' spacetime}\label{SimpViss}
%

The so-called ``Simpson--Visser'' spacetime, initially presented in reference~\cite{BTTW:Simpson:2019}, is given by the line element 
\begin{equation}\label{SV}
    \dd s^{2} = -\left(1-\frac{2m}{\sqrt{r^{2}+\ell^{2}}}\right)\,\dd t^{2} + \frac{\dd r^{2}}{\left(1-\frac{2m}{\sqrt{r^{2}+\ell^{2}}}\right)}+\left(r^{2}+\ell^{2}\right)\,\dd\Omega^{2}_{2} \ .
\end{equation}
It should be noted that the parameter ``$\ell$'' was in fact labelled ``$a$'' in the original article~\cite{BTTW:Simpson:2019}. This minor alteration is performed for consistency with the remainder of the discourse herein, as when in the axisymmetric environments of Chapters~\ref{C:SVCBB} and~\ref{C:AMCEOS}, it is prudent to use $\ell$ in order to avoid confusion with the spin parameter from Kerr spacetime, $a$.

\begin{figure}[!htb]
\begin{center}
\includegraphics[scale=0.5]{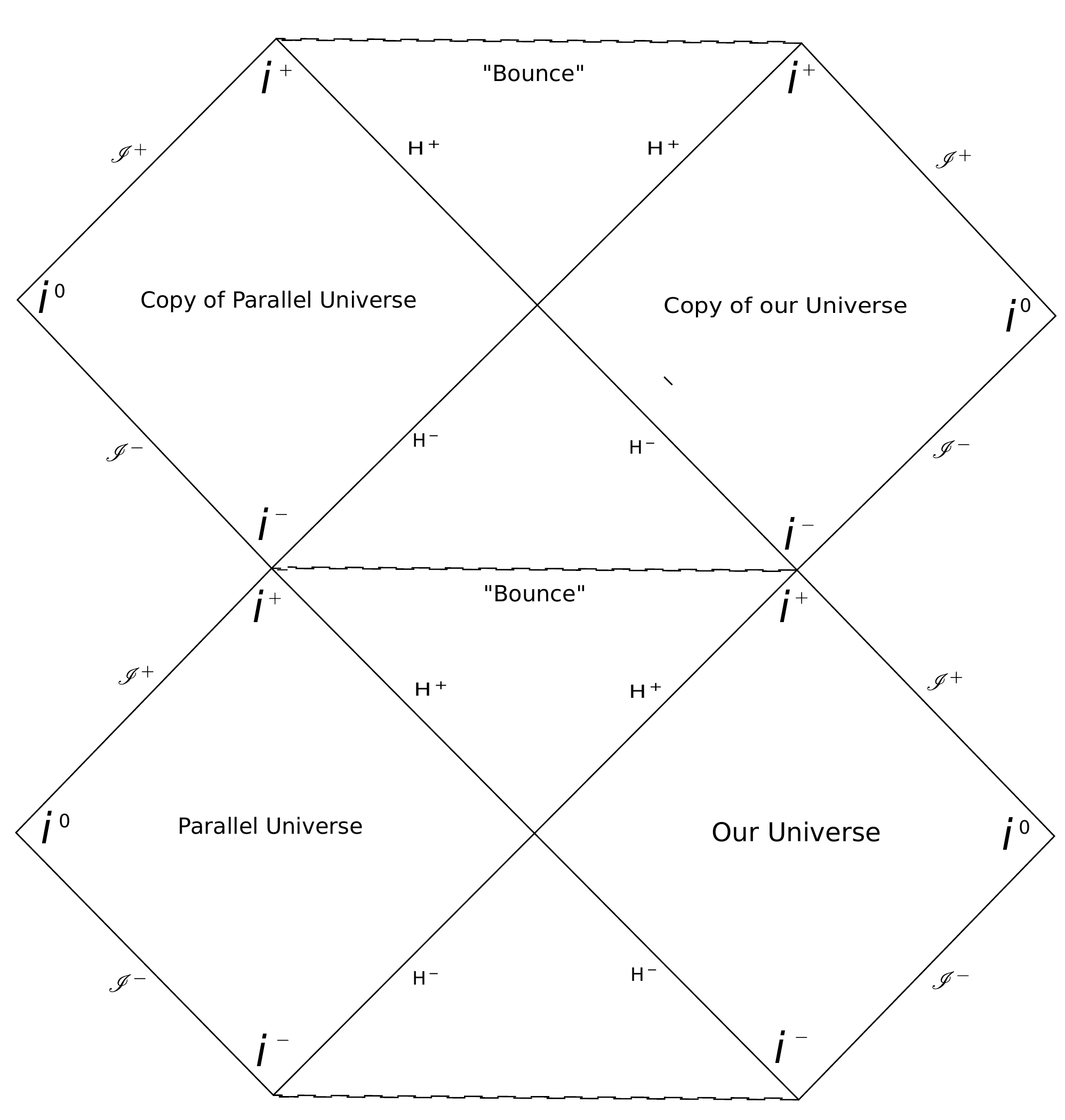}\qquad
\end{center}
{\caption[Carter--Penrose diagram for the maximally extended black-bounce case of Simpson--Visser spacetime]{{Carter--Penrose diagram for the maximally extended spacetime when $\vert\ell\vert\in(0,2m)$. In this example one ``bounces'' through the $r=0$  hypersurface in each black hole region into a future copy of the universe \textit{ad infinitum}.}}\label{F:bounce-1}}
\end{figure}

\begin{figure}[!htb]
\vspace{-0.5cm}
\begin{center}
\includegraphics[scale=0.5]{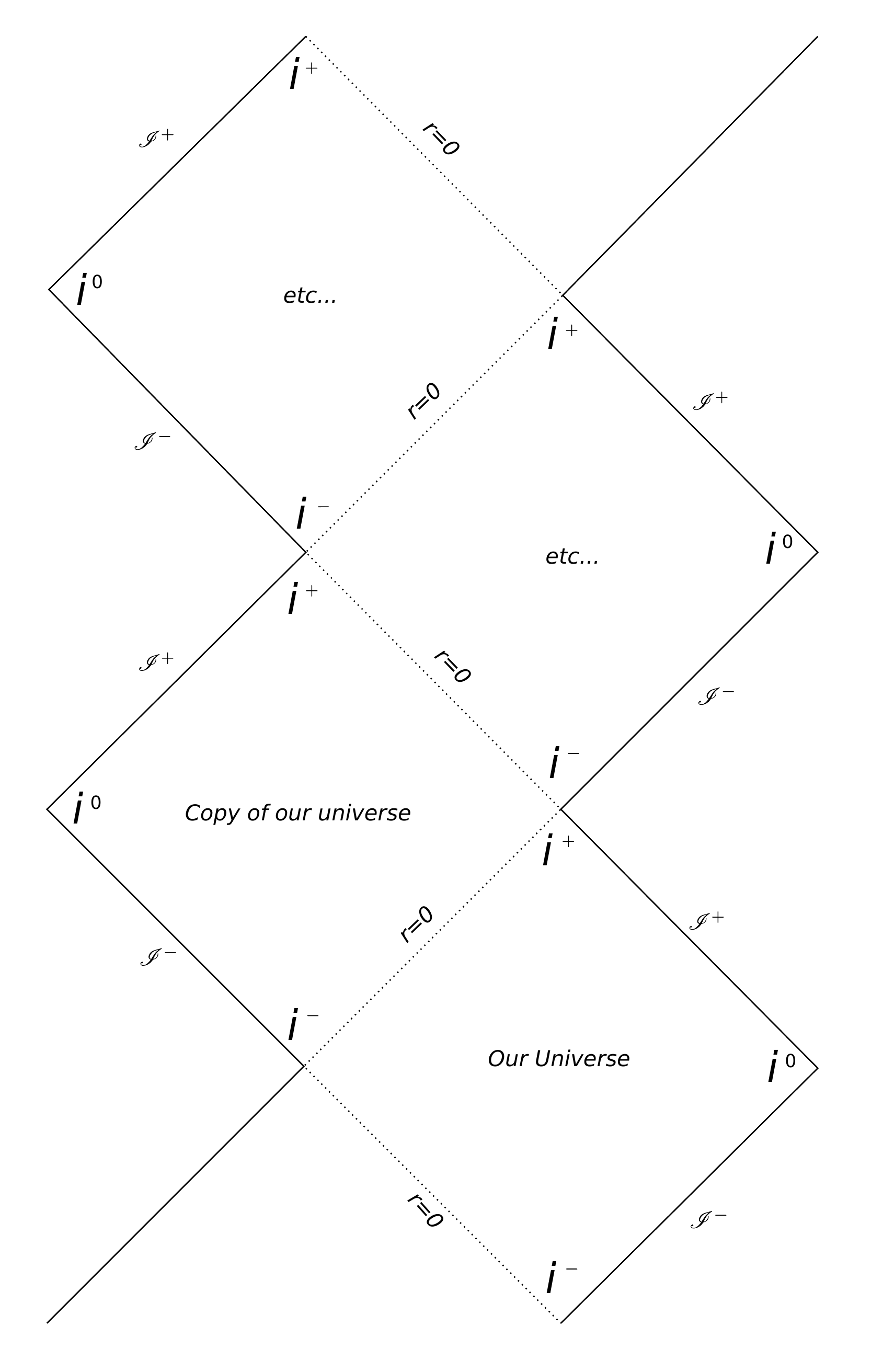}\qquad
\end{center}
{\caption[Carter--Penrose diagram for the maximally extended one-way wormhole case of Simpson--Visser spacetime]{{Carter--Penrose diagram for the maximally extended spacetime in the case when $\vert\ell\vert=2m$. In this example one has a one-way wormhole geometry with an extremal null throat}.}
\label{F:null-bounce-1}}
\end{figure}

The original motivation for the construction of the metric specified by Eq.~(\ref{SV}) was to minimally modify the Schwarzschild solution in common curvature coordinates such that the resulting candidate spacetime was globally nonsingular. It was considered that the most straightforward way to achieve this was to introduce a new scalar parameter to the line element in a tightly controlled manner; this is $\ell$ in Eq.~(\ref{SV}). By minimally modifying Schwarzschild, it was hoped the result would have a high degree of mathematical tractability. When viewed as a modification of Schwarzschild, there are the following two alterations:
\begin{itemize}
    \item $m \mapsto m(r) = \frac{mr}{\sqrt{r^{2}+\ell^{2}}}$ ;
    \item The coefficient of $\dd\Omega^{2}_{2}$ is modified from $r^2\rightarrow r^{2}+\ell^{2}$ .
\end{itemize}
Analysis of the nonzero curvature tensor components and Riemann curvature invariants concludes that for $\vert\ell\vert>0$, the resulting candidate spacetime is globally regular. Examination of the Kretschmann scalar, \textit{via} Theorem~\ref{Theorem:Kretsch}, concludes the same (see the discussion in \S~\ref{NBB:SV-bb} for specifics, as well as the discouse in the original article~\cite{BTTW:Simpson:2019}). It was also noticed that Eq.~(\ref{SV}) has very neat limiting behaviour. In the limit as $m\rightarrow0$, one obtains
    \begin{equation}
        \dd s^{2} = -\dd t^{2} + \dd r^{2} + \left(r^{2}+\ell^{2}\right)\,\dd\Omega^{2}_{2} \ .
    \end{equation}
    This is precisely the two-way traversable wormhole solution as presented in Morris and Thorne's aforementioned seminal paper~\cite{WISATUFIT:Morris:1988} (and arguably the most straightforward of all traversable wormhole geometries). In the limit as $\ell\rightarrow0$, Eq.~(\ref{SV}) becomes the Schwarzschild solution in the usual curvature coordinates.
The newly introduced scalar parameter $\ell$ hence quantifies the extent of the deviation away from Schwarzschild, and it invokes a rich, $\ell$-dependent horizon structure. The causal structure is characterised by (in the $r>0$ universe)
\begin{equation}
    r_{H} = \sqrt{(2m)^2-\ell^2} \ ,
\end{equation}
\clearpage
and the candidate spacetime neatly interpolates between the following qualitatively different geometries:
\begin{itemize}
    \item $\ell=0$ corresponds to the Schwarzschild black hole;
    \item $\vert\ell\vert\in\left(0, 2m\right)$ corresponds to a regular black hole in the sense of Bardeen~\cite{Tbilisi:Bardeen:1968};
    \item $\vert\ell\vert=2m$ corresponds to a one-way wormhole with an extremal null throat;
    \item $\vert\ell\vert>2m$ corresponds to a two-way traversable wormhole geometry in the canonical sense of Morris and Thorne.
\end{itemize}
The Carter--Penrose diagrams for the two lesser known cases are worth closer examination; this is when $\vert\ell\vert\in(0,2m)$ (see Fig.~\ref{F:bounce-1}), and when $\vert\ell\vert=2m$ (see Fig.~\ref{F:null-bounce-1}).
\vspace*{20pt}

\clearpage

Analysing the (non)satisfaction of the standard point-wise energy conditions of GR, one concludes that when $\vert\ell\vert>0$, the radial ``null energy condition'' (NEC) is manifestly violated in the region $\sqrt{r^2+\ell^2}>2m$. This is \textit{outside} any would-be horizons. In the context of static spherical symmetry, this is sufficient to conclude that all of the standard point-wise energy conditions shall be similarly violated for SV spacetime. If horizons are present, then the geometry has surface gravity
\begin{equation}
    \kappa = \frac{\sqrt{(2m)^2-\ell^2}}{8m^2} = \kappa_{Sch.}\,\sqrt{1-\frac{\ell^2}{(2m)^2}} \ ,
\end{equation}
and hence associated Hawking temperature
\begin{equation}
    T_{H} = \frac{\hbar\sqrt{(2m)^2-\ell^2}}{16\pi\,k_{B}\,m^2} = T_{H,Sch.}\,\sqrt{1-\frac{\ell^2}{(2m)^2}} \ .
\end{equation}
SV spacetime is amenable to the straightforward extraction of astrophysical observables, and indeed the coordinate locations of the photon sphere for null orbits and the ISCO for timelike orbits are simple, given by
\begin{equation}\label{SVorbits}
    r_{\gamma} = \sqrt{(3m)^2-\ell^2} \ , \qquad r_{ISCO} = \sqrt{(6m)^2-\ell^2} \ .
\end{equation}
Further research analysing SV spacetime has been performed in a plethora of other papers.\footnote{Please see~\href{https://inspirehep.net/literature?sort=mostrecent\&size=25\&page=1\&q=refersto\%3Arecid\%3A1709812\&ui-citation-summary=true}{SV-inspirehep.net} for a full list of pre-published and published papers citing reference~\cite{BTTW:Simpson:2019}; 132 total citations as of March 25th, 2023.} These analyses include many qualitatively different discussions, for instance examination of the quasi-normal modes and associated ringdown~\cite{EIBW:Bronnikov:2020, ROTRBHWT:Churilova:2020}, calculations pertaining to shadows and gravitational lensing effects~\cite{SGLBRSVBH:Ghosh:2021, GLBTPSIABSISDL:Tsukamoto:2021, SAOAOBBIBATAD:Guerrero:2021, CDBHCTSS:Crispino:2021, GLITSVBSIASDL:Tsukamoto:2021, GLIBS:Nascimento:2020}, as well as discourse surrounding precession phenomena~\cite{PAPMAABTW:Zhou:2020}.

One of the most important results pertaining to SV spacetime is its elevation to ``solution status'' in the recent (Dec. 2021) work by Bronnikov and Walia~\cite{FSFSVS:Bronnikov:2022}. Those authors show that SV spacetime can be obtained as an exact
solution to the Einstein field equations, sourced by a combination of a minimally coupled phantom
scalar field with a nonzero potential $V(\Phi)$, and a magnetic field in the framework of NLED (in the Lagrangian formalism given generally by some $\mathcal{L}\left(\mathcal{F}\right)$, $\mathcal{F} = F_{\mu\nu}F^{\mu\nu}$, where $F_{\mu\nu}$ is the appropriately modified Maxwell tensor from standard electromagnetism). From an action principle, in the Lagrangian formalism, one specifically has the following~\cite{FSFSVS:Bronnikov:2022}:
\begin{eqnarray}\label{SVaction}
    S &=& \int\,\sqrt{-g}\,\d^{4}r\left[R -2g^{\mu\nu}\partial_{\mu}\Phi\,\partial_{\nu}\Phi - 2\,V(\Phi) - \mathcal{L}\left(\mathcal{F}\right)\right] \ , \nonumber \\
    && \nonumber \\
    V(\Phi) &=& \frac{4m\cos^{6}\Phi\sec\Phi}{5\ell^3} = \frac{4m\ell^2}{5\left(r^2+\ell^2\right)^{\frac{5}{2}}} \ , \qquad \Phi = \pm\arctan\left(\frac{r}{\ell}\right) \ , \nonumber \\
    && \nonumber \\
    \mathcal{L}\left(\mathcal{F}\right) &=& \frac{12m\ell^2}{5\left(2q^2/\mathcal{F}\right)^{\frac{5}{4}}} = \frac{12m\ell^2}{5\left(r^2+\ell^2\right)^{\frac{5}{2}}} \ , \qquad \mathcal{F} = \frac{2q^2}{\left(r^2+\ell^2\right)^2} \ . \nonumber \\
    &&
\end{eqnarray}
Here $q$ is the charge associated with the electromagnetic field coupled to the background geometry \textit{via} NLED, and $R$ is the four-dimensional Ricci scalar. Extremising the action Eq.~(\ref{SVaction}) by varying $S$ with respect to the contrametric $g^{\mu\nu}$ leads to the Einstein equations, decomposable into a sum of stress-energy contributions from the scalar and electromagnetic fields respectively. Varying the action in $\Phi$ and $F_{\mu\nu}$ obtains a system of field equations for which the metric Eq.~(\ref{SV}) is the unique solution in spherical symmetry, promoting SV spacetime to an exact solution of the Einstein equations. For more detailed discussion please see reference~\cite{FSFSVS:Bronnikov:2022}.
\clearpage

The original construction was strictly a GR model: first a well-motivated geometry was built by minimally modifying Schwarzschild, then the geometry was coupled to the Einstein equations to examine the induced physics. The ``other'' direction where one fixes the stress-energy tensor and solves the resulting ten, highly nonlinear Einstein equations is comparably \textit{extremely} difficult. In fact, in general when given an underlying geometry coupled to a set of field equations it is either \textit{very} difficult or potentially impossible to recursively solve for the required action principle. With the appropriate Lagrangian for SV spacetime now in-hand, one can begin to examine whether there are any associations or patterns when compared with the quantised action of hypothetical candidate particles beyond the standard model (loosely similar to the procedure followed in reference~\cite{OQDOTSS:Kazakov:1993}). While SV spacetime is highly unlikely to be ``the answer'', the eternally optimistic theorist hopes that the general strategy of comparing the Lagrangians from which nonsingular candidate spacetimes are derived to the quantised actions of hypothetical particles will paint a clearer picture of where to head next (and may even reveal the specific experimental capacity required to make progress on quantum gravity).

%
%
%

\section{Vaidya black-bounce}\label{SVVaidya}

\begin{figure}[!htb]
\begin{center}
\includegraphics[scale=0.93]{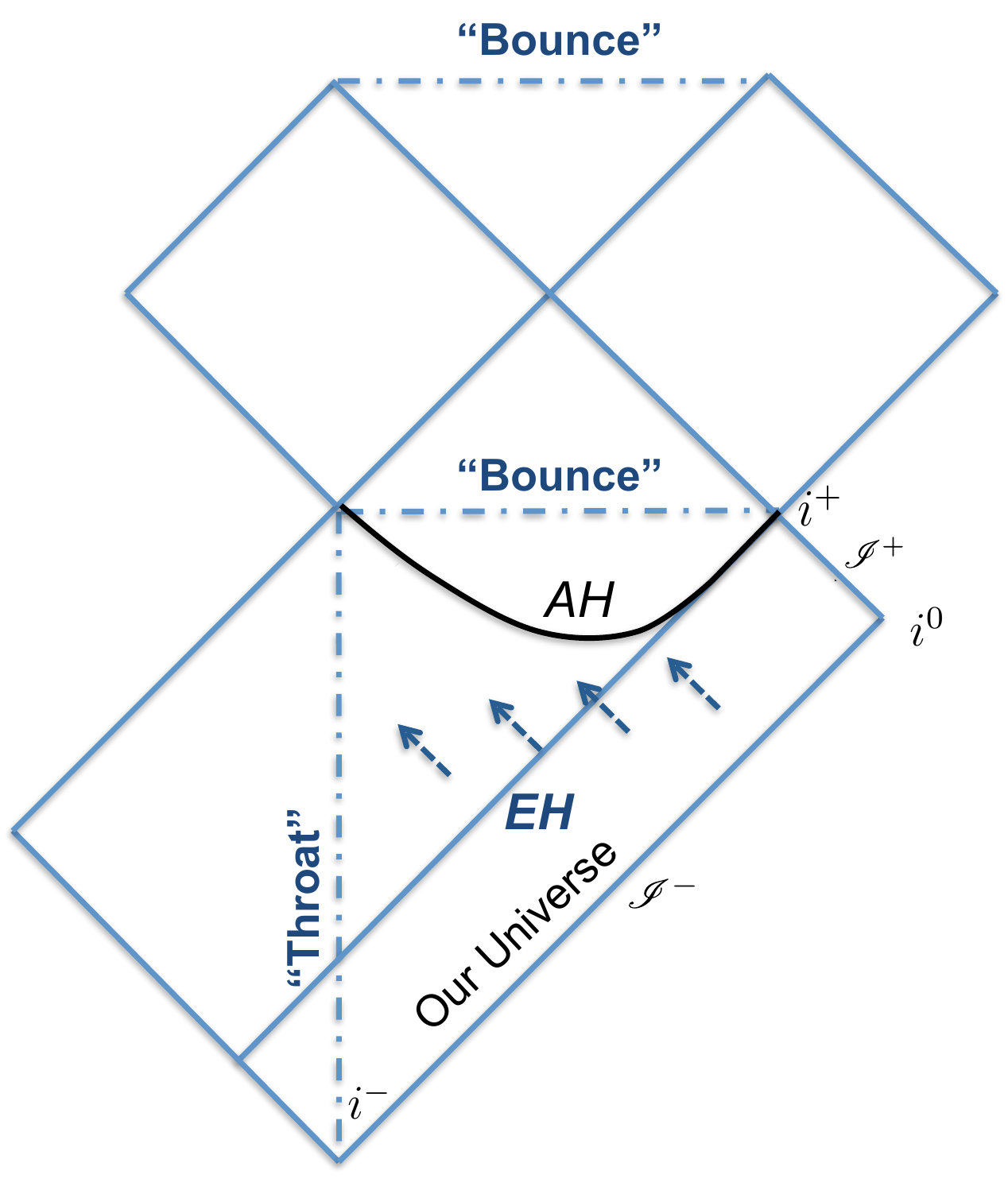}
\end{center}
\caption[Carter--Penrose diagram for a wormhole to black-bounce transition \textit{via} accretion of null dust]{Carter--Penrose diagram for a wormhole to black-bounce transition \textit{via} accretion of null dust.}
\label{F:WTB}
\end{figure}

\begin{figure}[!htb]
\begin{center}
\includegraphics[scale=0.91]{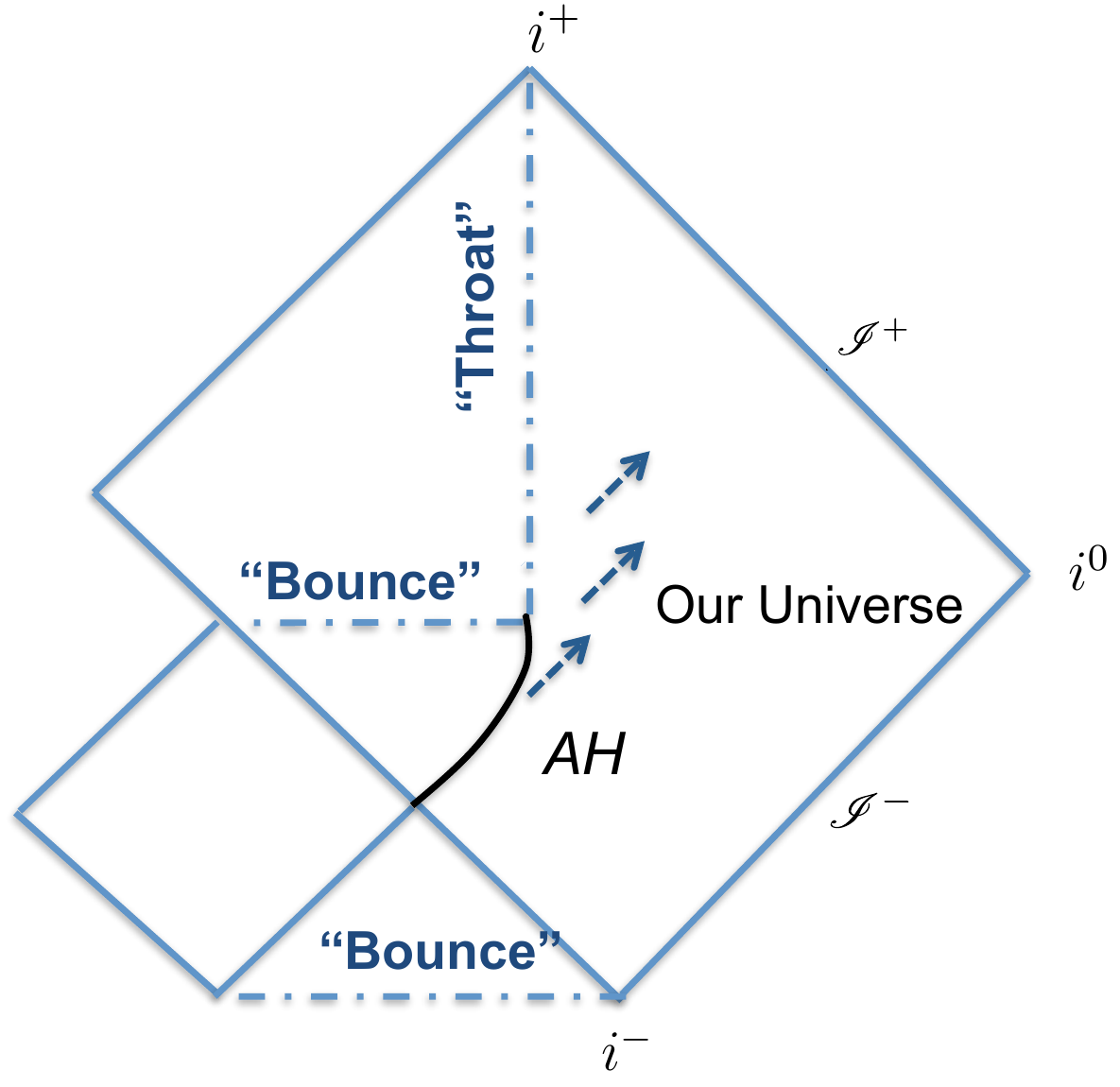}
\qquad
\end{center}
\caption[Carter--Penrose diagram for a black-bounce to traversable wormhole transition \textit{via} the emission of positive energy]{Carter--Penrose diagram for a black-bounce to wormhole transition due to the emission of positive energy. This is essentially the back reaction of the Hawking evaporation for a semi-classical regular black hole; in this case leaving a wormhole remnant.}
\label{F:BTW}
\end{figure}

The SV metric was elevated to the regime of dynamical spherical symmetry in reference~\cite{VSBATW:Moruno:2019}. One first rewrites the line element from Eq.~(\ref{SV}) in Eddington--Finkelstein coordinates, before playing a Vaidya-like ``trick'' by allowing the mass parameter $m$ to be a function of the null time coordinate $w$. The resulting candidate spacetime is given by the line element
\begin{equation}
    \dd s^2 = -\left(1-\frac{2m(w)}{\sqrt{r^2+\ell^2}}\right)\,\dd w^2 - (\pm2\,\dd w\dd r) + (r^2+\ell^2)\,\dd\Omega^2_2 \ ,
\end{equation}
where $w=\lbrace u,v\rbrace$ denotes the \textit{outgoing/ingoing} null time coordinate, representing \textit{retarded/advanced} time respectively. In the limit as $m(w)\rightarrow m$, SV spacetime is recovered (in Eddington--Finkelstein coordinates). Introducing dynamics in this specific manner allows for one to discuss simple phenomenological models of an evolving regular black hole/traversable wormhole geometry, either \textit{via} net accretion or net evaporation, whilst still keeping the discourse mathematically tractable. Analysis of the radial null curves yields a dynamical horizon location
\begin{equation}
    r_H(w) = \pm\sqrt{\left[2m(w)\right]^2-\ell^2} \ .
\end{equation}
\clearpage
While this permits analysis of numerous phenomenological models, the most qualitatively interesting are:
\begin{itemize}
    \item Increasing $m(v)$ crossing the $\ell/2$ limit describing the conversion of a traversable wormhole into a regular black hole \textit{via} the accretion of null dust. This qualitative scenario is depicted in Fig.~\ref{F:WTB}.
    \item Decreasing $m(u)$ crossing the $\ell/2$ limit describes the evaporation of a regular black hole leaving a traversable wormhole remnant. The causal structure is depicted in Fig.~\ref{F:BTW}. This phenomena is causally equivalent to a regular black hole transmuting into a traversable wormhole \textit{via} the accretion of phantom energy (though mathematically this scenario instead corresponds to decreasing $m(v)$ crossing the $\ell/2$ limit).
\end{itemize}

General discussion concerning evaporation/accretion models involving both regular black holes and traversable wormholes can be found in references~\cite{BHMDDTPEA:Babichev:2004, DECWGLEOS:Babichev:2004, WBHEETU:Moruno:2006, WITAU:Gonzalez-Diaz:2008, OTFODEAOBAW:Moruno:2008, OAODEOBAW:Madrid:2010, PSPATCTAESM:Lobo:2014, WGFGC:Chakrabarti:2021}. It is worth emphasising that one is able to classically describe the transmutation of a wormhole into a regular black hole, or \textit{vice versa}, only because the curvature-regularity of the black hole implies there is no topology change. For examination of other phenomenological models, and heightened detail, please see reference~\cite{VSBATW:Moruno:2019}.
\enlargethispage{20pt}

In what has become a somewhat typical result for nonsingular black hole mimickers in standard GR, the candidate spacetime is found to be in global violation of the NEC. For the ``Vaidya black-bounce'' spacetime, the existence of the radial null vector $k^\mu=(0,1,0,0)$ implies that one has
\begin{equation}
    T_{\mu\nu}k^{\mu}k^{\nu} \varpropto G_{\mu\nu}k^{\mu}k^{\nu} = -\frac{2\ell^2}{(r^2+\ell^2)^2} < 0 \ .
\end{equation}
For a full analysis of the nonzero curvature tensor components and Riemann curvature invariants for the Vaidya black-bounce geometry, as well as a more detailed discussion surrounding satisfaction/violation of the point-wise energy conditions, please see reference~\cite{VSBATW:Moruno:2019}.

\section{Black-bounce Reissner--Nordstr\"{o}m}\label{SVbbRN}
\vspace*{15pt}

While observationally the electric charges on astrophysical black holes are likely to be extremely low, $|Q|/m \ll 1$, introducing any nonzero electric charge has a significant theoretical impact. Consequently, given the demonstrated existence of the SV spacetime~\cite{BTTW:Simpson:2019, VSBATW:Moruno:2019, DTBTW:Lobo:2020, NBB:Lobo:2021, RPBH:Bronnikov:2006, RBHABU:Bronnikov:2007, FSFSVS:Bronnikov:2022} (recall: SV is the black-bounce analog to Schwarzschild), it is intuitive to suspect that the black-bounce analog to Reissner--Nordstr\"om (RN) spacetime also exists~\cite{THOTECWABB:Huang:2021}, and that it should be amenable to a reasonably tractable general relativistic analysis. This ``black-bounce Reissner--Nordstr\"{o}m'' (bbRN) spacetime was first constructed and analysed in reference~\cite{CBBS:Franzin:2021}.

Generally, it should be noted that the procedure by which one obtains a ``black-bounce'' variant from a known pre-existing solution in \textit{either} axisymmetry or spherical symmetry can be simplified. Given any spherically symmetric or axisymmetric geometry equipped with some metric $g_{\mu\nu}$, in possession of a curvature singularity at $r=0$ in standard $(t,r,\theta,\phi)$ curvature coordinates, the proposed procedure is explicitly designed to transmute the said geometry into a globally regular candidate spacetime whilst retaining the manifest symmetries. Applying this directly to RN spacetime yields the bbRN spacetime; a one-parameter class of geometry smoothly interpolating between standard GR electrovac black holes and traversable wormholes~\cite{WISATUFIT:Morris:1988, WTMATWEC:Morris:1988, LW:Visser:1995, TWSSE:Visser:1989, TWFSMSS:Visser:1989}. The black-bounce analog to Kerr--Newman is obtained similarly by applying this procedure to the Kerr--Newman (KN) spacetime. The ``black-bounce Kerr--Newman'' (bbKN) geometry is analysed in Chapter~\ref{C:SVCBB}.

The procedure is as follows:
\begin{itemize}
\item Leave the object $\d r$ in the line element undisturbed.
\item Whenever the metric components $g_{\mu\nu}$ have an explicit $r$-dependence, replace the $r$-coordinate by $\sqrt{r^2+\ell^2}$, where $\ell$ is some length scale (typically associated with the Planck length, and performing an identical role as the introduced scalar parameter for standard SV spacetime).
\end{itemize}
Leaving the object $\d r$ unchanged implies that the $r$ coordinate still performs an identical role to the curvature $r$ coordinate in terms of the spatial slicings of the spacetime, and ensures that one is not simply making a coordinate transformation. Crucially, replacing $r\rightarrow\sqrt{r^2+\ell^2}$ has the advantage of ``smoothing'' the geometry into something which is globally regular.
\vfill

\subsection{Preliminary geometric analysis}
\vspace*{10pt}

To begin with, consider the RN solution to the electrovac Einstein equations of GR, expressed in terms of standard $\left(t,r,\theta,\phi\right)$ curvature coordinates:
\begin{equation}
    \d s^2 = -f_\text{RN}(r)\d t^2 + \frac{\d r^2}{f_\text{RN}(r)}+r^2\d\Omega^{2}_{2} \ , \quad f_\text{RN}(r) = 1-\frac{2m}{r}+\frac{Q^2}{r^2}\ .
\end{equation}
Here $Q$ is the electrical charge of the centralised massive object controlling the curvature of the spacetime. Now perform the aforementioned ``regularising'' procedure; replace $r\rightarrow\sqrt{r^2+\ell^2}$ in the metric. The new candidate spacetime is described by the line element
\begin{equation}
    \d s^2 = -f(r)\d t^2 + \frac{\d r^2}{f(r)} + \left(r^2+\ell^2\right)\d\Omega^{2}_{2} \ , \quad f(r) = 1-\frac{2m}{\sqrt{r^2+\ell^2}}+\frac{Q^2}{r^2+\ell^2} \ .
\end{equation}
One can immediately see that the natural domains for the angular and temporal coordinates are unaffected by the regularisation procedure. In contrast the natural domain of the $r$ coordinate expands from $r\in[0,+\infty)$ to $r\in(-\infty,+\infty)$. Asymptotic flatness is preserved, as are the manifest spherical and time translation symmetries. 

Given the diagonal metric environment, it is trivial to establish the following tetrad:
\begin{align}
    \big(e_{\hat{t}}\big)^\mu &= \frac{1}{\sqrt{|f(r)|}} \left( 1, 0, 0, 0 \right) \ , \quad
    \big(e_{\hat{r}}\big)^\mu = \sqrt{|f(r)|} \left(0,1,0,0 \right) \ , \nonumber \\
    \big(e_{\hat{\theta}}\big)^\mu &= \frac{1}{\sqrt{r^2+\ell^2}}\left(0,0,1,0 \right) \ , \quad
    \big(e_{\hat{\phi}}\big)^\mu = \frac{1}{\sqrt{r^2+\ell^2}\,\sin\theta}\left(0,0,0,1 \right) \ ,\label{tetrad_bbRN}
\end{align}
and straightforward to verify that this is indeed a solution of $g_{\mu\nu}e_{\hat{\mu}}{}^{\mu}e_{\hat{\nu}}{}^{\nu}=\eta_{\hat{\mu}\hat{\nu}}$. Note however that the $-1$ in the Minkowski metric corresponds to the timelike direction, and is therefore in the $\hat t \hat t$ position when $f(r)>0$, and in the $\hat r \hat r$ position when instead $f(r)<0$. The analysis that follows is, when appropriate, performed with respect to this orthonormal basis.

Furthermore, note that as $\ell\to0$ one recovers the standard RN geometry, while when $m\to0$ and $Q\to0$ one recovers the standard Morris--Thorne wormhole~\cite{WISATUFIT:Morris:1988, WTMATWEC:Morris:1988, LW:Visser:1995, EMRATW:Boonserm:2018}:
\begin{equation}
    \d s^2 = -\d t^2 + \d r^2 + (r^2+\ell^2)\,\d\Omega^{2}_{2} \ .
\end{equation}
When only setting $Q\to0$ one recovers the standard SV spacetime of Eq.~(\ref{SV}), and setting \textit{both} $Q,\ell\to0$ returns Schwarzschild spacetime.
\clearpage

\subsubsection{Kretschmann scalar}\label{CBB:Kretschmann}

Firstly one must show that the bbRN spacetime is indeed globally regular. Conveniently, given that the spacetime is static, by Theorem~\ref{Theorem:Kretsch} examination of the Kretschmann scalar $K=R^{\mu\nu\rho\sigma}\,R_{\mu\nu\rho\sigma}$ will be sufficient to accomplish this task~\cite{NBB:Lobo:2021}.

Indeed, a simple computation shows that the Kretschmann scalar is quartic in $Q$ and given by
\begin{align}
    K &= \frac{4}{(r^2+\ell^2)^6}\bigg\lbrace\left(3\ell^4-10\ell^2r^2+14r^4\right)Q^4 \nonumber \\
    & \ \ +\sqrt{r^2+\ell^2}\left[2\ell^2(2\ell^2-3r^2)\sqrt{r^2+\ell^2}-2m(5\ell^4-11\ell^2r^2+12r^4)\right]Q^2 \nonumber \\[3pt]
    & \qquad \qquad \qquad \qquad \qquad \qquad \ \ +3\ell^8+(9m^2+6r^2)\ell^6+3r^2(r^2-m^2)\ell^4 \nonumber \\
    & \qquad \qquad \qquad \qquad \qquad \qquad \quad \ \ -8m\ell^2(\ell^4-r^4)\sqrt{r^2+\ell^2}+ 12m^2r^6\bigg\rbrace \ . \nonumber \\
    & \label{bbRNKretsch}
\end{align}
Examining the manifest finiteness of this quantity, it is clear one need only concern oneself with the behaviour of the denominator in the prefactor. In the limit as $r\rightarrow 0$ one has the result
\begin{equation}
    \lim_{r\rightarrow 0} K = \frac{12}{\ell^8}\,Q^4+\frac{8(2\ell-5m)}{\ell^7}\,Q^2+\frac{4\left(3\ell^2-8m\ell+9m^2\right)}{\ell^6} \ .
\end{equation}
Hence enforcing $\vert\ell\vert>0$, one can conclude that the Kretschmann scalar is globally finite. So in this manifestly static situation, by Theorem~\ref{Theorem:Kretsch}, one is guaranteed that all of the orthonormal Riemann curvature tensor components are automatically finite~\cite{NBB:Lobo:2021}. One may conclude that the geometry is indeed globally regular.

\subsubsection{Curvature tensors}\label{CBB:tensors}

For completeness, and in order to finish the investigation on curvature, here are the explicit forms for the various nonzero curvature tensor components and Riemann curvature invariants. This is mostly conveniently done in the introduced orthonormal basis of Eq.~(\ref{tetrad_bbRN}). The orthonormal components of the Riemann curvature tensor are given by
\begin{subequations}
\begin{align}
    R^{\hat{t}\hat{r}}{}_{\hat{t}\hat{r}} &= \frac{\ell^2-3r^2}{(r^2+\ell^2)^3}Q^2-\frac{m(\ell^2-2r^2)}{(r^2+\ell^2)^{5/2}} \ , \\
    R^{\hat{t}\hat{\theta}}{}_{\hat{t}\hat{\theta}} = R^{\hat{t}\hat{\phi}}{}_{\hat{t}\hat{\phi}} &= \frac{r^2}{(r^2+\ell^2)^3}Q^2-\frac{mr^2}{(r^2+\ell^2)^{5/2}} \ , \\
    R^{\hat{r}\hat{\theta}}{}
    _{\hat{r}\hat{\theta}} = R^{\hat{r}\hat{\phi}}{}_{\hat{r}\hat{\phi}} &= \frac{r^2-\ell^2}{(r^2+\ell^2)^3}Q^2+\frac{m(2\ell^2-r^2)-\ell^2\sqrt{r^2+\ell^2}}{(r^2+\ell^2)^{5/2}} \ , \\
    R^{\hat{\theta}\hat{\phi}}{}_{\hat{\theta}\hat{\phi}} &= -\frac{r^2}{(r^2+\ell^2)^3}Q^2+\frac{2mr^2+\ell^2\sqrt{r^2+\ell^2}}{(r^2+\ell^2)^{5/2}} \ ,
\end{align}
\end{subequations}
and in the limit as $r\rightarrow 0$ one has
\begin{equation}
    R^{\hat{t}\hat{r}}{}_{\hat{t}\hat{r}}\rightarrow \frac{Q^2-m\ell}{\ell^4} \ , \ \ R^{\hat{t}\hat{\theta}}{}_{\hat{t}\hat{\theta}}\rightarrow 0 \ , \ \ R^{\hat{r}\hat{\theta}}{}_{\hat{r}\hat{\theta}}\rightarrow -\frac{Q^2+\ell^2-2m\ell}{\ell^4} \ , \ \ R^{\hat{\theta}\hat{\phi}}{}_{\hat{\theta}\hat{\phi}}\rightarrow \frac{1}{\ell^2} \ .
\end{equation}
For the orthonormal components of the Ricci tensor one has
\begin{subequations}
\begin{align}
    R^{\hat{t}}{}_{\hat{t}} &= \frac{\ell^2-r^2}{(r^2+\ell^2)^{3}}Q^2 - \frac{m\ell^2}{(r^2+\ell^2)^{5/2}} \ , \\
    R^{\hat{r}}{}_{\hat{r}} &= -\frac{Q^2}{(r^2+\ell^2)^2} + \frac{\ell^2(3m-2\sqrt{r^2+\ell^2})}{(r^2+\ell^2)^{5/2}} \ , \\
    R^{\hat{\theta}}{}_{\hat{\theta}} &= R^{\hat{\phi}}{}_{\hat{\phi}} 
    = \frac{r^2-\ell^2}{(r^2+\ell^2)^3}Q^2+\frac{2m\ell^2}{(r^2+\ell^2)^{5/2}} \ ,
\end{align}
\end{subequations}
and in the limit as $r\rightarrow 0$
\begin{equation}
    R^{\hat{t}}{}_{\hat{t}}\rightarrow \frac{Q^2-m\ell}{\ell^4} \ , \quad R^{\hat{r}}{}_{\hat{r}}\rightarrow \frac{3m\ell-2\ell^2-Q^2}{\ell^4} \ , \quad R^{\hat{\theta}}{}_{\hat{\theta}}= R^{\hat{\phi}}{}_{\hat{\phi}}\rightarrow \frac{2m\ell-Q^2}{\ell^4} \ .
\end{equation}
Finally, for the Weyl tensor:
\begin{align}
    C^{\hat{t}\hat{r}}{}_{\hat{t}\hat{r}} &= -2C^{\hat{t}\hat{\theta}}{}_{\hat{t}\hat{\theta}} = -2C^{\hat{t}\hat{\phi}}{}_{\hat{t}\hat{\phi}} = -2C^{\hat{r}\hat{\theta}}{}_{\hat{r}\hat{\theta}} = -2C^{\hat{r}\hat{\phi}}{}_{\hat{r}\hat{\phi}} = C^{\hat{\theta}\hat{\phi}}{}_{\hat{\theta}\hat{\phi}} \nonumber \\[3pt]
    &= \frac{2(\ell^2-3r^2)}{3(r^2+\ell^2)^3}Q^2+\frac{2\ell^2\sqrt{r^2+\ell^2}-3m(\ell^2-2r^2)}{3(r^2+\ell^2)^{5/2}} \ ,
\end{align}
and in the limit as $r\rightarrow 0$
\begin{equation}
    C^{\hat{t}\hat{r}}{}_{\hat{t}\hat{r}}\rightarrow \frac{2Q^2+2\ell^2-3m\ell}{3\ell^4} \ .
\end{equation}

%
\subsubsection{Other Riemann curvature invariants}\label{CBB:invariants}

The explicit forms for the various nonzero Riemann curvature invariants for the bbRN geometry are presented below. Notably, all are globally finite, as was immediately guaranteed \textit{via} examination of the Kretschmann scalar from Eq.~(\ref{bbRNKretsch}).

The Ricci scalar is given by
\begin{equation}
    R = -\frac{2\ell^2}{(r^2+\ell^2)^{3}}\left(Q^2+r^2+\ell^2-3m\sqrt{r^2+\ell^2}\right) \ ,
\end{equation}
and as $r\rightarrow 0$
\begin{equation}
    \lim_{r\rightarrow 0} R = -\frac{2(Q^2+\ell^2-3m\ell)}{\ell^4} \ .
\end{equation}
The quadratic Ricci invariant $R^{\mu\nu}R_{\mu\nu}$ is given by
\begin{align}
    R^{\mu\nu}R_{\mu\nu} &= \frac{4\left(r^4-\ell^2r^2+\ell^4\right)}{(r^2+\ell^2)^{6}}Q^{4} + \frac{4\ell^2\left[m(r^2-4\ell^2)+(r^2+\ell^2)^{3/2}\right]}{(r^2+\ell^2)^{11/2}}Q^2 \nonumber \\[3pt]
    & \ \ + \frac{2\ell^4\left[2r^4-6m(r^2+\ell^2)^{3/2}+(9m^2+4\ell^2)(r^2+\ell^2)-2\ell^4\right]}{(r^2+\ell^2)^{6}} \ ,
\end{align}
and as $r\rightarrow 0$
\begin{equation}
    \lim_{r\rightarrow 0} R^{\mu\nu}R_{\mu\nu} = \frac{4}{\ell^8}\,Q^4+\frac{4(\ell-4m)}{\ell^7}\,Q^2+\frac{4\ell^2-12m\ell+18m^2}{\ell^6} \ .
\end{equation}
For the Weyl contraction $C^{\mu\nu\rho\sigma}C_{\mu\nu\rho\sigma}$ one has
\begin{align}
    & C^{\mu\nu\rho\sigma}C_{\mu\nu\rho\sigma} = \frac{16(3r^2-\ell^2)^{2}}{3(r^2+\ell^2)^{6}}Q^4 \nonumber \\[3pt]
    & \qquad \qquad \qquad + \frac{16(\ell^2-3r^2)\left[2\ell^2\sqrt{r^2+\ell^2}-3m(\ell^2-2r^2)\right]}{3(r^2+\ell^2)^{11/2}}Q^2 \nonumber \\
    & \qquad \qquad \qquad \quad +\frac{4\ell^4}{3(r^2+\ell^2)^6}\Bigg[4\ell^4 + (9m^2+8r^2)\ell^2 + 4r^4 -27m^2r^2 \nonumber \\
    & \qquad \qquad \qquad \qquad \ \ -12m\sqrt{r^2+\ell^2}\left(1+\frac{r^2}{\ell^2}\right)(\ell^2-2r^2) + \frac{36m^2r^6}{\ell^4}\Bigg] \ ,
\end{align}
and as $r\rightarrow 0$
\begin{equation}
    \lim_{r\rightarrow 0}C^{\mu\nu\rho\sigma}C_{\mu\nu\rho\sigma} = \frac{4\left(2Q^2+2\ell^2-3\ell m\right)^2}{3\ell^8} \ .
\end{equation}
So bbRN spacetime is globally regular, and amenable to tractable analysis of the Riemann curvature invariants and nonzero tensor components. It is prudent to discuss other geometrical properties; \textit{e.g.} horizons and characteristic orbits.

%
\subsection{Horizons and surface gravity}\label{CBB:horizon}
%
\vspace*{8pt}

In view of the diagonal metric environment, horizon locations are characterised by
\begin{align}
    g_{tt} &=
    0 \quad \Longrightarrow \quad r_{H} = S_1 \sqrt{ \left(m+S_2\sqrt{m^2-Q^2}\right)^2 -\ell^2 } \ .
\end{align}

Here $S_{1},S_{2}=\pm 1$, and choice of sign for $S_{1}$ dictates which universe one is in, whilst the choice of sign on $S_{2}$ corresponds to an outer/inner horizon respectively.
For horizons to exist one needs both $|Q|\leq m$ and $\ell \leq m \pm \sqrt{m^2-Q^2}$.
The case $|Q|=m$ while $\ell \leq m $ leads to extremal horizons at
$r_H = \pm \sqrt{m^2-\ell^2}$. When $|Q|>m$ there are no horizons, and the geometry is that of a traversable wormhole. Finally when $|Q|\leq m$ but $\ell$ is large enough, $\ell > m \pm \sqrt{m^2-Q^2}$, first the inner horizons vanish and then the outer horizons vanish.
\clearpage

\begin{figure}[!htb]
    \centering
    \includegraphics[width=.87\textwidth]{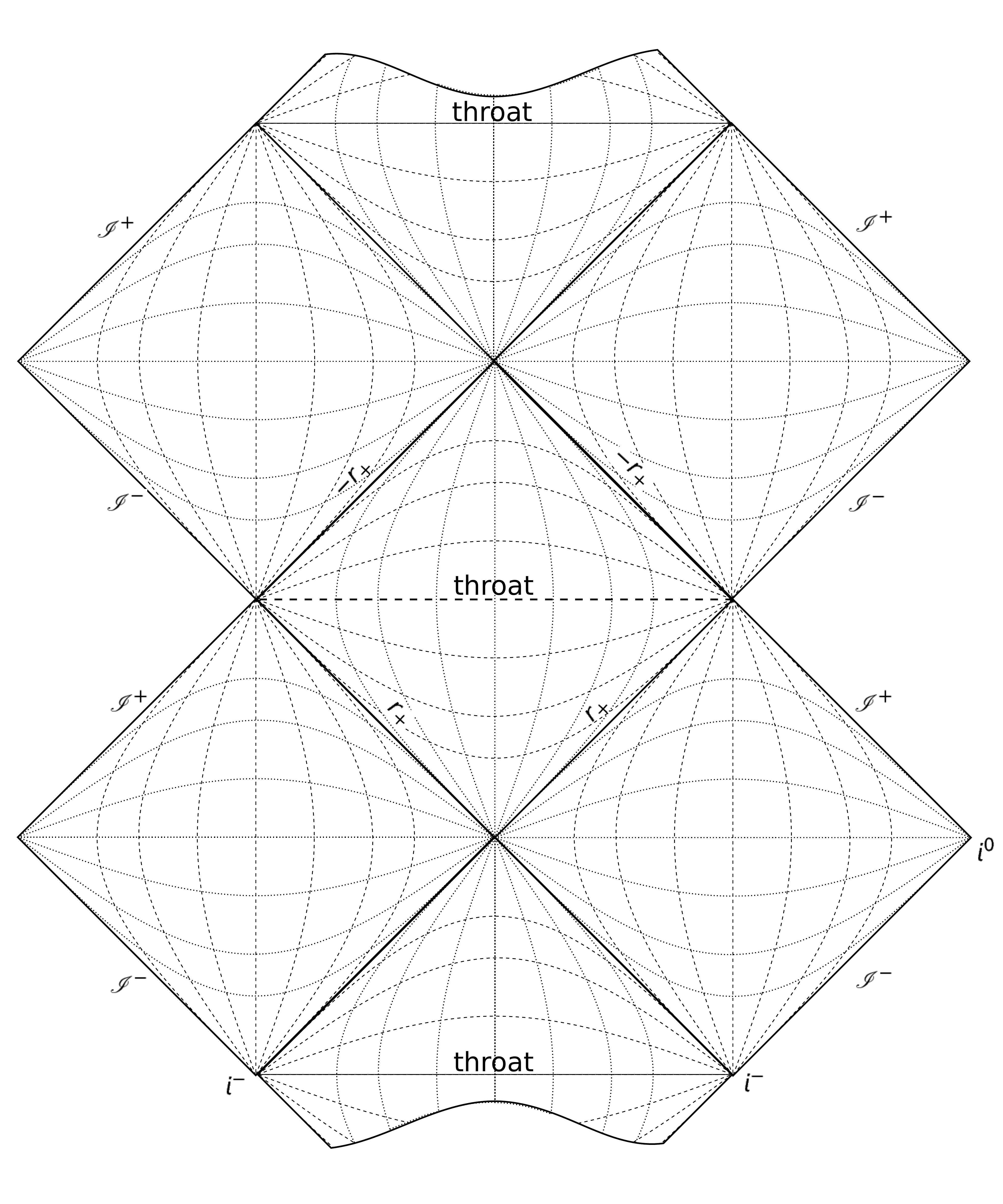}
    \caption[Carter--Penrose diagram for a maximally extended black-bounce Reissner--Nordstr\"{o}m regular black hole in the case where only outer horizons exist]{Carter--Penrose diagram for a regular black hole with only outer horizons, corresponding to $m-\sqrt{m^2-Q^2} < \ell < m+\sqrt{m^2-Q^2}$. The lower (upper) portion of the diagram corresponds to the $r>0$ ($r<0$) universe; the diagram continues indefinitely above and below the portion shown by repetition of this fundamental block. Here, $ r_+$ is $r_H$ with $S_2=+1$; the sign in front of it is $S_1$.\label{fig:penrose1}}
\end{figure}

\begin{figure}
    \centering
    \includegraphics[width=.999\textwidth]{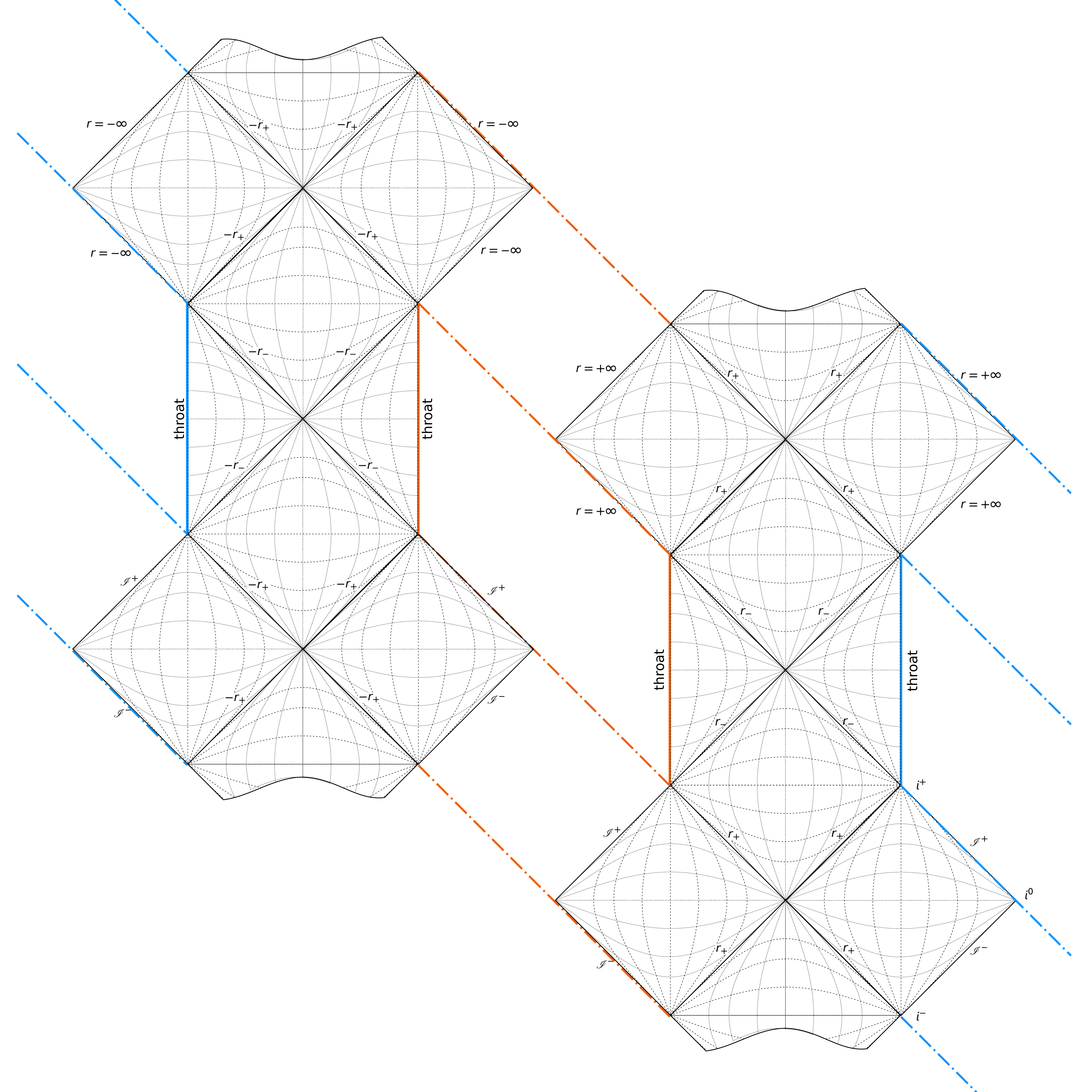}
    \caption[Carter--Penrose diagram for a maximally extended black-bounce Reissner--Nordstr\"{o}m regular black hole in the case where both inner and outer horizons exist]{Carter--Penrose diagram for a regular black hole with outer and inner horizons, corresponding to $\ell < m-\sqrt{m^2-Q^2}$. Vertical lines of the same colour are identified, as the right-hand (left-hand) part of the diagram represents the $r>0$ ($r<0$) universe; the diagram continues indefinitely above and below the portion shown. Here $r_+$ (resp.~$r_-$) is $r_H$ with $S_2=+1$ ($-1$); the sign in front of it is $S_1$.\label{fig:penrose2}}
\end{figure}

The structure of the maximally-extended spacetime can be visualised with the aid of the Carter--Penrose diagrams in Fig.~\ref{fig:penrose1} and Fig.~\ref{fig:penrose2} for the two qualitatively different regular black hole cases. Both diagrams will be relevant for the rotating generalisation of bbKN spacetime in Chapter~\ref{C:SVCBB} also; furthermore they are analogous to the Carter--Penrose diagrams of ``black-bounce Kerr'' (bbK) spacetime presented in reference~\cite{ANFORBHM:Franzin:2021}. Specifically, one should note that Fig.~\ref{fig:penrose1} is analogous to the first Carter--Penrose diagram explored in SV spacetime~\cite{BTTW:Simpson:2019}; \textit{i.e.} Fig.~\ref{F:bounce-1} already presented.
\vspace*{7pt}

It is straightforward to calculate the surface gravity at the event horizon in the $r>0$ universe for bbRN spacetime. Given one is working in standard curvature coordinates, the surface gravity $\kappa_{H}$ reduces to~\cite{DBHTAHS:Visser:1992}
\begin{equation}
    \kappa_{H} = \lim_{r\rightarrow r_{H}}\frac{1}{2}\frac{\partial_{r}g_{tt}}{\sqrt{-g_{tt}g_{rr}}} \ .
\end{equation}
For the metric one has $g_{tt} = -1/g_{rr}$ so this simplifies to
\begin{align}
    \kappa_{H} = \left.\frac{1}{2}\partial_{r}g_{tt}\right|_{r_H}
    = \frac{\sqrt{(m+\sqrt{m^2-Q^2})^2-\ell^2} \; \sqrt{m^2-Q^2}}{(m+\sqrt{m^2-Q^2})^3}
    = \kappa_H^\text{RN} \sqrt{\frac{r_H^2}{r_H^2+\ell^2}} \ ,
\end{align}
where $\kappa_H^\text{RN}$ is the surface gravity of a standard RN black hole of the same mass and charge, and as usual the associated Hawking temperature is given by $k_{B}T_{H} = \frac{\hbar}{2\pi}\kappa_{H}$~\cite{DBHTAHS:Visser:1992}. This gives the usual RN result as $\ell\to0$.

\subsection{ISCO and photon sphere}

With a look towards extraction of (potential) astrophysical observables for the bbRN spacetime, the coordinate locations of the notable orbits, the innermost stable circular orbit (ISCO) and the photon sphere~\cite{IAOSCOITPOAPCC:Boonserm:2020, PSISCOOSCO:Berry:2020, GCOQDRBH:Berry:2021}, are worth brief examination. Firstly, define the object
\begin{equation}
\epsilon = 
\begin{cases}
-1 & \quad\mbox{massive particle, \emph{i.e.}\ timelike worldline} \\
 0 & \quad\mbox{massless particle, \emph{i.e.}\ null worldline}.
\end{cases}
\end{equation}
Considering the affinely parameterised tangent vector to the worldline of a massive or massless particle, and fixing $\theta=\pi/2$ in view of spherical symmetry, one obtains the reduced equatorial problem
\begin{equation}
    \frac{\d s^2}{\d\lambda^2} = -g_{tt}\left(\frac{\d t}{\d\lambda}\right)^2+g_{rr}\left(\frac{\d r}{\d\lambda}\right)^2+(r^2+\ell^2)\left(\frac{\d\phi}{\d\lambda}\right)^2 = \epsilon \ .
\end{equation}
The Killing symmetries yield the following expressions for the conserved energy $E$, and angular momentum per unit mass $L$:
\begin{equation}
    \left(1-\frac{2m}{\sqrt{r^2+\ell^2}}+\frac{Q^2}{r^2+\ell^2}\right)\left(\frac{\d t}{\d\lambda}\right) = E \ , \quad (r^2+\ell^2)\left(\frac{\d\phi}{\d\lambda}\right) = L \ ,
\end{equation}
giving the ``effective potential'' for geodesic orbits
\begin{equation}
    V_{\epsilon}(r) = \left(1-\frac{2m}{\sqrt{r^2+\ell^2}}+\frac{Q^2}{r^2+\ell^2}\right)\left[-\epsilon+\frac{L^2}{r^2+\ell^2}\right] \ .
\end{equation}

\subsubsection*{Null orbits}

For the massless case of null orbits, \emph{e.g.}\ photon orbits, set $\epsilon=0$ and solve $V_{0}'(r) = 0$ for the location of the ``photon sphere''. First one has
\begin{align}
    V_{0}'(r) &= \frac{2L^2r\left[3m-\sqrt{r^2+\ell^2}\right]}{(r^2+\ell^2)^{5/2}} - \frac{4L^2r}{(r^2+\ell^2)^{3}}Q^2 \ ,
\end{align}
which gives the analytic location for the photon sphere in the $r>0$ universe \emph{outside} horizons:
\begin{equation}
    r_{\gamma} = \sqrt{\frac{m}{2}\left(9m+3\sqrt{9m^2-8Q^2}\right)-2Q^2-\ell^2} \ .
\end{equation}
In the limit as \textit{both} $Q,\ell\rightarrow 0$, the standard Schwarzschild result, $r_{\gamma}=3m$, is reproduced as expected. In the limit as $Q\rightarrow0$, the result for SV spacetime, Eq.~(\ref{SVorbits}), is returned, while in the limit as $\ell\rightarrow0$, the standard RN result is recovered, as expected.

\subsubsection*{Timelike orbits}

For the massive case of timelike orbits, the ISCO is extracted \emph{via} setting $V_{-1}'(r)=0$. Evaluating:
\begin{align}
    V_{-1}'(r) &= \frac{2r\left[m(r^2+\ell^2)+L^2(3m-\sqrt{r^2+\ell^2})\right]}{(r^2+\ell^2)^{5/2}} - \frac{2r(2L^2+r^2+\ell^2)}{(r^2+\ell^2)^{3}}Q^2 \ .
\end{align}
Equating $V_{-1}'(r)=0$ and solving for $r$ is not analytically feasible. Life can be made easier \emph{via} the change of variables $z=\sqrt{r^2+\ell^2}$, giving
\begin{equation}
    V_{-1}'(z) = -\frac{2\sqrt{z^2-\ell^2}}{z^6}\left\lbrace (2L^2+z^2)Q^2+z\left[L^2(z-3m)-mz^2\right]\right\rbrace \ .
\end{equation}
Assuming some fixed orbit at some $r_{c}$, hence fixing the corresponding $z_{c}=\sqrt{r_{c}^{2}+\ell^2}$, one may rearrange to find the required angular momentum per unit mass $L_{c}$ as a function of $z_{c}$ and the metric parameters. It follows that the ISCO will be located at the coordinate location where $L_{c}$ is minimised~\cite{IAOSCOITPOAPCC:Boonserm:2020, PSISCOOSCO:Berry:2020, GCOQDRBH:Berry:2021}. One has
\begin{align}
    L_{c} &= \frac{z_{c}\sqrt{mz_{c}-Q^2}}{\sqrt{z_{c}^{2}-3mz_{c}+2Q^2}} \,
    \Longrightarrow \, \frac{\partial L_{c}}{\partial z_{c}} = -\frac{6m^2z_{c}^2-mz_{c}(9Q^2+z_{c}^2)+4Q^4}{2\sqrt{mz_{c}-Q^2}(z_{c}^2-3mz_{c}+2Q^2)^{\frac{3}{2}}} \ .
\end{align}
Equating $\partial L_{c}/\partial z_{c}=0$ and solving for $z_{c}$, rearranging for $r_{c}$, and discounting complex roots leaves the analytic ISCO location for timelike particles in the $r>0$ universe:
\begin{equation}
    r_c = \frac{\sqrt{\left(A^2+2Am^2+4m^4-3m^2Q^2\right)^2-A^2\ell^2m^2}}{mA} \ ,
\end{equation}
%
%
where now the object $A$ is given by
\begin{align}
    A &= \left[2m^2Q^4+m^2\left(\pm B-9m^2\right)Q^2+8m^6\right]^{\frac{1}{3}} \ , \nonumber \\[3pt]
    B &= \sqrt{4Q^4-9m^2Q^2+5m^4} \ .
\end{align}
It is easily verified that in the limit as \textit{both} $Q,\ell\rightarrow 0$, $r_{c}\rightarrow 6m$, as expected for Schwarzschild. In the limit as $Q\rightarrow0$, the result for SV spacetime, Eq.~(\ref{SVorbits}), is regained, whilst in the limit as $\ell\rightarrow0$, the standard RN result is recovered, as expected.

\subsection{Stress-energy tensor and exact solution}\label{CBB:stress-energy}
\vspace*{10pt}

In the following discourse one assumes that the geometrodynamics is everywhere described by GR. Given the ``GR $+$ no singularities'' framework, this might seem a crude assumption for geometries associated to a possible regularisation of singularities by quantum gravity. However, once the regular spacetime settles down into an equilibrium state after the gravitational collapse, it is reasonable to expect this to be a good approximation everywhere if the final regularisation scale $\ell$ isn't much larger than the Planck scale (\textit{i.e.} for small $\ell$ the approximation only breaks down in a region where a mature theory of quantum gravity must necessarily take over). If $\ell$ is much larger than the Planck scale, it is still a good approximation sufficiently far away from the core region.

Given the above assumption, the determination of the stress-energy tensor associated to bbRN spacetime is easily accomplished by computing the nonzero components of the Einstein tensor. This leads to the following decomposition for the total stress-energy tensor $T^{\hat{\mu}}{}_{\hat{\nu}}$, valid \emph{outside} the outer horizon (and \emph{inside} the inner horizon):
\begin{equation}
    \frac{1}{8\pi}\,G^{\hat{\mu}}{}_{\hat{\nu}} = T^{\hat{\mu}}{}_{\hat{\nu}} 
    = \left[T_\text{bb}\right]^{\hat{\mu}}{}_{\hat{\nu}} + \left[T_{Q}\right]^{\hat{\mu}}{}_{\hat{\nu}} 
    = \diag\left(-\varepsilon, p_r, p_t, p_t\right) \ . 
\end{equation}
In contrast, \emph{between} the inner and outer horizons one has
\begin{equation}
    \frac{1}{8\pi}\,G^{\hat{\mu}}{}_{\hat{\nu}} 
    = T^{\hat{\mu}}{}_{\hat{\nu}} 
    = \left[T_\text{bb}\right]^{\hat{\mu}}{}_{\hat{\nu}} + \left[T_{Q}\right]^{\hat{\mu}}{}_{\hat{\nu}} 
    = \diag\left(p_r, -\varepsilon, p_t, p_t\right) \ . 
\end{equation}
Here $\left[T_\text{bb}\right]^{\hat{\mu}}{}_{\hat{\nu}}$ is the stress-energy tensor for the original electrically neutral SV spacetime~\cite{BTTW:Simpson:2019}, and $\left[T_{Q}\right]^{\hat{\mu}}{}_{\hat{\nu}}$ is the charge-dependent contribution to the stress-energy.
\clearpage

Examination of the nonzero components of the Einstein tensor, outside the outer horizon (and inside the inner horizon), yields the following:
\begin{subequations}
\begin{align}\label{rho}
    \varepsilon &= \frac{Q^2(r^2-2\ell^2)}{8\pi(r^2+\ell^2)^{3}} + \frac{\ell^2\left(4m-\sqrt{r^2+\ell^2}\right)}{8\pi(r^2+\ell^2)^{5/2}} \ , \\
    -p_{r} &= \frac{Q^2r^2}{8\pi(r^2+\ell^2)^{3}} + \frac{\ell^2}{8\pi(r^2+\ell^2)^{2}} \ , \\
    p_{t} &= \frac{Q^2r^2}{8\pi(r^2+\ell^2)^{3}} + \frac{\ell^2\left(\sqrt{r^2+\ell^2}-m\right)}{8\pi(r^2+\ell^2)^{5/2}} \ .
\end{align}
\end{subequations}
For the radial null energy condition (NEC)~\cite{LW:Visser:1995, TWSSE:Visser:1989, TWFSMSS:Visser:1989, TWWASECV:Dadhich:2003, QECVITW:Dadhich:2004, CASEC:Moruno:2017, GVP1:Visser:1996, GVP2:Visser:1996, GVP4:Visser:1997, SECFQVS:Moruno:2013, TFTEC:Barcelo:2002}, outside (inside) the outer (inner) horizon, one has:
\begin{align}
    \varepsilon + p_{r} 
    = -\frac{\ell^2\left[r^2+\ell^2-2m\sqrt{r^2+\ell^2}+Q^2\right]}{4\pi(r^2+\ell^2)^{3}}
    = -\frac{\ell^2 f(r)}{4\pi(r^2+\ell^2)^{2}} \ .
\end{align}
It is then clear that outside the outer horizon (or inside the inner one), where $f(r)>0$, the radial NEC is violated. At the horizons, where $f(r)=0$, one always has $\left.\left(\varepsilon+p_r\right)\right|_H = 0$. This on-horizon marginal satisfaction of the NEC is a quite generic phenomenon~\cite{DBHTAHS:Visser:1992, HCOS:Moruno:2021, DBHSGANHS:Martin:2004, DBHSASNH:Martin:2004}.

The explicit form for $\left[T_{Q}\right]^{\hat{\mu}}{}_{\hat{\nu}}$ can be trivially extracted:
\begin{align}\label{Tem}
    \left[T_{Q}\right]^{\hat{\mu}}{}_{\hat{\nu}} 
    &= \frac{Q^2r^2}{8\pi(r^2+\ell^2)^{3}}\, \diag\left( \frac{2\ell^2}{r^2}-1, -1, 1, 1\right) \nonumber \\
    &= \frac{Q^2r^2}{8\pi(r^2+\ell^2)^{3}}\left[\diag\left(-1, -1, 1, 1\right)+ \diag\left(\frac{2\ell^2}{r^2}, 0, 0, 0\right)\right] \ .
\end{align}
In the current situation, the first term above can be interpreted as the usual Maxwell stress-energy tensor~\cite{G:Misner:2000}
\begin{equation}\label{TMaxwell}
  \left[T_\text{Maxwell}\right]^{\hat{\mu}}{}_{\hat{\nu}} 
  = \frac{1}{4\pi}
  \left[
  -F^{\hat{\mu}}{}_{\hat{\alpha}}F^{\hat{\alpha}}{}_{\hat{\nu}}
  -\frac{1}{4}\delta^{\hat{\mu}}{}_{\hat{\nu}}F^{2}
  \right] \ ,
\end{equation}
while the second term can be interpreted as the stress-energy of ``charged dust'', with the density of the dust involving both the bounce parameter~$\ell$ and the total charge~$Q$. Overall:
\begin{equation}
  \left[T_{Q}\right]^{\hat{\mu}}{}_{\hat{\nu}} 
  = \left[T_\text{Maxwell}\right]^{\hat{\mu}}{}_{\hat{\nu}}
  +\Xi\; V^{\hat{\mu}}V_{\hat{\nu}} \ .
\end{equation}
The vector $V^{\hat\mu}$ is the normalised unit timelike eigenvector of the stress-energy, which in the current situation reduces to the normalised time-translation Killing vector, while the dust density $\Xi$ has to be determined. One obtains
\begin{equation}
    \left[T_{Q}\right]^{\hat{t}}{}_{\hat{t}} 
    = -\varepsilon_\text{em} = \left[T_\text{Maxwell}\right]^{\hat{t}}{}_{\hat{t}}  -\Xi= -\frac{1}{8\pi}E^2-\Xi \ .
\end{equation}
Comparing with Eq.~(\ref{Tem}), for the electric field strength $E$ it is readily seen that
\begin{align}\label{eq:ERN}
    E &= \frac{Qr}{(r^2+\ell^2)^{3/2}} = E_\text{RN}\left[\frac{r^3}{(r^2+\ell^2)^{3/2}}\right] \ ,
\end{align}
where $E_\text{RN}$ is the electric field strength of a typical RN black hole. The density of the dust $\Xi$ is given by
\begin{equation}
    \Xi = -\frac{1}{4\pi} \frac{Q^2\ell^2}{(r^2+\ell^2)^{3}} \ .
\end{equation}
As such, all told the electromagnetic stress-energy tensor for bbRN spacetime takes the following form
\begin{equation}
    \left[T_{Q}\right]^{\hat{\mu}}{}_{\hat{\nu}} 
    = \frac{1}{4\pi}\left[-F^{\hat{\mu}}{}_{\hat{\alpha}}F^{\hat{\alpha}}{}_{\hat{\nu}}-\frac{1}{4}\delta^{\hat{\mu}}{}_{\hat{\nu}}F^2\right] 
    -\frac{1}{4\pi}\frac{Q^2\ell^2}{(r^2+\ell^2)^3}V^{\hat{\mu}}V_{\hat{\nu}} \ .
\end{equation}
Finally, the electromagnetic potential is easily extracted \emph{via} integrating Eq.~\eqref{eq:ERN}, and in view of asymptotic flatness one may set the constant of integration to zero, yielding
\begin{equation}
    A_{\mu} = (\Phi_\text{em}(r), 0,0,0) = -\frac{Q}{\sqrt{r^2+\ell^2}}\;(1,0,0,0) \ .
    \label{eq:emstatic}
\end{equation}
Note that this really is simply the electromagnetic potential from standard RN spacetime, $-Q/r$, under the map $r\to \sqrt{r^2+\ell^2}$.

It is easy to verify that the electromagnetic field-strength tensor $F_{\mu \nu} = \nabla_{\mu} A_{\nu} - \nabla_{\nu} A_{\mu}$ satisfies $F_{[\mu\nu,\sigma]}=0$. The inhomogeneous Maxwell equation is, using $z=\sqrt{r^2+\ell^2}$: 
\begin{align}
    \nabla^{\hat {\mu}} F_{\hat{\mu} \hat {\nu}} &= {\frac{Q\ell^2}{z^6}\;\sqrt{z^2-2mz+Q^2}}\; \left(1,0,0,0\right) \ .\label{eq:maxinhomoRN}
\end{align}
The situation will get messier when adding rotation in Chapter~\ref{C:SVCBB}.

%
%
In the recent work (Dec. 2021) by Bronnikov and Walia~\cite{FSFSVS:Bronnikov:2022}, where SV spacetime was elevated to ``exact solution status'', the same was seen for bbRN spacetime. Those authors show that as an exact solution to the Einstein field equations, bbRN spacetime is sourced by a combination of a minimally coupled phantom scalar field with a nonzero potential $V(\Phi)$, and a magnetic field in the framework of NLED --- qualitatively the same as the SV case. In fact the only difference between this and the analogous result for SV spacetime is the presence of the charge-dependent terms coming from the nonzero charge $Q$ of the centralised massive object. From an action principle, in the Lagrangian formalism, one specifically has the following for bbRN spacetime~\cite{FSFSVS:Bronnikov:2022} (recall $\mathcal{F} = 2q^2/(r^2+\ell^2)^2\,,\newline \Phi = \pm\arctan\left(\frac{r}{\ell}\right)$):
\begin{eqnarray}\label{bbRNSolution}
    S &=& \int\,\sqrt{-g}\,\d^{4}r\left[R -2g^{\mu\nu}\partial_{\mu}\Phi\,\partial_{\nu}\Phi - 2\,V(\Phi) - \mathcal{L}\left(\mathcal{F}\right)\right] \ , \nonumber \\
    && \nonumber \\
    V(\Phi) &=& \frac{2\cos^6\Phi}{15\ell^4}\left(6m\ell\sec\Phi - 5Q^2\right) = \frac{2\ell^2\left[6m\sqrt{r^2+\ell^2}-5Q^2\right]}{15(r^2+\ell^2)^3} \ , \nonumber \\
    && \nonumber \\
    \mathcal{L}\left(\mathcal{F}\right) &=& \frac{12m\ell^2}{5\left(\frac{2q^2}{\mathcal{F}}\right)^{\frac{5}{4}}} + \frac{2Q^2\left[3\left(\frac{2q^2}{\mathcal{F}}\right)^{\frac{1}{2}}-4\ell^2\right]}{3\left(\frac{2q^2}{\mathcal{F}}\right)^{\frac{3}{2}}} \nonumber \\[4pt]
    &=& \frac{12m\ell^2}{5\left(r^2+\ell^2\right)^{\frac{5}{2}}} + \frac{2Q^2(3r^2-\ell^2)}{3(r^2+\ell^2)^3} \ . \nonumber \\
    &&
\end{eqnarray}
Here $q$ is the charge associated with the electromagnetic field coupled to the background geometry \textit{via} NLED, while $Q$ is the charge of the centralised massive object. $R$ is the four-dimensional Ricci scalar. Extremising the action Eq.~(\ref{bbRNSolution}) by varying $S$ with respect to the contrametric $g^{\mu\nu}$ leads to the Einstein equations, decomposable into a sum of stress-energy contributions from the scalar and electromagnetic fields respectively. Varying the action in $\Phi$ and $F_{\mu\nu}$ obtains a system of field equations for which the bbRN metric is the unique solution in spherical symmetry. For more detailed discussion please see reference~\cite{FSFSVS:Bronnikov:2022}. Notice specifically that in the limit as $Q\rightarrow0$, Eq.~(\ref{bbRNSolution}) becomes Eq.~(\ref{SVaction}), the analogous result for SV spacetime, as expected.

%

Having thoroughly explored the black-bounce analogs to Schwarzschild and RN spacetime, the SV and bbRN spacetimes respectively, the discourse is sufficiently mature to proceed to Chapter~\ref{C:SVCBB} and analyse the analog to Kerr--Newman by extending the analysis to the geometric environment of stationary axisymmetry.


%% file: 02-Black-bounce/4-SVCBB.tex
\allowdisplaybreaks
\definecolor{purple}{rgb}{1,0,1}
\definecolor{lime}{HTML}{A6CE39} 
\newcommand{\iu}{\mathrm{i}\mkern1mu}
\renewcommand{\r}{\hat{r}}
\renewcommand{\t}{\hat{t}}
\renewcommand{\th}{\hat{\theta}}
\newcommand{\ph}{\hat{\phi}}
\def\d{{\mathrm{d}}}
\def\O{{\mathcal{O}}}
\def\A{{\mathcal{A}}}
\chapter{Black-bounce Kerr--Newman}\label{C:SVCBB}

%
%
%

Having discussed the black-bounce Reissner--Nordstr\"{o}m (bbRN) spacetime~\cite{CBBS:Franzin:2021} in \S~\ref{SVbbRN}, and having knowledge of the black-bounce Kerr (bbK) spacetime constructed by Liberati \textit{et al} in reference~\cite{ANFORBHM:Franzin:2021}, it is immediate to suspect that it would be straightforward to discover and analyse the black-bounce Kerr--Newman (bbKN) geometry.

In reference~\cite{ANFORBHM:Franzin:2021}, those authors used the Newman--Janis procedure~\cite{NOTKSPM:Janis:1965, JAGRANCBH:Janis:2017, CKVOTNT:Rajan:2017} to transmute the original SV spacetime~\cite{BTTW:Simpson:2019} into an axisymmetric rotating version (the black-bounce analog to Kerr; bbK spacetime). Extending this, one can use the procedure by which one obtains a black-bounce variant from a known pre-existing solution as outlined in \S~\ref{SVbbRN}, and apply it to the standard Kerr--Newman spacetime, yielding the bbKN spacetime of reference~\cite{CBBS:Franzin:2021}. In particular, the existence of a nontrivial Killing tensor (and associated Carter-like constant) is verified for bbKN spacetime --- but without the existence of the full Killing tower~\cite{BHHSACI:Frolov:2017} of principal tensor and Killing--Yano tensor. Furthermore, the bbKN spacetime requires an interesting and nontrivial matter/energy content when viewed through the lens of standard GR.

Similarly to the previous cases, bbKN spacetime exhibits the interesting feature that it is an everywhere-regular one-parameter class of geometry smoothly interpolating between standard GR electrovac black holes and traversable wormholes~\cite{WISATUFIT:Morris:1988, WTMATWEC:Morris:1988, LW:Visser:1995, TWSSE:Visser:1989, TWFSMSS:Visser:1989}. In particular, the classical energy conditions will have nontrivial behaviour~\cite{LW:Visser:1995, TWSSE:Visser:1989, TWFSMSS:Visser:1989, TWWASECV:Dadhich:2003, QECVITW:Dadhich:2004, CASEC:Moruno:2017, GVP4:Visser:1997, GVP2:Visser:1996, GVP1:Visser:1996, SECFQVS:Moruno:2013, TFTEC:Barcelo:2002, TWVTSEC:Hochberg:1999, TQPOCP:Visser:2002}.
Furthermore, since the bbKN geometry can be fine-tuned to be arbitrarily close to Kerr--Newman, it serves as a ``black hole mimicker'' of direct interest to the observational community~\cite{LISA:Barausse:2020, PAOBHBGR:Carballo-Rubio:2018, WABHF:Damour:2007, OTVORBH:Carballo-Rubio:2018, GCBH:Carballo-Rubio:2020, OTPBATCOBH:Carballo-Rubio:2020, SDH:Barcelo:2009, BHIGR:Visser:2009, GCS:Mazur:2002, GVCS:Mazur:2004, SG:Visser:2004, POOH:Visser:2014}.

\section{Geometric analysis}\label{bbKNGeometry}

As expected, the bbKN geometry is qualitatively more complicated (but also physically richer) than that of the bbRN geometry from \S~\ref{SVbbRN}. Start with Kerr--Newman geometry in Boyer--Lindquist coordinates~\cite{TKSABI:Visser:2007, TKSRBHIGR:Scott:2009}:
\begin{eqnarray}
    \d s^2_\text{KN}  &=& -\frac{\Delta_\text{KN}}{\rho^2_\text{KN}}(a\sin^2\theta \d\phi - \d t)^2 
    + \frac{\sin^2\theta}{\rho^2_\text{KN}}\left[(r^2+a^2)\d\phi -a\d t\right]^2 \nonumber \\
    && \qquad \qquad \qquad \qquad \qquad \qquad \qquad
    + \frac{\rho^2_\text{KN}}{\Delta_\text{KN}}\d r^2+\rho^2_\text{KN}\d\theta^2 \ ,
\end{eqnarray}
where
\begin{equation}
    \rho^2_\text{KN} = r^2+a^2\cos^2\theta \ , \qquad \Delta_\text{KN} = r^2+a^2-2mr+Q^2 \ .
\end{equation}
Applying the ``regularisation'' procedure, the new candidate spacetime, bbKN spacetime, is given by the line element
\begin{equation}
    \d s^2 = -\frac{\Delta}{\rho^2}(a\sin^2\theta \d\phi - \d t)^2 + \frac{\sin^2\theta}{\rho^2}\left[(r^2+\ell^2+a^2)\d\phi -a\d t\right]^2 + \frac{\rho^2}{\Delta}\d r^2+\rho^2\d\theta^2 \ ,
    \label{eq:bbKN}
\end{equation}
where now $\rho^2$ and $\Delta$ are modified:
\begin{equation}
    \rho^2 = r^2+\ell^2+a^2\cos^2\theta \ , \qquad \Delta = r^2+\ell^2+a^2-2m\sqrt{r^2+\ell^2}+Q^2 \ .
\end{equation}
The natural domains of the angular and temporal coordinates are unaffected, while the radial coordinate again extends from the positive half line to the entire real line, and both manifest axisymmetry and asymptotic flatness are preserved. This spacetime is now \emph{stationary} but not static, hence one must examine more than just the Kretschmann scalar to draw any conclusion as to regularity.

\subsection{Curvature quantities}\label{CBB:tensorsKN}

Again using $z:= \sqrt{r^2 + \ell^2}$, where this shorthand now stands for the equatorial value of the parameter $\rho$, the Ricci scalar is given by
\begin{equation}
R = 2 \ell^2 \frac{m \left(\rho^4-2 z^4\right)+Q^2 z^3+z^3 \left(\rho^2-2 \Delta \right)}{\rho^6 z^3} \ .
\end{equation}
It is clearly finite in the limit $r\to 0$, \emph{i.e.}\ $z\to \ell$. So are the Kretschmann scalar and the invariants $R^{\mu \nu}R_{\mu \nu}$ and $C^{\mu \nu \rho \sigma}C_{\mu \nu \rho \sigma}$. The only potentially dangerous behaviour arises from the denominators, which, in the $r\to 0$ limit, take the form
\begin{equation}
    \frac{1}{\ell^2 (\ell^2 + a^2 \cos^2\theta)^6} \ .
\end{equation}
As long as $\ell \neq 0$, therefore, these quantities are never infinite.

Now turn the attention to the Einstein and Ricci tensors. Note that the (mixed) Einstein tensor $G^\mu _{\ \nu}$ can be diagonalised over the real numbers: its four eigenvectors $\{ e_{\hat{\mu}}\}_{\mu = t,r,\theta,\phi}$ form a globally defined tetrad~\cite{GPOPILS:Rajan:2016} and have explicit Boyer--Lindquist components:\enlargethispage{30pt}
\begin{align}
    \big(e_{\hat{t}}\big)^\mu &= \frac{1}{\sqrt{\rho^2 \abs{\Delta}}} \left( r^2+\ell^2+a^2, 0, 0, a \right) \ , \qquad
    \big(e_{\hat{r}}\big)^\mu = \sqrt{\frac{\abs{\Delta}}{\rho^2}} \left(0,1,0,0 \right)\ , \nonumber \\
    \big(e_{\hat{\theta}}\big)^\mu &= \frac{1}{\sqrt{\rho^2}}\left(0,0,1,0 \right) \ , \qquad
    \big(e_{\hat{\phi}}\big)^\mu = \frac{1}{ \sin \theta\sqrt{\rho^2}}\left(a \sin^2 \theta,0,0,1 \right) \ .
    \label{eq:tetrad}
\end{align}
Eigenvectors are defined up to multiplicative, possibly dimensionful constants. This choice of normalisation ensures that the tetrad $\{e_{\hat{\mu}}\}$ is orthonormal and reduces to Eq.~(\ref{tetrad_bbRN}) in the irrotational limit $a\to 0$. The tetrad Eq.~(\ref{eq:tetrad}) is hence used, along with the coordinate basis, to express components of curvature tensors.

In particular, the components of the Einstein tensor are
\begin{subequations}
\begin{align}
    G_{\hat t \hat t} &= \sign(\Delta) \frac{  \ell^2 \left[2 m \left(z^2-\rho^2\right)+z \left(\rho^2-2 \Delta \right)\right]+ Q^2 z \left(\rho^2-\ell^2\right)}{\rho^6 z} \ , \\
    G_{\hat r \hat r} &= - \sign(\Delta) \frac{ \ell^2 \left[2 m \left(z^2-\rho^2\right)+\rho^2 z\right]+Q^2 z \left(\rho^2-\ell^2\right) }{\rho^6 z} \ , \\
    G_{\hat \theta \hat \theta} &= \frac{\ell^2 \left[m \left(-\rho^4-2 \rho^2 z^2+2 z^4\right)+\rho^2 z^3\right]+Q^2 z^3 \left(\rho^2-\ell^2\right)}{\rho^6 z^3} \ , \\
    G_{\hat \phi \hat \phi} &= \frac{\ell^2 \left[m \left(-\rho^4-2 \rho^2 z^2+6 z^4\right)+z^3 \left(2 \Delta -\rho^2\right)\right]+Q^2 z^3 \left(\rho^2-3 \ell^2\right)}{\rho^6 z^3} \ .
\end{align}
\end{subequations}
Note that
\begin{equation}
  G_{\hat t \hat t} +  
  G_{\hat r \hat r} 
  = -\sign(\Delta)\; \frac{2\ell^2\Delta}{\rho^6} =-\frac{2\ell^2\vert\Delta\vert}{\rho^6} \ ,
\end{equation}
which is well-behaved at $\Delta=0$.

The Ricci tensor is clearly diagonal, too:
\begin{subequations}
\begin{align}
    R_{\hat t \hat t} &= \sign(\Delta) \frac{m \ell^2 \left(-\rho^4-2 \rho^2 z^2+4 z^4\right)+Q^2 z^3 \left(\rho^2-2 \ell^2\right)}{\rho^6 z^3} \ , \\
    R_{\hat r \hat r} &= \sign(\Delta) \frac{ \ell^2 \left[m \left(\rho^4+2 \rho^2 z^2-4 z^4\right)-2 \Delta  z^3\right]+Q^2 z^3 \left(2 \ell^2-\rho^2\right) }{\rho^6 z^3} \ , \\
    R_{\hat \theta \hat \theta} &= \frac{Q^2}{\rho^4} - \frac{2 \ell^2}{\rho^6} \left[\Delta +\frac{\rho^2 (m-z)}{z}\right] \ , \\
    R_{\hat \phi \hat \phi} &= \frac{2 m \ell^2 \left(2 z^2-\rho^2\right)+Q^2 z \left(\rho^2-2 \ell^2\right)}{\rho^6 z} \ .
\end{align}
\end{subequations}
Similarly, note that
\begin{equation}
  R_{\hat t \hat t} +  
  R_{\hat r \hat r} 
  = -\sign(\Delta)\; \frac{2\ell^2\Delta}{\rho^6}=-\frac{2\ell^2\vert\Delta\vert}{\rho^6} \ ,
\end{equation}
which is well-behaved at $\Delta=0$.

From these expressions one immediately notices that the curvature tensors are rational polynomials in the variable $z = \sqrt{r^2+\ell^2}$, which is strictly positive, and their denominators never vanish. The same is true for the Riemann and Weyl tensors; one may thus conclude that the spacetime is free of curvature singularities.

\subsection{Horizons, surface gravity, and ergosurfaces}
\vspace*{10pt}

Horizons are now associated to the roots of $\Delta$; specifically:
\begin{equation}
 r_H = S_1 \sqrt{\left(m+ S_2 \sqrt{m^2-Q^2-a^2} \right)^2-\ell^2} \ ,
\end{equation}
where $S_1$ and $S_2$ are defined as in \S~\ref{CBB:horizon}. The spacetime structures corresponding to $m-\sqrt{m^2-a^2-Q^2}<\ell<m+\sqrt{m^2-a^2-Q^2}$, and $\ell< m-\sqrt{m^2-a^2-Q^2}$, are analogous to their nonspinning counterparts from \S~\ref{SVbbRN}. As such, the causal structures are represented by the same Carter--Penrose diagrams as Fig.~\ref{fig:penrose1} and Fig.~\ref{fig:penrose2} respectively.

In the Kerr--Newman geometry, one demands $Q^2+a^2 \leq m^2$ to avoid the possibility of naked singularities. In the bbKN case, due to the global regularity, one need not worry about this eventuality and may consider arbitrary values of spin and charge. Thus, if $Q^2+a^2 > m^2$ or $\ell > m +S_2\sqrt{m^2-Q^2-a^2}$, the spacetime has no horizon.

If horizons are indeed present, their surface gravity is given by
\begin{equation}
    \kappa_{S_2} := \frac{1}{2} \dv{}{r} \eval{\bigg( \frac{\Delta}{r^2+\ell^2 + a^2} \bigg)}_{r_H} = \kappa^\text{KN}_{S_2} \sqrt{\frac{r_H^2}{r_H^2+\ell^2}} \ ,
\end{equation}
where $\kappa^\text{KN}_{S_2}$ is the surface gravity relative to the inner, when $S_2=-1$, or outer, when $S_2=+1$, horizon of a standard Kerr--Newman black hole with mass $m$, spin $a$, and charge $Q$.

The ergosurface is determined by $g_{tt}=0$, which is a quadratic equation in $r$. The roots are given by:
\begin{equation}
    r_\text{erg}=S_1 \sqrt{\left(m +S_2 \sqrt{m^2-Q^2-a^2\cos^2\theta} \right)^2-\ell^2} \ ,
\end{equation}
where $S_1$ and $S_2$ are as before.
\clearpage

\subsection{Geodesics and equatorial orbits}

%
Consider a test particle with mass $\mu$, energy $E$, component of angular momentum (per unit mass) along the rotation axis $L_z$, and zero electric charge. Its trajectory $x^\mu(\tau)$ is governed by the following set of first-order differential equations (see \emph{e.g.}\ reference~\cite{BHP:Frolov:1998}):
\begin{subequations}\begin{align}
    \rho^2 \dv{t}{\tau} &= a(L_z-a E\sin^2 \theta) + \frac{(r^2+\ell^2) + a^2}{\Delta}[E(r^2 + \ell^2 + a^2) - L_z a] \ , \\
    \rho^2 \dv{r}{\tau} &= \pm \sqrt{\mathcal{R}} \ ,\label{eq:eomrad} \\
    \rho^2 \dv{\theta}{\tau} &= \pm \sqrt{\Theta} \ , \\
    \rho^2 \dv{\phi}{\tau} &= \frac{L_z}{\sin^2 \theta} - a E + \frac{a}{\Delta} [E(r^2+ \ell^2 + a^2) - L_z a] \ ,
\end{align}\end{subequations}
where
\begin{align}
    \mathcal{R} &= [E (r^2 + \ell^2 + a^2) -L_z a]^2 - \Delta[\mu^2 (r^2+\ell^2) + (L_z -a E)^2 + \mathcal{Q}] \ ,\label{eq:eom1} \\
    \Theta &= \mathcal{Q} - \cos^2 \theta \bigg[ a^2 (\mu^2-E^2) + \frac{L_z^2}{\sin^2 \theta} \bigg] \ ,
    \label{eq:eom2}
\end{align}
and $\mathcal{Q}$ is a generalised Carter constant associated to the existence of a Killing tensor discussed in \S~\ref{CBB:Killing} below.

In view of the existence of the Killing tensor, there exist orbits that lie entirely on the equatorial plane $\theta=\pi/2$. Exploiting the conserved quantities, their motion is effectively one-dimensional and governed by the effective potential $\mathcal{R}$. Circular orbits in particular are given by
\begin{equation}
    \mathcal{R} =0 \qquad \text{and} \qquad \dv{\mathcal{R} }{r}=0 \ ;
    \label{eq:circ}
\end{equation}
in addition, when
\begin{equation}
    \dv[2]{\mathcal{R}}{r}>0 \ ,
\end{equation}
the orbits are stable. 

Solutions to Eq.~(\ref{eq:circ}) can be easily found by exploiting known results on the Kerr--Newman geometry~\cite{TKSRBHIGR:Scott:2009, TKSABI:Visser:2007, MOARCM:Couch:1965,GSOTKFOGF:Carter:1968}. Indeed, writing Eq.~(\ref{eq:eom1}) in terms of $z:= \sqrt{r^2+\ell^2}$, one immediately recognises the textbook result for a Kerr--Newman spacetime in which the Boyer--Lindquist radius has been given the uncommon name $z$. Moreover,
\begin{equation}
    \dv{\mathcal{R}}{r} = \dv{z}{r}\; \dv{\mathcal{R}}{z} \ ,
\end{equation}
so
\begin{equation}
    \dv{\mathcal{R}}{z} = 0 \quad \Longrightarrow \quad \dv{\mathcal{R}}{r} = 0 \ .
\end{equation}
Furthermore, at the critical point one has
\begin{equation}
    \frac{\d^2\mathcal{R}}{\d r^2} = 
    \left(\frac{\d z}{\d r}\right)^2 \;
    \frac{\d^2\mathcal{R}}{\d z^2} \ .
\end{equation}
So stability (or lack thereof) is unaffected by the substitution $r\longleftrightarrow z$. Therefore, suppose $z_0$ is such that
\begin{equation}
    \mathcal{R}(z_0) = 0 \qquad \text{and} \qquad \dv{\mathcal{R} (z_0)}{z} =0 \ ;
\end{equation}
that is, suppose the Kerr--Newman spacetime has a circular orbit at radius $z=z_0$, then the bbKN spacetime has a circular orbit at $r = r_0 := \sqrt{z_0^2-\ell^2}$. Clearly, this mapping is allowed only if $z_0 \geq \ell$. 

Noncircular and nonequatorial orbits, instead, require a more thorough analysis.

\subsection{Killing tensor and nonexistence of the Killing tower\label{CBB:Killing}}

The existence of the generalised Carter constant $\mathcal{Q}$ introduced in the previous section is guaranteed by the fact that the tensor
\begin{equation}
    K_{\mu \nu} = \rho^2 \left(l_\mu n_\nu + l_\nu n_\mu \right) + \left( r^2 + \ell^2 \right)g_{\mu \nu}
    \label{eq:killingtens}
\end{equation}
is a Killing tensor; it is easy to explicitly check that $K_{(\mu\nu;\lambda)}=0$. Here
\begin{equation}
    l^\mu = \left(\frac{r^2+\ell^2+a^2}{\Delta}, 1, 0, \frac{a}{\Delta}\right)
    \quad \text{and} \quad
    n^\mu = \frac{1}{2\rho^2}\left(r^2+\ell^2+a^2, - \Delta, 0, a \right)
\end{equation}
are a pair of geodesic null vectors belonging to a generalised Kinnersley tetrad --- see reference~\cite{ANFORBHM:Franzin:2021}.

Recall Theorem~\ref{Theorem:PGLT2}: Defining the Carter operator $\mathcal{K}\Phi = \nabla_\mu\left(K^{\mu\nu}\nabla_\nu\Phi\right)$, and the scalar wave operator $\Box\Phi = \nabla_\mu\left(g^{\mu\nu}\nabla_\nu\Phi\right)$, one has the result
\begin{equation}
    \left[\mathcal{K},\Box\right]\Phi = \frac{2}{3}\left(\nabla_\mu\left[R,K\right]^\mu{}_\nu\right)\nabla^\nu\Phi\ .
\end{equation}
This operator commutator will certainly vanish when the tensor commutator $\left[R,K\right]^\mu{}_\nu := R^\mu{}_\alpha K^\alpha{}_\nu - K^\mu{}_\alpha R^\alpha{}_\nu$ vanishes, and this tensor commutator certainly vanishes for bbKN spacetime. Hence by Theorem~\ref{Theorem:PGLT2}, the scalar wave equation (not just the Hamilton--Jacobi equation) separates on bbKN spacetime.

In the Kerr--Newman spacetime we started from, the Killing tensor is part of a ``Killing tower'' which ultimately descends from the existence of a closed conformal Killing--Yano tensor; a \emph{principal tensor} for short~\cite{BHHSACI:Frolov:2017}. Such a principal tensor is a rank-two, antisymmetric tensor $h_{\mu \nu}$ satisfying (in four spacetime dimensions) the equation:
\begin{equation}
    \nabla_\mu h_{\nu \alpha} = \frac{1}{3} \left[ g_{\mu \nu} \nabla^\beta h_{\beta \alpha} - g_{\mu \alpha} \nabla^\beta h_{\beta \nu} \right] \ .
    \label{eq:printenseq}
\end{equation}
In the language of forms, $\mathbf{h}$ is a non-degenerate two-form satisfying
\begin{equation}
    \mathbf{\nabla_Y h} = \mathbf{Y} \wedge \mathbf{X}, \qquad \mathbf{X} = \frac{1}{3} \div{\mathbf{h}} \ ,
\end{equation}
with $\mathbf{Y}$ any vector. The equation above implies, incidentally, that $\mathbf{h}$ is closed:
$\mathbf{\dd b}=0 $ so that locally
$\mathbf{h} = \mathbf{\dd b} $. The Hodge dual of a principal tensor is a Killing--Yano tensor, \emph{i.e.}
\begin{align}
    \mathbf{f} = \mathbf{\ast h } \qquad \text{is such that} \qquad
    \nabla_\mu f_{\nu \alpha} + \nabla_\nu f_{\mu \alpha} = 0 \ .
    \label{eq:KYeq}
\end{align}
A Killing--Yano tensor, in turn, squares to a tensor
\begin{equation}
    k_{\mu \nu}:= f_{\mu \alpha} f_\nu^{\ \alpha}
\end{equation}
that is a Killing tensor; $ k_{(\mu \nu;\lambda)}=0 $.

One may thus wonder whether the Killing tensor Eq.~(\ref{eq:killingtens}) derives from a principal tensor, as in the Kerr--Newman case. Naively, one may want to apply the usual trick $r \to \sqrt{r^2+\ell^2}$ to the Kerr--Newman principal tensor, or to the potential $\mathbf{b}$ (the two options are not equivalent). By adopting the first strategy, one finds a ``would-be'' Killing--Yano tensor that does indeed square to Eq.~(\ref{eq:killingtens}) but fails to satisfy Eq.~(\ref{eq:KYeq}). The second approach also fails.

In fact, one can prove that no principal tensor can exist in this spacetime. The system Eq.~(\ref{eq:printenseq}) is overdetermined and has a solution only if a certain integrability condition is satisfied: This condition implies that the corresponding spacetime be of Petrov type D. However, in reference~\cite{ANFORBHM:Franzin:2021}, Liberati \textit{et al} proved that the bbK spacetime is not algebraically special; it follows that neither can the bbKN spacetime be algebraically special. More prosaically, the nonexistence of the Killing tower can be seen as a side effect of the fact that the bbKN geometry does not fall into Carter's ``off-shell'' two-free-function distortion of Kerr~\cite{BHHSACI:Frolov:2017}. 

For reference, the would-be Killing--Yano tensor is given by:
\begin{align}
    f_{\mu \nu} &= \begin{pmatrix}
    0 & -a \cos \theta & 0 & 0 \\
    a \cos \theta & 0 & 0 & - a^2 \cos \theta \sin^2 \theta\\
    0& 0 & 0 & 0\\
    0 & a^2 \cos \theta \sin^2 \theta & 0 & 0
    \end{pmatrix}
\nonumber\\
    &+ \sqrt{r^2+\ell^2} \sin\theta \begin{pmatrix}
    0 & 0 & a  & 0 \\
    0 & 0 & 0 & 0\\
    - a & 0 & 0 & (r^2+\ell^2+a^2)\\
    0 & 0 & - (r^2+\ell^2+a^2) & 0
    \end{pmatrix} \ .
\end{align}
This would-be Killing--Yano tensor comes from~\cite[Eq.~(3.22), p.~47] {BHHSACI:Frolov:2017}, with coordinates changed to Boyer--Lindquist form, and with the substitution $r\to \sqrt{r^2+\ell^2}$ in the tensor components. 
It is easy to check that $f^2 = K$, but
\begin{equation}
\nabla_{(\mu} f_{\nu) \alpha} = \left(\sqrt{r^2+\ell^2}-r \right) \times \left[\text{tensor  that is finite as }\ell \to 0 \right] \ .
\end{equation}
This manifestly vanishes when $\ell\to0$, as it should to recover the Killing--Yano tensor of the standard Kerr--Newman spacetime.

Its divergence is in fact particularly simple:
\begin{equation}
    \nabla_\mu f^{\mu \nu} = 
    \left[\left( \sqrt{r^2+\ell^2}-r \right)\frac{2a \cos \theta}{\rho^2}\right]
    \left(1, 0, 0, 0 \right) \ .
\end{equation}
This again manifestly vanishes when $\ell\to0$, as it should.

Note that if one instead takes
\begin{equation}
    b_\mu \dd x^\mu = -\frac{1}{2} (r^2 +\ell^2-a^2\cos^2\theta) \dd t -\frac{1}{2} \left[-r^2-\ell^2 + (r^2+\ell^2+a^2)\cos^2\theta \right] a \dd \phi \ ,
\end{equation}
as in reference~\cite[eq.~(3.21), p.~47]{BHHSACI:Frolov:2017}, converted to Boyer--Lindquist coordinates, and subjected to the substitution $r\to\sqrt{r^2+\ell^2}$, one finds
\begin{equation}
f_{\mu \nu} \neq \nabla_\mu b_\nu - \nabla_\nu b_\mu \ .
\end{equation}
This is not surprising since derivatives are involved.

Having now completed a purely geometrical treatment of the properties of bbKN spacetime, one can move on to discuss the implications of fixing the definite geometrodynamics to be, as before, that of standard GR.

\section{Stress-energy tensor}
\vspace*{7pt}

%
One may again exploit the orthonormal tetrad from Eq.~(\ref{eq:tetrad}) to characterise the distribution of stress-energy in bbKN spacetime. Assuming standard GR holds, the Einstein tensor is proportional to the stress-energy tensor: one may thus interpret the one component of $G_{\hat{\mu} \hat{\nu}}$ that corresponds to the timelike direction as the energy density $\varepsilon$, and all the other nonzero components as the principal pressures $p_i$. As will be shortly seen, the interpretation of the bbKN stress-energy is considerably more subtle than the bbRN case --- in fact there are two qualitatively different interpretations presented, both with their set of motivations; choosing one is somewhat a matter of taste and context.
\clearpage

In particular, outside any horizon (technically, whenever $\Delta > 0$), one has
\begin{subequations}
\begin{align}
    \varepsilon &= -\frac{\ell^2 \left(2 a^2 z+2 m \rho^2-6 m z^2+2 z^3-\rho^2 z\right)}{8\pi \rho^6 z}-\frac{Q^2 \left(3 \ell^2-\rho^2\right)}{8\pi \rho^6} \ , \\
    p_1 &= \frac{\ell^2 \left[\rho^2 (2 m-z)-2 m z^2\right]}{8\pi \rho^6 z}+\frac{Q^2 \left(\ell^2-\rho^2\right)}{8\pi \rho^6} \ , \\
    p_2 &= \frac{\ell^2 \left[-m \rho^4+\rho^2 z^2 (z-2 m)+2 m z^4\right]}{8\pi \rho^6 z^3}+\frac{Q^2 \left(\rho^2-\ell^2\right)}{8\pi \rho^6} \ , \\
    p_3 &= \frac{\ell^2 \left[2 a^2 z^3+m \left(-\rho^4-2 \rho^2 z^2+2 z^4\right)-\rho^2 z^3+2 z^5\right]}{8\pi \rho^6 z^3}+\frac{Q^2 \left(\rho^2-\ell^2\right)}{8\pi \rho^6} \ .
\end{align}
\end{subequations}
The expressions above prove that bbKN spacetime is Hawking--Ellis type I~\cite{HCOS:Moruno:2021, TLSSOS:Ellis:1973, ECOTHT:Moruno:2018, GRCGSCATHC:Moruno:2017}.

Note that
\begin{equation}
    \varepsilon + p_1 = - \frac{\ell^2 \Delta}{{8\pi}\rho^6} \ .
\end{equation}
This is the same result one gets in the bbK spacetime~\cite{ANFORBHM:Franzin:2021}, modulo the redefinition of $\Delta$. Thus, in particular, the NEC is violated. Note that on the horizon $\Delta=0 $ so $\left.(\varepsilon+p_1)\right|_H =0$. 
This on-horizon simplification is a useful consistency check~\cite{DBHTAHS:Visser:1992, HCOS:Moruno:2021, DBHSGANHS:Martin:2004, DBHSASNH:Martin:2004}.

It is now prudent to try to characterise the spacetime by means of some variant of curved-spacetime Maxwell-like electromagnetism, that is, by assuming that some variant of Maxwell's equations hold. By doing so, as in the nonrotating case, one finds that the matter content is made up of two different components: one electrically neutral and one charged, with the charged component further subdividing into Maxwell-like and charged dust contributions. Isolating the $Q$-dependent contribution to the total stress-energy, one finds
\begin{equation}
\label{E:xxx}
  [T_Q]^{\hat\mu}{}_{\hat\nu}  = \frac{1}{8\pi} \frac{Q^2 \left(\rho^2-\ell^2\right)}{\rho^6}
  \left[ \diag\left(-1,-1,1,1\right) +\frac{2\ell^2}{\rho^2{-\ell^2}} \diag\left(1,0,0,0\right)
  \right] \ .
\end{equation}
This is structurally the same as what was observed for bbRN spacetime in Eq.~(\ref{Tem}), now with the substitutions
\begin{equation}
    \frac{Q^2 r^2} {(r^2+\ell^2)^3}
    \longleftrightarrow
    \frac{Q^2 \left(\rho^2-\ell^2\right)}{\rho^6}
    \qquad \hbox{and} \qquad
    \frac{2\ell^2}{r^2}
    \longleftrightarrow
    \frac{2\ell^2}{\rho^2{-\ell^2}} \ .
\end{equation}
The first term in Eq.~(\ref{E:xxx}) is structurally of the form of the Maxwell stress-energy tensor, and the second term is structurally of the form of charged dust. At first sight this seems to suggest that a similar treatment as the one presented for bbRN spacetime should lead to a consistent picture. 
However, as shall be seen below, this case is quite a bit trickier than the previous one.
%
\subsubsection{Electromagnetic potential and field-strength tensor}

The first step in carrying on the same interpretation for the stress-energy tensor as that applied to bbRN spacetime, is to introduce the electromagnetic potential. Of course, also in this case there is no obvious way to derive it, since one is not \emph{a priori} specifying the equations of motion for the electromagnetic sector. Therefore, one can choose to modify the Kerr--Newman potential in a minimal way, as was performed for the bbRN case, \emph{i.e.}\ by performing the usual substitution $r\to \sqrt{r^2+\ell^2}$. Thus, the proposal in Boyer--Lindquist coordinates reads
\begin{equation}
    A_{\mu} = -\frac{Q \sqrt{r^2+\ell^2}}{\rho^2} \left(1,0,0,-a \sin^2\theta \right) \ .
    \label{eq:empot}
\end{equation}
In the orthonormal basis one has
\begin{equation}
    A_{\hat \mu} = e_{\hat \mu}{}^\nu \; A_\nu = 
    -\frac{Q \sqrt{r^2+\ell^2}} {\sqrt{\rho^2|\Delta|}} \left( 1,0,0,0 \right) \ .
    \label{eq:empot2}
\end{equation}
This is a minimal modification in the sense that, when one puts $a\to0$, the corresponding electrostatic potential is that of bbRN spacetime from Eq.~(\ref{eq:emstatic}), and when $\ell\to 0$ one regains the usual result for standard Kerr--Newman. The potential Eq.~(\ref{eq:empot}) is also compatible with the Newman--Janis procedure as outlined in reference~\cite{JAGRANCBH:Janis:2017}, and as can be applied to the bbRN geometry.

One can now compute the electromagnetic field-strength tensor $F_{\mu \nu}$. In the orthonormal basis, its only nonzero components are
\begin{align}
   F_{\hat t \hat r} = -F_{\hat r \hat t} &=  - \frac{Q}{\rho^4} \sqrt{\frac{r^2 }{r^2+\ell^2}} (r^2 +\ell^2 - a^2 \cos^2\theta) \ , \\
   F_{\hat \theta \hat \phi} = -F_{\hat \phi \hat \theta} &=  \frac{2aQ\cos\theta \sqrt{r^2+\ell^2}}{\rho^4} \ .
\end{align}
The homogeneous Maxwell equation is trivially satisfied: $F_{[\mu\nu,\sigma]}=0$. For the inhomogeneous Maxwell equation, one has
\begin{align}
    \nabla^{\hat{\mu}} F_{\hat{\mu} \hat{\nu}} &= J_{\hat{\nu}} = \frac{Q\ell^2}{\rho^7 z} \left(-\frac{\Delta \left(\rho^4+2 \rho^2 z^2-4 z^4\right)}{ z^2 \sqrt{|\Delta|}}, 0, 0, {2 a \sin\theta \left(\rho^2-2 z^2\right)}{}\right) \ .\label{eq:maxinhomo}
\end{align}
One can interpret the right-hand side of Eq.~(\ref{eq:maxinhomo}) as an effective electromagnetic source. Note that in terms of the (orthonormal) components of the electric and magnetic fields one has $E_{\hat r} = F_{\hat t \hat r}$ and $B_{\hat r} = F_{\hat\theta \hat\phi}$. It is then easy to check that this implies that the Maxwell stress-energy tensor Eq.~(\ref{TMaxwell}) is diagonal in this orthonormal basis, and that
\begin{align}
    \left[T_\text{Maxwell} \right]^{\hat\mu}{}_{{\hat\nu}} =  \frac{E_{\hat r}^2 + B_{\hat r}^2}{8\pi}\, \diag\left(-1,-1,1,1\right) \ , \label{TMaxwell_bbKN}
\end{align}
independent of the specific values of $E_{\hat r}$ and $B_{\hat r}$. It is also useful to check that
\begin{equation}
    E_{\hat r}^2 +  B_{\hat r}^2
    = \frac{Q^2}{\rho^4} \frac{r^2}{r^2+\ell^2} + \frac{4 Q^2 \ell^2 a^2 \cos^2\theta}{\rho^8} \ .
\end{equation}

\subsubsection{Interpreting the bbKN stress-energy}

All of the above treatment is a relatively straightforward generalisation of the bbRN case and also provides the correct limits for $\ell\to 0$ and/or $a\to 0$ (recall $\ell\to0$ gives standard KN, while $a\to0$ gives bbRN, and applying both limits gives standard RN spacetime). However, when one attempts to interpret Eq.~(\ref{E:xxx}) as the sum of the Maxwell stress-energy tensor Eq.~(\ref{TMaxwell_bbKN}) together with a charged dust, an inconsistency appears in the form of extra terms. Assuming some generalisation of the energy density of the charged dust and working out the required electromagnetic potential also does not lead to satisfactory results.

In what follows, two alternative interpretations of the stress-energy tensor are presented, one based on a generalisation of the Maxwell dynamics to a nonlinear one, the other consisting of a generalisation of the charged dust fluid to one with anisotropic pressure.

\subsubsection{Nonlinear electrodynamics}

An alternative to identifying a Maxwell stress-energy tensor in Eq.~(\ref{E:xxx}) consists of generalising the decomposition of the charged part of the stress energy tensor adopted in the bbRN case to
\begin{equation}\label{TQorth_exp}
    \left[T_Q\right]^{\hat{\mu}}{}_{\hat{\nu}} 
    = \A  \left[T_\text{Maxwell}\right]^{\hat{\mu}}{}_{\hat{\nu}}
    + \Xi\; V^{\hat{\mu}}V_{\hat{\nu}} \ .
\end{equation}
The multiplicative factor $\A$ will soon be seen to be position-dependent, and to depend on the spin parameter $a$ and regularisation parameter $\ell$, but to be independent of the total charge $Q$. This sort of behaviour is strongly reminiscent of nonlinear electrodynamics (NLED), where quite generically one finds $ \left[T_\text{NLED}\right]^{\hat{\mu}}{}_{\hat{\nu}}\propto  \left[T_\text{Maxwell}\right]^{\hat{\mu}}{}_{\hat{\nu}}$. For various proposals regarding the use of NLED in regular black hole contexts, see~\cite{TBMAANMM:Ayon-Beato:2000, FPRBHS:Ayon-Beato:2005, GSBNE:Arellano:2007, RBHWANES:Balart:2014, BRBHWAES:Rodrigues:2018, RMBHAMFNE:Bronnikov:2001, MBUAWWAPS:Bolokhov:2012, NERBHAW:Bronnikov:2018}. The contribution $\Xi\, V^{\hat{\mu}}V_{\hat{\nu}}$ is again that appropriate to charged dust. The four-velocity $V^{\hat{\mu}}$ is now the (nongeodesic) unit vector parallel to the timelike leg of the tetrad.

If one now compares Eq.~(\ref{TMaxwell_bbKN}) with $ \left[T_Q\right]^{\hat\mu}{}_{{\hat\nu}}$ as defined in Eq.~(\ref{TQorth_exp}), one identifies
\begin{equation}
\A = \frac{Q^2(\rho^2-\ell^2)/\rho^6}{E_{\hat r}^2+B_{\hat r}^2} 
= \frac{\rho^2(\rho^2-\ell^2)(r^2+\ell^2)}{\rho^4 r^2 + 4 a^2\ell^2(r^2+\ell^2) \cos^2\theta} \ .
\end{equation}
Note that at small $\ell$
\begin{equation}
 \A = 1 - \frac{a^2\cos^2\theta \left(3 r^2-a^2\cos^2\theta\right)}{r^2(r^2+a^2\cos^2\theta)^2}\; \ell^2 +\O(\ell^4) \ . 
\end{equation}
So in the limit as $\ell\rightarrow 0$, $\mathcal{A}\rightarrow 1$, restoring standard Maxwell electromagnetism as would be expected for ordinary KN spacetime. Also, observe the large distance limit
\begin{equation}
\A = 1 -\frac{3\ell^2 a^2 \cos^2\theta}{r^4} + \O(r^{-6}) \ .
\end{equation}
That is, at sufficiently large distances, $ \left[T_Q\right]^{\hat{\mu}}{}_{\hat{\nu}}$ can safely be approximated as a Maxwell-like contribution plus a charged dust, while at small $r$ one has
\begin{equation}
\A = \frac{\ell^2+a^2\cos\theta^2}{4\ell^2}+ \O(r^2) \ .
\end{equation}
This indicates a simple rescaling of the Maxwell stress-energy, very similar to what happens in NLED, deep in the core of the black-bounce.

Indeed, it is possible to further characterise the departure from Maxwell-like behaviour by decomposing $\mathcal{A} = 1-a^2\ell^2\mathcal{F}$, where
\begin{equation}
    \mathcal{F} = \frac{\cos^2\theta\left[4(r^2+\ell^2)-{\rho^2}\right]}{\rho^4r^2+ 4a^2\ell^2\cos^2\theta(r^2+\ell^2)} \ .
\end{equation}
The motivation for doing so is to make utterly transparent the correct limiting behaviour for $\mathcal{A}$ both for $a\rightarrow 0$, and for $\ell\rightarrow 0$.

\subsubsection{Anisotropic fluid}
\enlargethispage{20pt}

Alternatively, one can instead generalise the pressureless dust fluid introduced in the bbRN case in \S~\ref{CBB:stress-energy}, and impose
\begin{equation}
\left[T_Q \right]_{\hat \mu \hat \nu} -    \left[T_\text{Maxwell} \right]_{\hat \mu \hat \nu} = \diag\left(\varepsilon_{f}, -p_{f}, p_{f}, p_{f}\right) \ ,
\label{eq:difference}
\end{equation}
which can be satisfied if
\begin{equation}
    \varepsilon_{f} = \frac{Q^2 \ell^2}{z^2 \rho^8} \left(4z^4-7z^2\rho^2+\rho^4 \right) \ , \quad p_{f} = \frac{Q^2 \ell^2}{z^2 \rho^8} \left(4z^4-5z^2\rho^2+\rho^4 \right) \ .
\end{equation}
This implies that the right-hand side of Eq.~(\ref{eq:difference}) can be interpreted, formally, as the stress-energy of an anisotropic fluid. Specifically, it can be written as
\begin{equation}
    \varepsilon_{f} V_{\hat \mu} V_{\hat \nu} + \frac{p_{f}}{3} \left(g_{\hat \mu \hat \nu} + V_{\hat \mu} V_{\hat \nu} \right) + \pi_{\hat \mu \hat \nu} \ ,
\end{equation}
with $V^{\hat \mu} = \left(1,0,0,0\right)$ (\emph{i.e.} $(e_{\hat t})^{\hat \mu}$) being the velocity of the fluid, and
\begin{equation}
    \pi_{\hat \mu \hat \nu} = \frac{2p_{f}}{3} \diag\left(0,-2,1,1\right)
\end{equation}
being the (traceless) anisotropic shear~\cite{TLSSOS:Ellis:1973}. Note that
\begin{equation}
    p_{f} \propto \left(4z^4-5z^2\rho^2+\rho^4 \right) 
    = \left(4z^2-\rho^2 \right) \left(z^2-\rho^2 \right) \propto a^2 \cos^2\theta \ .
\end{equation}
So when $a\to 0$ this anisotropic fluid reduces to the usual charged dust.

%

\subsubsection{Discussion so far}

Adding an electromagnetic charge to the SV and bbK spacetimes leads to the bbRN and bbKN spacetimes respectively, which are minimalist, one-parameter deformations of the entire Kerr--Newman family. They have the desirable properties that they simultaneously: (i)~pass all weak-field observational tests, (ii)~are globally regular (no curvature singularities), and (iii)~neatly interpolate between charged regular black holes and charged traversable wormholes.

While bestowing an electromagnetic charge on SV and bbK spacetimes in this manner is unlikely to be of direct astrophysical importance (since in any  plausible astrophysical situation  $|Q|/m \ll 1$), it is of considerable theoretical importance, as it generates an entirely new class of relatively clean everywhere-regular black holes to work with. Indeed, such geometries present interesting theoretical features, such as the existence of a nontrivial Killing tensor without the presence of the full Killing tower (principal tensor, Killing--Yano tensor), or the fact that the charge-dependent component of the stress-energy for the bbKN spacetime has a rather subtle physical interpretation. In particular, it can either be interpreted as charged dust together with a nonlinear modification to standard Maxwell electromagnetism, \emph{or} as standard Maxwell electromagnetism together with an anisotropic fluid.

While there is no simple way to remove this ambiguity, one should note that it appears somewhat problematic to justify from a physical point of view the introduction of a NLED for the bbKN spacetime, given that it is not required of the bbRN case, nor the bbK spacetime discussed in reference~\cite{ANFORBHM:Franzin:2021} for consistency with GR. This seems to suggest that the anisotropic fluid interpretation might be more natural.
%

The black-bounce analogs to the entire Kerr--Newman family have been explored thoroughly, as well as the Vaidya version in dynamic spherical symmetry allowing for nontrivial phenomenology. Extensions to these ``standard'' black-bounce spacetimes abound; for instance in Chapter~\ref{C:SVNBB}, several spherically symmetric extensions to SV spacetime are analysed which each invoke qualitative changes to the horizon structure and other features of the geometry.


%% file: 02-Black-bounce/3-SVNBB.tex
\chapter{Novel black-bounce spacetimes}\label{C:SVNBB}
\def\HMS{{\scriptscriptstyle{\rm HMS}}}

It is interesting to note that many kinds of regular black hole exist which are different to the ``black-bounce'' spacetimes discussed thus far, some of which share several qualitative similarities. For instance, there are geometries possessing a minimum of the areal radius in the T-region, where the radial coordinate is timelike, or on a horizon; these have been discussed in references~\cite{SASOCSBH:Bronnikov:1998, RPBH:Bronnikov:2006, RBHABU:Bronnikov:2007, MBUAWWAPS:Bolokhov:2012}. More specifically, the spacetimes described in reference~\cite{RPBH:Bronnikov:2006} have a de Sitter late-time asymptotic, making them in principle viable candidate cosmologies. In reference~\cite{MBUAWWAPS:Bolokhov:2012}, regular solutions with a phantom scalar and an electromagnetic field were obtained (note the qualitative similarity here to the source terms for SV spacetime from Eq.~(\ref{SVaction})), leading to a diversity of global structures, including those with up to four horizons. In addition to this, the stability of the solutions obtained in reference~\cite{RPBH:Bronnikov:2006} was analysed in reference~\cite{IOWARBHSBAPSF:Bronnikov:2012}, where it was shown that all of the configurations under study were unstable under spherically symmetric perturbations, except for a special class of black universes where the event horizon coincides with the minimum of the area function. Given the rich tapestry of regular spacetimes, it is natural to intuit that there should exist simple extensions to standard SV spacetime in spherical symmetry worthy of further exploration.

Thus, in Chapter~\ref{C:SVNBB}, a number of additional novel ``black-bounce'' spacetimes are developed and discussed. These candidate geometries were first presented and analysed in reference~\cite{NBB:Lobo:2021}. These are specific, everywhere-regular geometries where the ``area radius'' is always nonzero, thereby leading to a centralised ``throat'' that is either timelike (corresponding to a traversable wormhole), spacelike (corresponding to a ``bounce'' into a future universe), or null (corresponding to a ``one-way wormhole''). It is informative to embed the ``black-bounce'' family into a more general framework, by first performing a generalised analysis for candidate spacetimes in static spherical symmetry (this is done \textit{via} use of the ``Buchdahl'' gauge~\cite{AROTRSFS:Finch:1998, BTFPFS:Boonserm:2008, BTIGR:Boonserm:2012, OTNOTKS:Semiz:2020}), and to then consider a number of specific examples. The examples are constructed using a mass function similar to that of Fan--Wang~\cite{CORBHIGR:Fan:2016}, and fall into several particular cases, such as the original SV model, a Bardeen-type model, and other generalisations thereof. The regularity, (non)satisfaction of the classical energy conditions, and the causal structure of these models is discussed. The main results are several new geometries, more complex than the examples from Chapter~\ref{C:SVog}, with two or more horizons, and the possibility of an extremal case. Furthermore, a third general theorem is presented, regarding (non)satisfaction of the classical energy conditions in static spherical symmetry.

%
For this chapter \textit{only}, the metric signature $(+,-,-,-)$ is adopted.

\section{Generalising the black-bounce family}\label{NBB:general}
\vspace*{7pt}

\subsubsection{Metric and curvature}\label{NBBss:metric}

In standard $(t,r,\theta,\phi)$ curvature coordinates, the most general static spherically symmetric metric can always locally be cast into the form:
\begin{eqnarray}
\d s^2 = f(r)\,\d t^2 - {\d r^2 \over f(r)} -\Sigma^2(r)\,\d\Omega^{2}_{2} \ .\label{ele}
\end{eqnarray}
Here $f(r)$ and $\Sigma(r)$ are at this stage two freely specifiable functions. Horizons (if present) are located at the roots of $f(r)$, and the metric determinant is given by $g=-\Sigma^4(r) \sin^2\theta$. The area of a sphere at radial coordinate $r$ is $A(r) = 4\pi \Sigma^2(r)$. The coordinate choices implicit in Eq.~(\ref{ele}) are often called ``Buchdahl coordinates''~\cite{AROTRSFS:Finch:1998, BTFPFS:Boonserm:2008, BTIGR:Boonserm:2012, OTNOTKS:Semiz:2020}, or ``Buchdahl gauge''.

From this line element, one may easily calculate the nonzero components of the Riemann tensor:
\begin{eqnarray}
&& R^{tr}{}_{tr} = \frac{1}{2}f'' \ , \quad
R^{t\theta}{}_{t\theta} = R^{t\phi}{}_{t\phi} = \frac{f'\Sigma'}{2\Sigma} \ , \nonumber \\[3pt]
&& R^{r\theta}{}_{r\theta} = R^{r\phi}{}_{r\phi} = \frac{f'\Sigma'+2f\Sigma''}{2\Sigma} \ , \quad
R^{\theta\phi}{}_{\theta\phi} = \frac{f\Sigma'^2-1}{\Sigma^2} \ .\label{Rort}
\end{eqnarray}
To guarantee that the candidate spacetime is everywhere-regular, one must demand that:
\begin{itemize}
\itemsep-2pt
\item $\Sigma(r)$ is globally nonzero;
\item $\Sigma'(r)$ and $\Sigma''(r)$ are both globally finite;
\item All of $f(r)$, $f'(r)$, and $f''(r)$ are globally finite.
\end{itemize}
With a look towards appealing to Theorem~\ref{Theorem:Kretsch}, one may also calculate the Kretschmann scalar, $K=R_{\alpha\beta\mu\nu}R^{\alpha\beta\mu\nu}$, in terms of the semi-positive sum of squares~\cite{BHCAED:Bronnikov:2013} of the Riemann components from Eq.~(\ref{Rort}):
\clearpage
\begin{eqnarray}
K &=& 4\left(R^{tr}{}_{tr}\right)^2 + 4\left(R^{t\theta}{}_{t\theta}\right)^2 + 4\left(R^{t\phi}{}_{t\phi}\right)^2 \nonumber \\ 
&& \qquad +\,4\left(R^{r\theta}{}_{r\theta}\right)^2+4\left(R^{r\phi}{}_{r\phi}\right)^2+4\left(R^{\theta\phi}{}_{\theta\phi}\right)^2 \ .
\label{Kret1}
\end{eqnarray}
More explicitly, in view of the spherical symmetry, one has
\begin{eqnarray}
K = 4\left(R^{tr}{}_{tr}\right)^2 + 8\left(R^{t\theta}{}_{t\theta}\right)^2 + 8\left(R^{r\theta}{}_{r\theta}\right)^2 + 4\left(R^{\theta\phi}{}_{\theta\phi}\right)^2 \ .
\label{Kret2}
\end{eqnarray}
See Theorem~\ref{Theorem:Kretsch} for a more comprehensive justification of the fact that the Kretschmann scalar is semi-positive for the strictly static region of any static spacetime. Specifically, in the current situation there is the explicit sum of squares
\begin{eqnarray}
K &=& \frac{(\Sigma^2 f'')^2+2(\Sigma f' \Sigma')^2 +2\Sigma^2(f' \Sigma'+2f \Sigma'')^2
+4(1-f\Sigma'^2)^2}
{\Sigma^4} \ . \nonumber \\
&&
\label{Kret3}
\end{eqnarray}
By Theorem~\ref{Theorem:Kretsch}, verifying whether or not the Kretschmann scalar is globally finite will confirm the regularity of any static candidate spacetime.

Similarly one can consider the Weyl scalar $C_{\mu\nu\alpha\beta} C^{\mu\nu\alpha\beta}$, for which a minor variant of the argument in~\cite{BHCAED:Bronnikov:2013} yields:
\begin{eqnarray}
C_{\mu\nu\alpha\beta}C^{\mu\nu\alpha\beta} &=& 4\left(C^{tr}{}_{tr}\right)^2 + 4\left(C^{t\theta}{}_{t\theta}\right)^2 + 4\left(C^{t\phi}{}_{t\phi}\right)^2 \nonumber \\
&& \qquad + \,4\left(C^{r\theta}{}_{r\theta}\right)^2 + 4\left(C^{r\phi}{}_{r\phi}\right)^2 + 4\left(C^{\theta\phi}{}_{\theta\phi}\right)^2 \ .
\label{Csq1}
\end{eqnarray}
In view of spherical symmetry this reduces to
\begin{eqnarray}
C_{\mu\nu\alpha\beta}C^{\mu\nu\alpha\beta} = 4\left(C^{tr}{}_{tr}\right)^2 + 8\left(C^{t\theta}{}_{t\theta}\right)^2 + 8\left(C^{r\theta}{}_{r\theta}\right)^2 + 4\left(C^{\theta\phi}{}_{\theta\phi}\right)^2 \ .
\label{Csq2}
\end{eqnarray}
Indeed, explicit computation yields the perfect square
\begin{eqnarray}
C_{\mu\nu\alpha\beta}C^{\mu\nu\alpha\beta} = {1\over3} \left[f'' - {2 f'\Sigma'\over\Sigma} 
+ {2f (\Sigma'^2-\Sigma\Sigma'')\over\Sigma^2} - {2\over\Sigma^2} \right]^2 \ .
\label{Csq3}
\end{eqnarray}
Verifying whether or not the Weyl scalar is globally finite is a partial check on the regularity of any static spacetime.

\subsubsection{Stress-energy tensor}\label{NBBss:stress-energy}

In geometrodynamic units, the Einstein field equations are given by
\begin{eqnarray}
R_{\mu\nu}-\frac{1}{2}g_{\mu\nu}{R} = G_{\mu\nu} = 8\pi\,T_{\mu\nu} \ .\label{eqEinstein}
\end{eqnarray}
If one considers the matter sector to be an anisotropic fluid, then in regions where the $t$ coordinate is timelike (\textit{i.e.} $f(r)>0$), for instance in the domain of outer communication, the mixed components of the stress-energy tensor are given by
\begin{eqnarray}
T^{\mu}{}_{\nu} = {\diag}\left(\rho,-p_r,-p_t,-p_t\right) \ ,\label{EMT}
\end{eqnarray}
where $\rho$, $p_r$, and $p_t$ are the energy density and the two principal pressures, respectively. Taking into account the line element Eq.~(\ref{ele}), the Einstein equations Eq.~(\ref{eqEinstein}) provide the following stress-energy profile:
\begin{eqnarray}
\rho &=& -\frac{\Sigma \left(f' \Sigma'+2 f \Sigma''\right)+f \Sigma'^2-1}{8\pi \Sigma^2} \ ,\label{density+} \\[2pt]
p_r &=& \frac{\Sigma f'\Sigma'+f \Sigma'^2-1}{8\pi \Sigma^2} \ ,\label{p1+} \\[2pt]
p_t &=& \frac{\Sigma f''+2 f' \Sigma'+2 f \Sigma''}{16\pi \Sigma} \ .\label{p2+}
\end{eqnarray}
However, in regions where the $t$ coordinate is spacelike, \textit{i.e.} $f(r)<0$, one instead has
\begin{eqnarray}
T^{\mu}{}_{\nu} = {\diag}\left(-p_r,\rho,-p_t,-p_t\right) \ ,\label{EMT2}
\end{eqnarray}
where $p_r$ is the principal pressure in the now spacelike $t$ direction. Then in the sub-horizon regions where $t$ is spacelike there is the following stress-energy profile:
\begin{eqnarray}
\rho &=& -\frac{\Sigma f'\Sigma'+f \Sigma'^2-1}{8\pi \Sigma^2} \ ,\label{density-} \\[2pt]
p_r &=& \frac{\Sigma \left(f' \Sigma'+2 f \Sigma''\right)+f \Sigma'^2-1}{8\pi \Sigma^2} \ ,\label{p1-} \\[2pt]
p_t &=& \frac{\Sigma f''+2 f' \Sigma'+2 f \Sigma''}{16\pi \Sigma} \ .\label{p2-}
\end{eqnarray}
Furthermore, at any horizons that may be present where $f(r)=0$, one has
\begin{eqnarray}
\rho = -p_r =
-\frac{\Sigma f'\Sigma'-1}{8\pi \Sigma^2} \ ,\label{density0} 
\qquad
p_t = \frac{\Sigma f''+2 f' \Sigma'}{16\pi \Sigma} \ .\label{p20}
\end{eqnarray}
The on-horizon equality of $\rho = -p_r$ is well known~\cite{DBHSGANHS:Martin:2004, DBHTAHS:Visser:1992}, and is physically required to ensure that $\rho$ is continuous as one crosses the horizon.

Finally, the trace of the stress energy is given by
\begin{equation}
T^{\mu}{}_{\mu} = \rho - p_r - 2p_t = 
-{\Sigma^2 f'' +4 \Sigma(\Sigma'' f + \Sigma' f')  + 2 (\Sigma')^2 f - 2
\over 8\pi \Sigma^2} \ ,
\end{equation}
regardless of whether one is above or below any horizon that may be present.
\clearpage

To guarantee that the stress-energy is everywhere-regular, one must therefore demand that:
\begin{itemize}
\itemsep-2pt
\item $\Sigma(r)$ is globally nonzero;
\item $\Sigma'(r)$ and $\Sigma''(r)$ are both globally finite;
\item All of $f(r)$, $f'(r)$, and $f''(r)$ are globally finite.
\end{itemize}
Unsurprisingly, this is precisely the same set of conditions as was required for the Riemann tensor to be everywhere-regular.

\subsubsection{Hernandez--Misner--Sharp quasi-local mass}\label{NBBss:quasilocal-mass}

%
The Hernandez--Misner--Sharp quasi-local mass~\cite{OTAACIRSH:Hernandez:1966, REFASSGC:Misner:1964, GMQMIEG:Maeda:2008, SSTH:Nielsen:2008, KT:Abreu:2010, WPCF:Faraoni:2020} is most easily defined by inspecting the Riemann tensor component
\begin{eqnarray}
R^{\theta\phi}{}_{\theta\phi} = -\frac{2M_\HMS(r)}{\Sigma(r)^3}
= {f(r) \Sigma'(r)^2-1\over\Sigma(r)^2} \ .\label{Rmass}
\end{eqnarray}
Then
\begin{eqnarray}
M_\HMS(r)=\frac{1}{2}\Sigma(r)\left\{1-f(r)\Sigma'(r)^2\right\}\label{mass} \ .
\end{eqnarray}
Using this, $f(r)$ can be written as
\begin{eqnarray}
f(r) = 1 - \frac{2M_\HMS(r)-\Sigma(r)\{1-\Sigma'(r)^2\}}{\Sigma(r)\Sigma'(r)^2}\label{f} \ .
\end{eqnarray}
It will be useful to redefine $f(r)$ as
\begin{eqnarray}
f(r)=1-\frac{2M(r)}{\Sigma(r)}\label{fm} \ .
\end{eqnarray}
But now $M(r)$ is simply a function appearing in the metric, it is no longer the quasi-local mass obtained by integrating the energy density over the volume contained by a surface of radius $r$. Explicitly
\begin{eqnarray}
M(r) &=& \frac{M_\HMS(r)-{1\over2} \Sigma(r)\{1-\Sigma'(r)^2\}}{\Sigma'(r)^2} \ ; \nonumber \\[3pt]
\Longrightarrow \ M_\HMS(r) &=&  M(r) \Sigma'(r)^2+ {1\over2} \Sigma(r)\{1-\Sigma'(r)^2\} \ .
\end{eqnarray}
At any horizon that may or may not be present, where $f(r_H)=0$, in view of Eq's.~(\ref{mass}) and~(\ref{fm}) one has
\begin{equation}
M_\HMS(r_H) = M(r_H) = {\Sigma(r_H)\over2} \ .
\end{equation}
Finally, at any local extremum of $\Sigma(r)$ that may or may not be present, (that is, $\Sigma'(r_{ext})=0$, corresponding to a ``throat'', a ``bounce'', or an ``anti-throat''), in view of Eq's.~(\ref{mass}) and~(\ref{fm}) one has
\begin{equation}
M_\HMS(r_{ext}) = {\Sigma(r_{ext})\over2} \ ; \qquad  M(r_{ext}) = {\Sigma(r_{ext})\; \{1-f(r_{ext})\}\over2} \ .
\end{equation}
Either by differentiating $M_\HMS(r)$, or by substituting Eq.~(\ref{f}) into the Einstein equations, one may obtain the Hernandez--Misner--Sharp quasi-local mass in terms of the stress-energy component $T^t{}_t(r)$:
\begin{eqnarray}
M_\HMS'(r) &=& 4\pi \; T^t{}_t(r) \; \Sigma(r)^2 \; \Sigma'(r) \ ; \nonumber \\[3pt]
M_\HMS(r) &=& M_* + 4\pi\int_{r_*}^r T^t{}_t(\bar r) \; \Sigma(\bar r)^2 \; \Sigma'(\bar r) \; \d\bar{r} \ . 
\label{HMSmass}
\end{eqnarray}
Note that while the Hernandez--Misner--Sharp quasi-local mass can be defined for arbitrary values of $r$, it really only has its normal physical interpretation in the region where the $t$ coordinate is timelike, where $T^t{}_t \to \rho$. In this region, one has
\begin{eqnarray}
M_\HMS'(r) &=& 4\pi \rho(r) \; \Sigma(r)^2 \; \Sigma'(r) \ ; \nonumber \\[3pt]
M_\HMS(r) &=& M_\HMS(r_H) +4\pi\int_{r_H}^r\rho(\bar r)\;\Sigma(\bar r)^2\;\Sigma'(\bar r) \; \d\bar r \ . 
\label{HMSmass2}
\end{eqnarray}
That is, the ``mass function'' $M(r)$ defined in Eq.~(\ref{fm}) is not the energy contained within a surface of radius $r$;  one now sees that it is the Hernan\-dez--Misner--Sharp quasi-local mass $M_\HMS(r)$ that plays this role.

Note that in the limit $\Sigma(r)\rightarrow r$ one recovers the usual results
\begin{eqnarray}
M(r)= M_\HMS(r) \ , \quad f(r)=1-\frac{2M_\HMS(r)}{r} \ , \quad
R^{\theta\phi}{}_{\theta\phi}=-\frac{2M_\HMS(r)}{r^3} \ . \nonumber \\
&&
\end{eqnarray}
\enlargethispage{20pt}
\vspace*{-10pt}

\subsubsection{Energy conditions}\label{NBBss:ECs}

%
The standard energy conditions of classical GR are (mostly) linear in the stress-energy tensor, and have clear physical interpretations in terms of geodesic focusing, but suffer the drawback that they are often violated by semi-classical quantum effects. In contrast, it is possible to develop nonstandard energy conditions that are intrinsically nonlinear in the stress-energy tensor, and which exhibit better-controlled behaviour when semi-classical quantum effects are introduced, at the cost of a less direct applicability to geodesic focusing~\cite{SECFQVS:Moruno:2013, CAQFECFQVS:Moruno:2013, CASEC:Moruno:2017, SANEC:Moruno:2018}. The energy conditions have also found significant usage in cosmological settings~\cite{ECITEOGF:Visser:1997, GREC:Visser:1997, ECAGF:Visser:unknown, ECATCI:Barcelo:2000, TFTEC:Barcelo:2002, CMAEC:Cattoen:2007, CECHBDBTADB:Cattoen:2008}, in ``gravastars''~\cite{SG:Visser:2004, GMHAP:Cattoen:2005, GTG:Garcia:2012, NSAOTG:Garcia:2018, LSAOGTS:Garcia:2015, BSNH:Barcelo:2009}, and in various wormhole-related constructions~\cite{TWSSE:Visser:1989, TWFSMSS:Visser:1989, TSW:Poisson:1995, TWWASECV:Dadhich:2003, QECVITW:Dadhich:2004, R=0:Dadhich:2002}. See Appendix~\ref{B:GRbasics} for further detail.

The standard point-wise energy conditions~\cite{LW:Visser:1995} for the stress-energy tensor given by Eq.~(\ref{EMT}) are given by the following inequalities:
\begin{eqnarray}
&& NEC_{1,2} = WEC_{1,2} = SEC_{1,2} \Longleftrightarrow \rho + p_{r,t} \geq 0 \ ,\label{Econd1}\\[1pt]
&& SEC_3 \Longleftrightarrow \rho + p_r + 2p_t \geq 0 \ ,\label{Econd2}\\[1pt]
&& DEC_{1,2} \Longleftrightarrow 
(\rho + p_{r,t} \geq 0) \; \hbox{\textit{and}} \; (\rho - p_{r,t} \geq 0) \ ,\label{Econd3} \\[1pt]
&& DEC_3 = WEC_3 \Longleftrightarrow \rho \geq 0 \ .\label{Econd4}
\end{eqnarray}
This formulation has been carefully phrased such that it is true regardless of whether the $t$ coordinate is timelike or spacelike. Note that $DEC_{1,2} \Longleftrightarrow \left[(NEC_{1,2}) \hbox{ \textit{and} } (\rho - p_{r,t} \geq 0)\right]$. Enforcing satisfaction of the NEC, for all practical purposes one may as well subsume part of the DEC into the NEC and simply replace this with $DEC_{1,2} \Longrightarrow \rho - p_{r,t} \geq 0$.

Inserting the results from Eq's.~(\ref{density+})--~(\ref{p2+}), in regions where the $t$ coordinate is timelike, one has
\begin{eqnarray}
&& NEC_{1} = WEC_1 = SEC_1 \Longleftrightarrow
-\frac{2 f \Sigma''}{8\pi \Sigma} \geq 0 \ ,\label{cond1+}\\[2pt]
&& NEC_2 = WEC_2 = SEC_2 \Longleftrightarrow
\frac{\Sigma^2 f''-2 f \left(\Sigma \Sigma''+(\Sigma')^2\right)+2}{16\pi \Sigma^2} \geq 0 \ , \nonumber \\
&& \label{cond2+}\\
&& SEC_3 \Longleftrightarrow
\frac{\Sigma f''+2 f' \Sigma'}{8\pi \Sigma} \geq 0 \ ,\label{cond3+}\\[2pt]
&& DEC_{1} \Longrightarrow
{2\left(1 - f' \Sigma\Sigma' - f (\Sigma')^2 -f \Sigma\Sigma'' \right)
\over8\pi\Sigma^2} \geq 0 \ ,\label{cond4+}\\[2pt]
&& DEC_2 \Longrightarrow -\frac{\Sigma^2 f''+\Sigma \left(4 f' \Sigma'+6 f \Sigma''\right)+2 f \Sigma'^2-2}{16\pi \Sigma^2} \geq 0 \ ,\label{cond5+}\\[2pt]
&& DEC_3 = WEC_3 \Longleftrightarrow
-\frac{\Sigma \left(f' \Sigma'+2 f \Sigma''\right)+f (\Sigma')^2-1}{8\pi \Sigma^2} \geq 0 \ .\label{cond6+}
\end{eqnarray}
\enlargethispage{20pt}
Inserting the results from Eq's.~(\ref{density-})--~(\ref{p2-}), in regions where the $t$ coordinate is spacelike, one has
\begin{eqnarray}
&& NEC_{1} = WEC_1 = SEC_1 \Longleftrightarrow
+\frac{2 f \Sigma''}{8\pi \Sigma} \geq 0 \ ,\label{cond1-}\\[2pt]
&& NEC_2 = WEC_2 = SEC_2 \Longleftrightarrow
{ \Sigma^2 f'' - 2 (\Sigma')^2 f +2 \Sigma \Sigma'' f +2 \over16\pi\Sigma^2} \geq 0 \ , \nonumber \\
&& \label{cond2-}\\
&& SEC_3 \Longleftrightarrow
{ \Sigma f''+ 2 \Sigma' f' + 4 \Sigma'' f \over8\pi\Sigma} \geq 0 \ ,\label{cond3-}\\[2pt]
&& DEC_{1} \Longrightarrow
{2\left(1 - f' \Sigma\Sigma' - f (\Sigma')^2 -f \Sigma\Sigma'' \right)
\over8\pi\Sigma^2} \geq 0 \ ,\label{cond4-}\\[2pt]
&& DEC_2 \Longrightarrow
{-\Sigma^2 f'' - 2 \Sigma \Sigma'' f  -4 \Sigma\Sigma' f' -2 (\Sigma')^2 f +2
\over16\pi\Sigma^2} \geq 0 \ ,\label{cond5-}\\[2pt]
&& DEC_3 = WEC_3 \Longleftrightarrow
-{\Sigma \Sigma' f' + (\Sigma')^2 f - 1 
\over8\pi\Sigma^2} \geq 0 \ .
\label{cond6-}
\end{eqnarray}
That is, \emph{independent of whether one is above or below the horizon}, one has
\begin{equation}
NEC_{1} = WEC_1 = SEC_1 \Longleftrightarrow
-\frac{2 \vert f(r)\vert \;\Sigma''(r)}{8\pi \Sigma(r)} \geq 0 \ .\label{cond1b}
\end{equation}
So as long as one is not exactly \textit{on} any event horizon that might be present, one must have $f(r)\neq0$. Also $\Sigma(r)>0$ everywhere. So it is easily verified that, whenever $\Sigma''(r)>0$, $NEC_1, \ WEC_1$, and $SEC_1$ all exhibit negative values everywhere that is not precisely on the event horizon. Thus the NEC, and hence \emph{all} of the  standard point-wise energy conditions, are violated for black-bounce models whenever $\Sigma''(r)>0$.
\begin{theorem}\label{Theorem:EC}
For any static anisotropic fluid sphere with the line element as given by Eq.~(\ref{ele}), all of the standard point-wise energy conditions are violated whenever $f(r)\neq 0$, $\Sigma(r)>0$, and $\Sigma''(r)>0$.
\end{theorem}
%
%
Unfortunately, apart from $NEC_1$ and $DEC_1$, the other point-wise energy conditions do not transform nicely as one crosses any horizon that may be present.

The intention now is to look for models with positive energy density $\rho$, at least (insofar as possible) satisfying $WEC_3$. In addition to this, to extend beyond standard SV spacetime, one also looks for geometries that have a richer causal structure than the original model~\cite{BTTW:Simpson:2019}.

To quantify the amount of exotic matter present in the regions where the NEC is violated, one may apply a volume integral quantifier~\cite{TWWASECV:Dadhich:2003, QECVITW:Dadhich:2004}. With respect to Buchdahl coordinates, $\Sigma(r)$ defines the appropriate formula for the surface area of the spherical hypersurfaces \emph{via} $A = 4\pi \Sigma(r)^2$. It follows that if the NEC is violated when $r\in\left(r_{1},r_{2}\right)$, then the amount of exotic matter is quantified by the definite integral
\begin{eqnarray}\label{volumequantifier}
    \int_{r_{1}}^{r_{2}}\left(\rho+p_{r}\right)4\pi\Sigma^2 \, \d\Sigma &=& \int_{r_{1}}^{r_{2}}-\frac{2\vert f\vert\Sigma^{''}}{8\pi\Sigma}4\pi\Sigma^2 \, \d\Sigma = -\int_{r_{1}}^{r_{2}}\vert f\vert\Sigma^{''}\Sigma \, \d\Sigma \nonumber \\
    && \nonumber \\
    &=& -\int_{r_{1}}^{r_{2}}\vert f\vert\Sigma^{''}\Sigma^{'}\Sigma \, \d r = -\frac{1}{2}\int_{r_{1}}^{r_{2}} \vert f\vert\Sigma\left[\left(\Sigma^{'}\right)^{2}\right]^{'} \, \d r \nonumber \\
    && \nonumber \\
    &=& -\frac{1}{2}\vert f\vert\Sigma\left(\Sigma^{'}\right)^{2}\Bigg\vert_{r_{1}}^{r_{2}}+\frac{1}{2}\int_{r_{1}}^{r_{2}}\left(\vert f\vert\Sigma\right)^{'}\left(\Sigma^{'}\right)^{2} \, \d r \ . \nonumber \\
    &&
\end{eqnarray}
Given a specific candidate spacetime, \emph{i.e.} explicit forms for $f(r)$ and $\Sigma(r)$, one may compute this integral and obtain the amount of required exotic matter.
\clearpage

\subsection{Regaining Simpson--Visser spacetime}
\label{NBB:SV-bb}

%
The standard SV spacetime of \S~\ref{SimpViss} can be viewed as a special case of Eq.~(\ref{ele}). Specifically, take
\begin{eqnarray}
\Sigma(r) = \sqrt{r^2+\ell^2} \ , \quad M(r) = m \ , \quad f(r) = 1-\frac{2m}{\sqrt{r^2+\ell^2}} \ .\label{Visser}
\end{eqnarray}
%
%
%
To appeal directly to Theorem~\ref{Theorem:Kretsch}, the Kretschmann scalar can be written as the explicit sum of squares:
\begin{eqnarray}
K &=& \frac{4}{\left(r^2+\ell^2\right)^5}\Bigg\lbrace
m^2(2r^2-\ell^2)^2 + 2m^2r^4 + \left(2mr^2+\ell^2\sqrt{r^2+\ell^2}\right)^2 \nonumber \\
&& \qquad \qquad \qquad \qquad \qquad \quad + 2\left(mr^2-2m\ell^2+\ell^2\sqrt{r^2+\ell^2}\right)^2\Bigg\rbrace \ . \nonumber \\
&&
\end{eqnarray}
Provided $\ell\neq 0$ this is manifestly finite for all values of $r$ and $m$, confirming the regularity of SV spacetime, as already stated.

The standard point-wise energy conditions for SV spacetime, in the region where $t$ is timelike, can be written as
\begin{eqnarray}
NEC_1 &\Longleftrightarrow& -\frac{2 \ell^2 \left(\sqrt{r^2+\ell^2}-2 m\right)}{8\pi \left(r^2+\ell^2\right)^{5/2}} 
\geq 0 \ ,\label{NEC1out} \\[2pt]
NEC_2 &\Longleftrightarrow& \frac{3 \ell^2 m}{8\pi \left(r^2+\ell^2\right)^{5/2}} \geq 0 \ ,\label{NEC2out} \\[2pt]
WEC_3 &\Longleftrightarrow& -\frac{\ell^2 \left(\sqrt{r^2+\ell^2}- 4 m\right)}{8\pi \left(r^2+\ell^2\right)^{5/2}} \geq 0 \ , \\[2pt]
SEC_3 &\Longleftrightarrow& \frac{2 \ell^2 m}{8\pi \left(r^2+\ell^2\right)^{5/2}} \geq 0 \ , \\[2pt]
DEC_1 &\Longrightarrow& \frac{4m\ell^2}{8\pi \left(r^2+\ell^2\right)^{5/2}} \geq 0 \ , \\[2pt]
DEC_2 &\Longrightarrow& -\frac{\ell^2 \left(2\sqrt{r^2+\ell^2} - 5m\right)}{8\pi    \left(r^2+\ell^2\right)^{5/2}} \geq 0 \ .
\end{eqnarray}
Beginning with the case where the spacetime is a regular black hole, \textit{i.e.} $\ell\in\left(0,2m\right)$, it is immediately clear that the $NEC_1=WEC_1=SEC_1$ conditions are violated outside any horizons. Furthermore, $WEC_3$ and $DEC_2$ are violated whenever $\vert r\vert\gg r_{H}$.

For the case where one has a wormhole with a null throat, \textit{i.e.} $\ell=2m$, the $NEC_1$ condition is violated for all values of $r$; the $WEC_3$ and $DEC_2$ conditions are violated for $\vert r\vert\gg \ell$.

Finally, when one has a two-way wormhole, \textit{i.e.} $\ell>2m$, the $NEC_1$ and $WEC_3$ conditions are violated for all values of $r$, while the $DEC_2$ condition is violated when $\vert r\vert$ is sufficiently large. Notably, the energy density is always negative for this last case.

Recall that spherically symmetric regular black holes in GR coupled to NLED always violate $SEC_3$~\cite{RBHAEC:Zaslavskii:2010}, however, it is mathematically possible to satisfy this condition generically for the black-bounce spacetimes. However, even if the $SEC_3$ condition is satisfied, the $SEC_1$ condition is certainly violated --- at best one has \emph{partial} satisfaction of \emph{some} of the energy conditions.

In contrast, in the region where $t$ is spacelike (the existence of this region requires the existence of horizons, \textit{i.e.} $\ell<2m$), the point-wise energy conditions for SV spacetime can be written as
\begin{eqnarray}
NEC_1 &\Longleftrightarrow& \frac{2 \ell^2 \left(\sqrt{r^2+\ell^2}-2 m\right)}{8\pi \left(r^2+\ell^2\right)^{5/2}} \geq 0 \ , \\[2pt] 
NEC_2 &\Longleftrightarrow& \frac{(2\sqrt{r^2+\ell^2} -m) \ell^2}{8\pi \left(r^2+\ell^2\right)^{5/2}} \geq 0 \ , \\[2pt]
WEC_3 &\Longleftrightarrow& \frac{\ell^2 }{8\pi \left(r^2+\ell^2\right)^{2}} \geq 0 \ , \\[2pt]
SEC_3 &\Longleftrightarrow& \frac{2 \ell^2(2\sqrt{r^2+\ell^2} -3m)}{8\pi \left(r^2+\ell^2\right)^{5/2}} \geq 0 \ , \\[2pt]
DEC_1 &\Longrightarrow& \frac{4m\ell^2}{8\pi \left(r^2+\ell^2\right)^{5/2}} \geq 0 \ , \\[2pt]
DEC_2 &\Longrightarrow& \frac{\ell^2 m }{8\pi    \left(r^2+\ell^2\right)^{5/2}} \geq 0 \ .
\end{eqnarray}
Whenever $\sqrt{r^2+\ell^2}<2m$, \textit{i.e.} below the horizon, the $NEC_1=WEC_1=SEC_1$ conditions are certainly violated. This implies that below the horizon \emph{all} of the usual point-wise energy conditions are violated. For instance, even though the $WEC_3$ condition is satisfied below the horizon, the $WEC_1$ condition is not --- once again, at best one has \emph{partial} satisfaction of \emph{some} of the energy conditions.

To obtain the amount of exotic matter required for SV spacetime, simply apply the volume integral from Eq.~(\ref{volumequantifier}). In the $\ell>2m$ case with no horizons, one has the two-way traversable wormhole geometry. For this case, simply integrate the expression for $NEC_{1}$ above horizons from Eq.~(\ref{NEC1out}), all the way from $0$ to $+\infty$:
\enlargethispage{10pt}
\vspace*{3pt}
\begin{equation}
    \bigintssss_{0}^{+\infty}\frac{2\ell^2\left(2m-\sqrt{r^2+\ell^2}\right)}{8\pi\left(r^2+\ell^2\right)^{\frac{5}{2}}}\, \d V = \bigintssss_{0}^{+\infty}\frac{\ell^2r(2m-\sqrt{r^2+\ell^2})}{(r^2+\ell^2)^2}\, \d r = m - \ell \ .
\end{equation}
\vfill
\clearpage
Given $\ell>2m$, it follows that the amount of exotic matter present must be strictly greater than $m$ in order to stabilise the wormhole throat.

In the $\ell\in\left(0,2m\right)$ case where horizons are present, one finds the following for the amount of exotic matter \emph{inside} the horizon:
\begin{eqnarray}
    \bigintssss_{0}^{r_{H}}\frac{2\ell^2(\sqrt{r^2+\ell^2}-2m)}{8\pi(r^2+\ell^2)^{\frac{5}{2}}}\, \d V &=& \bigintssss_{0}^{\sqrt{\left(2m\right)^2-\ell^2}} \frac{\ell^2r(\sqrt{r^2+\ell^2}-2m)}{(r^2+\ell^2)^2}\, \d r \nonumber \\[3pt]
    &=& -\frac{(\ell-2m)^2}{4m} \ ,
\end{eqnarray}
and for the amount of exotic matter \emph{outside} the horizon:
\begin{eqnarray}
    \bigintssss_{r_{H}}^{+\infty} \frac{2\ell^2(2m-\sqrt{r^2+\ell^2})}{8\pi(r^2+\ell^2)^{\frac{5}{2}}}\, \d V &=& \int_{\sqrt{\left(2m\right)^2-\ell^2}}^{+\infty}\frac{\ell^2r(2m-\sqrt{r^2+\ell^2})}{(r^2+\ell^2)^2}\, \d r \nonumber \\[3pt]
    &=& -\frac{\ell^2}{4m} \ .
\end{eqnarray}
In all cases the amount of exotic matter required is strictly finite.

It is straightforward to extract the Hernandez--Misner--Sharp quasi-local mass, by suitably specifying the functions given in Eq.~(\ref{mass}):
\begin{eqnarray}
M_\HMS(r) =
\frac{\ell}{2}+\frac{8\pi}{2}\int_0^r T^t{}_t(r)\; r\sqrt{r^2+\ell^2}\,\d r = \frac{mr^2}{r^2+\ell^2}+\frac{\ell^2}{2\sqrt{r^2+\ell^2}} \ .\label{massVisser}
\end{eqnarray}
This mass is always positive. In comparison with Eq.~(\ref{HMSmass}), $r_*=0$, and $M_*=\ell/2$. There are also the following limits: $\lim_{r\rightarrow 0}M_\HMS(r)=\ell/2$, and $\lim_{r\rightarrow +\infty}M_\HMS(r)=m$.

Having explored SV spacetime more thoroughly in the Buchdahl framework, the primary goal below is to explore new black-bounce models that generalise the standard geometry. It is hoped that certain tractable modifications to the free functions from Eq.~(\ref{ele}), which act as extensions to standard SV, might somewhat ameliorate the violation of the point-wise energy conditions. In the remainder of Chapter~\ref{C:SVNBB}, $\Sigma(r)$ is fixed as $\Sigma(r)=\sqrt{r^2+\ell^2}$.

%
\section{New black-bounce spacetimes}\label{NBB:new}

Firstly, consider the following rather general class of black-bounce models which generalise standard SV spacetime, in which the functions $\Sigma(r)$, $M(r)$ and $f(r)$ are given by
\begin{eqnarray}
\Sigma(r)=\sqrt{r^2+\ell^2} \ , \ \ M(r)=\frac{m\,\Sigma(r)\, r^{k}}{\left(r^{2n}+\ell^{2n}\right)^{\frac{k+1}{2n}}} \ , \ \ f(r)=1-\frac{2M(r)}{\Sigma(r)} \ .\label{newBBS}
\end{eqnarray}
Here $n,k\in\mathbb{Z}^{+}$. This new model is inspired by the Fan--Wang mass function~\cite{CORBHIGR:Fan:2016} for regular black holes. SV spacetime is recovered by fixing $n=1$ and $k=0$, and for any $n$ and $k$, $\ell\rightarrow 0$ returns the Schwarzschild solution. The usual regular black hole solutions (Bardeen, Hayward, Frolov) are unable to be recovered by any limit due to the $\ell^2$ term present in $\Sigma(r)$. However, this model framework can generate several new classes of black-bounce worth further examination.

\subsection{Model $n=2$ and $k=0$}\label{NBBss:n=2+k=0}

Fixing $n=2$ and $k=0$ in Eq.~(\ref{newBBS}) yields
\begin{eqnarray}
\Sigma(r) = \sqrt{r^2+\ell^2} \ , \qquad f(r) = 1-\frac{2m}{\sqrt[4]{r^4+\ell^4}} \ .\label{mod1}
\end{eqnarray}
To verify the regularity of the model, by Theorem~\ref{Theorem:Kretsch}, one analyses the finiteness of the Kretschmann scalar by substituting these quantities into Eq.~(\ref{Kret3}):
\begin{eqnarray}
K &=& \frac{8m^2r^8}{\left(r^2+\ell^2\right)^2\left(r^4+\ell^4\right)^{5/2}} + \frac{{4m^2r^4\left(3\ell^4-2r^4\right)^2}}{\left(r^4+\ell^4\right)^{9/2}} + \frac{4\left(\frac{2mr^2}{\sqrt[4]{\ell^4+r^4}}+\ell^2\right)^2}{\left(r^2+\ell^2\right)^4} \nonumber \\[3pt]
&& \qquad \qquad \qquad + \frac{8\left[\ell^2\left(r^4+\ell^4\right)^{5/4}+m\left(r^6-2\ell^6-\ell^2r^4\right)\right]^2}{\left(r^2+\ell^2\right)^{4}\left(r^4+\ell^4\right)^{5/2}} \ .
\end{eqnarray}
The Kretschmann scalar is manifestly finite for all $r$, so by Theorem~\ref{Theorem:Kretsch} the model is indeed regular.

From~\eqref{mod1} it is clear that $f(r)=0$ provides two symmetric real values $r_{H}=\pm\sqrt[4]{(2m)^4-\ell^4}$. Provided $0<\ell<2m$, there is a regular black hole with two horizons, one in the positive-$r$ universe and another in the negative-$r$ universe. At $r=0$, the geometry is regular and can be readily extended to $r<0$, passing through a ``bounce'' to the negative-$r$ universe; this corresponds to a one-way spacelike throat in precisely the same manner as for standard SV spacetime. As $r\rightarrow0$, one observes that $f(r)\rightarrow(\ell-2m)/\ell<0$, with the signature changing to $(-,+,-,-)$ inside the horizon. If instead $\ell=2m$, the (maximally extended) spacetime has only extremal horizons, so one has a one-way wormhole geometry with an extremal null throat. For $\ell>2m$, there are no horizons and there is a traversable wormhole with a two-way timelike throat. As such, this particular class of black-bounce possesses qualitatively the same causal structures for the same values of $\ell$ as for standard SV spacetime. Indeed, this is the general case for any $n\in\mathbb{Z}^{+}$ when $k=0$.
\clearpage

In the region where the $t$ coordinate is timelike, the point-wise energy conditions are given by
\begin{eqnarray}
NEC_1 &\Longleftrightarrow& -\frac{2\ell^2\left(1-\frac{2m}{\sqrt[4]{\ell^4+r^4}}\right)}{8\pi\left(r^2+\ell^2\right)^2} \geq 0 \ , \\[3pt]
NEC_2 &\Longleftrightarrow& \frac{m\ell^2\left(2\ell^6+3\ell^4r^2+7\ell^2r^4-2r^6\right)}{8\pi\left(r^2+\ell^2\right)\left(r^4+\ell^4\right)^{9/4}} \geq 0 \ , \\[3pt]
WEC_3 &\Longleftrightarrow& -\frac{\ell^2\left[\left(r^4+\ell^4\right)^{5/4}-2m(2\ell^4+\ell^2r^2+r^4)\right]}{8\pi\left(r^2+\ell^2\right)^2\left(r^4+\ell^4\right)^{5/4}} \geq 0 \ , \\[3pt]
SEC_3 &\Longleftrightarrow& \frac{2m\ell^2r^2\left(3\ell^4+5\ell^2r^2-2r^4\right)}{8\pi\left(r^2+\ell^2\right)\left(r^4+\ell^4\right)^{9/4}} \geq 0 \ , \\[3pt]
DEC_1 &\Longrightarrow& \frac{4\ell^4m}{8\pi\left(r^2+\ell^2\right)\left(r^4+\ell^4\right)^{5/4}} \geq 0 \ , \\[3pt]
DEC_2 &\Longrightarrow& \frac{\ell^2\left[\frac{m\left(6\ell^8-\ell^6r^2+2\ell^4r^4-\ell^2r^6+6r^8\right)}{\left(r^4+\ell^4\right)^{9/4}}-2\right]}{8\pi\left(r^2+\ell^2\right)^2}\geq 0 \ .
\end{eqnarray}
The $NEC_1=WEC_1=SEC_1$ conditions are again violated for $\vert r\vert>r_{H}$, and the $NEC_2$, $SEC_3$, $DEC_{2}$, and $WEC_3$ conditions are all violated for $r\gg r_{H}$. Once again there is at least \textit{some} exotic matter present, maintaining the essential characteristics of the original SV spacetime.

In the sub-horizon region where the $t$ coordinate is spacelike (recall the existence of this region requires $\ell<2m$), one instead has:
%
\begin{eqnarray}
NEC_1 &\Longleftrightarrow& \frac{2\ell^2\left(1-\frac{2m}{\sqrt[4]{\ell^4+r^4}}\right)}{8\pi\left(r^2+\ell^2\right)^2} \geq 0 \ , \\[3pt]
NEC_2 &\Longleftrightarrow& \frac{{2}\ell^2\left(r^4+\ell^4\right)^{9/4}-m\ell^2(3r^2-\ell^2)\left(2r^6-r^4\ell^2-r^2\ell^4-{2}\ell^6\right)}{8\pi\left(r^2+\ell^2\right)^2\left(r^4+\ell^4\right)^{9/4}} \geq 0 \ , \nonumber \\
&& \\
WEC_3 &\Longleftrightarrow& \frac{\ell^2\left[\left(r^4+\ell^4\right)^{5/4}-2mr^2(r^2-\ell^2)\right]}{8\pi\left(r^2+\ell^2\right)^2\left(r^4+\ell^4\right)^{5/4}} \geq 0 \ , \\[3pt]
SEC_3 &\Longleftrightarrow& \frac{4\ell^2\left(r^4+\ell^4\right)^{9/4}-2m\ell^2\left(4\ell^8-3\ell^6r^2-3\ell^2r^6+6r^8\right)}{8\pi\left(r^2+\ell^2\right)^2\left(r^4+\ell^4\right)^{9/4}} \geq 0 \ , \\[3pt]
DEC_1 &\Longrightarrow& \frac{4\ell^4m}{8\pi\left(r^2+\ell^2\right)\left(r^4+\ell^4\right)^{5/4}} \geq 0 \ , \\[3pt]
DEC_2 &\Longrightarrow& {m\ell^2(2\ell^4-5r^2\ell^2+2r^4)\over8\pi (r^4+\ell^4)^{9/4}} \geq 0 \ .
\end{eqnarray}
So in between the horizons the $NEC_1=WEC_1=SEC_1$ conditions are once again violated, now for all $\vert r\vert<r_{H}$. This implies violation of all of the standard point-wise energy conditions, again maintaining the essential characteristics of the original SV spacetime.

The Hernandez--Misner--Sharp mass in the case of Eq.~(\ref{mod1}) is easily found \textit{via} substitution into Eq.~(\ref{mass}):
\begin{eqnarray}
M_\HMS(r) = \frac{\ell^2}{2\sqrt{r^2+\ell^2}}+\frac{mr^2}{\sqrt{r^2+\ell^2}\;\sqrt[4]{\ell^4+r^4}} \ .
\end{eqnarray}
The mass is always positive, with the following limiting behavour:\newline $\lim_{r\rightarrow 0}M_\HMS(r)=\ell/2$,  and $\lim_{r\rightarrow \infty}M_\HMS(r)=m$.

Consequently, one may conclude that constructing models of the form of Eq.~(\ref{newBBS}) by varying the (integer) value of $n$ and keeping $k$ fixed to be zero results in precisely the same qualitative characteristics across the board as found for the original SV spacetime. This immediately suggests to explore qualitatively different geometries one must change the value of $k$.

\subsection{Model $n=1$ and $k=2$}\label{NBBss:n=1+k=2}

%
If instead one fixes $n=1$ and $k=2$ in Eq.~(\ref{newBBS}), one has
\begin{eqnarray}
\Sigma(r) = \sqrt{r^2+\ell^2} \ , \qquad f(r) = 1-\frac{2mr^2}{(r^2+\ell^2)^{3/2}} \ .\label{mod2}
\end{eqnarray}
The function $f(r)$ is now identical to that which appears in the regular Bardeen black hole~\cite{Tbilisi:Bardeen:1968}, under the switch $\ell\rightarrow q$. However, the spacetime is completely different to Bardeen, due to the $\ell^2$ term appearing in $\Sigma^2$. Solving for the roots of $f(r)$, one finds: i) for $\ell<\ell_{ext}=4m/(3\sqrt{3})$, there are four real solutions, which are symmetrical to each other, namely, $(r_+,r_C,-r_C,-r_+)$, (here, $r_+$ corresponds to the event (outer) horizon, and $r_C$ corresponds to a Cauchy (inner) horizon); ii) for $\ell=\ell_{ext}$, there are two real solutions $(r_+,-r_+)$; iii) and for $\ell>\ell_{ext}$, no real value exists --- there are no horizons.

\begin{figure}[htb!]
\begin{center}
	\includegraphics[scale=1.08]{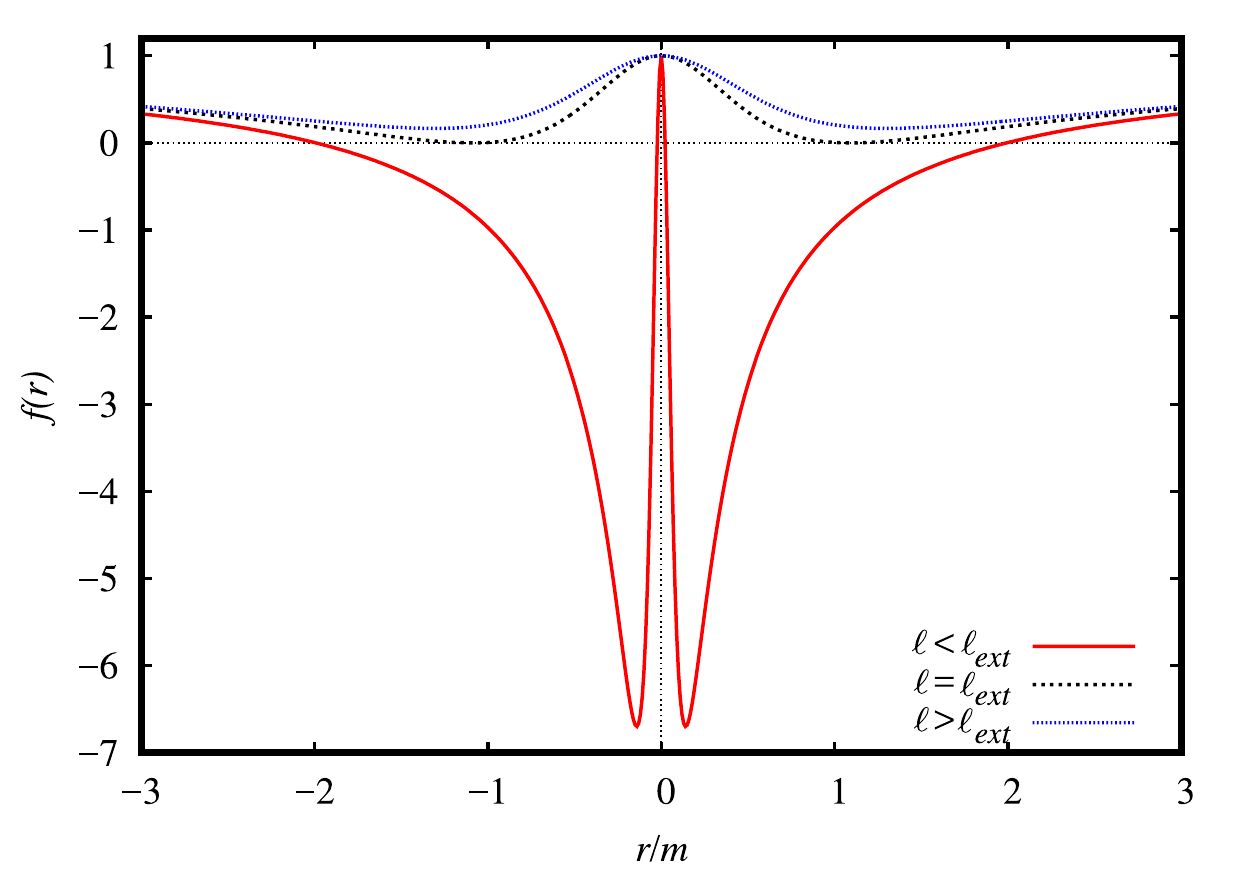}
	\caption[The various possibilities for the metric function $f(r)$, depending on the value of $\ell$, for the $n=1,\,k=2$ model]{Graphical representation of the possibilities of $f(r)$, given by Eq.~(\ref{mod2}). When $\ell<\ell_{ext}$, the signature changes four times, which translates as four horizons: two event (outer) and two Cauchy (inner) horizons. When $\ell=\ell_{ext}$, there is a black-bounce with two symmetric degenerate horizons. When $\ell>\ell_{ext}$, no horizon exists. See the text for more details.}
	\label{fig1}
\end{center}
\end{figure}

In this new model, there is the first drastic difference compared to standard SV. Specifically, when $\ell<\ell_{ext}$, by taking the limit $r\rightarrow 0$ in $f(r)$, one sees that $f(r)$ has a positive value with signature $(+,-,-,-)$, as opposed to signature $(-,+,-,-)$. This is due to the fact that SV spacetime only has a single horizon, changing the signature from $(+,-,-,-)$ outside, to $(-,+,-,-)$ inside the horizon (where $r=0$ is contained). However, for the model currently under examination, the signature changes four times; this is displayed in Fig.~\ref{fig1} which describes the behaviour of the metric function $f(r)$. Thus, there are two event (outer) and two Cauchy (inner) horizons in this case.

The Kretschmann scalar is given by
\begin{eqnarray}
K &=& \frac{8m^2r^4\left(r^2-2\ell^2\right)^2}{\left(r^2+\ell^2\right)^7}
+ \frac{{4}\left[\ell^2\left(r^2+\ell^2\right)^{3/2}+2mr^4\right]^2}{\left(r^2+\ell^2\right)^7} \nonumber \\
&& + \frac{{4}m^2\left(2\ell^4-11\ell^2r^2+2r^4\right)^2}{\left(r^2+\ell^2\right)^7} + \frac{8\left[\ell^2\left(r^2+\ell^2\right)^{3/2}+mr^2(r^2-4\ell^2)\right]^2}{\left(r^2+\ell^2\right)^{{7}}} \ . \nonumber \\
&&
\end{eqnarray}
It is easily verified that for $\ell>0$ this scalar is globally finite. Thus, by Theorem~\ref{Theorem:Kretsch}, the spacetime is always regular.
\enlargethispage{15pt}

\begin{figure}[!htb]
\begin{center}
	\includegraphics[scale=0.68]{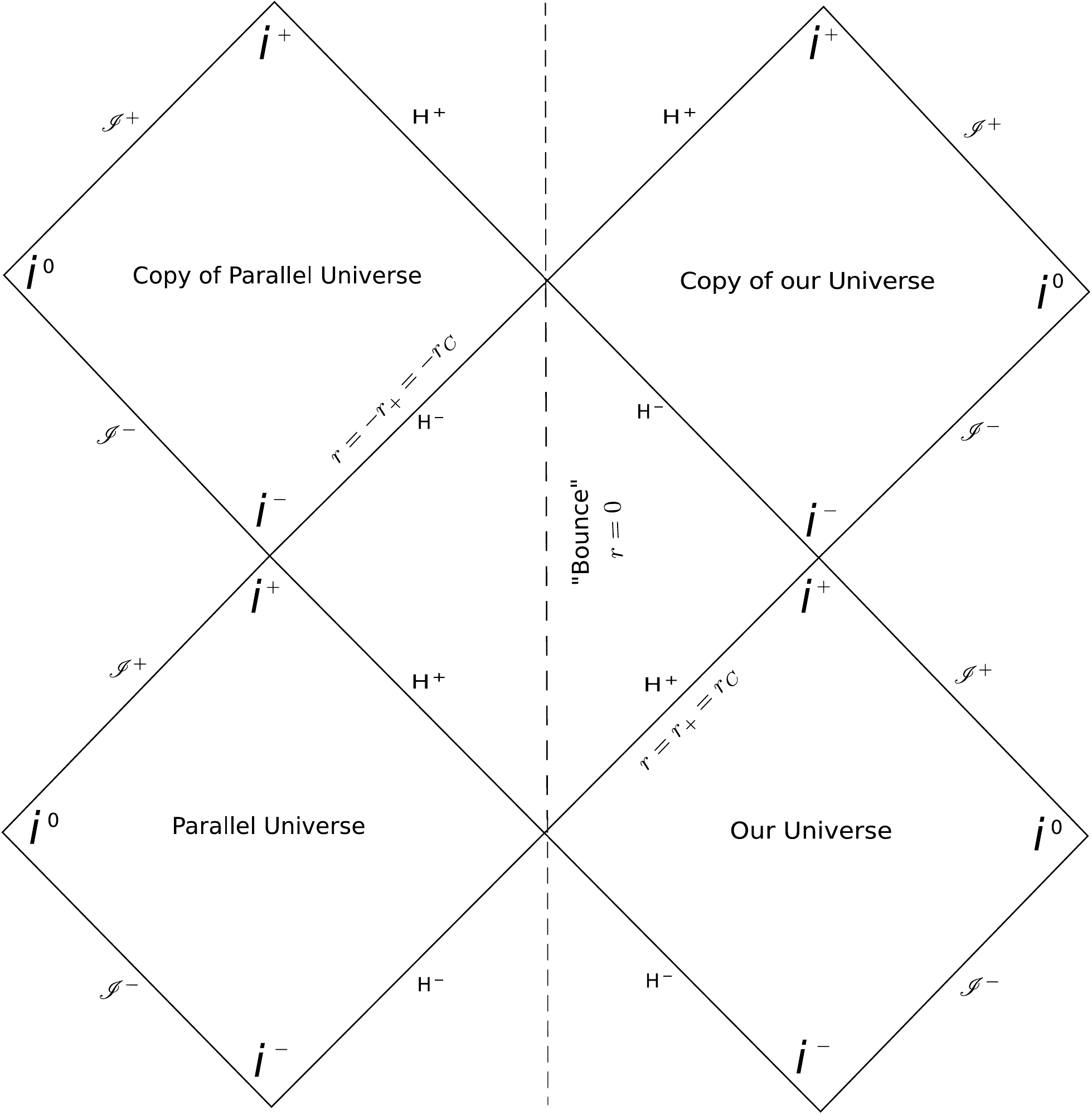}
	\caption[Carter--Penrose diagram for the extremal case, where only the extremal horizon exists, for the $n=1,\,k=2$ model]{In this example the horizon location is extremal. Mathematically there are repeated roots $r_+ = r_C$ of $f(r)$. Since the extremal horizon ($H_+ = C_+ = r_+$) is as usual an infinite proper distance from any point not on the extremal horizon, the Carter--Penrose diagram is somewhat misleading in that it would be infeasible to propagate through the extremal horizon to reach the hypersurface at $r = 0$. The bounce surface is now timelike, given $f(r)$ does not switch sign through the extremal horizon.}\label{figPenrose4}
\end{center}
\end{figure}

The causal structure is as follows: (i) When $\ell<\ell_{ext} = \frac{4m}{3\sqrt{3}}$, there are four horizons (two in each universe). The global causal structure is difficult to display in two dimensions; one has to ``cut the sheet''. This is the same picture as seen in the bbRN case from \S~\ref{SVbbRN}; see Fig.~\ref{fig:penrose2}. (ii) When $\ell=\ell_{ext} = \frac{4m}{3\sqrt{3}}$ there is a black-bounce with two symmetric degenerate horizons. The relevant Penrose diagram is depicted in Fig.~\ref{figPenrose4}. (iii) For the specific case of $\ell>\ell_{ext} = \frac{4m}{3\sqrt{3}}$, one has a horizonless traversable wormhole in the canonical sense of Morris and Thorne~\cite{WISATUFIT:Morris:1988}.
\clearpage
\vspace*{45pt}

\clearpage

In the region where the $t$ coordinate is timelike, the energy conditions for this model are given by the following relations:
\begin{eqnarray}
NEC_1 &\Longleftrightarrow& -\frac{2\ell^2\left(1-\frac{2mr^2}{\left(r^2+\ell^2\right)^{3/2}}\right)}{8\pi\left(r^2+\ell^2\right)^2} \geq 0 \ , \\[2pt]
NEC_2 &\Longleftrightarrow& \frac{m\ell^2(13r^2-2\ell^2)}{8\pi\left(r^2+\ell^2\right)^{7/2}} \geq 0 \ , \\[2pt]
WEC_3 &\Longleftrightarrow& -\frac{\ell^2\left[\left(r^2+\ell^2\right)^{3/2}-8mr^2\right]}{8\pi\left(r^2+\ell^2\right)^{7/2}} \geq 0 \ , \\[2pt]
SEC_3 &\Longleftrightarrow& \frac{2m\ell^2\left(7r^2-2\ell^2\right)}{8\pi\left(r^2+\ell^2\right)^{7/2}} \geq 0 \ , \\[2pt]
DEC_1 &\Longrightarrow& \frac{12\ell^2mr^2}{8\pi\left(r^2+\ell^2\right)^{7/2}} \geq 0 \ , \\[2pt]
DEC_2 &\Longrightarrow& -\frac{\ell^2\left[2(\ell^2+r^2)^{3/2}-m\left(2\ell^2+3r^2\right)\right]}{8\pi\left(r^2+\ell^2\right)^{7/2}} \ .
\end{eqnarray}
As mentioned above, the $NEC_1=WEC_1=SEC_1$ conditions are all violated when $\vert r\vert > r_+$, and the $DEC_{2}$ and $WEC_3$ conditions are violated when $\vert r\vert\gg r_+$. The $NEC_2$ condition is violated in the range $-\sqrt{2/13}\,\ell<r<\sqrt{2/13}\,\ell$, and finally the $SEC_3$ condition is violated for $-\sqrt{2/7}\,\ell<r<\sqrt{2/7}\,\ell$. As the $WEC_3=\rho$ condition is violated outside the event horizon, once again there are negative energy densities.
\enlargethispage{25pt}

In the region where the $t$ coordinate is spacelike, the energy conditions for this model are instead given by
\begin{eqnarray}
NEC_1 &\Longleftrightarrow& \frac{2\ell^2\left(1-\frac{2mr^2}{\left(r^2+\ell^2\right)^{3/2}}\right)}{8\pi\left(r^2+\ell^2\right)^2} \geq 0 \ , \\[2pt]
NEC_2 &\Longleftrightarrow& \frac{2\ell^2(r^2+\ell^2)^{3/2}+\ell^2m(9r^2-2\ell^2)}{8\pi\left(r^2+\ell^2\right)^{7/2}} \geq 0 \ , \\[2pt]
WEC_3 &\Longleftrightarrow& \frac{\ell^2\left[\left(r^2+\ell^2\right)^{3/2}+4mr^2\right]}{8\pi\left(r^2+\ell^2\right)^{7/2}} \geq 0 \ , \\[2pt]
SEC_3 &\Longleftrightarrow& \frac{4\ell^2(r^2+\ell^2)^{3/2}+2m\ell^2\left(3r^2-2\ell^2\right)}{8\pi\left(r^2+\ell^2\right)^{7/2}} \geq 0 \ , \\[2pt]
DEC_1 &\Longrightarrow& \frac{12\ell^2mr^2}{8\pi\left(r^2+\ell^2\right)^{7/2}} \geq 0 \ , \\[2pt]
DEC_2 &\Longrightarrow& -\frac{\ell^2m\left(r^2-2\ell^2\right)}{8\pi\left(r^2+\ell^2\right)^{7/2}} \geq 0  \ .
\end{eqnarray}
Once again, the $NEC_1=WEC_1=SEC_1$ conditions are violated for sub-horizon regions.
\clearpage

The Hernandez--Misner--Sharp mass in the case of Eq.~(\ref{mod2}) is straightforward to extract \textit{via} substitution into Eq.~(\ref{mass}):
\begin{eqnarray}
M_\HMS(r) = \frac{\ell^2}{2\sqrt{r^2+\ell^2}}+\frac{mr^4}{\left(r^2+\ell^2\right)^2} \ .
\end{eqnarray}
The mass is always positive and possesses the  limits $\lim_{r\rightarrow 0}M_\HMS(r)=\ell/2$,  and $\lim_{r\rightarrow \infty}M_\HMS(r)=m$.

In general, if one wishes to construct models by setting $k=2$ and varying $n$, it is straightforward to verify that the respective spacetimes all possess similar characteristics to the case considered above, and the energy density will always be negative for the region outside the event horizons.

\subsection{Model with zero energy density}\label{NBBss:zero}

While (as seen above) some of the energy conditions will always be violated, one wonders whether it is at least possible to satisfy the $WEC_3$ condition. This would require a non-negative energy density; it is prudent to first consider the marginal satisfaction case, where the energy density is identically zero. Using Eq.~(\ref{density+}), keeping $\Sigma(r)=\sqrt{r^2+\ell^2}$ fixed and setting $\rho(r)=0$ allows one to solve the resulting differential equation for the required $f(r)$:
\begin{equation}
f(r) = {(\sqrt{r^2+\ell^2}+K)\;\sqrt{r^2+\ell^2}\over r^2} \ ,
\end{equation}
where $K$ is some integration constant. However, the regularity condition that $f(0)$ be finite requires the integration constant $K$ to be set to $K=-\ell$, in which case
\begin{equation}
f(r) = {(\sqrt{r^2+\ell^2}-\ell)\;\sqrt{r^2+\ell^2}\over r^2} \ .
\end{equation}
This geometry is horizonless, and for small $r$ and large $r$ respectively one has
\begin{equation}
f(r) = {1\over2} + \O(r^2) \ , \quad \hbox{and} \quad f(r) = 1 - {\ell\over r} + \O(1/r^2) \ .
\end{equation}
Applying the Einstein equations to this metric, it is easy to verify that $\rho=0$, and that
\begin{equation}
p_r = -{2\ell^2(\sqrt{r^2+\ell^2}-\ell)\over r^2(r^2+\ell^2)^{3/2}} \ , \quad
p_t = {2\sqrt{r^2+\ell^2}(3\ell^2+2r^2)-\ell(7r^2+6\ell^2)\over2(r^2+\ell^2)^{3/2}r^4} \ .
\end{equation}
For this model $p_r<0$ for any nonzero $r$, and $p_t>0$ for any $r$, so certainly the $NEC_1$ condition is still violated throughout the spacetime. The $WEC_3$ condition is by construction marginally satisfied. So while one can force the WEC to be tolerably well-behaved, other energy conditions will still be violated. The Hernandez--Misner--Sharp mass for this spacetime is particularly simple; $M_\HMS(r) = \ell/2$ everywhere.

One could try to generalise this construction by choosing some positive function $\rho_*(r)>0$ and setting $\rho(r) = \rho_*(r)>0$. One would then solve the resulting differential equation for $f(r)$ arising from Eq.~(\ref{density+}), fixing the integration constant by demanding the finiteness of $f(0)$. Such a construction would satisfy the $WEC_3$ condition, at least in the domain of outer communication, but the status of the other energy conditions would remain unresolved.

However, when it comes to analysing the $NEC_1=WEC_1=SEC_1$ conditions more can be said. From Eq.~(\ref{cond1b}), substituting $\Sigma\to \sqrt{r^2+\ell^2}$, one can see that:
\begin{equation}
NEC_{1}=WEC_1=SEC_1 \Longleftrightarrow
-\frac{2\vert f(r)\vert\ell^2}{8\pi(r^2+\ell^2)} \geq 0 \ ,\label{cond1c}
\end{equation}
implying the $NEC_1=WEC_1=SEC_1$ conditions are violated everywhere except on the horizons themselves; indeed whenever $f(r)\neq 0$.

\subsection{Model $M(r)=m\cos^{2n}\left[r_0/\Sigma(r)\right]$}\label{NBBss:M(r)_1}

Introducing a completely new framework, if instead one chooses\newline $M(r)=m\cos^{2n}\left[r_0/\Sigma(r)\right]$, one has
\begin{eqnarray}
f(r) = 1-\frac{2M(r)}{\Sigma(r)} = 1-\frac{2m\cos^{2n}\left[r_0/\Sigma\right]}{\Sigma}\label{mod3} \ ,
\end{eqnarray}
where for $n\rightarrow 0$, retaining $\Sigma= \sqrt{r^2+\ell^2}$, one recovers standard SV spacetime. In the large-$r$ limit one has $\lim_{r\rightarrow \infty}f(r)=1$. Indeed, for $r\gg1$, $f(r)\sim 1-(2m/r)$. Also:
\begin{eqnarray}
\lim_{r\rightarrow 0}f(r)=1-\frac{2m\cos^{2n}\left(r_0/\ell\right)}{\ell} \ .
\end{eqnarray}
Through an appropriate choice for $r_0$, there are three possibilities. Namely: $f(0)>0$, with the signature $(+,-,-,-)$, $f(0)=0$ with $2m\geq \ell$, and $f(0)<0$, with signature $(-,+,-,-)$.

\begin{figure}[!htb]
\begin{center}
	\includegraphics[scale=.63]{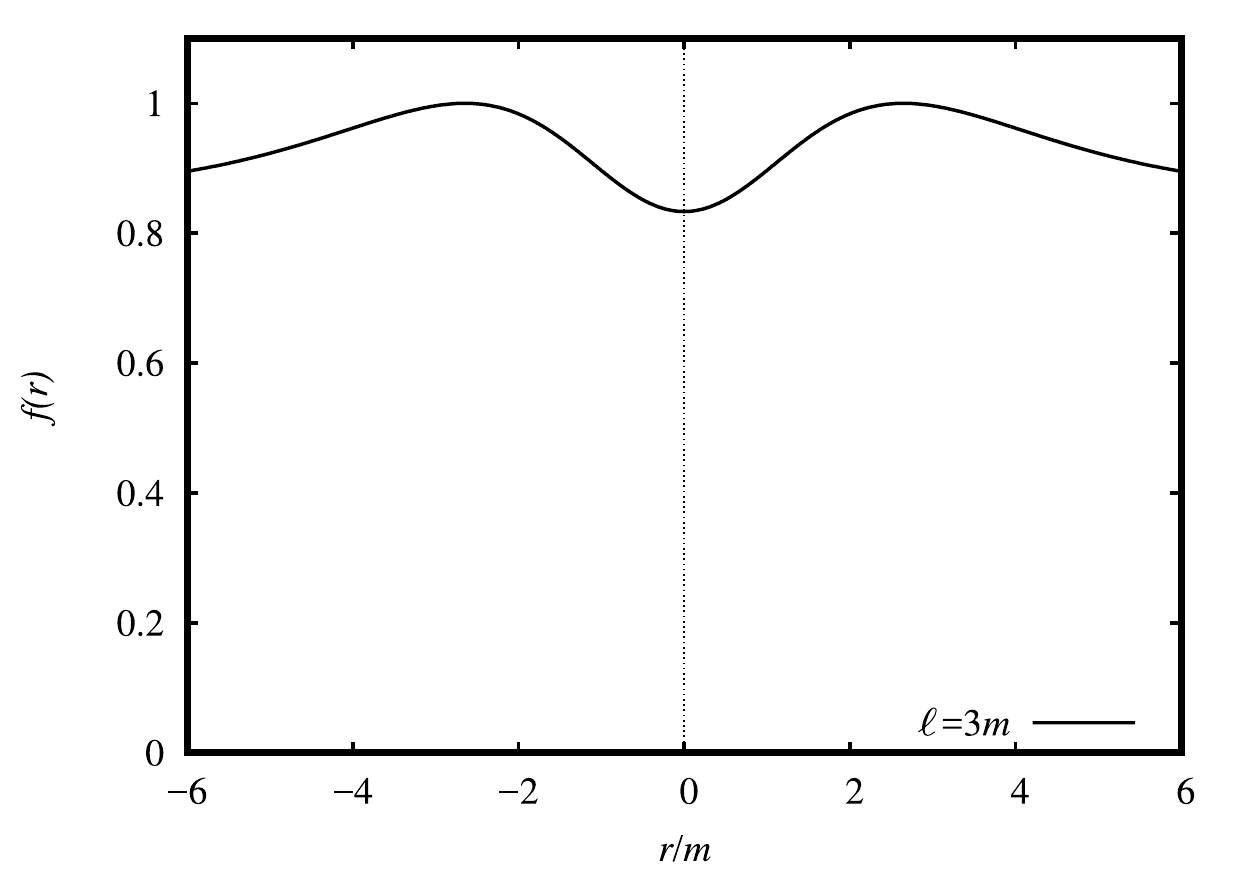}
	\includegraphics[scale=.63]{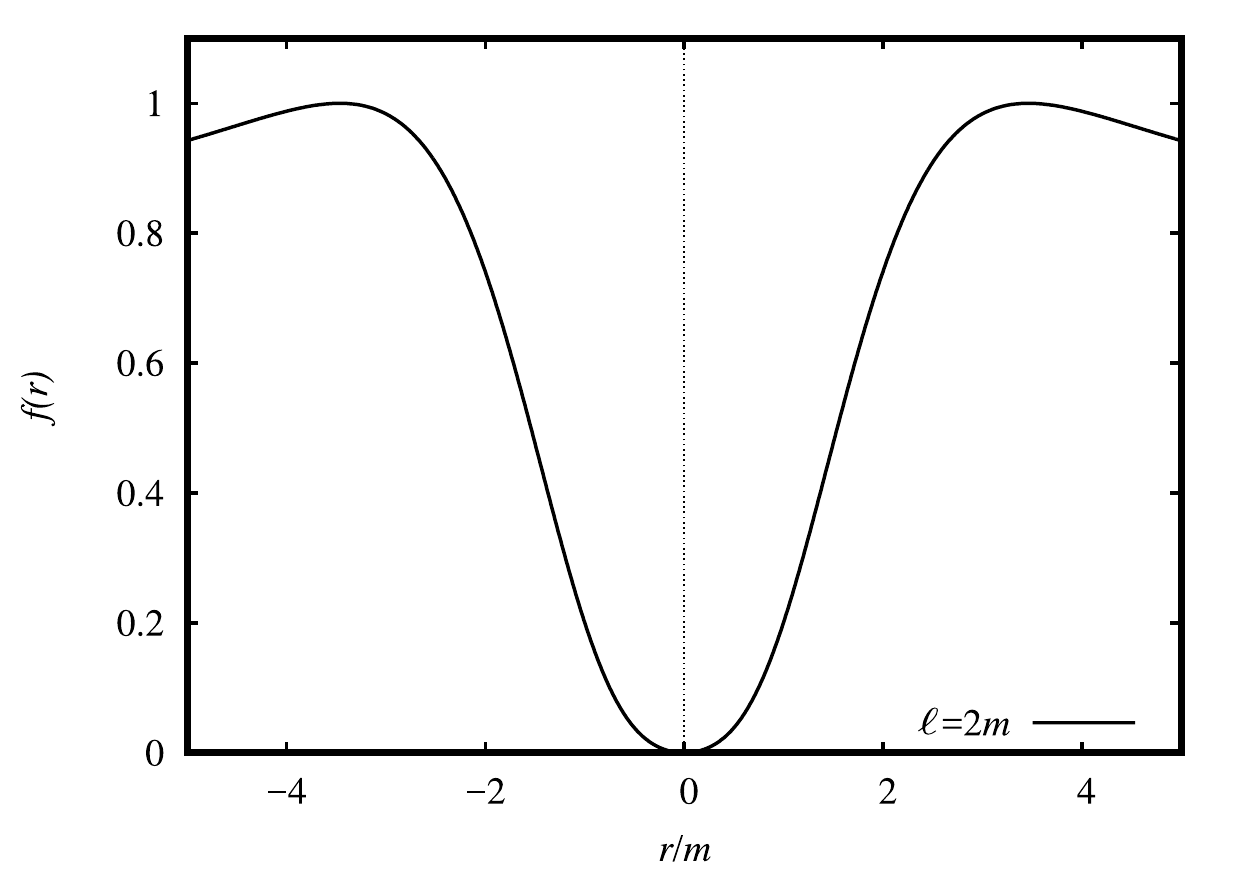}
	\includegraphics[scale=.63]{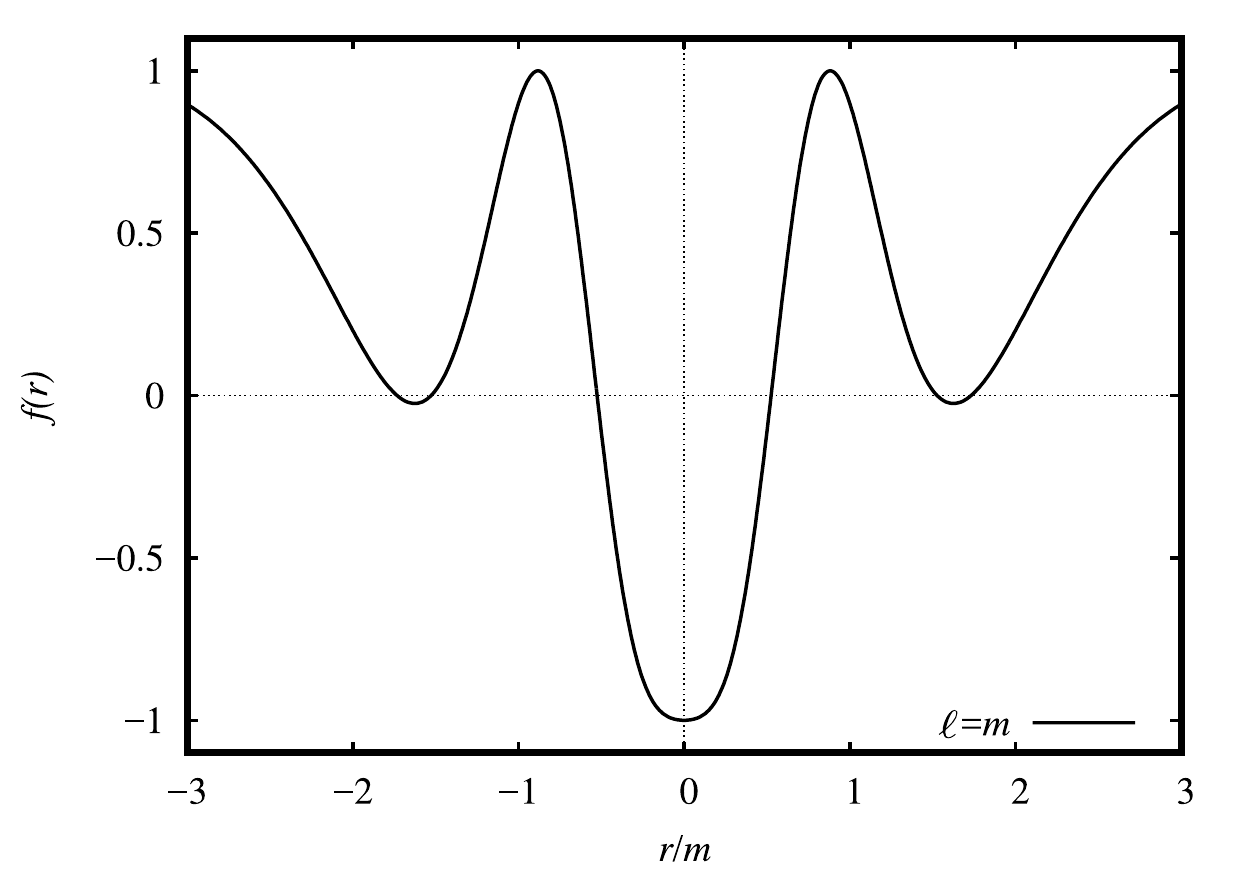}
	\includegraphics[scale=.63]{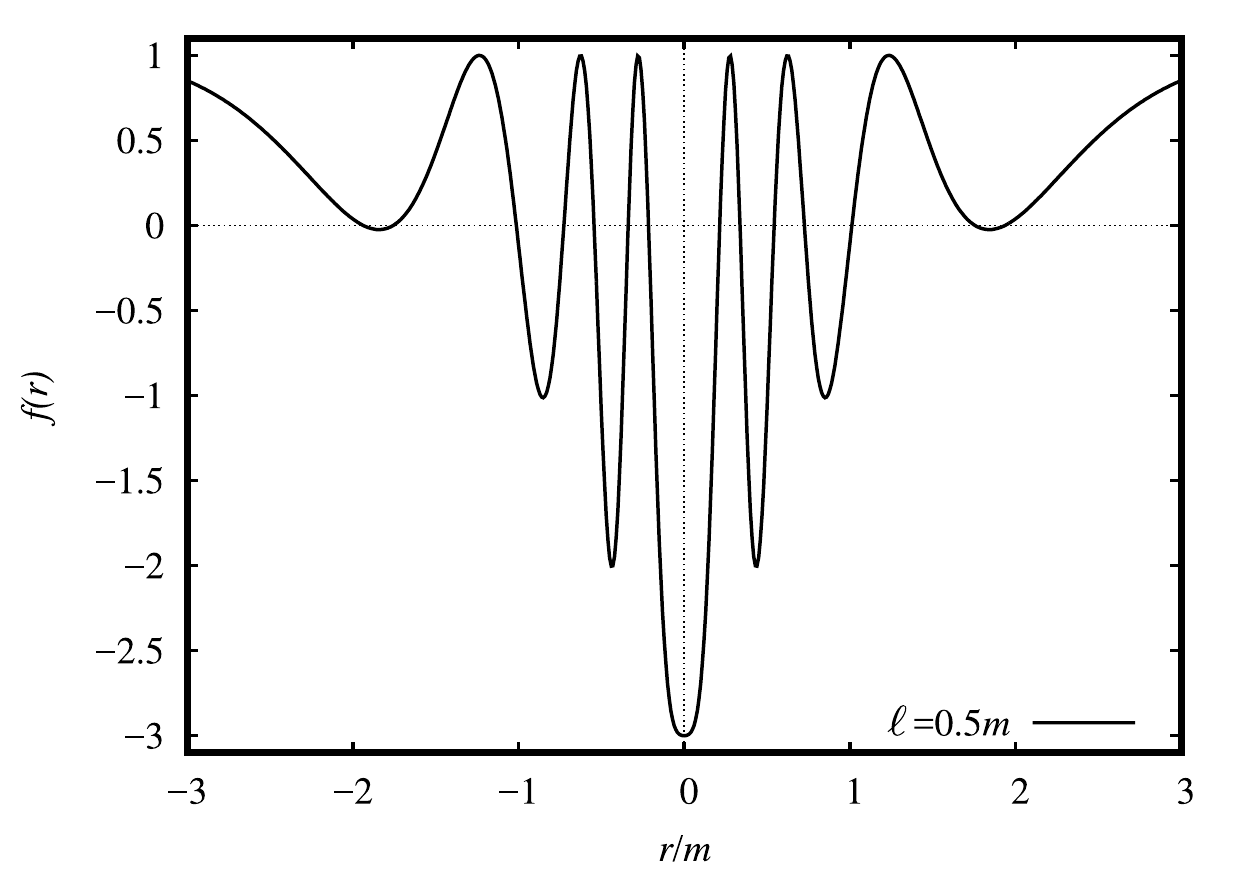}
	\caption[A subset of the various possibilities for $f(r)$ in the $r_0=2\pi\ell$ case of the $M(r)=m\,\cos^{2n}\Big(r_0/\Sigma(r)\Big)$ model]{Graphical representation of various possibilities for the $f(r)$ given by Eq.~(\ref{mod3}), fixing $r_0=2\pi\ell$.}\label{fig3}
\end{center}
\end{figure}

The number of horizons may also be very easily varied. In the plots of Fig.~\ref{fig3}, one may envision three structures of a black-bounce: (i) for $r_0=2\pi\ell$ and $\ell=3m$, there are no horizons, and consequently the geometry is a wormhole with a two-way timelike throat at $r=0$; (ii) for $r_0=2\pi\ell$ and $\ell=2m$, there is an extremal throat at $r=0$; (iii) for the specific example $r_0=2\pi\ell$ and $\ell=0.5m$, there is a regular black hole with precisely $14$ horizons, and where $r=0$ is a bounce. Thus, the number of horizons can in principle grow indefinitely. For the latter case, the causal structure cannot be represented in the typical fashion of a Carter--Penrose diagram.

The Kretschmann scalar can be easily extracted, and is globally finite for suitable values of the model's parameters, confirming regularity of the model \textit{via} Theorem~\ref{Theorem:Kretsch}.
\clearpage
\enlargethispage{30pt}

\clearpage

\begin{figure}[htb!]
\begin{center}
	\includegraphics[scale=.67]{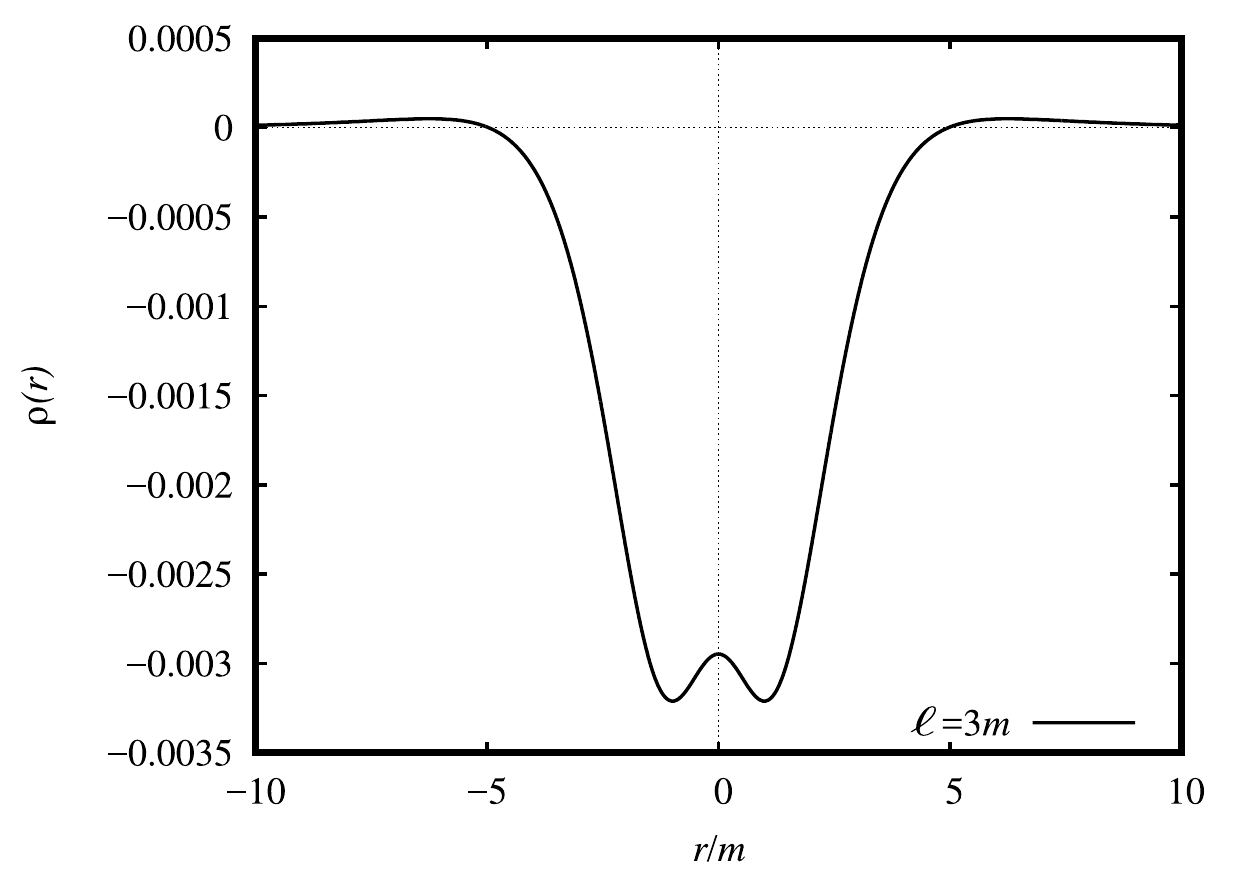}
	\includegraphics[scale=.67]{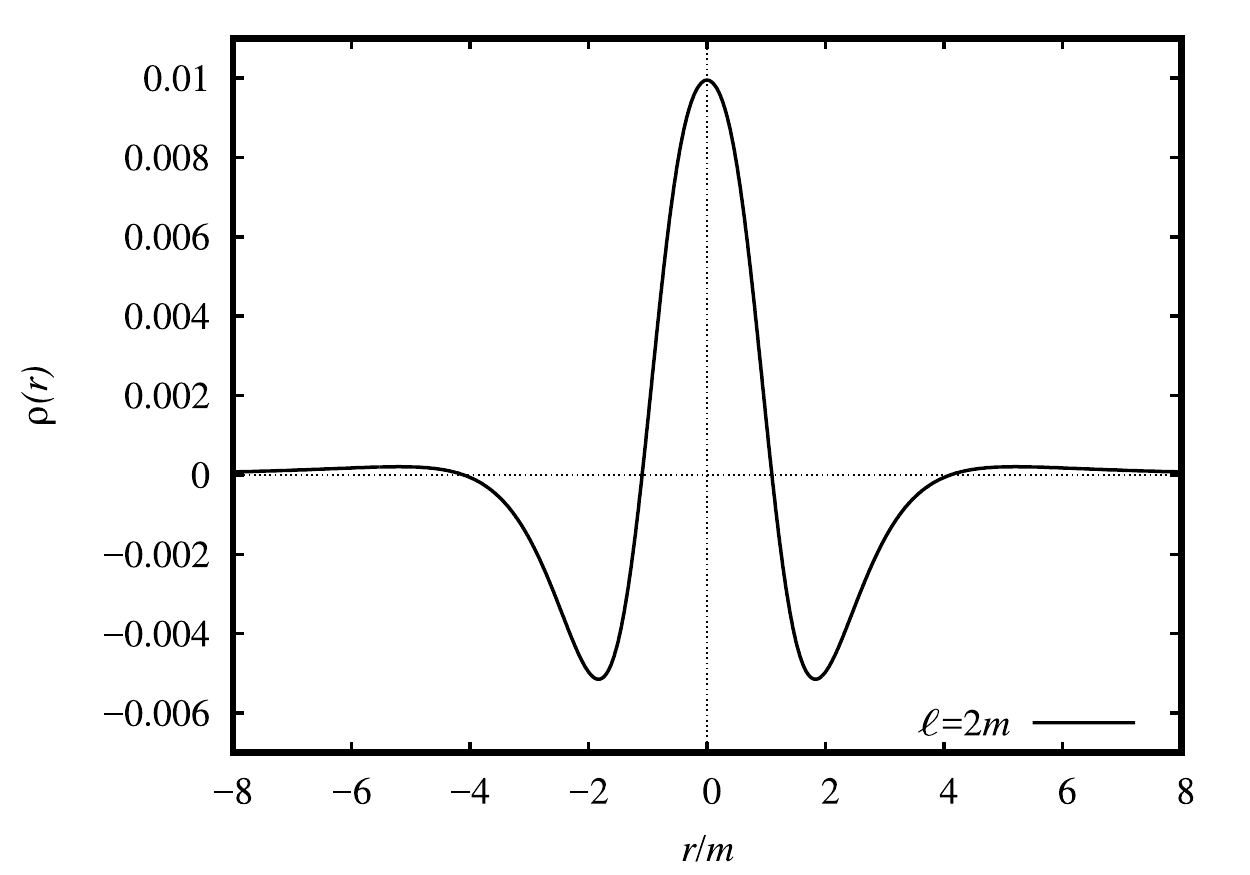}
	\caption[Graph of the energy density $\rho(r)$ in the $r_0=2\pi\ell$ case of the $M(r)=m\,\cos^{2n}\Big(r_0/\Sigma(r)\Big)$ model, for sample values of $\ell$]{Graphical representation of the energy density $\rho(r)$ for the model given by Eq.~(\ref{mod3}), for two sample values of $\ell$, fixing $r_0=2\pi\ell$.}\label{fig4}
\end{center}
\end{figure}

Analytically, the energy density is not particularly simple; it is represented in Fig.~\ref{fig4}. One sees that the energy density oscillates as the function $f(r)$. Asymptotically expanding the energy density as $r\rightarrow +\infty$, the dominant term is
\begin{equation}
\rho(r) \approx -\ell^2/8\pi r^4 \ .
\end{equation}
Therefore, the energy density is certainly negative for at least some regions outside the event horizon.
\vspace*{10pt}

The Hernandez--Misner--Sharp mass for this model is given by
\begin{eqnarray}
M_\HMS(r) = \frac{\ell^2}{2\sqrt{r^2+\ell^2}}+\frac{mr^2\cos^{2n}\left(r_0/\sqrt{r^2+\ell^2}\right)}{r^2+\ell^2} \ .
\end{eqnarray}
The mass is always positive with the limits $\lim_{r\rightarrow \infty}M_\HMS(r)=m$, and $\lim_{r\rightarrow 0}M_\HMS(r)=\ell/2$.
\clearpage

\subsection{Model $M(r)=m\arctan^n(r/\ell)\;(\Sigma/r) (2/\pi)^n$}\label{NBBss:M(r)_2}

\begin{figure}[htb!]
\begin{center}
	\includegraphics[scale=0.65]{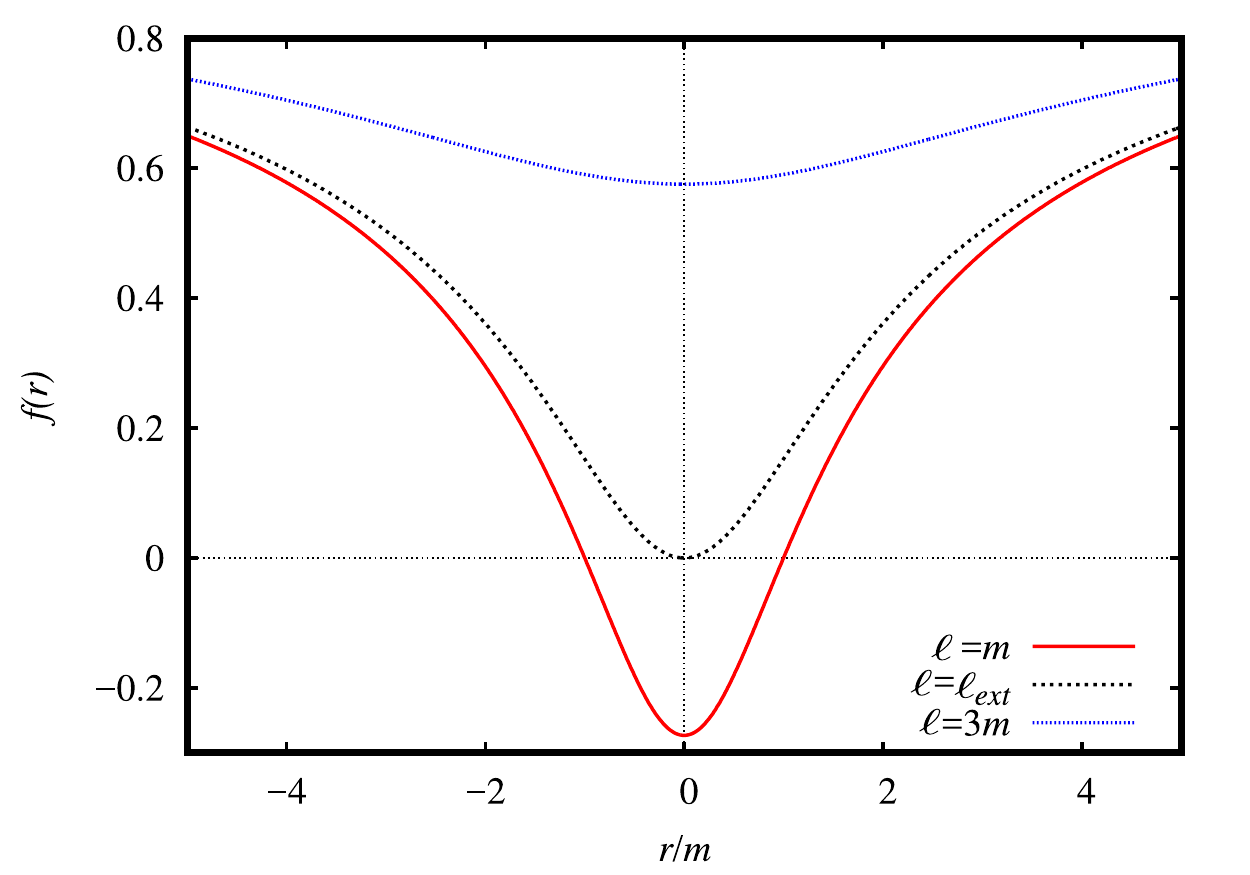} \includegraphics[scale=0.65]{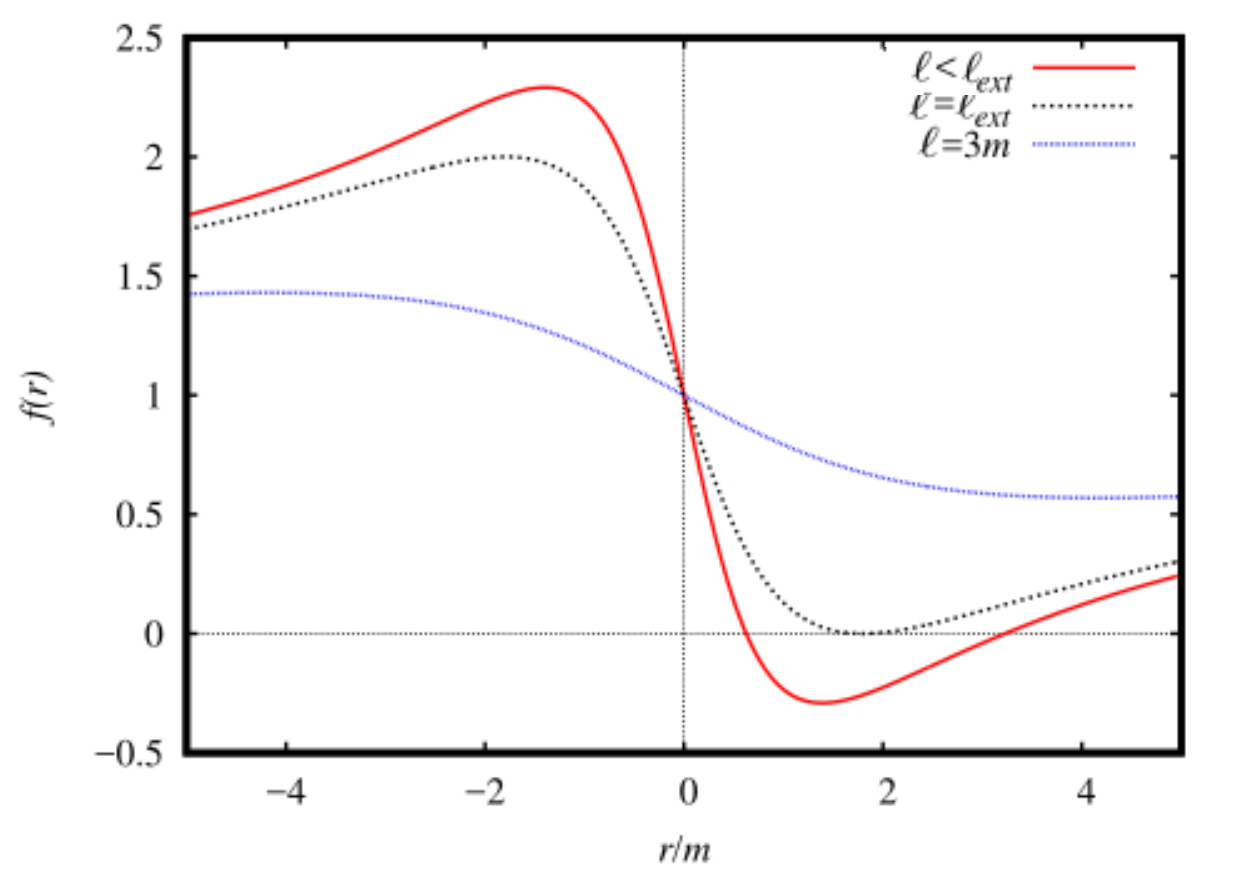}
	\caption[Two cases for $f(r)$ for the $M(r)=m\arctan^n(r/\ell)\;(\Sigma/r) (2/\pi)^n$ model]{Qualitative graphical representation of the $f(r)$ given by Eq.~(\ref{mod4}). In the upper plot $n=1$, and the lower plot $n=2$. For $n=2$, the extremal value is approximately $\ell_{ext}\approx5.16315560586775m/\pi^2$.}\label{fig5}
\end{center}
\end{figure}

So far, all models explored require a negative energy density at least \textit{somewhere} in the candidate spacetime. Using this as inspiration for the next model, one attempts to define a mass function that, for certain sub-cases, provides a globally positive energy density. Specifically, consider the case $M(r)=m\arctan^n(r/\ell)(\Sigma/r)(2/\pi)^n$. In this case the metric function $f(r)$ is given by
\begin{eqnarray}
f(r) = 1-\frac{2M(r)}{\Sigma(r)} = 1-\frac{2m\arctan^n(r/\ell)}{r}\left(\frac{2}{\pi}\right)^n \ .\label{mod4}
\end{eqnarray}
In the limits $(\ell,n)\rightarrow0$ one regains the Schwarzschild solution. It is simple to show that (for suitable ranges of values of the model's parameters) the Kretschmann scalar is globally finite, and so by Theorem~\ref{Theorem:Kretsch} the geometry is regular. Fixing $n$, one can regulate the presence of horizons by adjusting $\ell$; this is shown in Fig.~\ref{fig5}. For instance, consider $n=1,2$, where the extremal case for $n=1$ is given by $\ell_{ext}=4m/\pi$, and for $n=2$ is given by $\ell_{ext}\approx 5.16315560586775m/\pi^2$.
\enlargethispage{20pt}

\begin{figure}[htb!]
\begin{center}
	\includegraphics[scale=0.54]{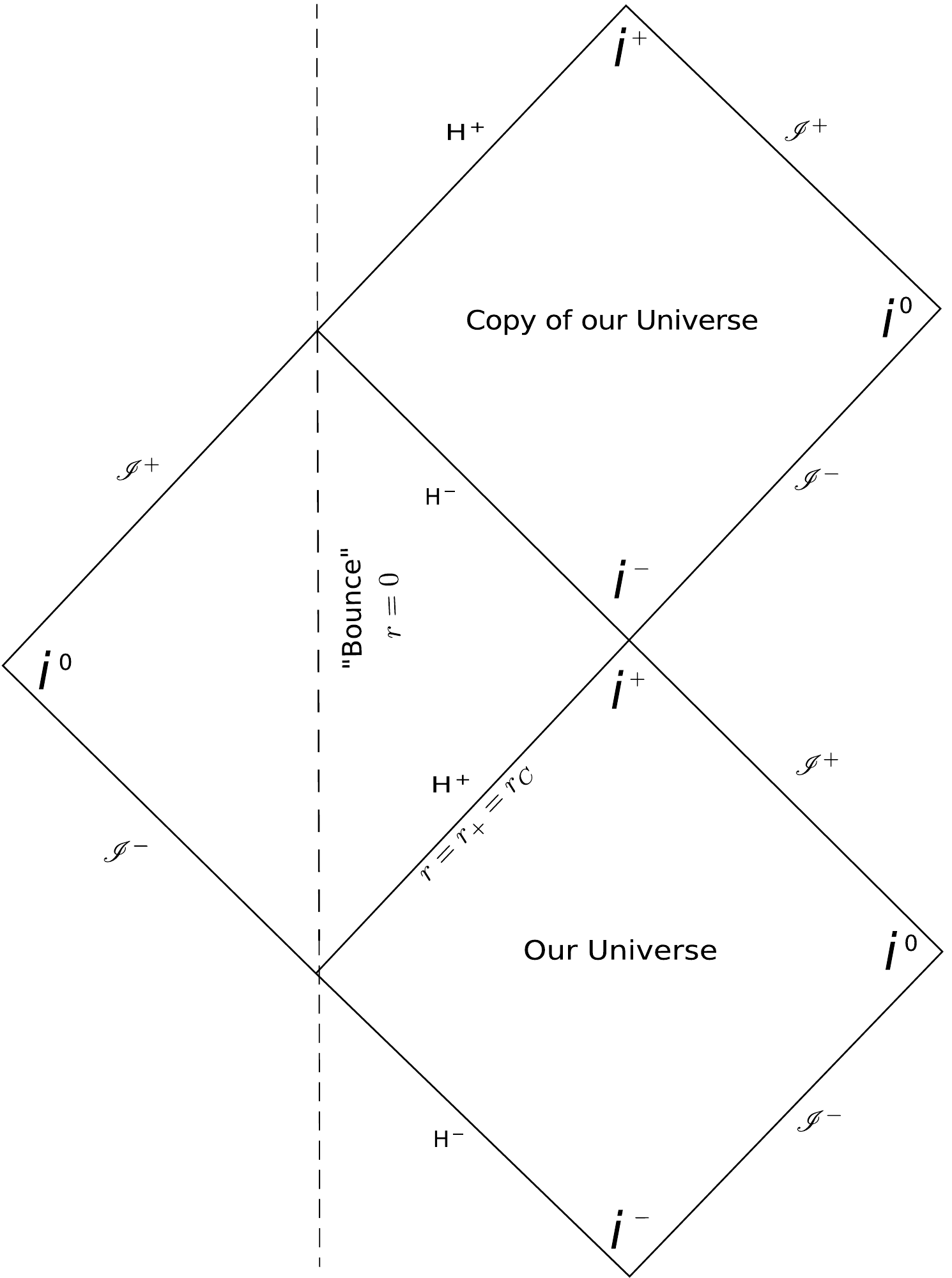}
	\caption[Carter--Penrose diagram for the case where there is an extremal horizon, followed by a timelike bounce, bouncing into a separate universe without horizons]{Carter--Penrose diagram for the case with an extremal horizon, followed by a timelike bounce at $r=0$, bouncing into a separate universe without horizons. Since the extremal horizon ($H+ = C+ = r+$) is as usual an infinite proper distance from any point not on the extremal horizon, the Carter--Penrose diagram is misleading in that it would be infeasible to propagate through the extremal horizon to reach the hypersurface at $r=0$.}
	\label{figPenrose5}
\end{center}
\end{figure}

\begin{figure}[htb!]
\begin{center}
	\includegraphics[scale=0.85]{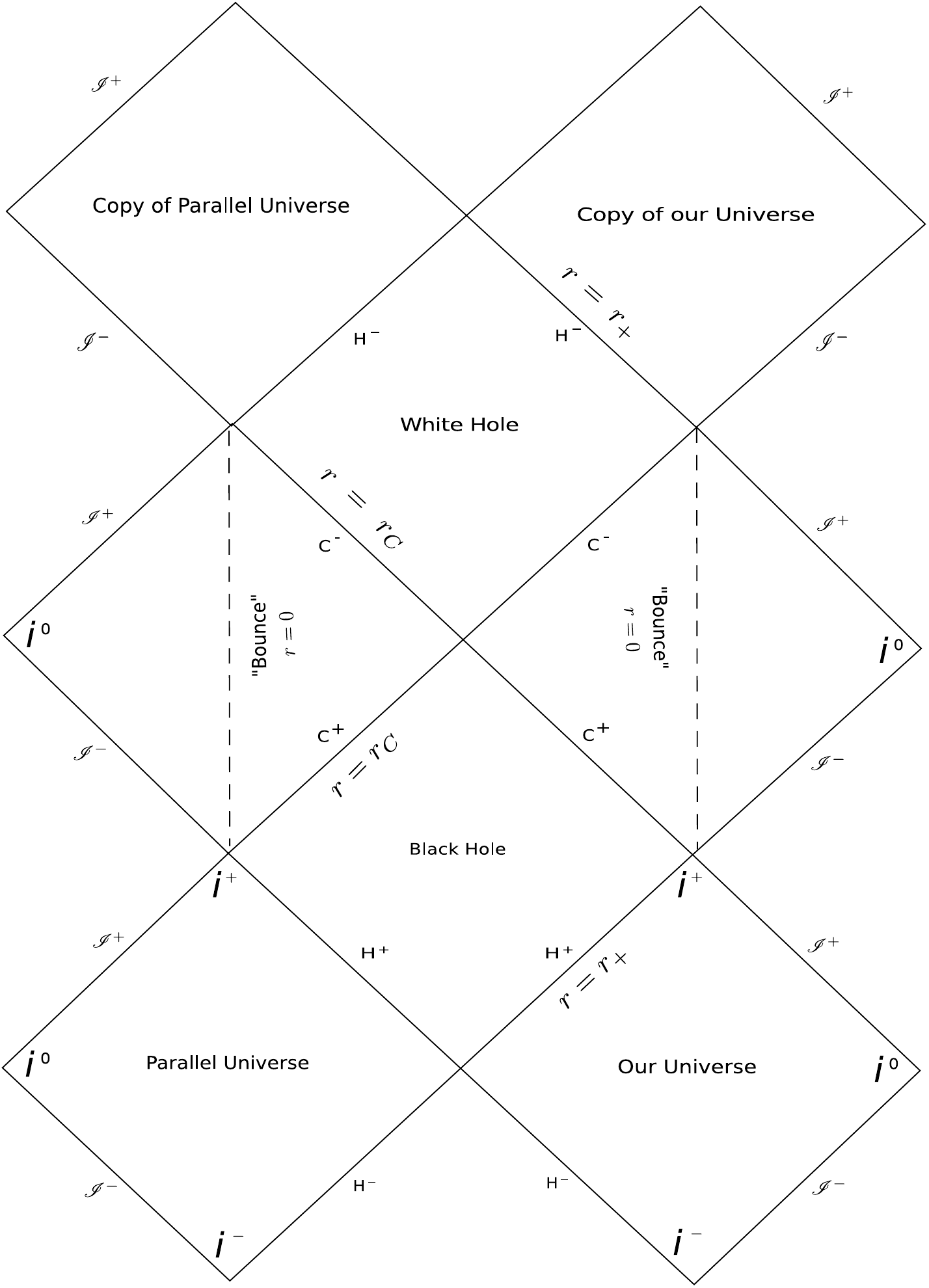}
	\caption[Carter--Penrose diagram for the case where there is both an inner and outer horizon, followed by a timelike bounce, bouncing into a separate universe without horizons]{Carter--Penrose diagram for the case where there is both an inner and outer horizon, followed by a timelike ``bounce'' at $r=0$, bouncing into a separate universe without horizons.}
	\label{figPenrose6}
\end{center}
\end{figure}

It is straightforward to represent the various causal structures for certain sub-cases using Carter--Penrose diagrams. For example:
\begin{itemize}
\itemsep-3pt
    \item Both the cases $n=1$ and $\ell>\ell_{ext}=4m/\pi$, as well as $n=2$ and $\ell>\ell_{ext}\approx5.16m/\pi^2$, are two-way traversable wormholes in the canonical sense of Morris and Thorne~\cite{WISATUFIT:Morris:1988};
    \item When $n=1$ and $\ell=\ell_{ext}=4m/\pi$, there is the one-way wormhole geometry with extremal null throat of Fig.~\ref{F:null-bounce-1};
    \item When $n=1$ and $\ell<\ell_{ext}=4m/\pi$, there is the classic spacelike bounce case with one horizon in each universe as in Fig.~\ref{fig:penrose1};
    \item The case when $n=2$ and $\ell=\ell_{ext}\approx5.16m/\pi^2$ is depicted in Fig.~\ref{figPenrose5}. There is an extremal horizon, followed by a timelike bounce at $r=0$, bouncing into a separate universe without horizons;
    \item The case when $n=2$ and $\ell<\ell_{ext}\approx5.16m/\pi^2$ is depicted in  Fig.~\ref{figPenrose6}. There is both an inner and outer horizon, followed by a timelike bounce at $r = 0$, bouncing into a separate universe without horizons.
\end{itemize}
\enlargethispage{30pt}

\clearpage

\vspace*{50pt}
\clearpage

\begin{figure}[htb!]
\begin{center}
	\includegraphics[scale=0.7]{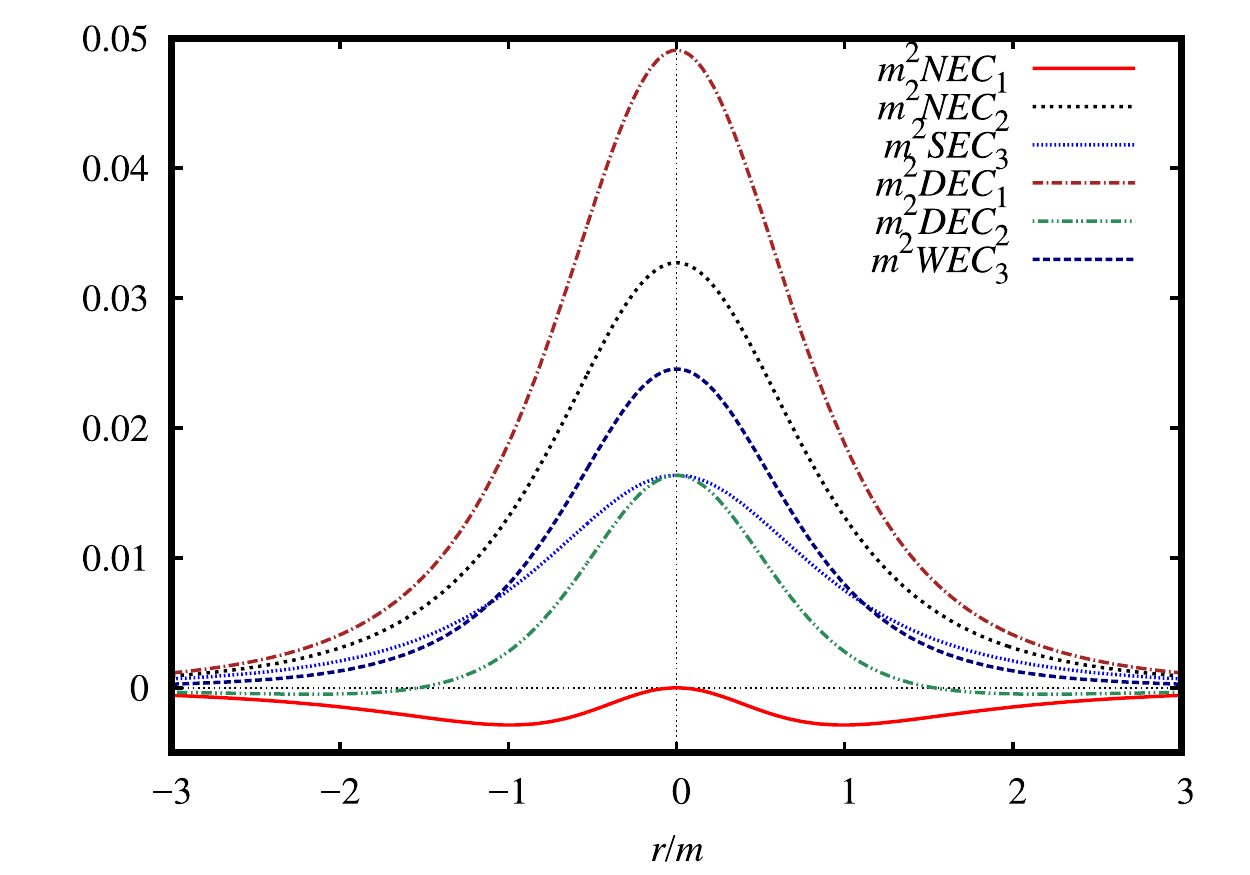} \includegraphics[scale=0.7]{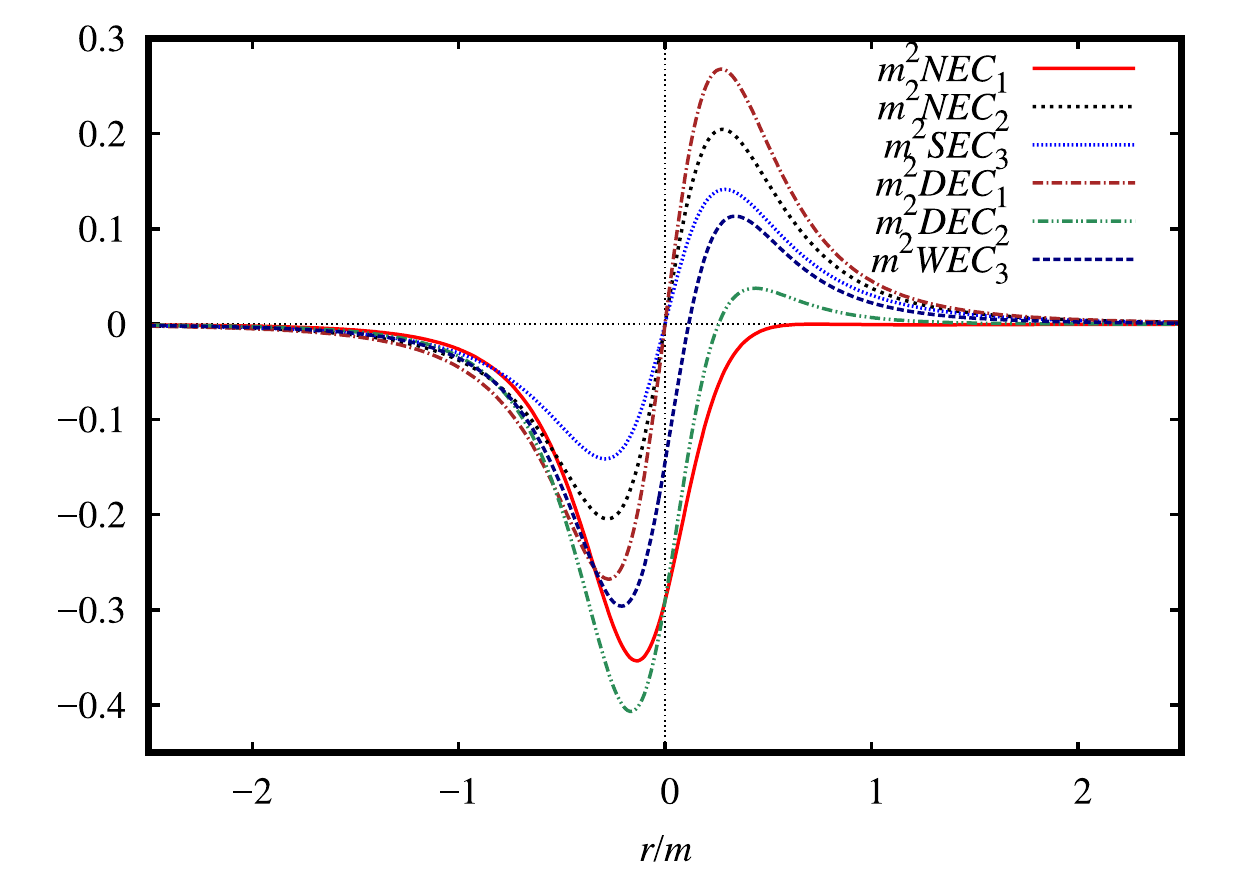}
	\caption[Graphical representation of the energy conditions for the $M(r)=m\arctan^n(r/\ell)\;(\Sigma/r) (2/\pi)^n$ model]{Graphical representation of the energy conditions for sub-cases of the model given by Eq.~(\ref{mod4}). In the upper plot $n=1$ and $\ell=\ell_{ext}=4m/\pi$, while in the lower plot $n=2$ and $\ell=\ell_{ext}\approx5.16m/\pi
^2$.}\label{fig6}
\end{center}
\end{figure}

The energy conditions are depicted in Fig.~\ref{fig6}. In both sub-cases presented, one verifies that all of the energy conditions are violated. However, in particular, the $WEC_3$ condition is always satisfied when $n=1$, and either $\ell<\ell_{ext}$ \textit{or} $\ell=\ell_{ext}$; the energy density is globally positive. A new feature that highlights the difference between black-bounce candidates and the typical regular black holes considered is that the condition $SEC_3$, defined in Eq.~(\ref{Econd1}), can be satisfied; verified \textit{via} examination of Fig.~\ref{fig6}.

The Hernandez--Misner--Sharp mass is given by
\begin{eqnarray}
M_\HMS(r) = \frac{\pi^{-n}\left(\ell^2\pi^n+2^{n+1}mr\arctan\left(\frac{r}{\ell}\right)^n\right)}{2\sqrt{r^2+\ell^2}} \ .\label{massmod4}
\end{eqnarray}
Note that for odd $n$ the mass is always positive, and there are the following limits: $\lim_{r\rightarrow \infty}M_\HMS(r)=m$, and $\lim_{r\rightarrow 0}M_\HMS(r)=\ell/2$. 
%

\subsection{Model $M(r)=m\arctan^n(r/\ell)(2/\pi)^n$}\label{NBBss:M(r)_3}

\begin{figure}[htb!]
\begin{center}
	\includegraphics[scale=0.7]{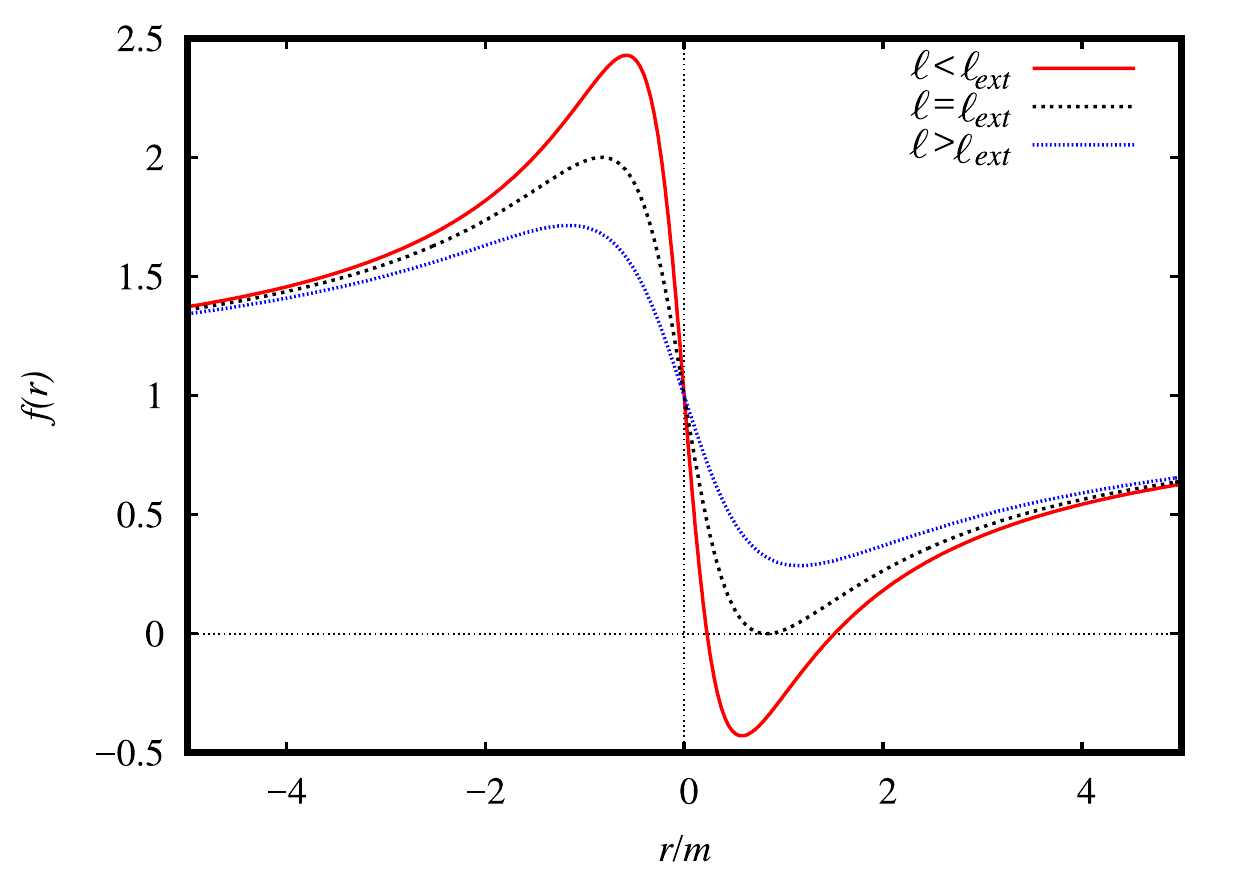}
	\includegraphics[scale=0.7]{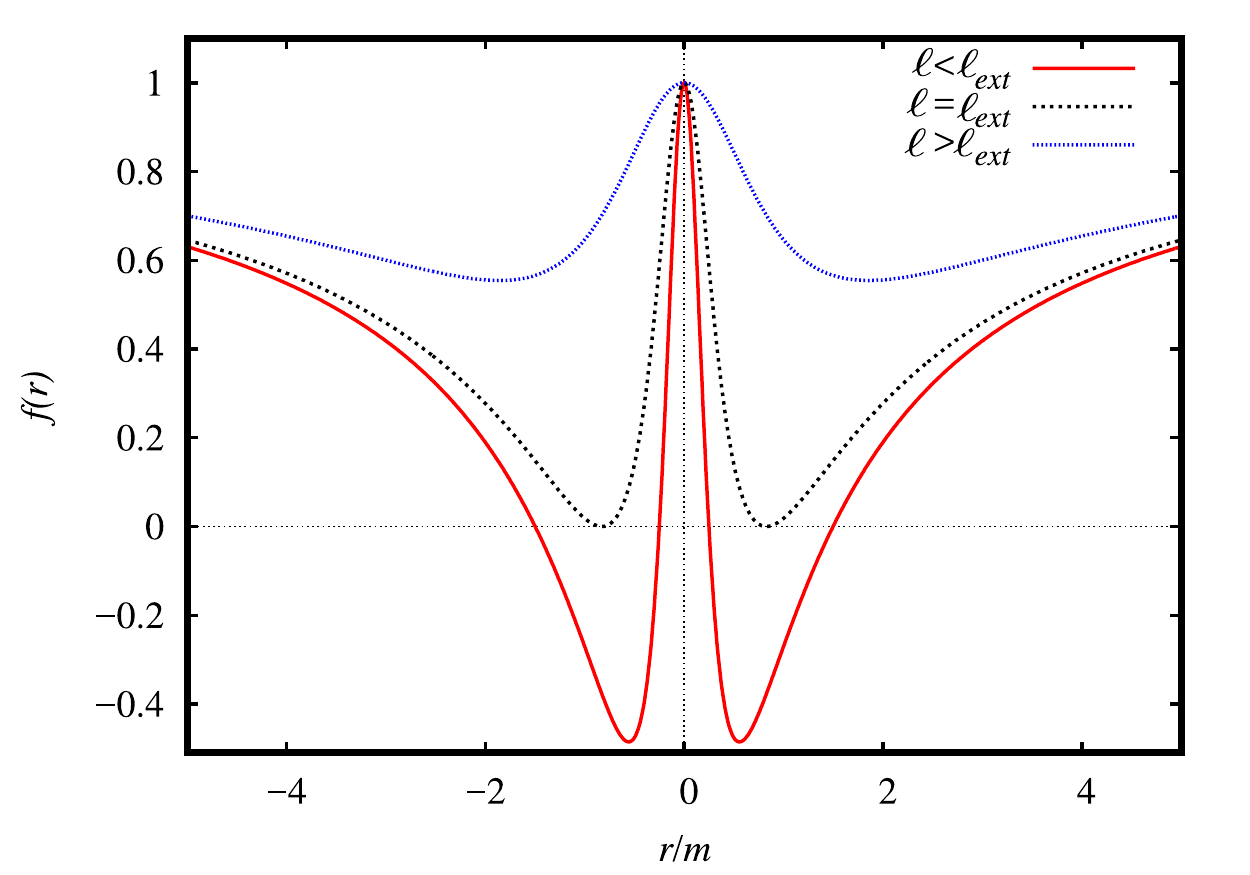}
	\caption[Graphical representation of $f(r)$, for various values of $\ell$, for the $M(r)=m\arctan^n(r/\ell)(2/\pi)^n$ model]{Graphical representation of $f(r)$ for the model given by Eq.~(\ref{mod5}), depicting $n=1$ (upper) and $n=2$ (lower) for different values of $\ell$.}\label{fig9}
\end{center}
\end{figure}

Another mass function which can provide a positive energy density is given by $M(r)=m\arctan^n(r/\ell)(2/\pi)^n$, so that $f(r)$ takes the form
\begin{eqnarray}
f(r) = 1-\frac{2M(r)}{\Sigma(r)} = 1-\frac{2m\arctan^n(r/\ell)}{\sqrt{r^2+\ell^2}}\left(\frac{2}{\pi}\right)^n \ .\label{mod5}
\end{eqnarray}
$n\rightarrow 0$ returns standard SV spacetime, and when $(a,n)\rightarrow 0$, one recovers the Schwarzschild solution. The Kretschmann scalar is easily verified to be globally finite. Fig.~\ref{fig9} shows $f(r)$, where for the cases $n=1$ and $n=2$, there are horizons dependent on the values of $\ell$: when $n=1$, $\ell_{ext}=0.714410046190945m$, and when $n=2$, $\ell_{ext}=0.4456300400812961661m$.

\enlargethispage{20pt}

The causal structures are: (i) both the cases $n=1$ with $\ell>\ell_{ext}\approx0.714m$, and $n=2$ with $\ell>\ell_{ext}\approx0.446m$, are standard two-way traversable wormholes; (ii) when $n=2$ and $\ell=\ell_{ext}\approx0.446m$, see Fig.~\ref{figPenrose4}; (iii) when $n=1$ and $\ell=\ell_{ext}\approx0.714m$, see Fig.~\ref{figPenrose5}; (iv) when $n=1$ and $\ell<\ell_{ext}\approx0.714m$, see Fig.~\ref{figPenrose6}.

\begin{figure}[htb!]
\begin{center}
    \includegraphics[scale=0.7]{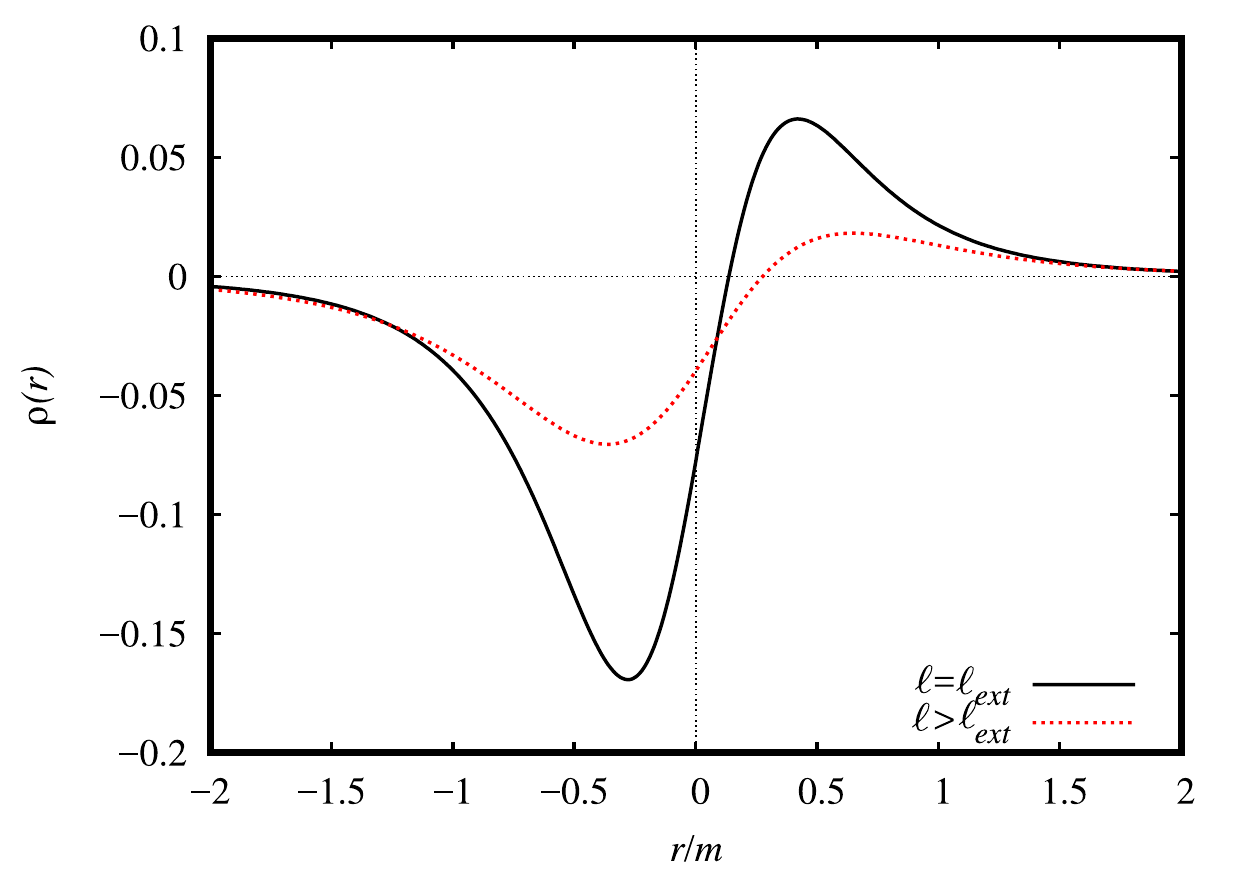}
	\includegraphics[scale=0.7]{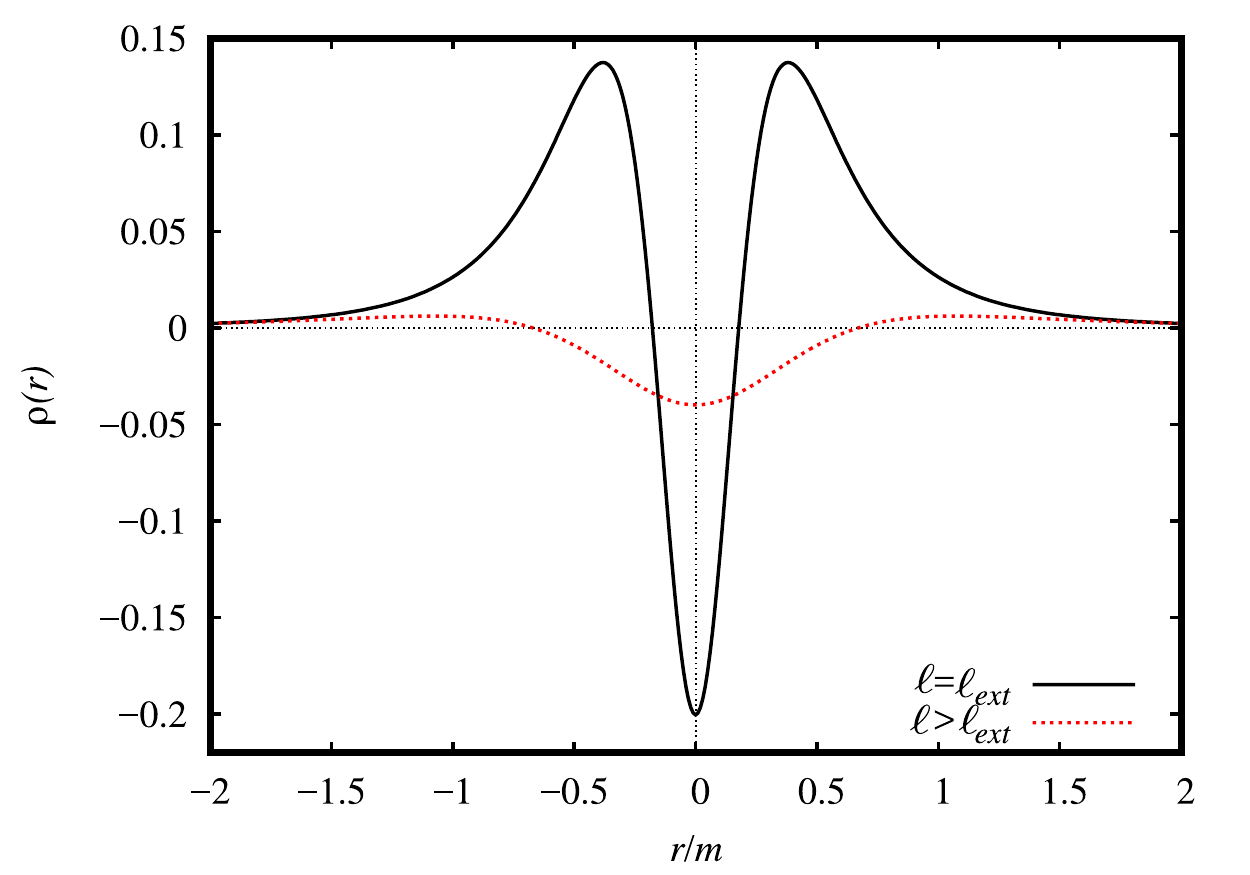}
	\caption[Graph of the energy density for certain sub-cases of the\newline $M(r)=m\arctan^n(r/\ell)(2/\pi)^n$ model]{Energy density for the model given by Eq.~(\ref{mod5}), showing $n=1$ (upper) and $n=2$ (lower).}\label{fig10}
\end{center}
\end{figure}

Notice that for odd $n$, the positive and negative regions of $r$ are not symmetric, contrary to the situation for even $n$. In the $n=1$ case, the energy density is positive for $r>r_+$, and negative for $-\infty<r<r_+$; this is depicted in the upper plot of Fig.~\ref{fig10}. In the lower plot of Fig.~\ref{fig10}, the $n=2$ case is shown, where the energy density is positive for $r>r_+$, and negative inside of the horizon.

The Hernandez--Misner--Sharp mass is given by
\begin{eqnarray}
M_\HMS(r) = \frac{\ell^2}{2\sqrt{r^2+\ell^2}}+\frac{m\left(\frac{2}{\pi}\right)^nr^2\arctan\left(\frac{r}{\ell}\right)^n}{r^2+\ell^2} \ .\label{massmod5}
\end{eqnarray}
For even $n$ the mass is always positive, and possesses the limits:\newline $\lim_{r\rightarrow \infty}M_\HMS(r)=m$, and $\lim_{r\rightarrow 0}M_\HMS(r)=\ell/2$.
\clearpage

\section{Conclusion of Part I}\label{NBB:conclusion}
\enlargethispage{10pt}

%
%
%
\enlargethispage{30pt}
In Chapter~\ref{C:SVNBB}, several new classes of black-bounce which generalise the geometry of original SV spacetime have been presented and analysed. All models have been shown to be everywhere-regular, and to have an area of the angular part of the metric that is always positive and nonzero. All new models have an everywhere-positive Hernandez--Misner--Sharp mass (also explicitly shown for the original SV case). The analyses reveal that it is straightforward to create extensions to SV spacetime which contain not only an arbitrary number of horizons, but also various combinations of event, Cauchy, and degenerate extremal horizons. This extends the discourse surrounding causal structures in regular spacetimes beyond those usually considered. Analysis of the point-wise energy conditions for each model demonstrates that it is theoretically possible to force satisfaction of the WEC in the domain of outer communication, and is also possible to satisfy the condition $SEC_3$, although the SEC will generically still be violated \emph{somewhere} for regular black holes in static spherical symmetry~\cite{RBHAEC:Zaslavskii:2010}. Furthermore, Theorem~\ref{Theorem:EC} concerning (non)satisfaction of the energy conditions in static spherical symmetry is now in-hand. Overall, these are just a small subset of the salient features of these new black-bounce models. One could also study geodesics, dynamic thin-shells, thermodynamics, the scattering and absorption of quantum fields, shadows (silhouettes) and quasinormal modes for each candidate spacetime, or indeed in a more general setting.

In Chapters~\ref{C:SVog} and~\ref{C:SVCBB}, the black-bounce analogs to Schwarzschild, Reissner--Nordstr\"{o}m, Kerr, and Kerr--Newman spacetimes have been presented and explored. These are the SV, bbRN, bbK, and bbKN spacetimes respectively. All are smooth interpolations between (charged or uncharged) regular black holes and traversable wormholes, depending on the value of the bounce parameter $\ell$, and all are amenable to highly tractable analysis and the extraction of (potential) astrophysical observables. These include the notable orbits for each candidate spacetime. One is able to easily discuss the relevant causal structures for each geometry, and to extract the surface gravity if appropriate. Notably, both the SV and bbRN spacetimes are now exact solutions of the Einstein equations, courtesy of Bronnikov and Walia's work~\cite{FSFSVS:Bronnikov:2022} finding suitable source terms in the Lagrangian formalism.

%
It is worth noting that in the past two-three years, as at March 25th, 2023, the five major articles considered and discussed in Part~\ref{PartI} (these are references~\cite{BTTW:Simpson:2019, VSBATW:Moruno:2019, ANFORBHM:Franzin:2021, CBBS:Franzin:2021, NBB:Lobo:2021}) have been cited a cumulative total of 393 times between them. An exhaustive post-literature review for the entire family of black-bounce spacetimes is an enormous task; the intention of Part~\ref{PartI} is to capture only a subset of all available results, specifically those which are most fundamental and pertain to the root spacetimes from which the subsequent analyses stem. Of the multitude of future research opportunities involving the black-bounce family, perhaps the most crucial are the following:
\begin{itemize}
    \item Elevating the bbK and bbKN spacetimes to ``exact solution status'' by finding the correct combination of suitable source terms in the Lagrangian formalism is likely the most important result still outstanding. Given these are the black-bounce analogs to Kerr and Kerr--Newman respectively, they are likely to be the most astrophysically appropriate candidate spacetimes from the family. Associating the bounce parameter $\ell$ with the Planck length is to conjecture the existence of some hypothetical quantum mechanism which acts as the regulariser for the candidate spacetimes; once the appropriate Lagrangian is discovered, comparison with candidate Lagrangians arising from quantum field theory could help shed light on whether such a mechanism is more than just hypothetical. It should be noted that while the ``eureka'' result is highly unlikely, this strategy is a tractable path to progress, and is consistent with the framework of ``turning one knob at a time'' as explored in the Introduction (Chapter~\ref{C:intro}, \S~\ref{Intro:knob}).
    \item Continuing to streamline the discourse between theory and experiment is crucial. While significant research has already been performed which begins to quantify the potential astrophysical observables for various black-bounce candidates (\textit{e.g.} for standard SV spacetime the quasi-normal modes~\cite{EIBW:Bronnikov:2020, ROTRBHWT:Churilova:2020}, shadows and gravitaional lensing effects~\cite{SGLBRSVBH:Ghosh:2021, GLBTPSIABSISDL:Tsukamoto:2021, SAOAOBBIBATAD:Guerrero:2021, CDBHCTSS:Crispino:2021, GLITSVBSIASDL:Tsukamoto:2021, GLIBS:Nascimento:2020}, and precession phenomena~\cite{PAPMAABTW:Zhou:2020}), it is critical that engagement between the theoretical, numerical, experimental, and observational relativity communities is prioritised. Taking the stance of a strict empiricist, understanding the specific sensitivities of the instruments involved in the field's current ``golden era'' of experimental and observational capacity is paramount. The associated margins of error for specific measurements can be used to inform the most prudent line of inquiry for both the numerical and theoretical researchers currently operating in gravitational physics; there is little point to constructing toy models without an end goal which appeals to the traditional scientific method.
\end{itemize}
\vspace*{60pt}

\begin{center}
    This concludes Part~\ref{PartI}.
\end{center}
\vspace*{8pt}

%% file: 03-AMC/1-AMCog.tex
\part[The family of black holes with asymptotically \\\hspace*{.8cm} Minkowski cores]{The family of black holes with asymptotically Minkowski cores}\label{PartII}

\chapter{``aMc'' spacetime}\label{C:AMCog}

%
%
%
%
%
\vspace*{7pt}

Another family of candidate geometries modelling alternatives to classical black holes is the family of regular black holes with asymptotically Minkowski cores~\cite{RBHWAMC:Simpson:2019, PSISCOOSCO:Berry:2020, TTWCFARBHWAMC:Berry:2020, EOS:Simpson:2021, AVKS:Simpson:2022}. The family stems from the analog to Schwarzschild spacetime, analysed in reference~\cite{RBHWAMC:Simpson:2019}, simply dubbed ``the regular black hole with asymptotically Minkowski core''. For compactness of notation, this specific geometry is henceforth labelled ``aMc'' spacetime. All members of this family model regular black holes in the sense of Bardeen~\cite{Tbilisi:Bardeen:1968}, are curvature-regular, and pass all weak-field observational tests of standard GR. This family has seen slightly less popularity in the community than the black-bounce family of Part~\ref{PartI}, but has still commanded attention and interest in the past two years; for instance as at March 25th, 2023, reference~\cite{RBHWAMC:Simpson:2019} has received 76 citations since publication in December 2019. It is the current author's opinion that this family, in particular the ``eye of the storm'' (eos) geometry of Chapter~\ref{C:AMCEOS}, is more highly desirable than the black-bounce family for several key reasons --- this is discussed at length throughout Part~\ref{PartII}.

In Chapter~\ref{C:AMCog}, the static spherically symmetric aMc spacetime is presented and discussed, and its key features analysed. The main results are largely drawn from references~\cite{RBHWAMC:Simpson:2019, PSISCOOSCO:Berry:2020}. These include analysis of curvature quantities, causal structure, surface gravity, notable orbits, and the stress-energy tensor and associated energy conditions. At the conclusion of Chapter~\ref{C:AMCog}, a proposal is presented for a geometry which acts as the aMc analog to Reissner--Nordstr\"{o}m spacetime, ``aMcRN'' spacetime; this is thematically similar to the bbRN spacetime of \S~\ref{SVbbRN}. Its global regularity is demonstrated, while investigation of other features is left as an avenue for further research. This candidate geometry is seemingly new to the literature, first presented in this thesis.
\clearpage

In Chapter~\ref{C:AMCQNM} an analysis of the quasi-normal modes of aMc spacetime is performed; the fundamental modes of spin zero and spin one perturbations are extracted to first-order using the WKB approximation (for details on these pieces of terminology, see Chapter~\ref{C:AMCQNM}, or the original work in reference~\cite{ROTRBHWAMC:Simpson:2021}). Qualitative features of the quasi-normal mode profiles are then compared with known results, with the long-term aim of speaking directly to the ringdown calculation for the LIGO/Virgo collaboration~\cite{BBHMITFALIGOOR:Abbott:2016, OOGWFABBHM:Abbott:2016, OOA50SMBBH:Abbott:2017, AIOTBBHM:Abbott:2016, POTBBHM:Abbott:2016, GW23SM:Abbott:2020, GWTC1:Abbott:2019}, or indeed LISA~\cite{LISA:Barausse:2020}. Remaining in the geometrical environment of spherical symmetry, in Part~\ref{PartIII}, Chapter~\ref{C:SVthinshell}, a dynamical thin-shell variant of aMc spacetime is briefly discussed which goes beyond the static case; this research was first conducted in reference~\cite{TTWCFARBHWAMC:Berry:2020}.

Chapter~\ref{C:AMCEOS} migrates the discourse to the geometric environment of stationary axisymmetry. It is here that the highly desirable eos spacetime~\cite{EOS:Simpson:2021, AVKS:Simpson:2022} is explored, constructed according to a set of carefully chosen criteria for the metric. This model is the analog to the Kerr solution. It is extraordinarily tractable, possesses many desirable geometric features, is extremely well-behaved in the context of the point-wise energy conditions, and amenable to the straightforward extraction of astrophysical observables potentially falsifiable/verifiable by the experimental and observational communities. Argument is made that this is the most well-motivated model regular black hole currently available in the GR literature. At the end of Chapter~\ref{C:AMCEOS}, in \S~\ref{eosKN}, a proposal is presented for a geometry which would act as the analog to Kerr--Newman spacetime, ``eosKN'' spacetime; this is thematically similar to the bbKN spacetime of Chapter~\ref{C:SVCBB}. This is a new geometry presented for the first time in this thesis. Further research on eosKN spacetime is relegated to the domain of the future.
\enlargethispage{20pt}

In the domain of outer communication, all candidate spacetimes discussed henceforth have metric signature $(-,+,+,+)$.
\enlargethispage{10pt}

\section{Preliminary geometric analysis}\label{AMC1}

It is well established that all static spherically symmetric spacetimes have a line element which can without loss of generality be represented as~\cite{LW:Visser:1995}
\begin{equation}\label{Eq1}
    \d s^2 = -\e^{-2\Phi(r)}\left(1-\frac{2m(r)}{r}\right)\,\d t^2 + \frac{\d r^2}{1-\frac{2m(r)}{r}} + r^2\,\d\Omega^2_{2} \ .
\end{equation}
Here $\Phi(r)$ and  $m(r)$ are \emph{a priori} arbitrary functions of $r$. The subsequent analysis could equivalently be performed in the Buchdahl gauge of Chapter~\ref{C:SVNBB}, where the two free functions $f(r), \Sigma(r)$ are used instead, in slightly different places in the metric. Ultimately, with standard curvature coordinates in generic static spherical symmetry one must have two independent free functions of the radial coordinate $r$ \textit{somewhere} in the metric. The chosen gauge ought to be that which renders the relevant discussion the most tractable.

Historically, the majority of prominent regular black hole models are of the form of Eq.~(\ref{Eq1}), and they exhibit an asymptotically de Sitter core with finite central energy density and an equal-but-opposite central pressure (see for instance the commonly studied Bardeen/Hayward/Frolov~\cite{Tbilisi:Bardeen:1968, FAEONBH:Hayward:2006, ILPAABHMWACAH:Frolov:2014} regular black holes, as well as the Mazur--Mottola gravastars~\cite{GCS:Mazur:2002, GVCS:Mazur:2004, DEACS:Mazur:2005, SG:Visser:2004}, and Dymnikova's models~\cite{Dymnikova1, Dymnikova2, Dymnikova3, Dymnikova4, Dymnikova5, Dymnikova6, Dymnikova7}). The aMc spacetime under consideration instead possesses an asymptotically Minkowski core (the energy density and associated pressures asymptote to zero). This physical difference can be made mathematically explicit by examining the required conditions on the two functions $\Phi(r)$ and $m(r)$ appearing in the metric given by Eq.~(\ref{Eq1}):
\begin{itemize}
    \item \emph{de Sitter core:} One would require that $\lim_{r\rightarrow 0}\rho(r) = \rho_0 \neq 0$; that is, $\rho(r) = \O(1)$. For static spherically symmetric spacetimes, $\rho(r)=m'(r)/4\pi r^{2}$~\cite{LW:Visser:1995, GR:Wald:2010, G:Misner:2000}, so this implies $m(r) = \O(r^3)$. For $\Phi(r)$ it is sufficient to demand $\Phi(r) = \O(1)$. The standard Bardeen/Hayward/\newline Frolov/gravastar/Dymnikova models all fall into this class.
    \item \emph{asymptotically Minkowski core:} One would require that $\lim_{r\rightarrow 0}\rho(r) = 0$; that is,  $\rho(r) = o(r)$. This in turn implies $m(r) = o(r^3)$. For $\Phi(r)$ it is once again sufficient to demand $\Phi(r) = \O(1)$.
\end{itemize}
%
To keep the discourse as close to Schwarzschild as possible, one may as well immediately set $\Phi(r) = 0$. This preserves all of the key features in the desired physics while heightening mathematical tractability. $\Phi(r)=0$ spacetimes have a long and respected history. Specifically, see references~\cite{LW:Visser:1995, WIGTT:Jacobson:2007, Kiselev:Boonserm:2020, Kiselev:Visser:2020}.

The choice for $m(r)$ is more subtle. By including a new scalar parameter $\ell$ \textit{via} choosing $m(r)=\e^{-\ell/r}$,\footnote{Similarly to the discourse in Chapter~\ref{C:SVog}, while the original articles~\cite{RBHWAMC:Simpson:2019, PSISCOOSCO:Berry:2020, TTWCFARBHWAMC:Berry:2020} use $a$ for the scalar parameter, $\ell$ is employed so as not to create confusion with the spin parameter $a$ from Kerr spacetime. This will be particularly important in Chapter~\ref{C:AMCEOS}.} aMc spacetime has the effect of ``exponentially suppressing'' the mass of the centralised object as one nears the coordinate location $r = 0$ (that is, for $\ell>0$). This guarantees that the object has an asymptotically Minkowski core. Many other choices for $m(r)$ are possible; this is discussed further in \S~\ref{EOSSS}. Note that the ``suppression'' parameter $\ell$ is different from the ``bounce'' parameter $\ell$ of Part~\ref{PartI}, though the usual interpretation of both would be to associate them with the Planck scale, and hence the existence of some hypothetical quantum mechanism. Furthermore, note that while the exponential function $\e^{-\ell/r}$ guarantees that for $r>0$ everything is $C^{\infty}$-smooth, analyticity is lost, and there is a severe discontinuity at $r=0$, rendering the negative-$r$ domain grossly unphysical. One may safely omit $r<0$ from the analysis.

The aMc line element is given explicitly by~\cite{RBHWAMC:Simpson:2019}
\begin{equation}\label{amcmetric}
    \d s^2 = -\left(1-\frac{2m\,\e^{-\ell/r}}{r}\right)\d t^2 + \frac{\d r^2}{\left(1-\frac{2m\,\e^{-\ell/r}}{r}\right)} + r^2 \; \d\Omega^2_{2} \ .
\end{equation}
Note that $\ell\to0$ returns Schwarzschild spacetime.

Conducting a standard GR analysis of the resulting  spacetime, it is straightforward to demonstrate that this metric does indeed correspond to a regular black hole in the sense of Bardeen. Furthermore, for this toy model the resulting curvature quantities are significantly simpler than the analogous results for Bardeen/Hayward/Frolov, or indeed SV spacetime, at the cost of having many physically interesting features defined by the Lambert $W$ function, one of the special functions of mathematics~\cite{SAOTLWFTP:Valluri:2000, TLWFAQS:Valluri:2009, BTGFFSBH:Boonserm:2008, QF:Boonserm:2011, RELSAGFFDBH:Boonserm:2013, EMRATW:Boonserm:2018, SRGEWTLWF:Sonoda:2013, AFOTEPITLNL...:Sonoda:2013, OTLWF:Corless:1996, FWLDAWDL:Vial:2012, WPATLWF:Stewart:2011, SPAWDL:Stewart:2012, PATLWF:Visser:2018}. This special function is finding increasingly common applications in physics.

A rather different (extremal) version of the aMc model, based on nonlinear electrodynamics, has been previously discussed by Culetu~\cite{OARMSS:Culetu:2013}, with follow-up on some aspects of the non-extremal case in references~\cite{NBHWANES:Culetu:2015, OARCBHWANES:Culetu:2015, SAEBHWATSOEM:Culetu:2016}. See also references~\cite{RBHWANES:Balart:2014, RBHIf(T):Houndjo:2015, RBHIf(R)GCTNE:Junior:2016}. Part of the GRF essay~\cite{SATFOBHE:Xiang:2013} is based on a mass function of the form $m(r)= m\,\e^{-\ell^2/r^2}$ (this choice of mass function is discussed further in \S~\ref{EOSSS}). Another very different type of exponential modification of Schwarzschild and Kerr geometries, obtained by inserting factors of the form $1-a\exp(-b^3/r^3)$ into the spacetime metric, has been discussed by Takeuchi~\cite{HFAAIRRBHWAT:Takeuchi:2016}. The overall framework in these references is significantly different from that of reference~\cite{RBHWAMC:Simpson:2019}, and of Chapter~\ref{C:AMCog}.

\subsection{Curvature quantities}\label{suppressedtensors}
\vspace*{8pt}

Before proceeding it is prudent to introduce a relevant piece of mathematical detail: For any polynomial function $p(r)$, when $\ell>0$, ${\e^{-\ell/r}}/{p(r)} \rightarrow 0$ as $ r\rightarrow 0^{\mbox{+}}$. That is, the exponential expression dominates the asymptotic behaviour for small $r$. The relevant curvature quantities for standard aMc spacetime are displayed below.

Riemann tensor nonzero components:
\begin{eqnarray}
    R^{tr}{}_{tr} &=& \frac{m\left(\ell^{2}-4\ell r+2r^{2}\right)\e^{-\ell/r}}{r^{5}} \ , \nonumber \\
    R^{t\theta}{}_{t\theta} &=& R^{t\phi}{}_{t\phi} = R^{r\theta}{}_{r\theta} = R^{r\phi}{}_{r\phi} = \frac{m\left(\ell-r\right)\e^{-\ell/r}}{r^{4}} \ , \nonumber \\
    R^{\theta\phi}{}_{\theta\phi} &=& \frac{2m\,\e^{-\ell/r}}{r^{3}} \ .
\end{eqnarray}
Ricci tensor nonzero components:
\begin{equation}
    R^{t}{}_{t} = R^{r}{}_{r} = \frac{m\,\ell\left(\ell-2r\right)\e^{-\ell/r}}{r^{5}} \ , \qquad 
    R^{\theta}{}_{\theta} = R^{\phi}{}_{\phi} = \frac{2m\,\ell\,\e^{-\ell/r}}{r^{4}} \ .
\end{equation}
Einstein tensor nonzero components:
\begin{equation}\label{einsteinsuppressed}
    G^{t}{}_{t} = G^{r}{}_{r} = -\frac{2m\,\ell\,\e^{-\ell/r}}{r^{4}} \ , \qquad 
    G^{\theta}{}_{\theta} = G^{\phi}{}_{\phi} = -\frac{m\ell\left(\ell-2r\right)\e^{-\ell/r}}{r^{5}} \ .
\end{equation}
Weyl tensor nonzero components:
\begin{eqnarray}
    -\frac{1}{2}C^{tr}{}_{tr} = -\frac{1}{2}C^{\theta\phi}{}_{\theta\phi} = C^{t\theta}{}_{t\theta} &=& C^{t\phi}{}_{t\phi} = C^{r\theta}{}_{r\theta} = C^{r\phi}{}_{r\phi} \nonumber \\
    &=& -\frac{m\left(\ell^{2}-6\ell r+6r^{2}\right)\e^{-\ell/r}}{6r^{5}} \ .
\end{eqnarray}
Ricci scalar:
\begin{equation}
    R = \frac{2m\,\ell^{2}\,\e^{-\ell/r}}{r^{5}} \ .
\end{equation}
The Ricci contraction $R_{\mu\nu}\,R^{\mu\nu}$:
\begin{equation}
    R_{\mu\nu}\,R^{\mu\nu} = \frac{2m^{2}\ell^{2}\left(\ell^{2}-4\ell r+8r^{2}\right)\e^{-2\ell/r}}{r^{10}} \ .
\end{equation}
The Kretschmann scalar $R_{\mu\nu\alpha\beta}\,R^{\mu\nu\alpha\beta}$:
\begin{equation}
    K = \frac{4m^{2}\left(\ell^{4}-8\ell^{3}r+24\ell^{2}r^{2}-24\ell r^{3}+12r^{4}\right)\e^{-2\ell/r}}{r^{10}} \ .
\end{equation}
%
The Weyl contraction $C_{\mu\nu\alpha\beta}\,C^{\mu\nu\alpha\beta}$:
\begin{equation}
    C_{\mu\nu\alpha\beta}\,C^{\mu\nu\alpha\beta} = \frac{4m^{2}\left(\ell^{2}-6\ell r+6r^{2}\right)^{2}\e^{-2\ell/r}}{3r^{10}} \ .
\end{equation}
Analysis of the Kretschmann scalar demonstrates it is globally finite, hence by Theorem~\ref{Theorem:Kretsch}, the spacetime is curvature-regular. Notably, as $r\to0$, all curvature quantities tend to zero, reflecting the fact that the core is asymptotically Minkowski. As $r\to+\infty$, all curvature quantities tend to zero; asymptotic flatness is preserved.

One of the primary benefits of the aMc model is its relative tractability when compared with the majority of models in the literature. For instance, in reference~\cite{RBHWAMC:Simpson:2019}, an explicit comparison of the nonzero components of the Einstein tensor is performed, comparing aMc with the usual Bardeen, Hayward, and Frolov models. One can also compare with the results from SV spacetime from reference~\cite{BTTW:Simpson:2019}. The expressions for aMc are \emph{significantly} more tractable; this comes at the cost of defining qualitatively important features of the geometry in terms of the Lambert $W$ function, such as the horizon locations (see below).

\subsection{Horizons and surface gravity}
\label{sec:3+1}

%
The Lambert $W$ function is defined as $W(x)\e^{W(x)}=x$. This special function of mathematics is becoming increasingly important in theoretical phy\-sics, and is a critical tool in the analysis of aMc spacetime.

In view of the diagonal metric environment of Eq.~(\ref{amcmetric}), one may examine horizon locations by simply setting $g_{tt}=0$:
\begin{equation}
    g_{tt} = 0 \quad \Longrightarrow \quad 
    r = -\frac{\ell}{W\!\!\left(-\frac{\ell}{2m}\right)} \quad \Longrightarrow \quad 
    r = 2m\,\e^{W\!\left(-\frac{\ell}{2m}\right)} \ .
\end{equation}
There is a coordinate location for the horizon defined \emph{explicitly} in terms of the Lambert $W$ function. Given both $\ell,r>0$, in order for horizons to be defined there are two possibilities:
\begin{itemize}
\item 
Taking the $W_{0}\left(x\right)$ branch of the real-valued Lambert $W$ function:
\begin{equation}
 W_{0}\!\left(-\frac{\ell}{2m}\right)<0 \quad \Longrightarrow \quad \ell\in\,\left(0, \ \frac{2m}{\e}\right] \ .
\end{equation}
Fixing $\ell$ in this interval causes $W_{0}\left(-\frac{\ell}{2m}\right)\in\,\left[-1, \ 0\right)$, hence the possible coordinate locations of \emph{this} horizon are given by $r_{H}\in\,[\ell, +\infty)$.
\item 
Taking the $W_{-1}\left(x\right)$ branch of the real-valued Lambert $W$ function: The $W_{-1}(x)$ branch only returns outputs for $x\in\,\left[-\frac{1}{\e},0\right)$, hence there are the same restrictions on $\ell$ as for the previous case; $\ell\in\,\left(0,\frac{2m}{\e}\right]$. The difference is that fixing $\ell$ in this interval causes $W_{-1}\left(-\frac{\ell}{2m}\right)\in\,[-1, -\infty)$, hence the possible coordinate locations for \emph{this} horizon are given by $r_{H}\in\,(0, \ell]$.
\end{itemize}
It follows that when fixing $\ell\in\,\left(0,\frac{2m}{\e}\right)$, there is an event (outer) horizon at $r_{H_{+}}=2m\,\e^{W_{0}\left(-\frac{\ell}{2m}\right)}$, and a Cauchy (inner) horizon at $r_{H_{-}}=2m\,\e^{W_{-1}\left(-\frac{\ell}{2m}\right)}$. When $\ell=2m/\e$, the two horizons merge to single degenerate extremal horizon. When $\ell>2m/\e$, the geometry is horizonless, modelling some compact massive object other than a black hole; \textit{e.g.} a candidate gravastar, star, or planet.

Perturbatively, for small $\ell$ one has
\begin{equation}
r_{H_{+}} = 2m -\ell +\O(\ell^2) \ ,\label{E:outer-pert}
\end{equation}
cleanly reproducing Schwarzschild in the $\ell\to0$ limit. For the inner horizon, since $r_{H_-}<2m$, then
\begin{equation}
r_{H_-} = {\ell\over \ln(2m/r_{H_-})}
\end{equation}
implies that $r_{H_-}<\ell$. Hence there is a strict upper bound given by the simple analytic expression:
\begin{equation}
r_{H_-} < {\ell\over \ln(2m/\ell)} \ .\label{E:inner-bound}
\end{equation}
Certainly $\lim_{\ell\to0} r_{H_-}(m,\ell)=0$, as expected, but the form of $r_{H_-}(m,\ell)$ is not analytic. This bound can also be viewed as the first term in an asymptotic expansion~\cite{OTLWF:Corless:1996}, based on (as $x\to 0^+$)
\begin{equation}
W_{-1}(-x) = \ln(x) + \O\lbrace\ln\left[-\ln(x)\right]\rbrace = -\ln(1/x) + \O\lbrace\ln\left[\ln(1/x)\right]\rbrace \ ,
\end{equation}
leading to
\begin{equation}
r_{H_-} = {\ell\over \ln(2m/\ell) +\O\lbrace\ln\left[\ln(2m/\ell)\right]\rbrace} 
= {\ell\over \ln(2m/\ell)} +
\O\left\lbrace\ell\ln\left[\ln(2m/\ell)\right]\over \left[\ln(2m/\ell)\right]^2 \right\rbrace \ .\label{E:inner-asymp}
\end{equation}
More specifically (as $\ell/m\to 0$ or $m/\ell\to+\infty$)
\begin{equation}
{r_{H_-} \over \ell} = {1\over \ln(2m/\ell)} +
\O\left\lbrace\ln\left[\ln(2m/\ell)\right]\over \left[\ln(2m/\ell)\right]^2 \right\rbrace \ .\label{E:inner-asymp2}
\end{equation}
%
The surface gravity for both horizons in the $\ell\in(0,\frac{2m}{\e})$ case is easily computed. The Killing vector which is null at the event horizon is $\xi=\partial_{t}$, that is 
$\xi^\mu=(1,0,0,0)^\mu$. This gives the norm:
\begin{equation}
    \xi^{\mu}\xi_{\mu} = g_{\mu\nu}\xi^{\mu}\xi^{\nu} = g_{tt} = -\left(1-\frac{2m\,\e^{-\frac{\ell}{r}}}{r}\right) \ .
\end{equation}
Given the following relation for the surface gravity $\kappa$ (see for instance references~\cite{GR:Wald:2010, G:Misner:2000}):
\begin{equation}
    \nabla_{\nu}\left(-\xi^{\mu}\xi_{\mu}\right) = 2\kappa\,\xi_{\nu} \ ,
\end{equation}
one finds that
\begin{equation}
 \nabla_{\nu}\left(1-\frac{2m\,\e^{-\frac{\ell}{r}}}{r}\right) = 2\kappa\,\xi_{\nu} \ .
\end{equation}
Now perform a case-by-case analysis for each horizon:
\begin{itemize}
    \item Outer: $r_{H_{+}} = 2m\,\e^{W_{0}\left(-\frac{\ell}{2m}\right)}$, implying
    \begin{eqnarray}
    \kappa_{out} &=& \frac{1}{2}\partial_{r}\left(1-\frac{2m\,\e^{-\frac{\ell}{r}}}{r}\right)\Bigg\vert_{r=2m\,\e^{W_{0}\left(-\frac{\ell}{2m}\right)}} \nonumber \\[3pt]
    &=& \kappa_{Sch.}\left\lbrace-\frac{2m\,W_{0}\left(-\frac{\ell}{2m}\right)}{\ell}\left[1+W_{0}\left(-\frac{\ell}{2m}\right)\right]\right\rbrace \ .
\end{eqnarray}
The associated Hawking temperature is given by~\cite{GR:Wald:2010, G:Misner:2000}:
\begin{equation}
   T_{out} = \frac{\hslash\kappa_{out}}{2\pi k_{B}} = T_{Sch.}\left\lbrace-\frac{2m\,W_{0}\left(-\frac{\ell}{2m}\right)}{\ell}\left[1+W_{0}\left(-\frac{\ell}{2m}\right)\right]\right\rbrace \ .
\end{equation}
The area of the outer horizon is given by
\begin{equation}
    A_{out} = 4\pi\left[2m\,\e^{W_{0}\left(-\frac{\ell}{2m}\right)}\right]^{2} = \frac{4\pi \ell^2}{\left[W_{0}\!\left(-\frac{\ell}{2m}\right)\right]^{2}} \ .
\end{equation}
%
\item Inner: $r_{H_{-}} = 2m\,\e^{W_{-1}\left(-\frac{\ell}{2m}\right)}$; through similar analysis one finds: 
\begin{eqnarray}
    \kappa_{inn} &=& \kappa_{Sch.}\left\lbrace-\frac{2m\,W_{-1}\!\left(-\frac{\ell}{2m}\right)}{\ell}\left[1+W_{-1}\!\left(-\frac{\ell}{2m}\right)\right]\right\rbrace \ , \\[3pt]
    T_{inn} &=& T_{Sch.}\left\lbrace-\frac{2m\,W_{-1}\!\left(-\frac{\ell}{2m}\right)}{\ell}\left[1+W_{-1}\!\left(-\frac{\ell}{2m}\right)\right]\right\rbrace \ , \\[3pt]
    A_{inn} &=& \frac{4\pi \ell^{2}}{\left[W_{-1}\!\left(-\frac{\ell}{2m}\right)\right]^{2}} \ .
\end{eqnarray}
\item Extremal case: Extremality arises when $\ell=2m/\e$, at coordinate location $r_{H_{ext}} = 2m/\e = \ell$, where $W_0(-{1\over \e})=-1=W_{-1}(-{1\over \e})$. Consequently $\kappa_{ext}, T_{ext}=0$, as expected, and trivially $A_{ext} = 4\pi \ell^2$.
\end{itemize}

\section{Notable orbits}\label{AMCISCOOSCO}
%
%
%

The notable orbits for aMc spacetime can be extracted, and present far more curious physics that one might first suspect. Mathematically, the analysis is also significantly more intricate than for many candidate spacetimes in static spherical symmetry.

\subsubsection{Geodesics and the effective potential}

In order to calculate the location of the photon sphere and extremal stable circular orbit (ESCO) for aMc spacetime, one must first establish the relevant effective potential. Begin by considering the affinely parameterised tangent vector to the worldline of a massive or massless particle in aMc spacetime:
\begin{multline}
g_{\mu\nu}\dv{x^\mu}{\lambda}\dv{x^\nu}{\lambda} = -\left(1-\frac{2m\,\e^{-\ell/r}}{r}\right)\left(\dv{t}{\lambda}\right)^2 + \left(\frac{1}{1-\frac{2m\,\e^{-\ell/r}}{r}}\right)\left(\dv{r}{\lambda}\right)^2  \\
+ r^2 \left[ \left(\dv{\theta}{\lambda}\right)^2 + \sin^2\theta \left(\dv{\phi}{\lambda}\right)^2 \right] = \epsilon \ ,
\label{tangent}
\end{multline}
where \( \epsilon \in \{-1,0\} \), with $-1$ corresponding to a massive (timelike) particle and 0 corresponding to a massless (null) particle. In view of spherical symmetry, one can set \( \theta = \pi/2 \) without any loss of generality and reduce Eq.~(\ref{tangent}) to
\begin{equation}
\epsilon = -\left(1-\frac{2m\,\e^{-\ell/r}}{r}\right)\left(\dv{t}{\lambda}\right)^2 + \left(\frac{1}{1-\frac{2m\,\e^{-\ell/r}}{r}}\right)\left(\dv{r}{\lambda}\right)^2 + r^2 \left(\dv{\phi}{\lambda}\right)^2 \ .
\label{tangentreduced}
\end{equation}
The presence of time-translation and angular Killing vectors implies the conserved quantities
\begin{equation}
E = \left(1-\frac{2m\,\e^{-\ell/r}}{r}\right)\left(\dv{t}{\lambda}\right) \qq{and} L = r^2\left(\dv{\phi}{\lambda}\right) \ ,
\end{equation}
corresponding to the energy and angular momentum of the particle, respectively. Thus, Eq.~(\ref{tangentreduced}) implies
\begin{equation}
E^2 = \left(\dv{r}{\lambda}\right)^2  + \left(1-\frac{2m\,\e^{-\ell/r}}{r}\right)\left(\frac{L^2}{r^2}-\epsilon\right) \ .
\end{equation}
This defines an ``effective potential'' for geodesic orbits
\begin{equation}
V_\epsilon (r) = \left(1-\frac{2m\,\e^{-\ell/r}}{r}\right)\left(\frac{L^2}{r^2}-\epsilon\right) \ ,
\end{equation}
with the circular orbits corresponding to extrema of this potential.

\subsection{Photon spheres}\label{aMcphotonspheres}

It is prudent to subdivide the discussion into two topics: First the \emph{existence} of circular photon orbits (photon spheres), and then the \emph{stability} of circular photon orbits. The discussion is considerably more complex than for the Schwarzschild spacetime, where there is only one unstable circular photon orbit, at $r=3m$. Once the extra parameter $\ell$ is nonzero, and in particular sufficiently large, the set of photon orbits exhibits more diversity.

\subsubsection{Existence of photon spheres}

\begin{figure}[!htbp]
\centering
\begin{subfigure}{.5\textwidth}
  \centering
  \includegraphics[width=.96\linewidth]{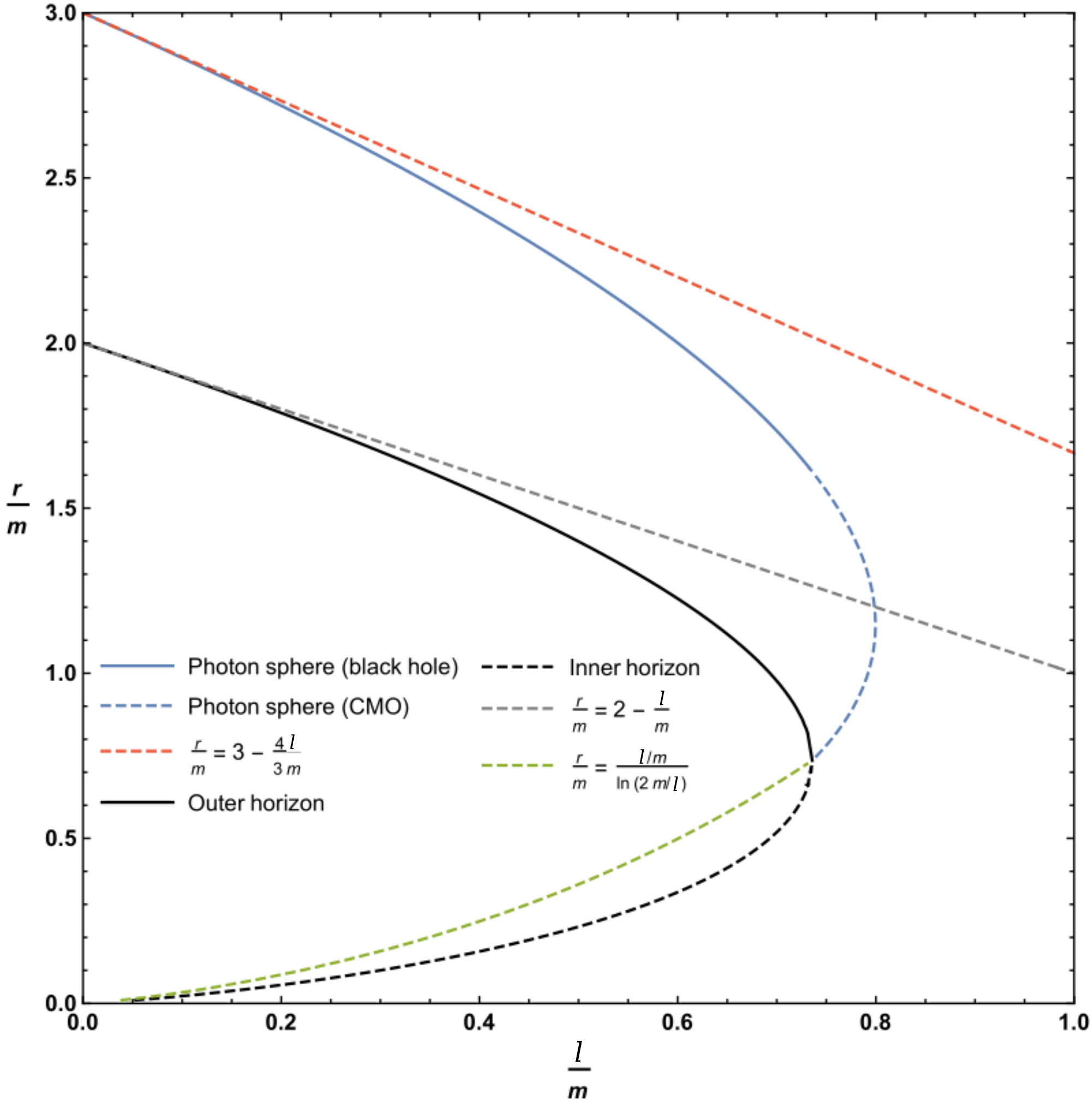}
  \caption{}
  \label{fig:iscom}
\end{subfigure}%
\begin{subfigure}{.5\textwidth}
  \centering
  \includegraphics[width=.96\linewidth]{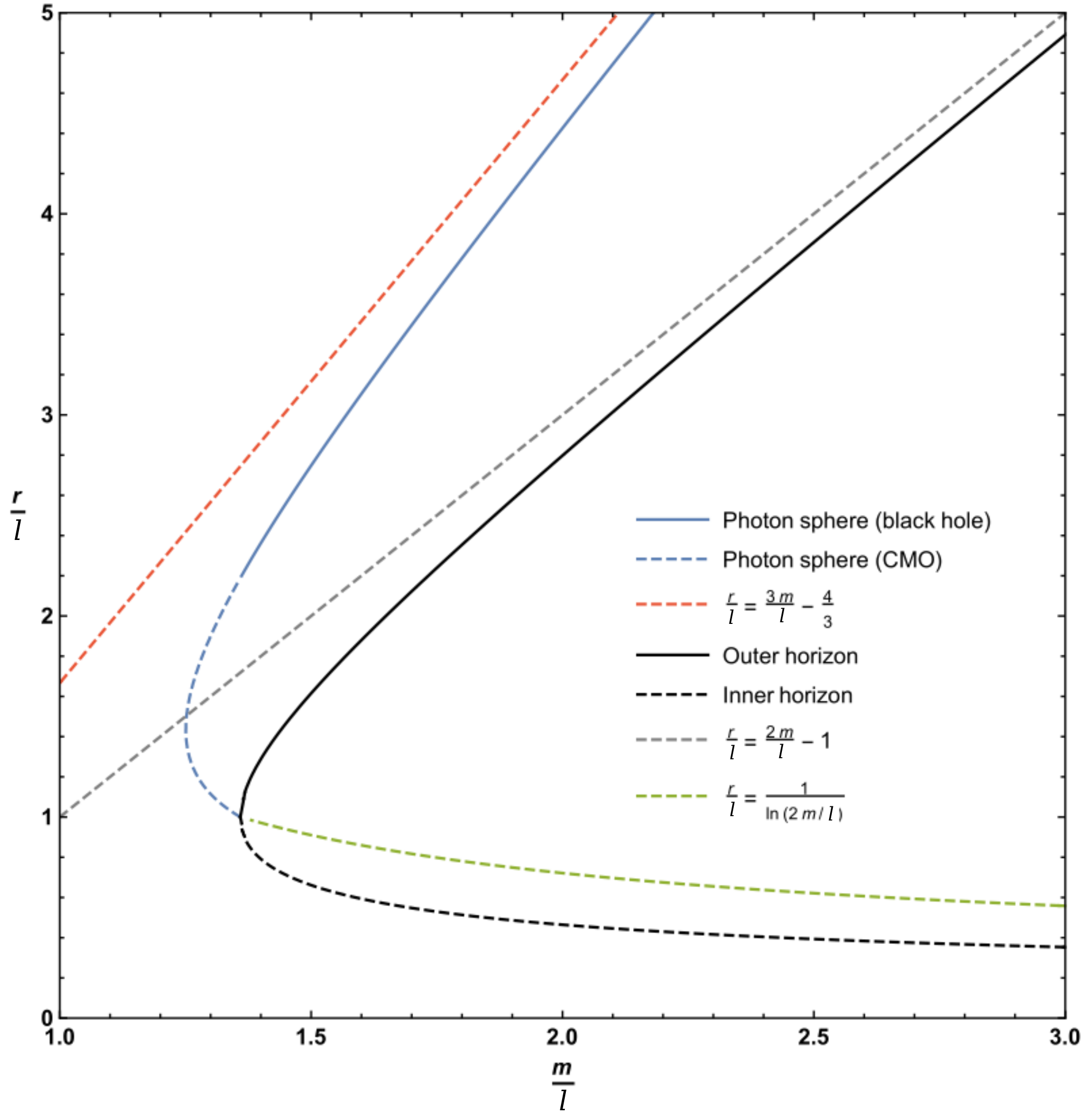}
  \caption{}
  \label{fig:iscoa}
\end{subfigure}
\caption[Location of the photon sphere, inner horizon, and outer horizon for aMc spacetime]{Location of the photon sphere, inner horizon,  and outer horizon as a function of the parameters \( \ell \) and \( m \). The dashed blue line represents the extension of the photon sphere to horizonless compact massive objects (CMOs), whilst the dashed red line is the asymptotic solution for small values of the parameter \( \ell \), given by Eq.~\eqref{E:small-a}. The dashed grey line is the asymptotic solution to the outer horizon for small values of \( \ell \), given by Eq.~\eqref{E:outer-pert}. The dashed green line is the simple analytic bound and asymptotic estimate for the location of the inner horizon, given by Eq's.~\eqref{E:inner-bound} and~\eqref{E:inner-asymp2}.}
\label{fig:photonsphere}
\end{figure}

For null trajectories one has
\begin{equation}
V_0(r) = \left(1-\frac{2m\,\e^{-\ell/r}}{r}\right)\frac{L^2}{r^2} \ .
\end{equation}
So for circular photon orbits
\begin{equation}
V_0'(r_\gamma) = \frac{2L^2}{r_\gamma^5} \left[ m\,\e^{-\ell/r_\gamma}(3r_\gamma-\ell)-r_\gamma^2 \right] = 0 \ .
\end{equation}
%
To be explicit about this, the location of a circular photon orbit, \( r_\gamma \), is given implicitly by the equation
\begin{equation}
r_\gamma^2 = m\,\e^{-\ell/r_\gamma}(3r_\gamma-\ell) \ ,\label{eq;photonsphere}
\end{equation}
where \( \ell \) and \( m \) are fixed by the geometry of the spacetime.\footnote{As $\ell\to 0$, $r_\gamma\to 3m$, as expected for Schwarzschild spacetime.} The curve described by the loci of these circular photon orbits has been plotted in two distinct ways in Fig.~\ref{fig:photonsphere}.

For clarity, defining  \( w = r_\gamma/\ell \) and \( z = m/\ell \), one can rewrite the condition for circular photon orbits as
\begin{equation}
w^2 = z \, \e^{-1/w} (3w-1) \ ;
\quad \implies \quad 
 z = {w^2 \; \e^{1/w}\over3w-1} \ .\label{E:z-for-photon}
\end{equation}
Fig.~\ref{fig:photonsphere} also contains plots of the locations of both inner and outer horizons.

The inner and outer horizons merge at $\ell/m=2/\e = 0.7357588824...$, that is, at $m/\ell = \e/2=1.359140914...$. For $\ell/m>2/\e$, that is for $m/\ell < \e/2$, one is dealing with a horizonless compact object and there is a region where there are \emph{two} circular photon orbits. Note that the curve described by the loci of circular photon orbits terminates once one hits a horizon, that is, at $w=1$. Sub-horizon curves of constant $r$ are spacelike (tachyonic), and \emph{cannot} be  lightlike, so they are explicitly excluded. That is, photon spheres can only \emph{exist} in the region $w\in(1,+\infty)$.

Given this, the next step is to analyse stability \emph{versus} instability, and find the exact location of the various turning points.

\subsubsection{Stability \emph{versus} instability for circular photon orbits} 

To check the \emph{stability} of these circular photon orbits, one needs to investigate
\begin{equation}
V_0''(r_\gamma) = \frac{2L^2}{r_\gamma^7} \left[ 3r_\gamma^3 - m\,\e^{-\ell/r_\gamma}(6r_\gamma-\ell)(2r_\gamma-\ell) \right] \ .
\label{E:V''}
\end{equation}

\textbf{Perturbative analysis (small $\ell$):} 

Note that determining $r_\gamma(m,\ell)$ from Eq.~(\ref{eq;photonsphere}) is not analytically feasible, but $r_\gamma(m,\ell)$ can certainly be estimated perturbatively for small $\ell$. One finds
\begin{equation}
r_\gamma(m,\ell) = 3m - \frac{4m\ell}{r_\gamma} + \mathcal{O}(\ell^2) \quad \implies \quad r_\gamma(m,\ell) = 3m - {4\over3}\ell + \mathcal{O}(\ell^2) \ .\label{E:small-a}
\end{equation}
So, for small values of \( \ell \), the standard result for $r_\gamma$ from Schwarzschild spacetime is recovered.

Estimating $V_0''(r_\gamma)$ by  now substituting the approximate location of the photon sphere as \( r_\gamma(m,\ell) = 3m - 4\ell/3 +\mathcal{O}(\ell^2)\) gives
\begin{equation}
V_0''(r_\gamma(m,\ell)) =  -{2L^2\over81 m^4}\left(1 +  {4\over3}\, {\ell\over m} +\mathcal{O}(\ell^2)\right) \ .\label{E:V''-small-a}
\end{equation}
This quantity is manifestly negative for small $\ell$. That is, within the limits of the current small-$\ell$ approximation, photons are in an unstable orbit at the small-$\ell$ photon sphere.

\textbf{Non-perturbative analysis:} 

Whilst determining $r_\gamma(m,\ell)$ is analytically infeasible, it should be noted that in contrast both $\ell(m,r_\gamma)$ and $m(r_\gamma,\ell)$ are easily determined analytically:
\begin{equation}
\ell(m,r_\gamma) = r_\gamma (3 - W(r_\gamma \e^3/m)) \ , \quad m(r_\gamma,\ell) = {r_\gamma^2 \; \e^{\ell/r_\gamma} \over(3r_\gamma-\ell)} \ .\label{E:non-pert}
\end{equation}
Consequently, at the peak one can write
\begin{equation}
V_0(r_\gamma,m) = {L^2\over r_\gamma^2} \left( 1- {2\over W(r_\gamma \e^3/m)}\right) \ ,
\quad
V_0(r_\gamma,\ell) = {L^2\over r_\gamma^2} \; {r_\gamma-\ell\over 3r_\gamma-\ell} \ . 
\end{equation}
Regarding stability, in the first case, substituting Eq.~(\ref{E:non-pert}\,a) into Eq.~\eqref{E:V''} gives
\begin{equation}
V_0''(r_\gamma,m) = -{2L^2
\left( W(r_\gamma \e^3/m)^2- W(r_\gamma \e^3/m)-3\right)\over r_\gamma^4  W(r_\gamma \e^3/m)} \ .
\end{equation}
\clearpage
Using properties of the Lambert $W$ function, one sees that this is negative for $r_\gamma/m >  {1\over 2} (1+\sqrt{13}) \; \e^{-5/2 +\sqrt{13}/2} = 1.146702958...$, implying instability of the circular photon orbits in this region, and stability outside this region.

That is, on the curve of circular photon orbits, $V''(r_\gamma)=0$ at the point
\begin{equation}
(r_\gamma/m, \ell/m)_* = ( 1.146702958..., 0.7995092385...) \ .
\end{equation}
In the second case, substituting Eq.~(\ref{E:non-pert}\;b) into Eq.~\eqref{E:V''} gives
\begin{equation}
V_0''(r_\gamma,\ell) = -{2L^2\over r_\gamma^5} \; {3r_\gamma^2-5\ell r_\gamma+\ell^2\over 3r_\gamma-\ell} \ . 
\end{equation}
This will certainly be negative for $r_\gamma/\ell > (5+\sqrt{13})/6 = 1.434258546...$, implying instability of the circular photon orbits in this region, and stability outside this region.

That is, on the curve of circular photon orbits, $V''(r_\gamma)=0$ at the point
\begin{equation}
(r_\gamma/\ell,m/\ell)_* = ( 1.434258546..., 1.250767286...) \ .
\end{equation}
Consequently, on the curve of circular photon orbits one has \emph{existence} and \emph{stability} in the region $w\in(1,1.434258546...)$, and \emph{existence} and \emph{instability} in the region $w\in(1.434258546...,\infty)$. Precisely at the point $w=1.434258546...$ the photon sphere exhibits marginal stability.

\subsubsection{Turning points} 

To evaluate the exact location of the turning points on the curve described by the loci of circular photon orbits, recall that using  \( w = r_\gamma/\ell \) and \( z = m/\ell \) allows one to write this curve as
\begin{equation}
w^2 = z \, \e^{-1/w} (3w-1) \quad \implies \quad z = {w^2\, \e^{1/w}\over (3w-1)} \ .\label{E:z-for-photon2}
\end{equation}
%
This allows the calculation of
\begin{equation}
\dv{z}{w} = \e^{1/w} \, \frac{3w^2 - 5w +1}{(3w-1)^2} \ ,
\end{equation}
which has a zero at \( w = (5+\sqrt{13})/6 \), where $V_0''(r_\gamma,\ell)=V_0''(w)=0$.

At this point $z$ takes on its maximum value
\begin{equation}
z = \e^{6/(5+\sqrt{13})} \frac{(5+\sqrt{13})^2}{18(3+\sqrt{13})}  =
\e^{(5-\sqrt{13})/2}\;{(2+\sqrt{13})\over 9} \ .
\end{equation}
Consequently, no photon sphere can exist if
\begin{equation}
\frac{\ell}{m} > \e^{-(5-\sqrt{13})/2}\;(\sqrt{13}-2) = 0.7995092385... \ ,\label{eq;photonlimit1}
\end{equation}
or equivalently
\begin{equation}
\frac{m}{\ell} < \e^{(5-\sqrt{13})/2}\;{(2+\sqrt{13})\over 9} =1.250767286.... \ .\label{eq;photonlimit2}
\end{equation}
Note that this happens when
\begin{equation}
{r_\gamma\over m} > {1\over2} (1+\sqrt{13}) \e^{-(5-\sqrt{13})/2} \ , \quad
{r_\gamma\over \ell} > {5+\sqrt{13}\over 6} \ ,
\end{equation}
which was where $V_0''(r_c,m)=0$.

\begin{figure}[!htbp]
\centering
\begin{subfigure}{.5\textwidth}
  \centering
  \includegraphics[width=.96\linewidth]{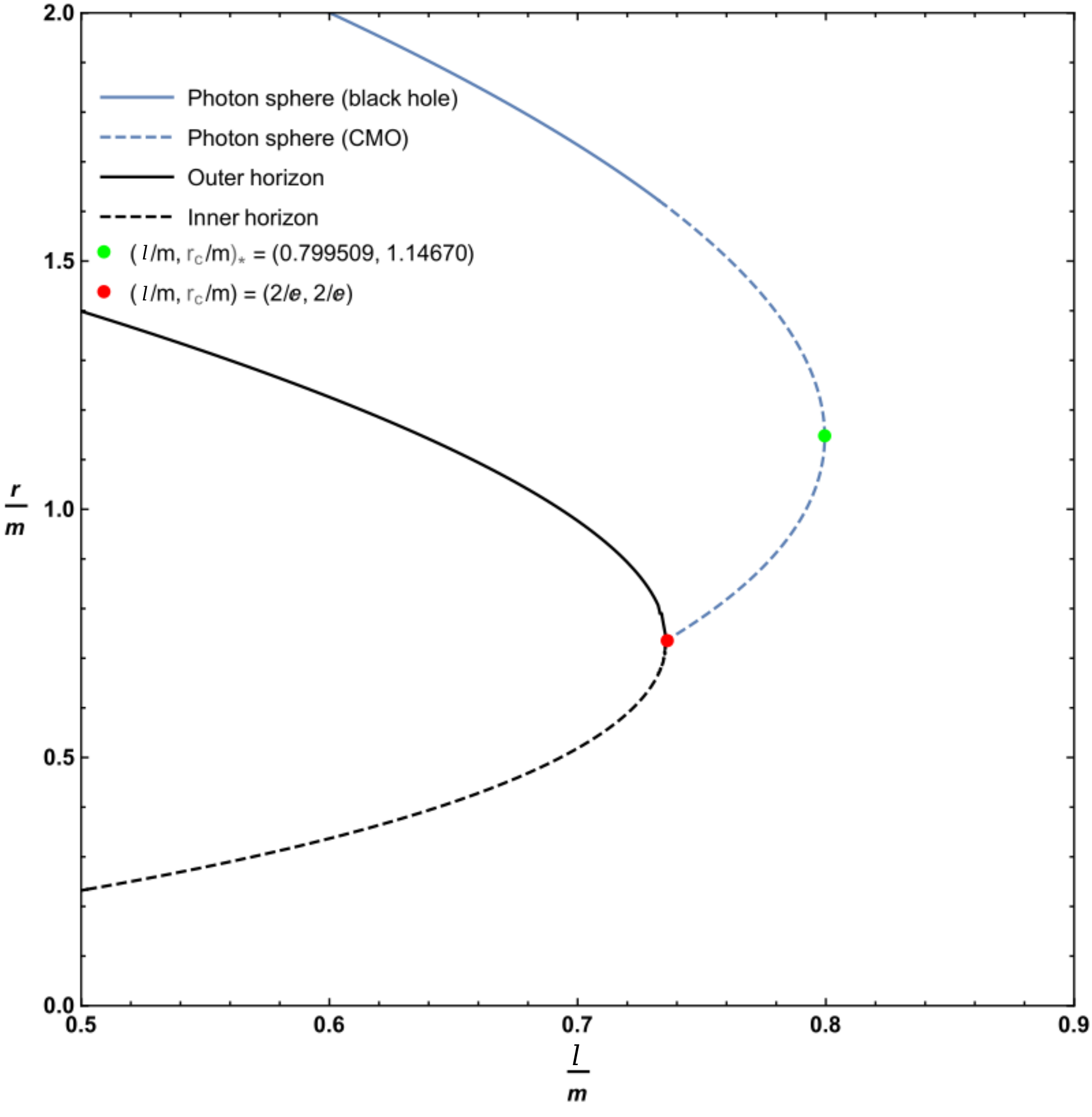}
  \caption{}
  \label{fig:iscom}
\end{subfigure}%
\begin{subfigure}{.5\textwidth}
  \centering
  \includegraphics[width=.96\linewidth]{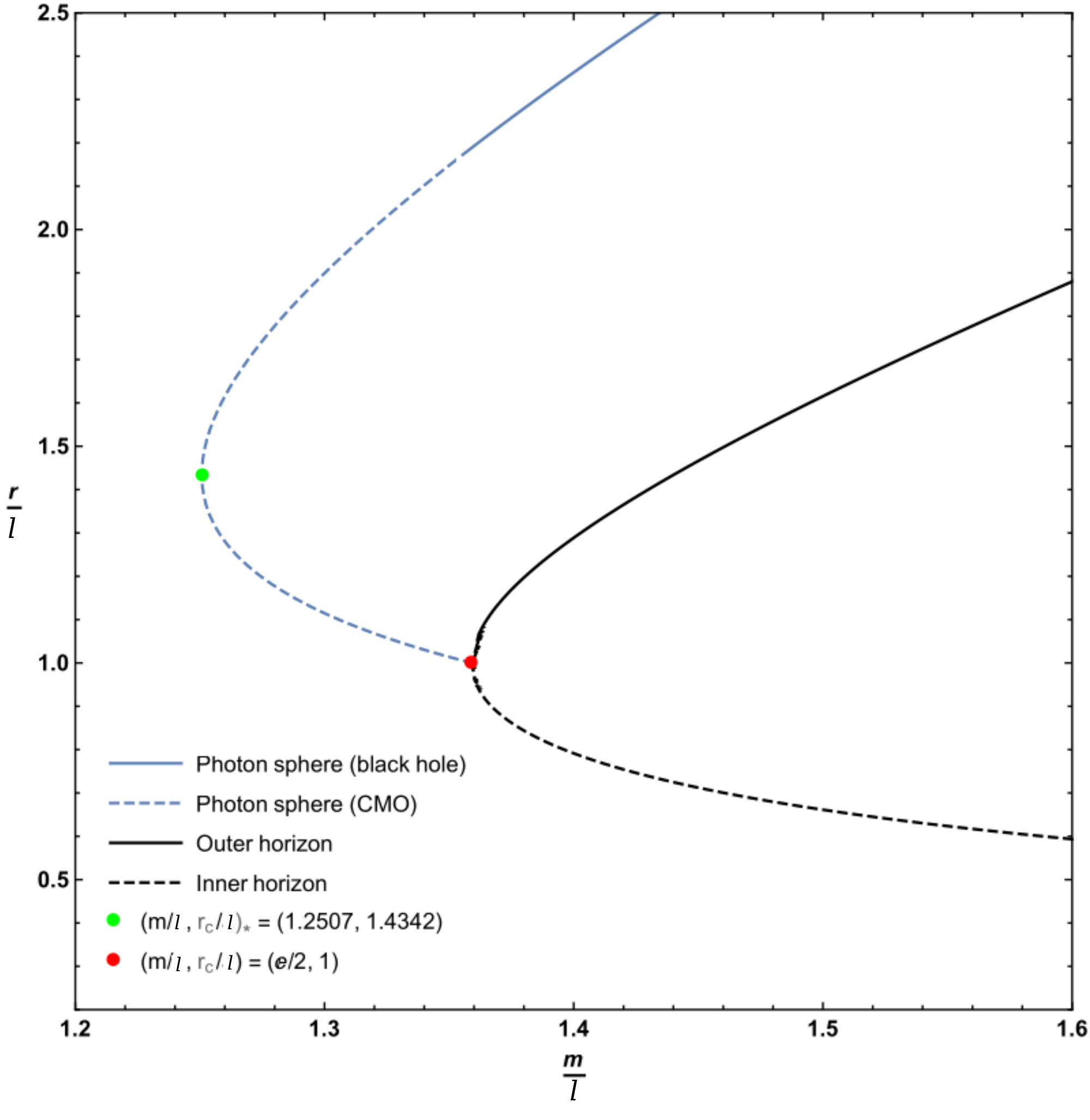}
  \caption{}
  \label{fig:iscoa}
\end{subfigure}
\caption[Zoomed in plots of the location of the photon sphere, inner horizon, and outer horizon for aMc spacetime]{Zoomed in plots of the location of the photon sphere, inner horizon, and outer horizon as a function of the parameters \( \ell \) and \( m \), focusing on the extremal and merger regions. The dashed blue line represents the extension of the photon sphere to horizonless compact massive objects (CMOs). Whenever the location of the photon sphere is double-valued the upper branch corresponds to an unstable photon orbit while the lower branch corresponds to a stable photon orbit.}\label{fig:photonsphere2}
\end{figure}

As can be seen, originally in Fig.~\ref{fig:photonsphere}, and now in more detail in the zoomed-in plot of Fig.~\ref{fig:photonsphere2}, for horizonless compact massive objects there is a region where there are two possible locations for the photon sphere for fixed values of \( m \) and \( \ell \). Furthermore when this happens it is the upper branch that corresponds to an unstable photon orbit, while the lower branch is a stable photon orbit.
\vspace*{8pt}

\newpage

\subsection{Timelike circular orbits}
\def\ISCO{{\hbox{\tiny ISCO}}}
\def\ESCO{{\hbox{\tiny ESCO}}}

First check the \emph{existence}, and then the \emph{stability}, of timelike circular orbits. Even in Schwarzschild spacetime ($\ell\to0$) this is not entirely trivial: Timelike circular orbits, at some fixed $r_c$, \emph{exist} for all $r_c\in (3m,+\infty)$; they are unstable for $r_c\in (3m,6m)$, exhibit marginal stability at $r_c=6m$, and are stable for $r_c\in (6m,+\infty)$. Once the parameter $\ell$ is nonzero the situation is much more complex.

\subsubsection{Existence of circular timelike orbits} 

For timelike trajectories, the effective potential is given by
\begin{equation}
V_{-1}(r) = \left(1-\frac{2m\,\e^{-\ell/r}}{r}\right)\left(1+\frac{L^2}{r^2}\right) \ ,
\end{equation}
and so the locations of the circular orbits can be found from
\begin{equation}
V_{-1}'(r_c) = -\frac{2}{r_c^5}\left\{ L^2r_c^2+m\,\e^{-\ell/r_c} [ \ell(L^2+r_c^2)-r_c(3L^2+r_c^2) ] \right\} = 0 \ .
\end{equation}
That is, all timelike circular orbits (there will be infinitely many of them) must satisfy
\begin{equation}
 L^2r_c^2+m\,\e^{-\ell/r_c} [ \ell(L^2+r_c^2)-r_c(3L^2+r_c^2) ]  = 0 \ .
\end{equation}
This is not analytically solvable for \( r_c(L,m,\ell) \), however one \emph{can} solve for the required angular momentum $L_c(r_c,m,\ell)$ of these circular orbits:
\begin{equation}
L_c(r_c,m,\ell)^2 = {\frac{r_c^2 \, m(r_c-\ell)}{m\ell-3mr_c+r_c^2\,\e^{\ell/r_c}}} \ .
\label{E:Lsq}
\end{equation}
Physically one must demand $0\leq L_c^2 <+\infty$, so the boundaries for the \emph{existence} region of circular orbits (whether stable or unstable) are given by
\begin{equation}
r_c = \ell \ , \quad {m\ell-3mr_c+r_c^2\,\e^{\ell/r_c}} = 0 \ .
\end{equation}
The first of these conditions $r_c=\ell$, comes from the fact that in aMc spacetime gravity is effectively repulsive for $r<\ell$. Remember that $g_{tt} = -(1-2m\e^{-\ell/r}/r)$, and that the pseudo-force due to gravity depends on $\partial_r g_{tt}$. Specifically
\begin{equation}
\partial_r g_{tt} = - {2m\over r^2} \; \e^{-\ell/r} \; \left(1-{\ell\over r}\right) \ ,
\end{equation}
and this changes sign at $r=\ell$. So for $r>\ell$ gravity attracts to the centre, but for $r<\ell$ gravity repels from the centre.

If gravity is repulsive, there is no way to counter-balance it with a centrifugal pseudo-force, and so there is simply no way to get a circular orbit, regardless of whether it be stable or unstable. Precisely at $r=\ell$ there are stable ``orbits'' where the test particle just sits there, with zero angular momentum, no sideways motion required. Since by construction $r_c>r_{H_{+}} \geq \ell$, this constraint is relevant only for horizonless CMOs.

 The second of these conditions is exactly the location of the photon orbits considered in \S~\ref{aMcphotonspheres}. Physically what is going on is this: At large distances it is easy to put a massive particle into a circular orbit with $L_c\propto\sqrt{mr_c}$. As one moves inwards and approaches the photon orbit, the massive particle must move more and more rapidly, and the angular momentum per unit mass must diverge when a particle with nonzero invariant mass tries to orbit at the photon orbit.
 
\begin{figure}[!htbp]
\centering
\begin{subfigure}{.5\textwidth}
  \centering
  \includegraphics[width=.96\linewidth]{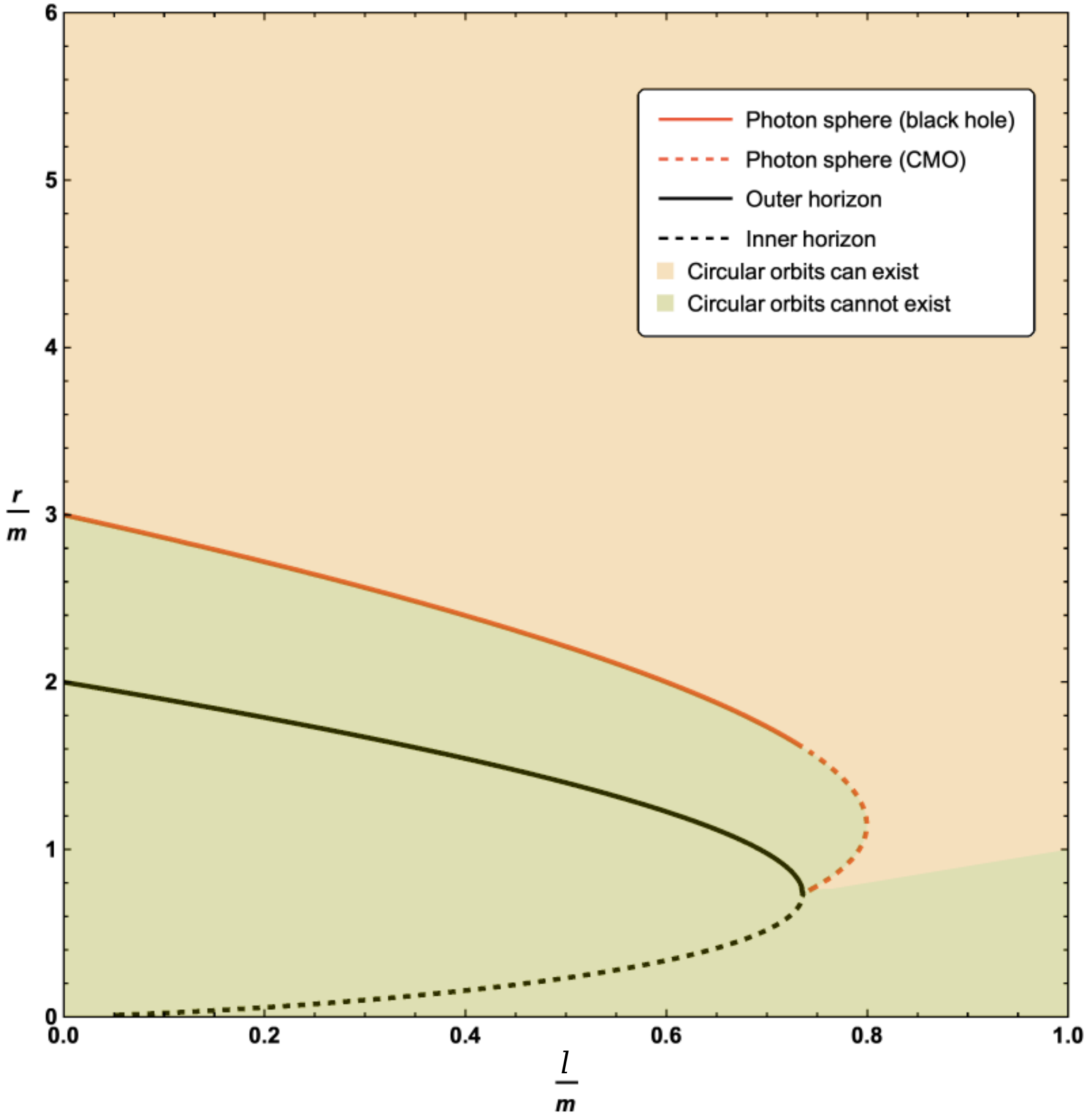}
  \caption{}
  \label{fig:iscom}
\end{subfigure}%
\begin{subfigure}{.5\textwidth}
  \centering
  \includegraphics[width=.96\linewidth]{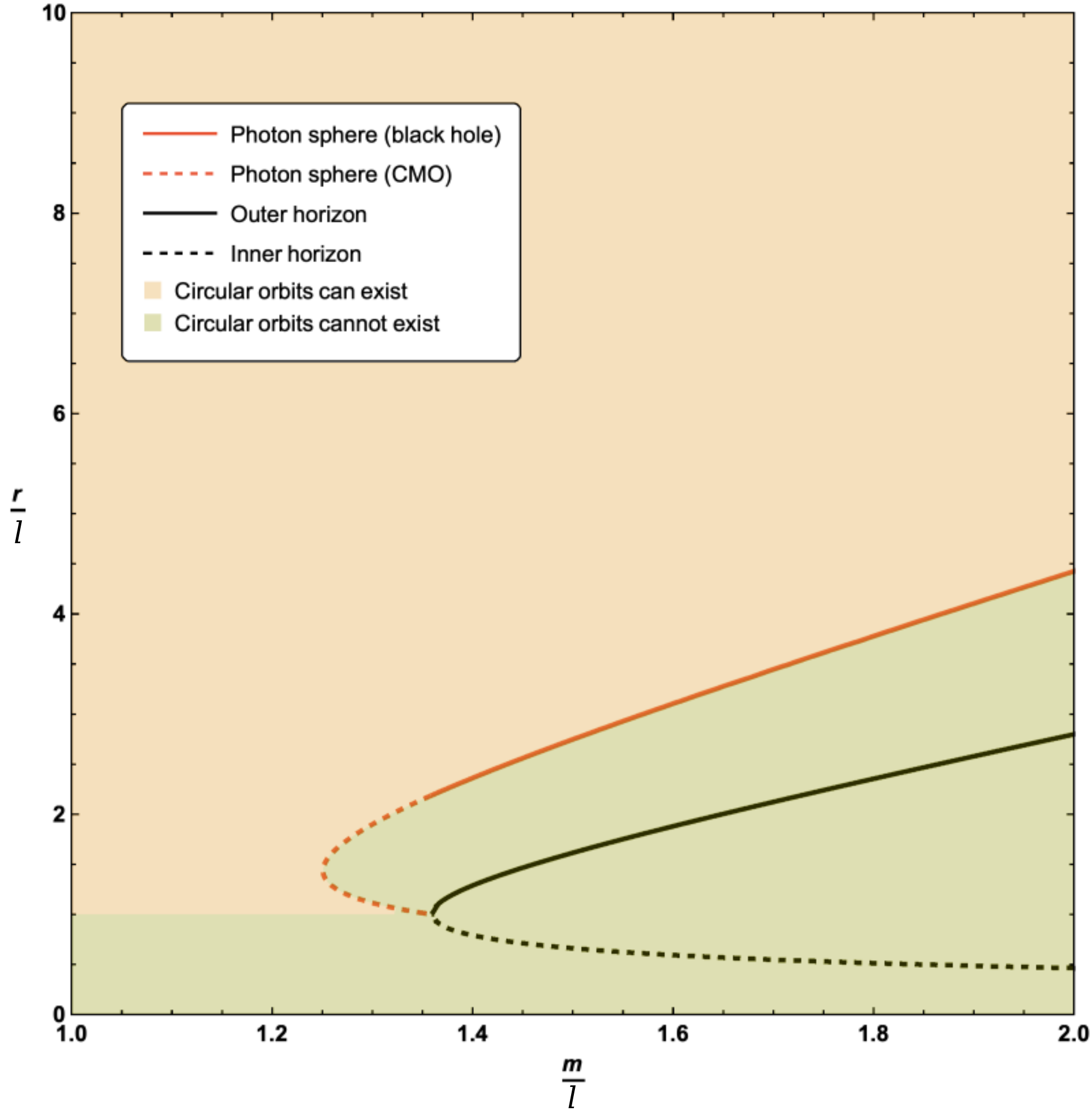}
  \caption{}
  \label{fig:iscoa}
\end{subfigure}
\caption[Locations of the \emph{existence} region for timelike circular orbits for aMc spacetime]{Locations of the \emph{existence} region for timelike circular orbits in terms of the circular null geodesics, outer horizon, and inner horizon for various values of the parameters \( \ell \) and \( m \).}
\label{fig:isco-existence}
\end{figure}

 Thus the existence region (rather than just its boundary) for timelike circular orbits is given by
 \begin{equation}
r_c > \ell \ , \quad {m\ell-3mr_c+r_c^2\,\e^{\ell/r_c}}>0 \ .
\end{equation}
The situation is depicted in Fig.~\ref{fig:isco-existence}.

\subsubsection{Stability \emph{versus} instability for  circular timelike orbits} 
\enlargethispage{30pt}

Now consider the general expression
\begin{equation}
V_{-1}''(r) = {6L^2 r^3- 2m(2r^4-4\ell r^3+(12L^2+\ell^2)r^2-8L^2\ell r+L^2\ell^2)\e^{-\ell/r}\over r^7} \ ,
\end{equation}
and substitute the known value of $L\to L_c(r_c)$ for circular orbits; see Eq.~\eqref{E:Lsq}. Then
\begin{equation}
V_{-1}''(r_c) = -{2m \e^{-\ell/r_c}(2m(3r_c^2-3\ell r_c+\ell^2) \e^{-\ell/r_c} -r_c(r_c^2+\ell r_c-\ell^2))\over
(r_c^2 - m(3r_c-\ell) \e^{-\ell/r_c})r^4} \ .
\end{equation}
Note that $V_{-1}''(r_c) \to+\infty$ at the photon orbit, where the denominator vanishes.

To locate the \emph{boundary} of the region of \emph{stable} circular orbits, the ESCO (extremal stable circular orbit), one needs to set $V_{-1}''(r_c)=0$, leading to
\begin{equation}
2m(3r_c^2-3\ell r_c+\ell^2) \e^{-\ell/r_c} = r_c(r_c^2+\ell r_c-\ell^2) \ .\label{E:stable}
\end{equation}
Note that locating this boundary is equivalent to extremising $L_c(r_c)$. To see this, consider the quantity $[V_{-1}'(L(r),r)]=0$ and differentiate:
\begin{equation}
{\dd{} [V_{-1}'(L(r),r)]\over \dd r }  = 
\left.{\partial V_{-1}'(L,r)\over\partial L}\right|_{L=L(r)} \times {\dd L(r)\over \dd r}
+ \left.V''_{-1}(L,r)\right|_{L=L(r)} \ .
\end{equation}
This implies
\begin{equation}
0  = 
\left.{\partial V_{-1}'(L,r)\over\partial L}\right|_{L=L(r)} \times {\dd L(r)\over \dd r}
\;+\; \left.V_{-1}''(L,r)\right|_{L=L(r)} \ .
\end{equation}
Thence
\begin{equation}
 \left.V_{-1}''(L,r)\right|_{L=L(r)}  = - \left.{\partial V_{-1}'(L,r)\over\partial L}\right|_{L=L(r)} 
 \times {\dd L(r)\over \dd r} \ .
 \end{equation}
But it is easily checked that ${\partial V_{-1}'(L,r)/\partial L}$ is nonzero outside the photon sphere, that is, in the existence region for circular timelike geodesics. Thence:
\begin{equation}
 \left.V_{-1}''(L,r)\right|_{L=L(r)}  = 0
 \quad \Longleftrightarrow  \quad
 {\dd L(r)\over \dd r}=0 \ .
 \end{equation}
So one might as well extremise $L^2_c(r_c)$, as in Eq.~(\ref{E:Lsq}), and one again finds Eq.~(\ref{E:stable}).

\begin{figure}[!htbp]
\centering
\begin{subfigure}{.5\textwidth}
  \centering
  \includegraphics[width=.96\linewidth]{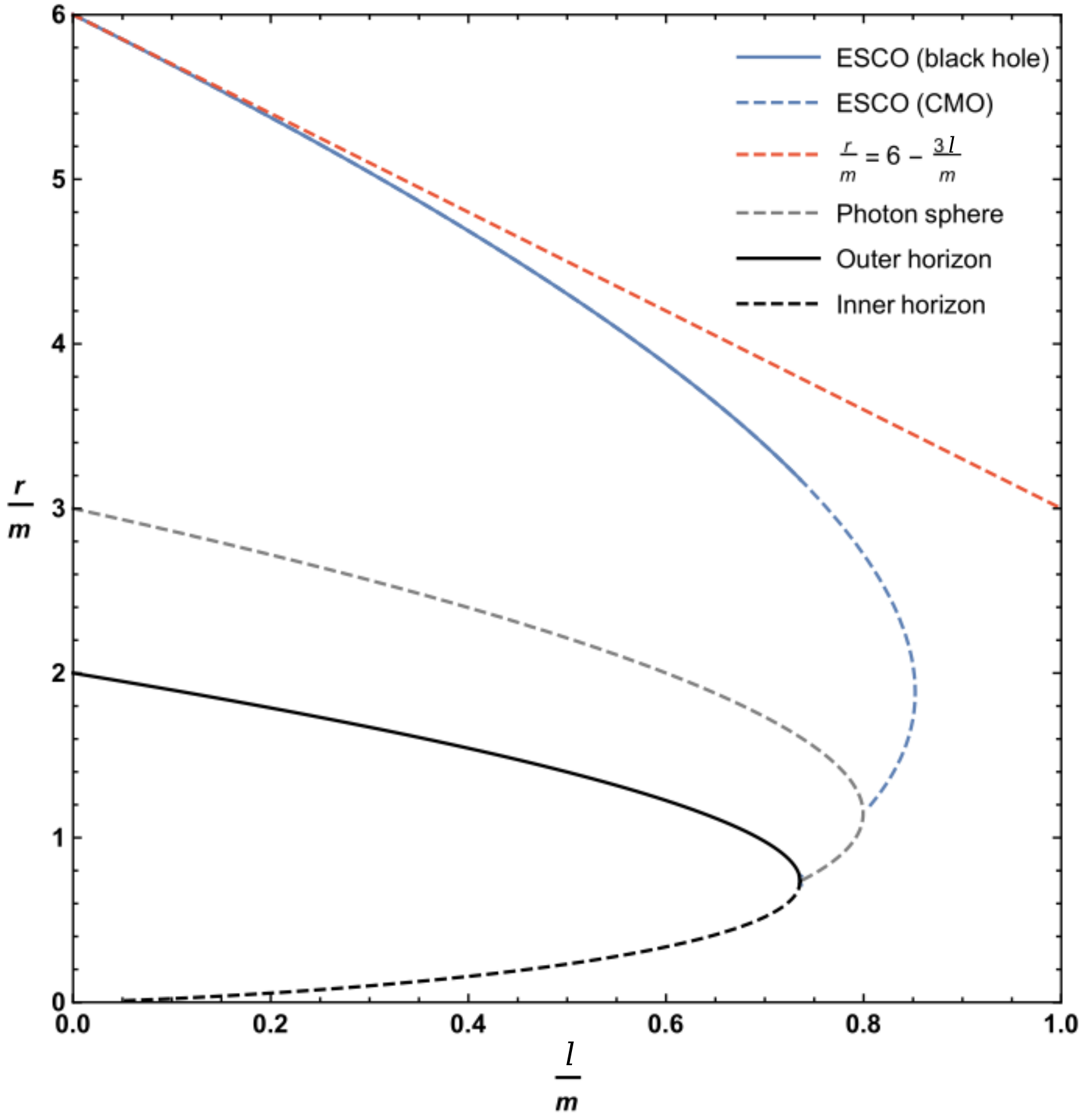}
  \caption{}
  \label{fig:iscom}
\end{subfigure}%
\begin{subfigure}{.5\textwidth}
  \centering
  \includegraphics[width=.95\linewidth]{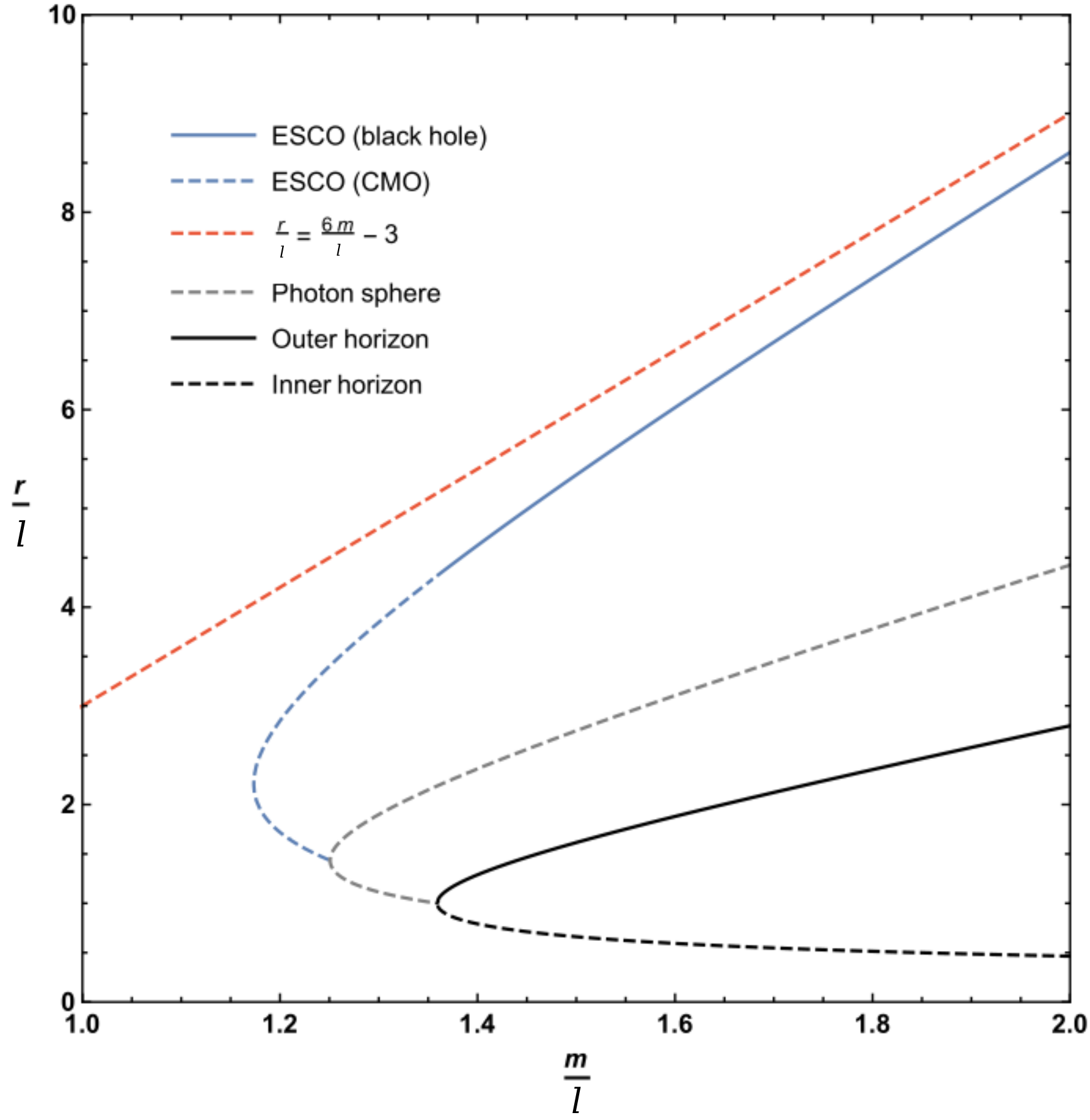}
  \caption{}
  \label{fig:iscoa}
\end{subfigure}
\caption[Example locations of the ESCO, photon sphere, outer horizon, and inner horizon for aMc spacetime]{Locations of the ESCO, photon sphere, outer horizon, and inner horizon for various values of the parameters \( \ell \) and \( m \).
The dashed blue line represents the extension of the ESCO to CMOs. The dashed red curves in each case are the asymptotic locations of the ISCO for small values of \( \ell \).}
\label{fig:esco}
\end{figure}

\begin{figure}[!htbp]
\centering
\begin{subfigure}{.5\textwidth}
  \centering
  \includegraphics[width=.96\linewidth]{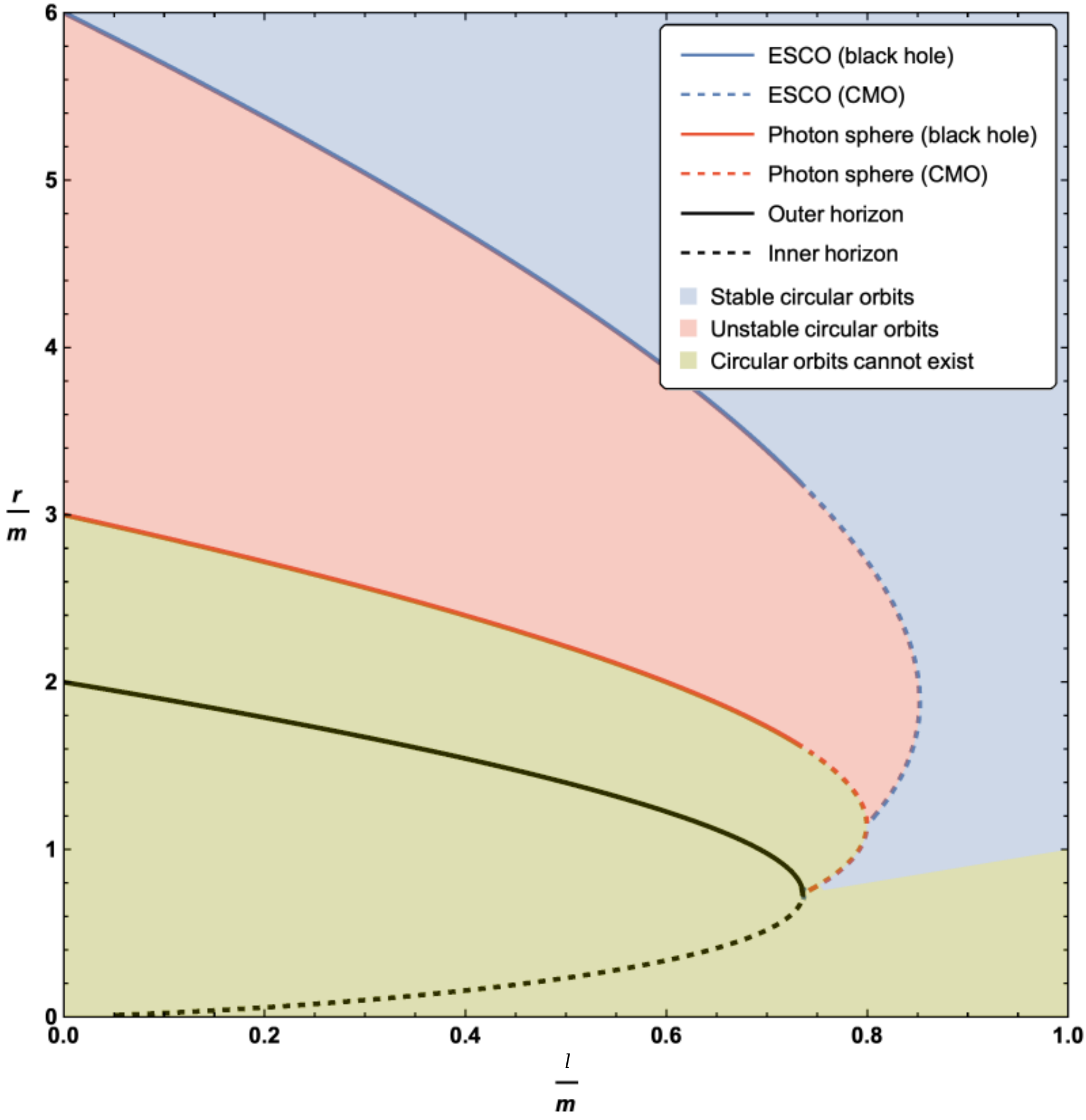}
  \caption{}
  \label{fig:iscom}
\end{subfigure}%
\begin{subfigure}{.5\textwidth}
  \centering
  \includegraphics[width=.96\linewidth]{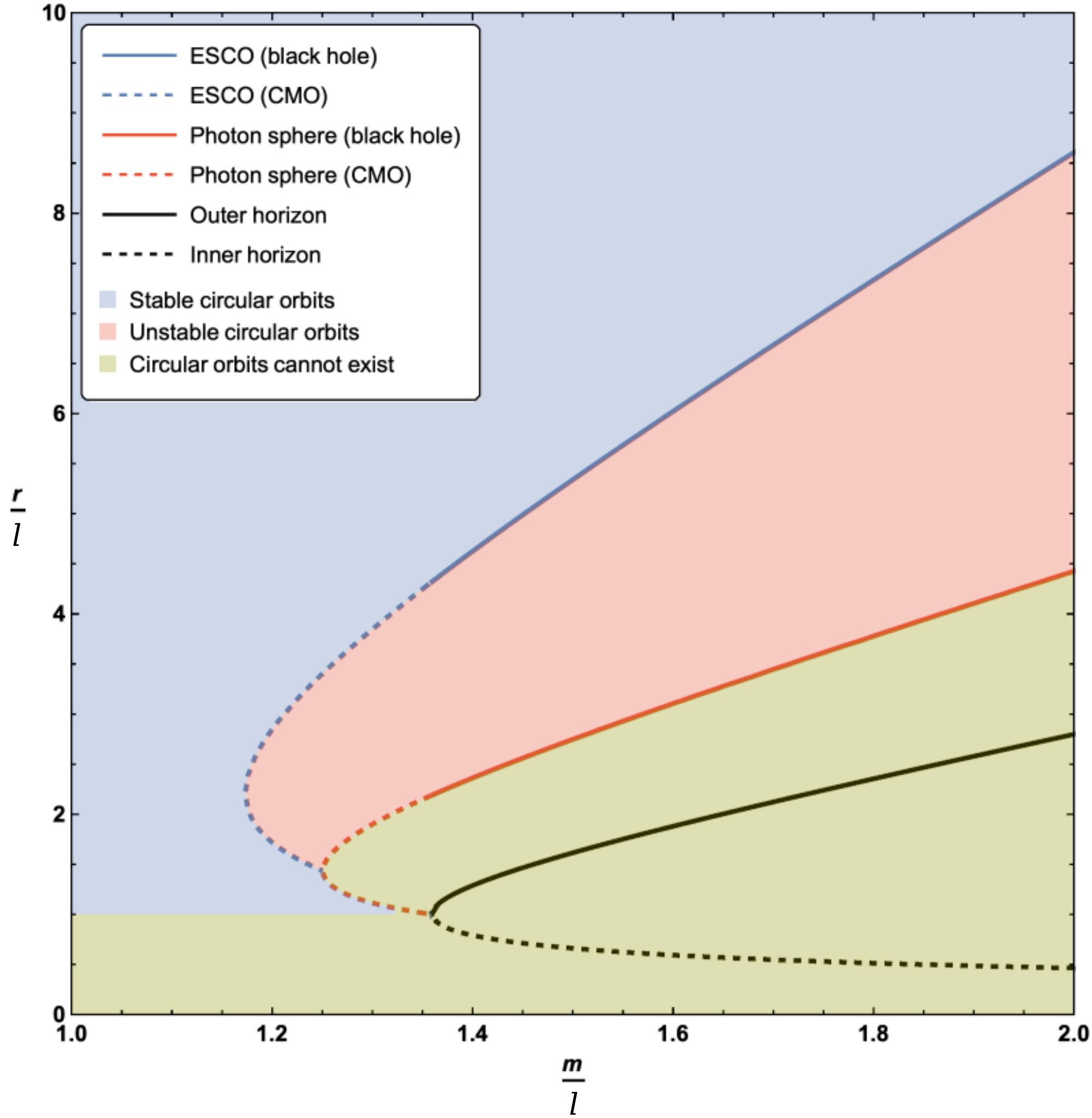}
  \caption{}
  \label{fig:iscoa}
\end{subfigure}
\caption[More example locations of the ESCO, photon sphere, outer horizon, and inner horizon for aMc spacetime]{Locations of the ESCO, photon sphere, outer horizon, and inner horizon for various values of the parameters \( \ell \) and \( m \).
The dashed blue line represents the extension of the ESCO to CMOs. The dashed red line represents the extension of the photon sphere to CMOs. The blue region denotes stable timelike circular orbits, while the red region denotes unstable timelike circular orbits, and the green region denotes the nonexistence of timelike circular orbits. Where the ESCO is single-valued it is an ISCO. Where the ESCO is double-valued the upper branch is an ISCO and the lower branch is an OSCO (outermost stable circular orbit).}\label{fig:esco-2}
\end{figure}

Defining  \( w = r_c/\ell \) and \( z = m/\ell \), the curve describing the boundary of the region of stable timelike circular orbits can be rewritten as
\begin{equation}
2z (3w^2-3w+1)\e^{-1/w} = w(w^2+w-1) \ .\label{E:dd}
\end{equation}
Plots of the boundary implied by Eq.~\eqref{E:stable}, or equivalently Eq.~\eqref{E:dd}, can be seen in Fig.~\ref{fig:esco}. As for the photon sphere, there is the interesting result that the extension of the ESCO to horizonless compact massive objects results in up to two possible ESCO locations for fixed values of \( \ell \) and \( m \). Perhaps unexpectedly, the curve of ESCOs does not terminate at the horizon --- it terminates once it hits the curve of circular photon orbits at a very special point. One now turns to analysing the qualitative behaviours and the various turning points presented in Figs.~\ref{fig:esco} and~\ref{fig:esco-2}. Note that where the ESCO is single-valued it is an ISCO (innermost stable circular orbit). Where the ESCO is double-valued, the upper branch is an ISCO and the lower branch is an OSCO (outermost stable circular orbit)~\cite{IAOSCOITPOAPCC:Boonserm:2020}.

\enlargethispage{60pt}

\clearpage

\textbf{Perturbative analysis (small $\ell$):} 

First, investigate the existence region perturbatively for small $\ell$. One has
\begin{equation}
L_c(r_c,m,\ell)^2 = {mr_c^2\over r_c-3m} -{2mr_c(r_c-m)\over(r_c-3m)^2} \; \ell + \mathcal{O}(\ell^2) \ .
\end{equation}
Note that this approximation diverges at the Schwarzschild photon sphere $r=3m$. So for small $\ell$ the boundary for the region of \emph{existence} of timelike circular orbits is still $r=3m$.

Now investigate the \emph{stability} region perturbatively for small $\ell$. Rearranging Eq.~(\ref{E:stable}), one sees that
\begin{equation}
r_c = {6m(r_c^2-\ell r_c+\ell^2/3) \e^{-\ell/r_c}\over r_c^2+\ell r_c-\ell^2} 
= 6m\left( 1-{3\ell\over r_c} +\O(\ell^2) \right) \ .
\end{equation}
Thence
\begin{equation}
r_c = 6m -3\ell +\O(\ell^2) \ ,
\end{equation}
which sensibly reproduces the Schwarzschild ISCO to lowest order in $\ell$, and explains the asymptote in Fig.~\ref{fig:esco} (b).

Furthermore, for small $\ell$, substituting $L_c(r_c)$ into $V''_{-1}(L,r_c)$ and expanding gives
\begin{equation}
V_{-1}''(r_c) = {2m(r_c-6m)\over r_c^3(r_c-3m)} 
+  {4m^2(7r_c-15m)\over r^4(r_c-3m)^2} \; \ell  +\O(\ell^2) \ .
\end{equation}
Demanding that this quantity be zero self-consistently yields $r_c = 6m -3\ell +\O(\ell^2)$. 

\textbf{Non-perturbative analysis:} 

\begin{figure}[h]
\centering
\begin{subfigure}{.5\textwidth}
  \centering
  \includegraphics[width=.96\linewidth]{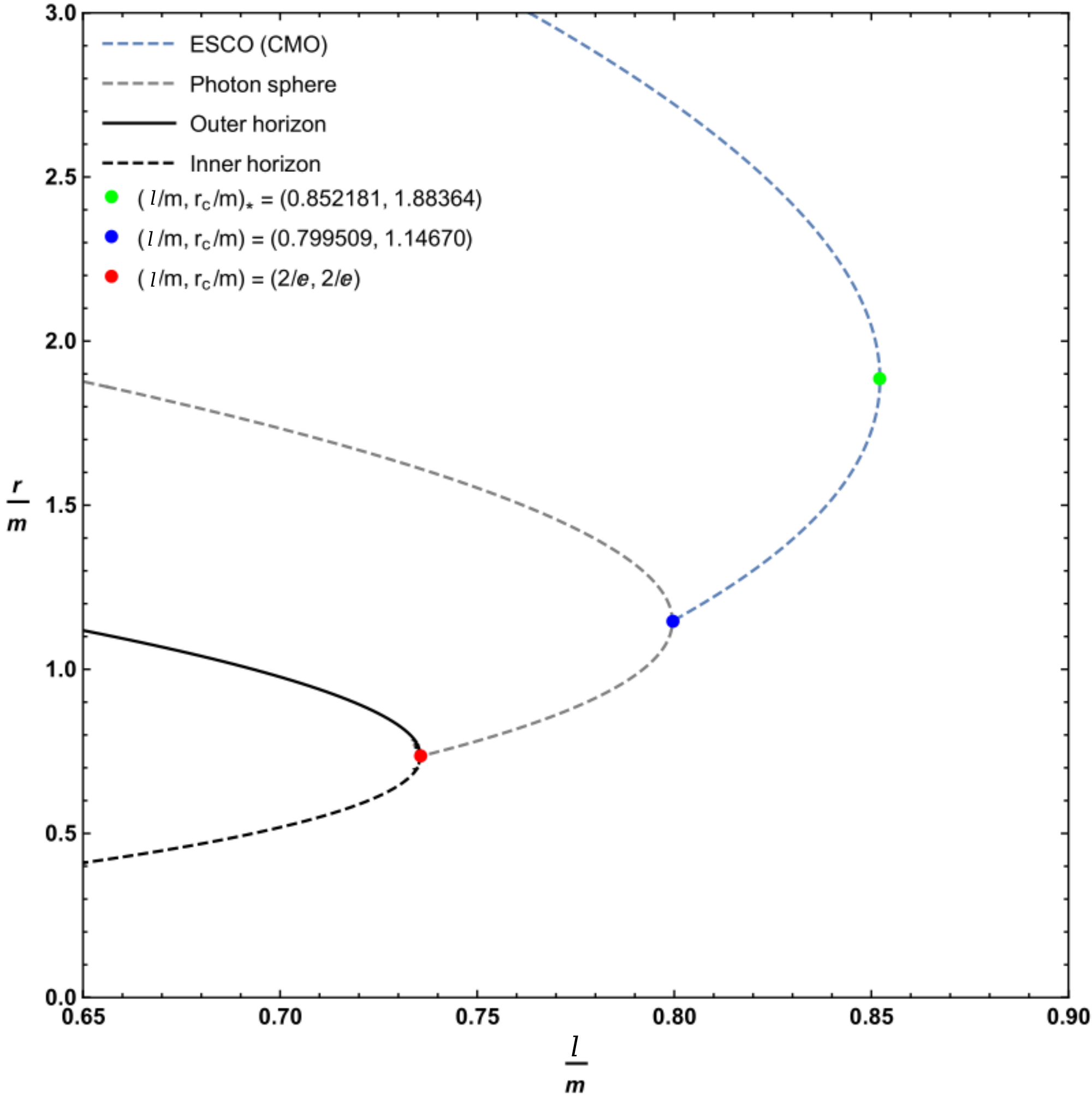}
  \caption{}
  \label{fig:iscom}
\end{subfigure}%
\begin{subfigure}{.5\textwidth}
  \centering
  \includegraphics[width=.96\linewidth]{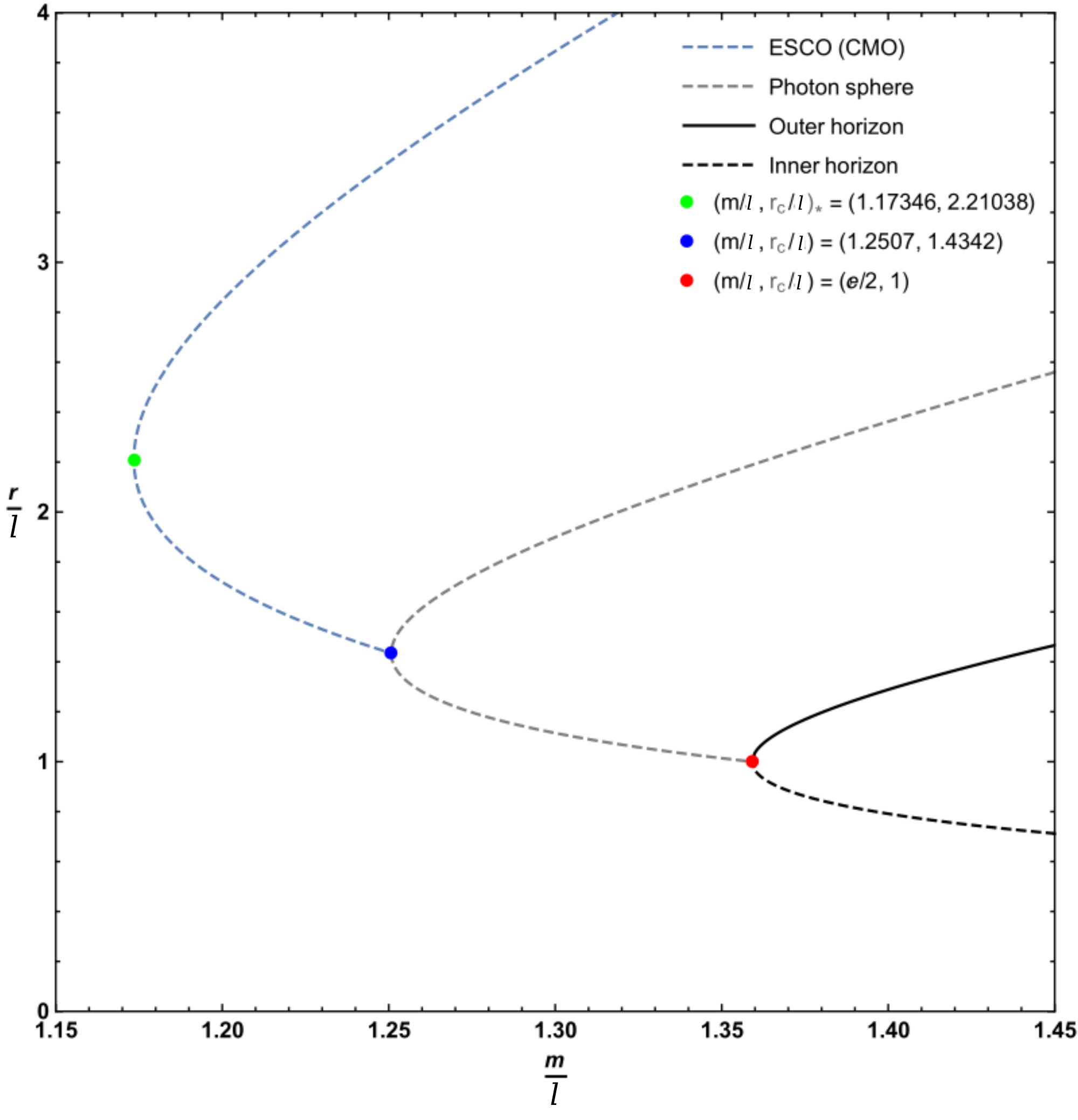}
  \caption{}
  \label{fig:iscoa}
\end{subfigure}
\caption[Zoomed in plot of example locations for the ESCO, outer horizon, and inner horizon for aMc spacetime]{Zoomed in plot of the locations of the ESCO, outer horizon, and inner horizon for various values of the parameters \( \ell \) and \( m \), focusing on the turning points. The dashed blue line represents the extension of the ESCO to CMOs. Where the ESCO is single-valued it is an ISCO. Where the ESCO is double-valued the upper branch is an ISCO and the lower branch is an OSCO.}
\label{fig:esco-3}
\end{figure}

Already it has been seen that, defining  \( w = r_c/\ell \) and \( z = m/a \), the curve describing the boundary of the region of stable timelike circular orbits can be rewritten as
\begin{equation}
2z (3w^2-3w+1)\,\e^{-1/w} = w(w^2+w-1) \ .\label{E:dd3}
\end{equation}
Thence
\begin{equation}
    z = {w(w^2+w-1)\,\e^{1/w}\over 2(3w^2-3w+1)} \ .
    \label{E:z-for-esco}
\end{equation}
Examining the turning points of $z(w)$, the derivative is
\begin{equation}
{\dd z\over\dd w} = {(w-1)(3w^4-6w^3-3w^2+4w-1)\,\e^{1/w} \over 2w(3w^2-3w+1)^2} \ .
\end{equation}
\clearpage
There is one obvious local extrema at $w=1$, corresponding to $z=\e/2$. Physically this corresponds to the point where inner and outer horizons merge and become extremal --- but from inspection of Fig.~\ref{fig:esco}, the descriptive plots of Fig.~\ref{fig:esco-2}, and the zoomed-in plots of Fig.~\ref{fig:esco-3}, one sees that the curve of ESCOs hits the photon orbit (and becomes unphysical) before getting to this point. In terms of the variables used when plotting Figs.~\ref{fig:esco}--\ref{fig:esco-3} this unphysical (from the point of view of ESCOs) point corresponds to
\begin{equation}
(r_c/\ell, m/\ell)_* = (1, \e/2) \ ; \qquad (r_c/m, \ell/m)_* = (2/\e, 2/\e) \ . 
\end{equation}
The other local extrema is located at the only physical root of the quartic polynomial
\begin{equation}
3w^4-6w^3-3w^2+4w-1 =0 \ .
\end{equation}
While this can be solved analytically, the results are highly intractable; it is better to use numerics. Two roots are complex, one is negative, and the only physical root is $w= 2.210375896...$, \textit{i.e.} $z=1.173459017...$. Physically this implies that the ESCO curve should exhibit a nontrivial local extremum --- and from inspection of Fig.~\ref{fig:esco} one sees that the curve of ESCOs does indeed have a local extremum at this point. In terms of the variables used when plotting Fig.~\ref{fig:esco} this extremal point corresponds to
\begin{equation}
(r_c/\ell, m/\ell)_* = (2.210375896, 1.173459017) \ , 
\end{equation}
and
\begin{equation}
(r_c/m, \ell/m)_* = (1.883641323, 0.8521814444) \ . 
\end{equation}

\subsubsection{Intersection of ESCO and photon sphere } 

One can rewrite the curve for the loci of the photon spheres, given by Eq.~\eqref{E:z-for-photon}, as
\begin{equation}
\e^{-1/w} z = {w^2\over (3w-1)} \ .
\end{equation}
Similarly, for the loci of ESCOs rewrite Eq.~\eqref{E:z-for-esco} as
\begin{equation}
\e^{-1/w} z = {w(w^2+w-1)\over 2 (3w^2-3w+1)} \ .
\end{equation}
These curves cross at
\begin{equation}
{w\over (3w-1)} = {(w^2+w-1)\over 2 (3w^2-3w+1)} \ .
\end{equation}
That is, at
\begin{equation}
(w-1)(3w^2-5w+1)=0 \ ,
\end{equation}
with explicit roots at
\begin{equation}
1, \quad {5\pm\sqrt{13}\over6} \ .
\end{equation}
The physically relevant root is $w = {5+\sqrt{13}\over6}= 1.434258546...$, which was where it was previously determined that the photon sphere becomes stable, and at the point where the curve of photon spheres maximises the value of $z=m/\ell$.

\subsubsection{Explicit result for the angular momentum} 

One can rewrite the curve for the angular momentum Eq.~\eqref{E:Lsq} as
\begin{equation}
L_c^2 = 
\ell^2  \left( \e^{-1/w} z \; w^2 (w-1) \over  w^2 - \e^{-1/w} z (3w-1)\right) \ .
\end{equation}
Similarly, for the loci of ESCOs one can rewrite Eq.~\eqref{E:z-for-esco} as
\begin{equation}
\e^{-1/w} z = {w(w^2+w-1)\over 2 (3w^2-3w+1)} \ .
\end{equation}
Then substitute this into back into $L_c$:
\begin{equation}
L_c^2 = \ell^2\; {w^2(w^2+w-1)\over 3w^2-5w+1} \ .
\end{equation}
\clearpage
This has a pole at $w = {5+\sqrt{13}\over6}= 1.434258546...$, and is then positive and finite for all $w >{5+\sqrt{13}\over6}$. Of course the point $w = {5+\sqrt{13}\over6}$ on the ESCO curve is exactly where the ESCO curve hits the photon curve, so one would expect the angular momentum to diverge there. Asymptotically for large $r$ (large $w=r_c/\ell$) one has $L_c^2 \sim \ell^2 w^2/3$ and $m/\ell = z \sim w/6$, so $L_c^2 \sim 2 m r_c$ as expected from the large-distance Newtonian limit.

\subsubsection{Summary of notable orbits} 

Overall, the boundary of the stability region for timelike circular orbits is rather complicated. In terms of the variable $w=r_c/\ell$:
\begin{itemize}
\item For $w\in ({5+\sqrt{13}\over6}, +\infty)$ there is an ESCO. This ESCO then subdivides as follows:
\begin{itemize}
\item For $w \in( 2.210375896, +\infty)$ there is an ISCO.
\item For $w \in ({5+\sqrt{13}\over6}, 2.210375896)$ there is an OSCO.
\end{itemize}
\item For $w\in (1, {5+\sqrt{13}\over6})$ the stability region is bounded by a stable photon orbit.
\item The line $w=1$ bounds the stability and existence region for timelike circular orbits from below.
\end{itemize}
This is considerably more complicated than might reasonably have been expected.

%
\section{Stress-energy tensor}\label{AMCStress}

Consider the Einstein field equations for this spacetime; $G^{\mu}{}_{\nu}=8\pi \, T^{\mu}{}_{\nu}$. From Eq.~(\ref{einsteinsuppressed}), in the domain of outer communication one has
\begin{equation}\label{stresssuppressed}
    T^{\mu}{}_{\nu} = \begin{bmatrix}
    -\rho & 0 & 0 & 0 \\
    0 & p_r & 0 & 0 \\
    0 & 0 & p_t & 0 \\
    0 & 0 & 0 & p_t
    \end{bmatrix} \ ,
\end{equation}
with
\begin{eqnarray}
    \rho &=& - p_r = \frac{2m\,\ell\,\e^{-\ell/r}}{8\pi r^{4}} \ , \nonumber \\
    p_t &=& -\frac{m\,\ell\left(\ell-2r\right)\e^{-\ell/r}}{8\pi r^{5}} \ .
\end{eqnarray}
Examining where the energy density is maximised for this spacetime is of specific interest due to the exponential suppression of the mass:
\begin{equation}
    \frac{\partial{\rho}}{\partial{r}} = \frac{2m\,\ell\,\e^{-\ell/r}}{8\pi r^{6}}\left(\ell-4r\right) \ .
\end{equation}
It follows that $\rho$ is maximised at coordinate location $r=\frac{\ell}{4}$. It is worth re-emphasising the high tractability of these expressions.
\enlargethispage{20pt}

\subsubsection{Null energy condition}

Recall that satisfaction of the NEC implies that both $\rho+p_r\geq 0$ and $\rho+p_t\geq 0$ globally. First, consider $\rho+p_r$ and note that it is identically zero. This result is common to all $\Phi(r)=0$ spacetimes with respect to the current static spherically symmetric gauge; see for example references~\cite{LW:Visser:1995, WIGTT:Jacobson:2007, Kiselev:Visser:2020, Kiselev:Boonserm:2020}. Now, noting that for all $\Phi(r)=0$ spacetimes $\rho+p_t = \frac{r}{2}\,\rho'$ (see reference~\cite{Kiselev:Boonserm:2020}), one has:
\begin{equation}
    \rho+p_t = \frac{r}{2}\,\rho' = \frac{m\,\ell\,\e^{-\ell/r}}{8\pi r^5}\left(\ell-4r\right) \ .
\end{equation}
This changes sign when $r=\frac{\ell}{4}$. There is therefore the somewhat nontypical situation where the radial NEC is satisfied by the geometry whilst the transverse NEC is violated in the deep core (note the boundary of the violated region is precisely the locus for which $\rho$ is maximised).

\subsubsection{Strong energy condition}

In order to satisfy the SEC, one of the required conditions is that $\rho+p_r+2p_t\geq 0$ globally. Evaluating:
\begin{equation}
    \rho+p_r+2p_t = 2p_t = -\frac{m\ell\left(\ell-2r\right)\e^{-\ell/r}}{4\pi r^{5}} \ .
\end{equation}
This is negative for the region $r<\frac{\ell}{2}$; the SEC is violated in the deep core --- this is consistent with the general situation, as proven in reference~\cite{RBHAEC:Zaslavskii:2010}. Notably, there is satisfaction of the energy conditions everywhere in the domain of outer communication.

\section{Introducing the ``aMcRN'' model}\label{S:amcRN}
\enlargethispage{15pt}

As discussed above, aMc spacetime, the analog to Schwarzschild, is already in-hand. Along similar lines to the construction of bbRN spacetime in \S~\ref{SVbbRN}, it is intuitive to imagine that an asymptotically Minkowski core analog to Reissner--Nordstr\"{o}m (RN) also exists, and is amenable to highly tractable analysis. It pays to view aMc spacetime as a minimal modification of Schwarzschild; one simply performs the following ``regularising'' procedure in standard curvature coordinates:
\begin{itemize}
    \item Make the modification $m\to m(r)=m\,\e^{-\ell/r}$.
\end{itemize}
It should be emphasised that this is \emph{not} a coordinate transformation.

Extending this process by minimally modifying RN is very straightforward; given standard RN spacetime in curvature coordinates:
\begin{equation}
    \d s^2 = -f_\text{RN}(r)\d t^2 + \frac{\d r^2}{f_\text{RN}(r)}+r^2\d\Omega^{2}_{2} \ , \quad f_\text{RN}(r) = 1-\frac{2m}{r}+\frac{Q^2}{r^2} \ ,
\end{equation}
\clearpage
one simply performs the following ``regularising'' procedure:
\enlargethispage{20pt}
\begin{itemize}
    \item Make the modification $m\to m(r)=m\,\e^{-\ell/r}$;
    \item Make the modification $Q^2\to Q^2(r)=Q^2\,\e^{-\ell/r}$.
\end{itemize}
It should be emphasised that this is \emph{not} a coordinate transformation. One could also choose $Q^2(r)=Q^2\,\e^{-\ell^2/r^2}$, at the cost of tractability in the resulting curvature quantities. For now, generalisations of this process are left for future research.

This procedure yields the following static spherically symmetric line element, for now dubbed ``aMcRN'' spacetime:
\begin{eqnarray}
    \d s^2 &=& -f_\text{aMcRN}(r)\d t^2 + \frac{\d r^2}{f_\text{aMcRN}(r)}+r^2\d\Omega^{2}_{2} \ , \nonumber \\[3pt]
    f_\text{aMcRN}(r) &=& 1-\frac{2m\,\e^{-\ell/r}}{r}+\frac{Q^2\,\e^{-\ell/r}}{r^2} \ .\label{amcRN}
\end{eqnarray}
$\ell\to0$ recovers standard RN spacetime, $Q\to0$ returns aMc spacetime, and both $\ell,Q\to0$ regains Schwarzschild. As far as the author is aware, this is the first presentation of Eq.~(\ref{amcRN}) in the literature.

Extraction of the Kretschmann scalar is straightforward, given by
\begin{eqnarray}
    K &=& \frac{\e^{-2\ell/r}}{r^{12}}\Bigg\lbrace\left(\ell^4-12\ell^3r+52\ell^2r^2-88\ell r^3+56r^4\right)Q^4 \nonumber \\
    && \ -\left(4\ell^4mr -40\ell^3mr^2+144\ell^2mr^3-192\ell mr^4+96mr^5\right)Q^2 \nonumber \\
    && \quad +4\ell^4m^2r^2-32\ell^3m^2r^3+96\ell^2m^2r^4-96\ell m^2r^5+48m^2r^6\Bigg\rbrace \ . \nonumber \\
    && \label{amcRNK}
\end{eqnarray}
%
Examination of Eq.~(\ref{amcRNK}) concludes that the Kretschmann scalar is globally finite, and hence by Theorem~\ref{Theorem:Kretsch}, aMcRN spacetime is curvature-regular.

The horizon locations are messy, given implicitly in terms of the Lambert $W$ function \emph{via} examination of the roots of $f_{\text{aMcRN}}(r)$:
\begin{equation}
    r_{H_{+,-}} = -\frac{\ell}{W_{i\in\lbrace0,-1\rbrace}\left(-\frac{r_{H}\ell}{2mr_{H}-Q^2}\right)} \ .
\end{equation}
For an alternative representation, one can demand satisfaction of
\begin{equation}
    \e^{-\ell/r}\left[\frac{2m}{r}-\frac{Q^2}{r^2}\right] = 1 \ ,
\end{equation}
and provided $Q<m$, and $\ell\ll m$, both an event (outer) and Cauchy (inner) horizon will exist in the $r>0$ universe, corresponding to a choice of the $W_0$ or $W_{-1}$ branch of the Lambert $W$ function respectively. For certain values of $Q,\ell$, the horizons will merge to one extremal horizon; proper examination of this, and of the other salient features of aMcRN spacetime, is for now left as an avenue for further research.

\subsubsection{Discussion so far}\label{sec:dis}

The aMc model presented is a regular black hole geometry for $\ell\in\left(0, \frac{2m}{\e}\right]$, and accordingly violates the SEC~\cite{RBHAEC:Zaslavskii:2010}. There is one degenerate extremal horizon for $\ell=\frac{2m}{\e}$, and both an event (outer) and Cauchy (inner) horizon when $0<\ell<\frac{2m}{\e}$. For $\ell>\frac{2m}{\e}$, there are no horizons and aMc spacetime models a compact massive object that isn't a black hole; \textit{e.g.} a gravastar, star, or planet. The notable orbits for the model are highly nontrivial, but amenable to extraction and analysis, and present several atypical features of qualitative interest. The model satisfies the radial NEC (there is no wormhole throat), but in the black hole case violates the transverse NEC in the deep core for $r<\frac{\ell}{4}$. The SEC violation is also only in the deep core, where $r<\frac{\ell}{2}$; one has satisfaction of the point-wise energy conditions in the domain of outer communication. The energy density $\rho$ is maximised at $r=\frac{\ell}{4}$, and the exponential expression present in the metric implies that the regular black hole has an asymptotically Minkowski core as $r\rightarrow 0$. The curvature of the geometry is asymptotically flat at infinity, at the core, and has some maximal peak in between. Surface gravity is easily extracted at each horizon, and consequently the Hawking temperature. Notably, the relevant curvature quantities are significantly simpler than for canonical regular black hole solutions (Bardeen/Hayward/Frolov \emph{etc.}), at the cost of having some important physical features defined explicitly in terms of the Lambert $W$ function. The newly constructed aMcRN candidate spacetime has also been presented, and demonstrated to be an everywhere-regular model for charged regular black holes, with precise examination of its important features left as an opportunity for future research.

%

%% file: 03-AMC/2-AMCQNM.tex
\chapter{Quasi-normal modes and the ringdown}\label{C:AMCQNM}
Given the conditions that a propagating waveform is purely ingoing at the horizon and purely outgoing at spatial infinity, the proper oscillation frequencies of a candidate black hole spacetime are determined \textit{via} analysis of the permitted quasi-normal modes (QNMs). QNM analysis is by now utterly standard, with a wealth of literature containing QNM analyses in many varied contexts~\cite{SOGRBASBH:Vishveshwara:1970, GFOAPFIASGAITH:Zerilli:1970, LWTOGWFAVBH:Press:1971, GRFAPFRIASBH:Davis:1971, NPORGC1:Price:1972, NPORGC2:Price:1972, RFITSB:Bardeen:1973, PAFGAERIARG:Zerilli:1974, TQMOTSBH:Chandrasekhar:1975, ROOARRBH:Detweiler:1977, BHAGW2:Detweiler:1979, QOOASBH:Blome:1984, NATTQMOABH:Ferrari:1984, TROASBH:Bachelot:1993, ESOREAQMOCO:Fiziev:2006, QMOBH:Konoplya:2011, IOWARBHSBAPSF:Bronnikov:2012, QMORBH:Flachi:2013, QMOBBH:Correa:2012, OQMFGPOBBH:Ulhoa:2014, QMOTFARBH:Abdujabbarov:2015, GWFQMOACOLW:Aneesh:2018, QMOBHANS:Santos:2019, ADROQMISASBH:Aragon:2020, ACBSRABHQF:Cuadros-Melgar:2020, ROTRBHWT:Churilova:2020}, as well as the QNMs of propagating waveforms emanating from an astrophysical source being directly observed \textit{via} the LIGO/Virgo binary merger events~\cite{BBHMITFALIGOOR:Abbott:2016, OOGWFABBHM:Abbott:2016, OOA50SMBBH:Abbott:2017, AIOTBBHM:Abbott:2016, POTBBHM:Abbott:2016, GW23SM:Abbott:2020, GWTC1:Abbott:2019}. Given the hope that LIGO/Virgo (or more likely LISA~\cite{LISA:Barausse:2020}) will eventually be able to delineate the fingerprint of classical black holes from possible black hole mimickers, it is increasingly relevant to analyse well-motivated candidate spacetimes which model black hole mimickers and to compile results that speak to the advances made in observational and gravitational wave astronomy. It should be noted that such analysis is purely classical, as is consistent with the relevant ringdown calculation for LIGO/Virgo. As already discussed, it is well-known that classical curvature singularities in GR typically occur at a distance scale that necessitates a complete theory of quantum gravity. More specifically, treatments of both the classical analysis and aspects from quantum mechanics can lead to the amelioration of curvature singularities in certain configurations~\cite{QMFATAOFIGR:Fulling:1973, QEOANS:Goswami:2006}. However, in the absence of a phenomenologically falsifiable/verifiable theory of quantum gravity, appealing to the strategy of ``turning one knob at a time'' (see \S~\ref{Intro:knob}), it is well-motivated to construct nonsingular model spacetimes and extract (potential) astrophysical observables which are in principle falsifiable/verifiable by the observational community through the lens of GR.
\clearpage

%
Consequently, the analysis in Chapter~\ref{C:AMCQNM} contains some of the first steps required to speak directly to the calculations relevant for the gravitational wave observatories. These results were first presented in reference~\cite{ROTRBHWAMC:Simpson:2021}. In \S~\ref{S:RW_potential} a Regge--Wheeler analysis is performed for the aMc spacetime discussed in Chapter~\ref{C:AMCog}, followed by a first-order WKB approximation of the permitted QNMs for propagating waveforms in \S~\ref{S:QNMs}. The approximation is computed for spin zero scalar and spin one vector perturbations. Subsequently, in \S~\ref{S:Results} numerical results analysing the respective fundamental modes are compiled for various values of the $\ell$ parameter (which quantifies the deviation from Schwarzschild spacetime), and for various multipole numbers $k$. For the electromagnetic spin one fluctuations, and for the scalar spin zero fluctuations, the signals are found to be both shorter-lived and higher-energy than their Schwarzschild counterparts (for a specific range of relevant values of $\ell$). To conclude Chapter~\ref{C:AMCQNM}, in \S~\ref{S:RWperturb} a completely general analysis of first-order perturbations of the Regge--Wheeler potential is conducted, before specialising to Schwarzschild spacetime. A general result is presented explicating the shift in QNMs under perturbation of the Regge--Wheeler potential.

Recall from Eq.~(\ref{amcmetric}) that aMc spacetime is given by the line element
\begin{equation}
\dd s^2 = -\left(1-\frac{2m\,\e^{-\ell/r}}{r}\right)\dd{t}^2 + \frac{\dd{r}^2}{1-\frac{2m\,\e^{-\ell/r}}{r}} + r^2\,\d\Omega^{2}_{2} \ .\label{amcmetric2}
\end{equation}
For thorough discussions of the important features of this geometry, see references~\cite{RBHWAMC:Simpson:2019, PSISCOOSCO:Berry:2020}, or review Chapter~\ref{C:AMCog}.

\section{Regge--Wheeler potential}\label{S:RW_potential}

%
In this section the spin-dependent Regge--Wheeler potentials are explored. The spin two axial mode involves somewhat messier perturbations, and hence do not lend themselves nicely to the WKB approximation and subsequent computation of QNM profiles without the assistance of numerical code. Due to this ensuing intractability, the relevant Regge--Wheeler potential for the spin two axial mode is explored for completeness, before specialising the QNM discourse to spin zero (scalar) and spin one (electromagnetic, \textit{e.g.}) perturbations only. The QNMs of spin two axial perturbations are relegated to the domain of future research. Given one does not know the spacetime dynamics \textit{a priori}, the relativistic Cowling approximation~\cite{Cowling} is invoked, where one allows the scalar/vector field of interest to oscillate whilst keeping the candidate geometry fixed. This formalism closely follows that of reference~\cite{RELSAGFFDBH:Boonserm:2013}.

To proceed, one implicitly defines the tortoise coordinate \textit{via}
\begin{equation}
\dd{r^*} = \frac{\dd{r}}{1-\frac{2m\,\e^{-\ell/r}}{r}} \ .\label{tortoise}
\end{equation}
Although this equation is not analytically integrable, one can still conduct analysis of the Regge--Wheeler potential through this implicit definition of the tortoise coordinate. The coordinate transformation Eq.~(\ref{tortoise}) allows one to write the spacetime metric Eq.~(\ref{amcmetric2}) in the following form:
\begin{equation}
\dd{s}^2 = \left( 1- \frac{2m\,\e^{-\ell/r}}{r} \right) \bigg\{ -\dd{t}^2 + \dd{r_*}^2 \bigg\} + r^2 \left( \dd{\theta}^2 + \sin^2\theta \dd{\phi}^2 \right) \ ,
\end{equation}
which can then be rewritten as
\begin{equation}\label{metricstar}
\dd{s}^2 = A(r_*)^2 \big\{ -\dd{t}^2 + \dd{r_*}^2 \big\} + B(r_*)^2 \left( \dd{\theta}^2 + \sin^2\theta \dd{\phi}^2 \right) \ .
\end{equation}
In Regge and Wheeler's original work~\cite{SOASS:Regge:1957}, they show that for perturbations in a black hole spacetime, assuming a separable wave form of the type
\begin{equation}
\Psi(t,r_*,\theta,\phi) = \e^{i\omega t} \psi(r_*) Y(\theta,\phi)
\label{sepwaveform}
\end{equation}
results in the following differential equation (now called the Regge--Wheeler equation):
\begin{equation}
\pdv[2]{\psi(r_*)}{r_*} + \big\{ \omega^2 - \V_S \big\}\,\psi(r_*) = 0 \ .
\label{RWeq}
\end{equation}
Here \( Y(\theta,\phi) \) represents the spherical harmonic functions, \( \psi(r_*) \) is a propagating scalar, vector, or spin two axial bivector field in the candidate spacetime, \( \V_S \) is the spin-dependent Regge--Wheeler potential, and \( \omega \) is some (possibly complex) temporal frequency in the Fourier domain~\cite{ESOREAQMOCO:Fiziev:2006, GWFQMOACOLW:Aneesh:2018, QMOBHANS:Santos:2019, BTTW:Simpson:2019, RELSAGFFDBH:Boonserm:2013, SOASS:Regge:1957, EMRATW:Boonserm:2018}. The method for solving Eq.~\eqref{RWeq} is dependent on the spin of the perturbations and on the background spacetime. For instance, for vector perturbations (\(S=1\)), specialising to electromagnetic fluctuations, one analyses the electromagnetic four-potential subject to Maxwell's equations:
\enlargethispage{20pt}
\begin{equation}
    \frac{1}{\sqrt{-g}}\,\partial_{\mu}\left(F^{\mu\nu}\,\sqrt{-g}\right) = \bf{0} \ ,
\end{equation}
whilst for scalar perturbations ($S=0$), one solves the minimally coupled massless Klein--Gordon equation
\begin{equation}
    \square\psi(r) = \frac{1}{\sqrt{-g}}\,\partial_{\mu}\left(\sqrt{-g}\,\partial^{\mu}\,\psi\right) = 0 \ .
\end{equation}
Further details can be found in references~\cite{QMOBHANS:Santos:2019, ADROQMISASBH:Aragon:2020, SOASS:Regge:1957, RELSAGFFDBH:Boonserm:2013}. For spins $S\in\lbrace0,1,2\rbrace$, this yields the general result in static spherical symmetry~\cite{RELSAGFFDBH:Boonserm:2013, EMRATW:Boonserm:2018}:
\begin{equation}
\V_{0,1,2} = \left\{\frac{A^2}{B^2}\right\} \left[k(k+1)+S(S-1)(g^{rr}-1)\right] + (1-S) \frac{\partial^2_{r_*}B}{B} \ ,
\end{equation}
where $A$ and $B$ are the relevant functions as specified by Eq.~(\ref{metricstar}), $k$ is the multipole number (with $k\geq S$), and $g^{rr}$ is the relevant contrametric component with respect to standard curvature coordinates (for which the covariant components are presented in Eq.~(\ref{amcmetric2})).

For the spacetime under consideration, one has $A(r)=\sqrt{1-\frac{2m\,\e^{-\ell/r}}{r}}$,\newline \( B(r) = r \), $g^{rr} = 1-\frac{2m\,\e^{-\ell/r}}{r}$, and \( \partial_{r_*} = \left(1-\frac{2m\,\e^{-\ell/r}}{r}\right) \partial_r \). Hence
\begin{eqnarray}
\frac{\partial^2_{r_*}B}{B} &=& \frac{\left(1-\frac{2m\,\e^{-\ell/r}}{r}\right)\partial_r\left[1-\frac{2m\,\e^{-\ell/r}}{r}\right]}{r} \nonumber \\[3pt]
&=& \left( \frac{r - 2m\,\e^{-\ell/r}}{r^3} \right) \left[ \frac{2m\,\e^{-\ell/r}(r-\ell)}{r^2} \right] \ ,
\end{eqnarray}
and so one has the exact result that
\begin{equation}
\V_{0,1,2} = \left( \frac{r - 2m\,\e^{-\ell/r}}{r^3} \right) \left\{ k(k+1) + \frac{2m\,\e^{-\ell/r}}{r}(1-S)\left[S + 1 - \frac{\ell}{r}\right] \right\} \ .
\end{equation}
That is,
\begin{equation}
\V_{0,1,2} = \left( 1- \frac{2m\,\e^{-\ell/r}}{r} \right) \left\{ \frac{k(k+1)}{r^2}  + \frac{2m\,\e^{-\ell/r}}{r^3}(1-S)\left[S+1-\frac{\ell}{r}\right] \right\} \ .
\end{equation}
Note that at the outer horizon, where one has \( r_{H_{+}} = 2m \,\e^{W_{0}\left(-\frac{\ell}{2m}\right)} \), with $W$ being the special function now commonly called the Lambert $W$ function~\cite{RELSAGFFDBH:Boonserm:2013, EMRATW:Boonserm:2018, SAOTLWFTP:Valluri:2000, TLWFAQS:Valluri:2009, BTGFFSBH:Boonserm:2008, QF:Boonserm:2011, SRGEWTLWF:Sonoda:2013, AFOTEPITLNL...:Sonoda:2013, OTLWF:Corless:1996, FWLDAWDL:Vial:2012, WPATLWF:Stewart:2011, SPAWDL:Stewart:2012, PATLWF:Visser:2018}, the Regge--Wheeler potential vanishes. Taking the limit as \( \ell\rightarrow0 \) recovers the known Regge--Wheeler potentials for spin zero, spin one, and spin two axial perturbations in the Schwarzschild spacetime:
\begin{equation}
\V_{Sch.,0,1,2} = \lim_{\ell\rightarrow0} \V_{0,1,2} = \left(1-\frac{2m}{r}\right) \left\{ \frac{k(k+1)}{r^2} + \frac{2m}{r^3}(1-S^2) \right\} \ .
\end{equation}
Note that in Regge and Wheeler's original work~\cite{SOASS:Regge:1957}, only the spin two axial mode was analysed. However, this result agrees both with the original work, as well as with later results extending to spin zero and spin one perturbations~\cite{QMOBHANS:Santos:2019}. It is informative to explicate the exact form for the Regge--Wheeler potential for each spin case, and to then plot the qualitative behaviour of the potential as a function of the dimensionless variables $r/m$ and $\ell/m$ for the respective dominant multipole numbers ($k=S$).

\begin{itemize}
    \item Spin one vector field: The conformal invariance of spin one massless particles in $(3+1)$ dimensions implies that the $\frac{\partial_{r_*}^2B}{B}$ term vanishes, and indeed mathematically the potential reduces to the highly tractable
    \begin{equation}
        \V_1 = \left(1-\frac{2m\,\e^{-\ell/r}}{r}\right)\frac{k(k+1)}{r^2} \ .
    \end{equation}
    Specialising to the dominant multipole number $k=1$ then gives
    \begin{equation}\label{ell1}
        \eval{\V_1}_{k=1} = \frac{2}{r^2}\left(1-\frac{2m\,\e^{-\ell/r}}{r}\right) \ .
    \end{equation}
    Now, in order to examine the qualitative features of the potential it is of mathematical convenience to define the new dimensionless variables $x=r/m$, and $y=\ell/m$. It is worth re-iterating that convention in the historical literature would be to set $\ell\sim m_{p}$, such that the newly introduced scalar parameter appeals to the quantum gravity regime. This would imply that $y=\ell/m\sim m_{p}/m_{sun}\ll 1$. In view of the redefinition of parameters, Eq.~(\ref{ell1}) may be re-expressed as
    \begin{equation}
        \eval{m^2\V_1}_{k=1} = \frac{2}{x^2}\left(1-\frac{2\,\e^{-y/x}}{x}\right) \ .
    \end{equation}
    The qualitative features of $\V_1$ are then plotted in Fig.~\ref{F:1}, for the full range of $y$ such that the spacetime still possesses a nontrivial horizon structure, and the domain for $x$ such that one is strictly outside the horizon.
    \enlargethispage{35pt}
\begin{figure}[htb!]
\begin{center}
\includegraphics[scale=0.32]{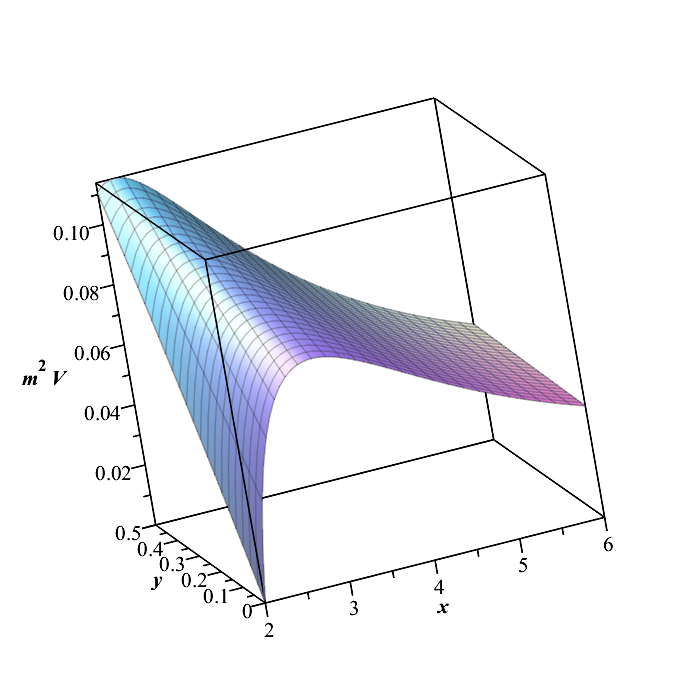}
\includegraphics[scale=0.33]{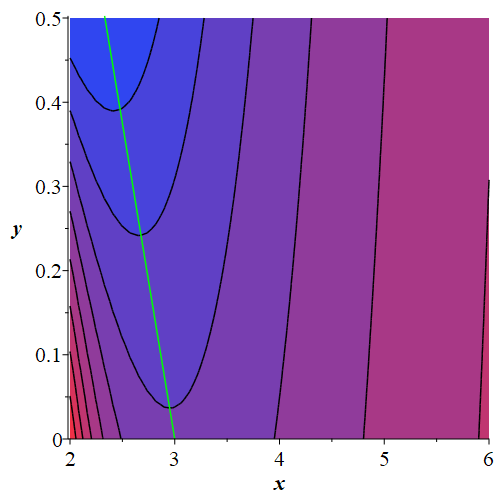}
\caption[Qualitative features of the spin one Regge--Wheeler potential for aMc spacetime for the dominant multipole number $k=1$]{The qualitative features of the spin one Regge--Wheeler potential for the dominant multipole number $k=1$ are depicted. The upper plot is a three-dimensional plot of $m^2\V_1$. The lower plot is a contour plot; `blue'$\rightarrow$`red' corresponds to `high'$\rightarrow$`low'.}
\label{F:1}
\end{center}
\end{figure}
\vfill
\clearpage

An immediate sanity check from Fig.~\ref{F:1} is that for $\ell=0$, where the candidate spacetime reduces to Schwarzschild, one observes a peak at $r=3m$. This is the expected location of the photon sphere for Schwarzschild, and is indeed the corresponding location of the peak of the relevant spin one Regge--Wheeler potential. As $\ell$ increases, the $r$-coordinate location of the peak decreases. For all values of $\ell$, there is falloff at spacial infinity, and once the peak is crested there is rapid falloff as one approaches the horizon location (where the potential vanishes completely). The green line present in the lower plot of Fig.~\ref{F:1} corresponds to the approximate location of the photon sphere from Eq.~(\ref{E:small-a}); $r_\gamma\approx 3m-\frac{4}{3}\ell$. This approximation is used for the location of the peak of the spin one potential in order to extract the QNM profile approximations in \S~\ref{S:QNMs}. One can see that for the given domain and range this approximation has high accuracy, closely matching with the locations of the peaks.
\item Spin zero scalar field: The potential now becomes
    \begin{equation}
        \V_0 = \left(1-\frac{2m\,\e^{-\ell/r}}{r}\right)\left\lbrace \frac{k(k+1)}{r^2}+\frac{2m\,\e^{-\ell/r}}{r^3}\left(1-\frac{\ell}{r}\right)\right\rbrace \ ,
    \end{equation}
    and, fixing the dominant multipole number $k=0$, one specialises to the scalar $s$-wave, which is of particular importance, yielding
    \begin{equation}
        \eval{\V_0}_{k=0} = \frac{2m\,\e^{-\ell/r}}{r^3}\left(1-\frac{2m\,\e^{-\ell/r}}{r}\right)\left(1-\frac{\ell}{r}\right) \ .
    \end{equation}
    Once again, to examine the qualitative features of the potential it is convenient to re-express this in terms of the dimensionless variables $x=r/m, \ y=\ell/m$:
    \begin{equation}
        \eval{m^2\V_0}_{k=0} = \frac{2\,\e^{-y/x}}{x^3}\left(1-\frac{2\,\e^{-y/x}}{x}\right)\left(1-\frac{y}{x}\right) \ .
    \end{equation}
    The qualitative features of $\V_0$ are then displayed in Fig.~\ref{F:2}.
    
    The most notable feature is the spin zero peak; there is a slight shift in the peak locations between the spin one and spin zero potentials. The green line in the lower plot of Fig.~\ref{F:2} is a ``line of best fit'', obtained \textit{via} manual corrections starting from the approximate location of the photon sphere from Eq.~(\ref{E:small-a}). This marks the $\ell$-dependent coordinate location $r_0\approx \frac{41}{15}m-\frac{4}{3}\ell$. Given one does not have information concerning how the peak shifts when comparing the spin one and spin zero potentials \textit{a priori}, and the peak location is not analytically solvable (see \S~\ref{S:QNMs}), this approximation is the best one can do in order to retain the desired level of mathematical tractability. Consequently, in \S~\ref{S:QNMs}, the approximation for $r_0$ as above is used in the computation of the relevant QNM profiles. The remaining features of the plot are qualitatively similar to those for the spin one case.
    %
\begin{figure}[htb!]
\begin{center}
\includegraphics[scale=0.32]{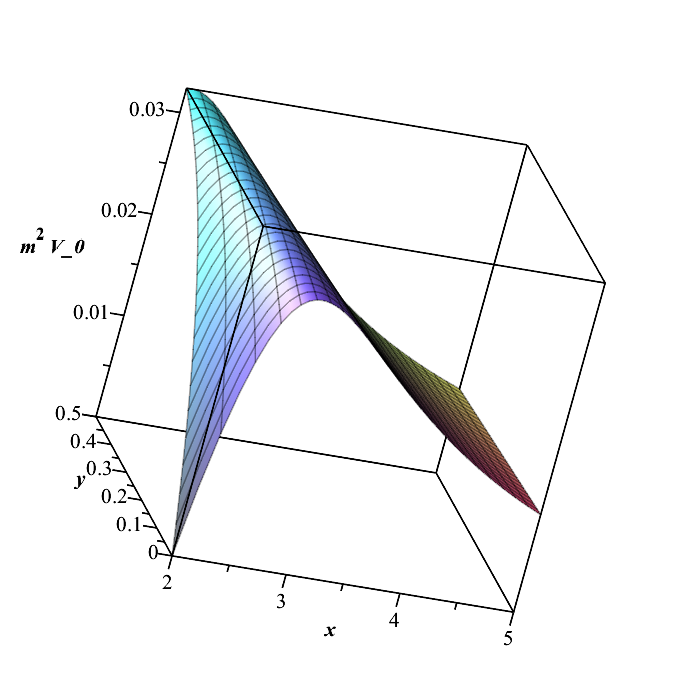}
\includegraphics[scale=0.33]{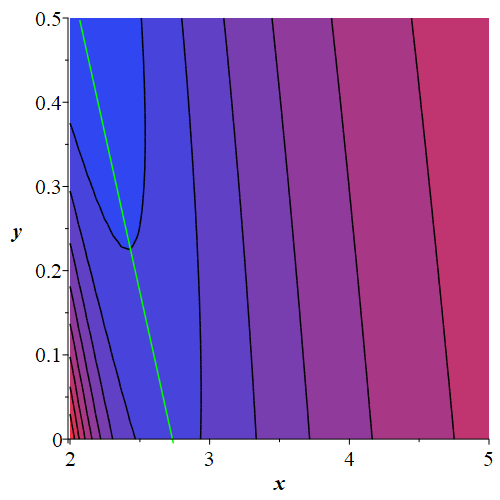}
\caption[Qualitative features of the spin zero Regge--Wheeler potential for aMc spacetime for the dominant multipole number $k=0$]{The qualitative features of the spin zero Regge--Wheeler potential for the dominant multipole number $k=0$ are depicted. The upper plot is a three-dimensional plot of $m^2\V_0$. The lower plot is a contour plot; `blue'$\rightarrow$`red' corresponds to `high'$\rightarrow$`low'.}
\label{F:2}
\end{center}
\end{figure}
    %
    \item Spin two bivector field (axial mode): The potential becomes
    \begin{equation}
        \V_2 = \left(1-\frac{2m\,\e^{-\ell/r}}{r}\right)\left\lbrace \frac{k(k+1)}{r^2}-\frac{2m\,\e^{-\ell/r}}{r^3}\left(3-\frac{\ell}{r}\right)\right\rbrace \ ,
    \end{equation}
    and, fixing the dominant multipole number $k=2$, one finds:
    \begin{equation}
        \eval{\V_2}_{k=2} = \frac{1}{r^2}\left(1-\frac{2m\,\e^{-\ell/r}}{r}\right)\left\lbrace 6-\frac{2m\,\e^{-\ell/r}}{r}\left(3-\frac{\ell}{r}\right)\right\rbrace \ .
    \end{equation}
    Once again, it is informative to re-express this in terms of the dimensionless variables $x=r/m$, $y=\ell/m$, giving
    \begin{equation}
        \eval{m^2\V_2}_{k=2} = \frac{1}{x^2}\left(1-\frac{2\,\e^{-y/x}}{x}\right)\left\lbrace 6 -\frac{2\,\e^{-y/x}}{x}\left(3-\frac{y}{x}\right)\right\rbrace \ .
    \end{equation}
    The qualitative features of $\V_2$ are then displayed in Fig.~\ref{F:3}.
    
    Once again the approximate location for the peak of the spin two (axial) potential is obtained \textit{via} application of manual corrections to the approximate location of the photon sphere from Eq.~(\ref{E:small-a}), and is found to be $r_2\approx\frac{10}{3}m-\frac{5}{3}\ell$ (this is the green line in the lower plot of Fig.~\ref{F:3}). This approximation would serve as a starting point to extract QNM profile approximations for the spin two axial mode, similarly to the processes performed for spins one and zero in \S~\ref{S:QNMs}. However, for a combination of readability and tractability, this is for now a topic for future research. The remaining qualitative features of the spin two (axial) potential are similar to those for spins one and zero.
\begin{figure}[htb!]
\begin{center}
\includegraphics[scale=0.32]{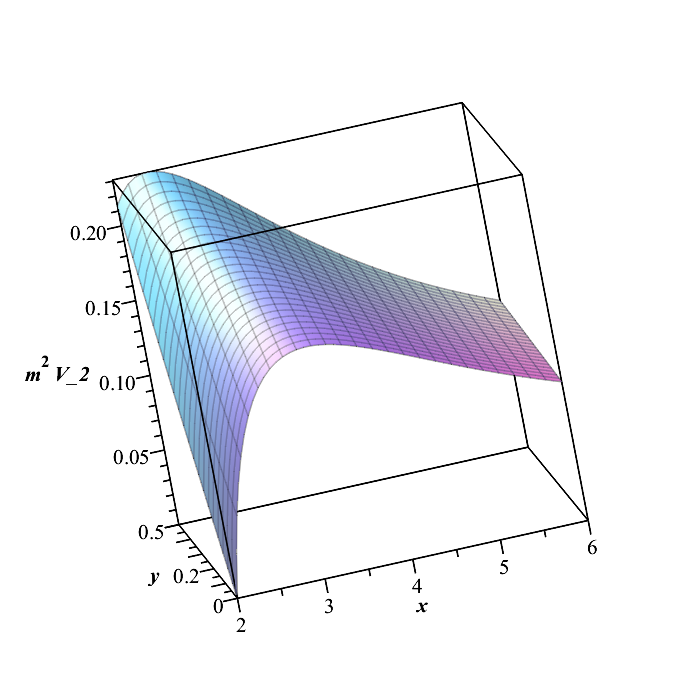}
\includegraphics[scale=0.33]{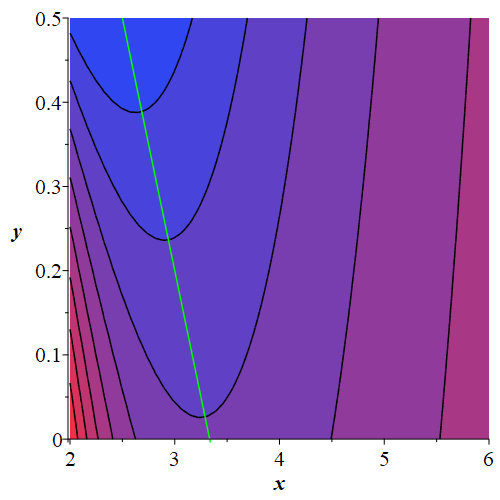}
\caption[Qualitative features of the spin two axial Regge--Wheeler potential for aMc spacetime for the dominant multipole number $k=2$]{The qualitative features of the spin two axial Regge--Wheeler potential for the dominant multipole number $k=2$ are depicted. The upper plot is a three-dimensional plot of $m^2\V_2$. The lower plot is a contour plot; `blue'$\rightarrow$`red' corresponds to `high'$\rightarrow$`low'.}
\label{F:3}
\end{center}
\end{figure}
%
\end{itemize}
\clearpage

\section{First-order WKB approximation}\label{S:QNMs}

In order to calculate the QNM profiles for the candidate spacetime, one first defines them in the standard way: they are the \( \omega \) present in the right-hand-side of Eq.~\eqref{sepwaveform}, and they satisfy the ``radiation'' boundary conditions that \( \Psi \) is purely outgoing at spacial infinity and purely ingoing at the horizon~\cite{QOOASBH:Blome:1984, QMOBHANS:Santos:2019}. Due to the inherent difficulty of analytically solving the Regge--Wheeler equation, a standard approach in the literature is to use the WKB approximation. Although the WKB method was originally constructed to solve Schr\"odinger-type equations in quantum mechanics, the close resemblance between the Regge--Wheeler equation Eq.~\eqref{RWeq} and the Schr\"odinger equation allows for it to be readily adapted to the general relativistic setting. Given the use of the WKB approximation, one cannot extend the analysis of the QNMs for the candidate spacetime to the case when $\ell>2m/\e$, as for this case there are no horizons in the geometry. The existence of the outer horizon (or at the very least an extremal horizon) is critical to setting up the correct radiative boundary conditions. Other techniques for approximating the QNMs, \textit{e.g.} time domain integration (see reference~\cite{ROTRBHWT:Churilova:2020} for an example), are likely to be applicable in this context. For now, this research is relegated to the domain of the future.

In order to proceed with the WKB method, one makes the stationary ansatz $\Psi\sim \e^{i\omega t}$, such that all of the qualitative behaviour for $\Psi$ is encoded in the profiles of the respective $\omega$. Computing a WKB approximation to first-order yields a relatively simple and tractable approximation to the QNM profiles for a black hole spacetime~\cite{QOOASBH:Blome:1984, QMOBHANS:Santos:2019, ACBSRABHQF:Cuadros-Melgar:2020}:
\begin{equation}\label{BLAH}
\omega^2 \approx \bigg[ \V(r_*) - i \left(n+\frac{1}{2}\right) \sqrt{-2 \, \partial_{r_*}^2 \V(r_*)} \bigg]_{r_*=r_{max}},
\end{equation}
where \( n \in \mathbb{N} \) is the overtone number, and where \( r_* = r_{max} \) is the tortoise coordinate location which maximises the relevant Regge--Wheeler potential. It is worth noting that $\eval{\V(r_*)}_{r_*=r_{max}}=\eval{\V(r)}_{r=r_{max}}$; this will be used in the subsequent analysis.

In-depth calculations of the WKB approximation up to higher orders in a general setting can also be found in references~\cite{QOOASBH:Blome:1984, QMOBHANS:Santos:2019, ACBSRABHQF:Cuadros-Melgar:2020}. Furthermore, various improvements and refinements to the WKB approximation have been explored in references~\cite{BHNM:Schutz:1985, BHNM:Iyer:1987, QBOTDSBHAHOWKBA:Konoplya:2003, QMOSDSBH:Zhidenko:2004, QMOBH:Matyjasek:2017, HOWKBFFQMAGF:Konoplya:2019}. These include the derivation of a generalised higher-order formula, and the exploration of an improved variant of the original WKB approximation using Pad\'{e} approximants. These formulae present various intractabilities by hand and are best handled by numerical code. Consequently, a first-order calculation is performed to holistically analyse the qualitative aspects of the QNMs; refinement in accuracy is left to the numerical relativity community to explore further.

\subsection{Spin one}

For spin one particles recall that the relevant Regge--Wheeler potential is given by
\begin{equation}
    \V_1(r) = \left(1-\frac{2m\,\e^{-\ell/r}}{r}\right)\frac{k(k+1)}{r^2} \ .
\end{equation}
The $\V_1(r)$ Regge--Wheeler potential is proportional to the $V_0(r)$ effective potential used for determining the location of the photon sphere for massless particles~\cite{PSISCOOSCO:Berry:2020}. Specifically, one has
\begin{equation}
\pdv{\, \V_1}{r} = \frac{2k(k+1)}{r^3}\bigg\{ \frac{m \, \e^{-\ell/r}}{r}\left(3-\frac{\ell}{r}\right) - 1 \bigg\} \ .
\end{equation}
The resulting stationary points are not analytically solvable, and \textit{via} comparison with Eq.~(\ref{eq;photonsphere}) one sees that the spin one Regge--Wheeler potential is maximised precisely at the location of the photon sphere; \( r_1 = r_\gamma = \sqrt{ m\,\e^{-\ell/r_\gamma}(3r_\gamma-\ell) } \approx 3m - \frac{4}{3}\ell \). Thus, one immediately obtains the spin one, first-order WKB approximation for the real part of the QNMs (recall $y=\ell/m$):
\begin{eqnarray}
\Re(\omega^2) \approx \eval{\V_1}_{r=r_1} &\approx& \frac{9 \, k(k+1)}{(9m-4\ell)^2} \left\lbrace 1 - \frac{6m\,\e^{\frac{3\ell}{4\ell-9m}}}{9m-4\ell} \right\rbrace \nonumber \\[3pt]
&=& \frac{9k(k+1)}{m^2}\left\lbrace\frac{1}{(9-4y)^2}\left[1-\frac{6\,\e^{\frac{3y}{4y-9}}}{9-4y}\right]\right\rbrace \ .
\end{eqnarray}
Letting $r_c=3m-\frac{4}{3}\ell$, recalling $x=r/m$ (\textit{i.e.} $x_c=r_c/m$), and defining $z=\ell/r_c$, alternative representations include (eliminating $m$):
\begin{eqnarray}
    \Re(\omega^2) &\approx& \frac{k(k+1)}{r_c^2}\left\lbrace 1 - \frac{2\,\e^{-\ell/r_c}(r_c+\frac{4}{3}\ell)}{3r_c}\right\rbrace \nonumber \\[3pt]
    &=& \frac{k(k+1)}{r_c^2}\left\lbrace 1 - \frac{2\,\e^{-z}(1+\frac{4}{3}z)}{3}\right\rbrace \ ,
\end{eqnarray}
or (eliminating $\ell$):
\begin{equation}
    \Re(\omega^2) \approx \frac{k(k+1)}{r_c^2}\left\lbrace 1 - \frac{2m\,\e^{\frac{3(r_c-3m)}{4r_c}}}{r_c}\right\rbrace = \frac{k(k+1)}{r_c^2}\left\lbrace 1 - \frac{2\,\e^{\frac{3}{4}\left(1-\frac{3}{x_c}\right)}}{x_c}\right\rbrace \ .
\end{equation}
\vspace*{7pt}

Which expression is preferred is a matter of taste and context.
\clearpage

Now, to compute the imaginary part of the QNMs, note that
\begin{eqnarray}
\eval{ \pdv[2]{\V_1}{r_*} }_{r=r_1} &=& \left(1-\frac{2m\,\e^{-\ell/r}}{r}\right) \Bigg\lbrace\left(1-\frac{2m\,\e^{-\ell/r}}{r}\right)\pdv[2]{\V_1}{r} \nonumber \\[3pt]
&& \qquad \qquad \qquad \qquad \quad + \frac{2m\,\e^{-\ell/r}(r-\ell)}{r^3} \pdv{\V_1}{r} \Bigg\rbrace \Bigg\vert_{r=r_1} \ . \nonumber \\
&&
\end{eqnarray}
But already it is known that \( \partial\V_1/\partial r \eval_{r=r_1} = 0 \), and so this reduces to
\begin{equation}
\eval{ \pdv[2]{\V_1}{r_*} }_{r=r_1} = \eval{ \left(1-\frac{2m\,\e^{-\ell/r}}{r}\right)^2 \pdv[2]{\V_1}{r} }_{r=r_1} \ .
\end{equation}
Thus, for spin one particles one finds
\begin{align}
\eval{ \pdv[2]{\V_1}{r_*} }_{r=r_1} &= \eval{ k(k+1) \left(1-\frac{2m\,\e^{-\frac{\ell}{r}}}{r}\right)^2 \left\{ \frac{6}{r^4} - (\ell-2r)(\ell-6r)\frac{2m\,\e^{-\frac{\ell}{r}}}{r^7} \right\} }_{r=r_1}	\notag \\[3pt]
&\approx \frac{486 \, k(k+1)}{(9m-4\ell)^4} \left(1 - \frac{6m\,\e^{3\ell\over4\ell-9m}}{9m-4\ell} \right)^2 	\notag \\
&\hspace{1.5cm}\times\left\{ 1-\frac{27m\,\e^{3\ell\over4\ell-9m} (18m-11\ell)(2m-\ell)}{(9m-4\ell)^3} \right\} \ ,
\end{align}
thereby giving
\begin{align}
\Im(\omega^2) &\approx \eval{ -\left(n+\frac{1}{2}\right)\sqrt{-2 \, \partial_{r_*}^2 \V_1(r)} }_{r=r_1} 	\notag \\[3pt]
&\approx -\frac{18\sqrt{3}}{(9m-4\ell)^2} \, \left(n+\frac{1}{2}\right) \left(1 - \frac{6m\,\e^{3\ell\over4\ell-9m}}{9m-4\ell} \right) \notag \\
&\hspace{1cm}\times\sqrt{k(k+1) \left\lbrace \frac{27m\,\e^{3\ell\over4\ell-9m} (18m-11\ell)(2m-\ell)}{(9m-4\ell)^3} - 1 \right\rbrace} \notag \\[3pt]
&= -\frac{\left(n+\frac{1}{2}\right)}{m^2}18\sqrt{3}\left(1-\frac{6\,\e^{\frac{3y}{4y-9}}}{9-4y}\right) \notag \\
&\hspace{1cm}\times\sqrt{\frac{k(k+1)}{(9-4y)^4}\left\lbrace\frac{27\,\e^{\frac{3y}{4y-9}}(18-11y)(2-y)}{(9-4y)^3}-1\right\rbrace} \ .
\end{align}
\clearpage
Alternative expressions include (eliminating $m$):
\begin{eqnarray}
    \Im(\omega^2) &\approx& -\frac{2\sqrt{3}\left(n+\frac{1}{2}\right)}{r_c^2}\left(1-\frac{2\,\e^{-\ell/r_c}(3r_c+4\ell)}{9r_c}\right) \nonumber \\[3pt]
    && \hspace{1cm}\times \sqrt{k(k+1)\left\lbrace\frac{(3r_c+4\ell)(2r_c-\ell)(6r_c-\ell)\,\e^{-\ell/r_c}}{27r_c^3}-1\right\rbrace} \nonumber \\[3pt]
    &=& -\frac{2\sqrt{3}\left(n+\frac{1}{2}\right)}{r_c^2}\left(1-\frac{2\,\e^{-z}(3+4z)}{9}\right) \nonumber \\[3pt]
    &&\hspace{1cm}\times\sqrt{k(k+1)\left\lbrace\frac{(3+4z)(2-z)(6-z)\,\e^{-z}}{27}-1\right\rbrace} \ ,
\end{eqnarray}
or (eliminating $\ell$):
\begin{eqnarray}
    \Im(\omega^2) &\approx& -\frac{2\sqrt{3}\left(n+\frac{1}{2}\right)}{r_c^2}\left(1-\frac{2m\,\e^{\frac{3(r_c-3m)}{4r_c}}}{r_c}\right) \nonumber \\[3pt]
    && \hspace{1cm}\times \sqrt{k(k+1)\left\lbrace\frac{3m\,\e^{\frac{3(r_c-3m)}{4r_c}}(11r_c-9m)(3r_c-m)}{16r_c^3}-1\right\rbrace} \nonumber \\[3pt]
    &=& -\frac{2\sqrt{3}\left(n+\frac{1}{2}\right)}{r_c^2}\left(1-\frac{2\,\e^{\frac{3}{4}\left(1-\frac{3}{x_c}\right)}}{x_c}\right) \nonumber \\[3pt]
    &&\hspace{1cm}\times\sqrt{k(k+1)\left\lbrace\frac{3\,\e^{\frac{3}{4}\left(1-\frac{3}{x_c}\right)}\left(11x_c-9\right)\left(3x_c-1\right)}{16x_c^3} - 1 \right\rbrace} \ . \nonumber \\
    &&
\end{eqnarray}
Thus, the first-order WKB approximation of the spin one QNMs, for general multipole numbers $k$ and oscillation modes $n$, is given by
\enlargethispage{10pt}
\begin{eqnarray}\label{spinoneapprox}
    \omega^2 &\approx& \frac{9k(k+1)}{(9m-4\ell)^2}\left(1-\frac{6m\,\e^{\frac{3\ell}{4\ell-9m}}}{9m-4\ell}\right) \Bigg\lbrace 1- 2\sqrt{3}i\left(n+\frac{1}{2}\right) \nonumber \\[3pt]
    && \qquad \qquad \times\sqrt{\frac{1}{k(k+1)}\left[\frac{27m\,\e^{\frac{3\ell}{4\ell-9m}}(18m-11\ell)(2m-\ell)}{(9m-4\ell)^3}-1\right]}\Bigg\rbrace \nonumber \\
    && \nonumber \\
    &=& \frac{9k(k+1)}{m^2(9-4y)^2}\left(1-\frac{6\,\e^{\frac{3y}{4y-9}}}{9-4y}\right)\Bigg\lbrace 1 - 2\sqrt{3}i\left(n+\frac{1}{2}\right) \nonumber \\[3pt]
    && \qquad \qquad \qquad \times\sqrt{\frac{1}{k(k+1)}\left[\frac{27\,\e^{\frac{3y}{4y-9}}(18-11y)(2-y)}{(9-4y)^3}-1\right]}\Bigg\rbrace \ . \nonumber \\
    &&
\end{eqnarray}
\clearpage
For the purposes of extracting numerical results in \S~\ref{S:Results}, it is useful to redefine this as the dimensionless quantity
\begin{eqnarray}\label{eqforspinoneres}
    m^2\omega^2 &\approx& \frac{9k(k+1)}{(9-4y)^2}\left(1-\frac{6\,\e^{\frac{3y}{4y-9}}}{9-4y}\right)\Bigg\lbrace 1 - 2\sqrt{3}i\left(n+\frac{1}{2}\right) \nonumber \\[3pt]
    && \qquad \qquad \qquad \times\sqrt{\frac{1}{k(k+1)}\left[\frac{27\,\e^{\frac{3y}{4y-9}}(18-11y)(2-y)}{(9-4y)^3}-1\right]}\Bigg\rbrace \ . \nonumber \\
    &&
\end{eqnarray}
Alternative expressions include (eliminating $m$):
\begin{eqnarray}
    \omega^2 &\approx& \frac{k(k+1)}{r_c^2}\left(1-\frac{2(3r_c+4\ell)\,\e^{-\ell/r_c}}{9r_c}\right)\Bigg\lbrace 1-2\sqrt{3}i\left(n+\frac{1}{2}\right) \nonumber \\[3pt]
    && \qquad \times\sqrt{\frac{1}{k(k+1)}\left[\frac{(2r_c-\ell)(6r_c-\ell)(3r_c+4\ell)\,\e^{-\ell/r_c}}{27r_c^3}-1\right]}\Bigg\rbrace \nonumber \\
    && \nonumber \\
    &=& \frac{k(k+1)}{r_c^2}\left(1-\frac{2(3+4z)\,\e^{-z}}{9}\right)\Bigg\lbrace 1 - 2\sqrt{3}i\left(n+\frac{1}{2}\right) \nonumber \\[3pt]
    && \qquad \qquad \times\sqrt{\frac{1}{k(k+1)}\left[\frac{(3+4z)(2-z)(6-z)\e^{-z}}{27}-1\right]}\Bigg\rbrace \ , \nonumber \\
    &&
\end{eqnarray}
or (eliminating $\ell$):
\begin{eqnarray}
    \omega^2 &\approx& \frac{k(k+1)}{r_c^2}\left(1-\frac{2m\,\e^{\frac{3(r_c-3m)}{4r_c}}}{r_c}\right)\Bigg\lbrace 1-2\sqrt{3}i\left(n+\frac{1}{2}\right) \nonumber \\[3pt]
    && \qquad \times\sqrt{\frac{1}{k(k+1)}\left[\frac{3m\,\e^{\frac{3(r_c-3m)}{4r_c}}(11r_c-9m)(3r_c-m)}{16r_c^3}-1\right]}\Bigg\rbrace \nonumber \\
    && \nonumber \\
    &=& \frac{k(k+1)}{r_c^2}\left(1-\frac{2\,\e^{\frac{3}{4}\left(1-\frac{3}{x_c}\right)}}{x_c}\right)\Bigg\lbrace 1 - 2\sqrt{3}i\left(n+\frac{1}{2}\right) \nonumber \\[3pt]
    && \qquad \times\sqrt{\frac{1}{k(k+1)}\left[\frac{3\,\e^{\frac{3}{4}\left(1-\frac{3}{x_c}\right)}(11x_c-9)(3x_c-1)}{16x_c^3}-1\right]}\Bigg\rbrace \ . \nonumber \\
    &&
\end{eqnarray}
In the Schwarzschild limit one obtains
\begin{equation}
\omega^2_{Sch.} = \lim_{\ell\rightarrow0} \left(\omega^2\right) \approx \frac{k(k+1)}{27m^2}\left( 1 - \frac{i(2n+1)}{\sqrt{k(k+1)}} \right) \ ,
\end{equation}
which agrees with existing work in the literature~\cite{QOOASBH:Blome:1984, QMOBHANS:Santos:2019}.


\subsection{Spin zero}

For spin zero particles recall that one has the following specific form for the Regge--Wheeler potential:
\begin{equation}
    \mathcal{V}_{0}(r) = \left(1-\frac{2m\,\e^{-\ell/r}}{r}\right)\left\lbrace\frac{k(k+1)}{r^{2}}+\frac{2m\,\e^{-\ell/r}\left(1-\frac{\ell}{r}\right)}{r^{3}}\right\rbrace \ .
\end{equation}
It is immediately clear that the peak of this potential is going to be slightly shifted from the location of the photon sphere, which for the spin one case maximised $\mathcal{V}_{1}(r)$. Computing:
\begin{eqnarray}
    \frac{\partial{\mathcal{V}_{0}}}{\partial{r}} &=& \frac{1}{r^3}\Bigg\lbrace\left[\frac{2m\,\e^{-\ell/r}}{r}\left(3-\frac{\ell}{r}\right)-2\right]\left[k(k+1)+\frac{2m\,\e^{-\ell/r}}{r}\left(1-\frac{\ell}{r}\right)\right] \nonumber \\[3pt]
    && \qquad \quad + \left(1-\frac{2m\,\e^{-\ell/r}}{r}\right)\left[\frac{-2m\,\e^{-\ell/r}}{r}\left(\left(\frac{\ell}{r}\right)^2-3\left(\frac{\ell}{r}\right)+1\right)\right]\Bigg\rbrace \ . \nonumber \\
    &&
\end{eqnarray}
The associated stationary points are not analytically solvable for $r$. It is worth noting that in general, the stationary points of $\V_0$ are $k$-dependent, unlike in the case for $\V_1$. Without knowledge of the location of the peak for the spin zero potential \textit{a priori}, the best line of inquiry which retains the desired level of tractability is to specialise to the scalar $s$-wave (corresponding to $k=0$), which is of particular importance, playing a dominant role in the signal. This constraint is fit for purpose in extracting the relevant results in \S~\ref{S:Results}. Specialising to the $s$-wave, one finds
\begin{eqnarray}
    \eval{\frac{\partial\V_0}{\partial r}}_{k=0} &=& \frac{2m\,\e^{-\ell/r}}{r^4}\Bigg\lbrace\frac{m\,\e^{-\ell/r}}{r}\left[4\left(\frac{\ell}{r}\right)^2-14\left(\frac{\ell}{r}\right)+8\right] \nonumber \\[3pt]
    && \qquad \qquad \qquad \qquad -\left[\left(\frac{\ell}{r}\right)^2-5\left(\frac{\ell}{r}\right)+3\right]\Bigg\rbrace \ ,
\end{eqnarray}
and still the associated stationary points are not analytically solvable. As such, to make progress one uses the approximate location of the peak as found in \S~\ref{S:RW_potential}; $r_0\approx\frac{41}{15}m-\frac{4}{3}\ell$. See Fig.~\ref{F:2} for details. One obtains the following approximation for the real part of the spin zero QNMs for the scalar $s$-wave:
\begin{eqnarray}\label{Respin0}
    \Re(\omega^2) &\approx& \frac{6750m\,\e^{\frac{15\ell}{20\ell-41m}}(41m-35\ell)\left(30m\,\e^{\frac{15\ell}{20\ell-41m}}+20\ell-41m\right)}{(20\ell-41m)^5} \nonumber \\[3pt]
    &=& \frac{6750}{m^2}\left[\frac{\e^{\frac{15y}{20y-41}}(41-35y)(30\,\e^{\frac{15y}{20y-41}}+20y-41)}{(20y-41)^5}\right] \ .
\end{eqnarray}
For the imaginary part of the spin zero QNMs, firstly one has
\begin{eqnarray}
    \eval{\frac{\partial^2\V_0}{\partial r_*^2}}_{r=r_0} = \eval{\left(1-\frac{2m\,\e^{-\ell/r}}{r}\right)^2\frac{\partial^2\V_0}{\partial r^2}}_{r=r_0} \ ,
\end{eqnarray}
with
\begin{eqnarray}
    \frac{\partial^2\V_0}{\partial r^2} &=& \frac{2}{r^4}\Bigg\lbrace 3k(k+1) + \frac{4m^2\,\e^{-2\ell/r}}{r^2}\left[2\left(\frac{\ell}{r}\right)^2-10\left(\frac{\ell}{r}\right)+5\right]\left(\frac{\ell}{r}-2\right) \nonumber \\[2pt]
    && \qquad \qquad \quad -\frac{m\,\e^{-\ell/r}}{r}\Bigg[12(k^2+k-1)-4\left(\frac{\ell}{r}\right)(2k^2+2k-7) \nonumber \\[2pt]
    && \qquad \qquad \qquad \qquad \qquad \qquad +\left(\frac{\ell}{r}\right)^2(k^2+k-11)+\left(\frac{\ell}{r}\right)^3\Bigg]\Bigg\rbrace \ , \nonumber \\
    &&
\end{eqnarray}
giving
\begin{eqnarray}
    \frac{\partial^2\V_0}{\partial r_*^2} &=& \frac{2}{r^4}\left(1-\frac{2m\,\e^{-\frac{\ell}{r}}}{r}\right)^2\Bigg\lbrace \frac{4m^2\,\e^{-\frac{2\ell}{r}}}{r^2}\left[2\left(\frac{\ell}{r}\right)^2-10\left(\frac{\ell}{r}\right)+5\right]\left(\frac{\ell}{r}-2\right) \nonumber \\[2pt]
    && \quad +3k(k+1)-\frac{m\,\e^{-\frac{\ell}{r}}}{r}\Bigg[12(k^2+k-1)-4\left(\frac{\ell}{r}\right)(2k^2+2k-7) \nonumber \\[2pt]
    && \qquad \qquad \qquad \qquad \qquad \qquad \ \ +\left(\frac{\ell}{r}\right)^2(k^2+k-11)+\left(\frac{\ell}{r}\right)^3\Bigg]\Bigg\rbrace \ . \nonumber \\
    &&
\end{eqnarray}
For tractability, it is now prudent to specialise to $k=0$. This yields
\begin{eqnarray}
    \eval{ \frac{\partial^2\V_0}{\partial r_*^2}}_{k=0} &=& \frac{2}{r^4}\left(1-\frac{2m\,\e^{-\ell/r}}{r}\right)^2\Bigg\lbrace\left(\frac{2m\,\e^{-\ell/r}}{r}\right)^2\left(\frac{\ell}{r}-2\right) \nonumber \\[2pt]
    && \times\left[2\left(\frac{\ell}{r}\right)^2-10\left(\frac{\ell}{r}\right)+5\right] - \frac{m\,\e^{-\ell/r}}{r}\Bigg[-12+28\left(\frac{\ell}{r}\right) \nonumber \\[2pt]
    && \qquad \qquad \qquad \qquad \qquad \qquad \qquad -11\left(\frac{\ell}{r}\right)^2+\left(\frac{\ell}{r}\right)^3\Bigg]\Bigg\rbrace \ , \nonumber \\
    &&
\end{eqnarray}
and one obtains the following approximation for the imaginary part of the $s$-wave spin zero QNMs (expressed only as a function of $y$ for readability):
\clearpage
\begin{eqnarray}\label{Imspin0}
    \Im(\omega^2) &\approx& \eval{ -\left(n+\frac{1}{2}\right)\sqrt{-2\,\left[\partial_{r_*}^{2}\V_0\left(r_0\approx\frac{41}{15}m-\frac{4}{3}\ell\right)\right]}}_{k=0} \nonumber \\[3pt]
    &=& -\frac{1350}{m^2}\left(n+\frac{1}{2}\right)\left[\frac{\sqrt{5}\,\e^{\frac{15y}{20y-41}}(30\,\e^{\frac{15y}{20y-41}}+20y-41)}{(20y-41)^{\frac{11}{2}}}\right] \nonumber \\[2pt]
    &&\times\Bigg\lbrace 100\,\e^{\frac{15y}{20y-41}}(59950y^3-247230y^2+327795y-137842) \nonumber \\[2pt]
    && \qquad \qquad \qquad \quad +2112500y^4-13535125y^3+31644825y^2 \nonumber \\[2pt]
    && \qquad \qquad \qquad \qquad \qquad \qquad \quad -31703660y+11303044\Bigg\rbrace^{\frac{1}{2}} \ . \nonumber \\
    &&
\end{eqnarray}
Combining Eq.~(\ref{Respin0}) and Eq.~(\ref{Imspin0}) gives the approximation for the dimensionful $\omega^2$, however the expression is unwieldy and not particularly useful to display explicitly here. Using Eq's.~(\ref{Respin0}) and (\ref{Imspin0}) to compute the real and imaginary approximations for the dimensionless $m^2\omega^2$ respectively is sufficient for extracting the relevant results in \S~\ref{S:Results}.

\section{Numerical results}\label{S:Results}

Given that the behaviour of the waveform is aggressively governed by the fundamental mode, to extract profile approximations it is both physically well-motivated and mathematically tractable to specialise to $n=0$. Given the WKB approximation is only valid in the presence of a nontrivial horizon structure, and that the candidate spacetime only possesses horizons for $\ell\in\left(0,\frac{2m}{\e}\right)$, it is prudent for one to define the dimensionless object $\hat{\ell}=\ell/\ell_{max}=\frac{\ell\e}{2m}=\frac{\e}{2}y$, such that $\hat{\ell}\in(0,1)$. One may then define the dimensionless $\hat{\omega}=\omega m$, such that all of the qualitative information for the dimensionful $\omega$ as a function of the dimensionful $\ell$ is now encoded in the dimensionless $\hat{\omega}$ as a function of the dimensionless $\hat{\ell}$. As such, one then examines $\hat{\omega}$ by plugging in values of $y=\frac{2}{\e}\hat{\ell}$ into the relevant equations from \S~\ref{S:QNMs} on a case-by-case basis. Lastly, it will be of most use to analyse the dominant multipole number in each spin case. Given $k\geq S$, for electromagnetic spin one fluctuations this will correspond to analysing $k=1$, for scalar spin zero fluctuations one analyses the $s$-wave corresponding to fixing $k=0$ (as already stipulated), and finally for spin two axial perturbations one would fix $k=2$. Notably this implies that the approximate locations for the peaks of the relevant Regge--Wheeler potentials computed in \S~\ref{S:RW_potential} are directly applicable here.

\subsection{Spin one}

Consequently, to analyse QNM profile approximations for electromagnetic spin one fluctuations on the background spacetime, fix the fundamental mode $n=0$, and analyse the special case of the dominant multipole number $k=1$. Substituting these values into Eq.~(\ref{eqforspinoneres}) and computing the resulting square root gives the results from Table~\ref{tab:1} for the approximation of $\hat{\omega}$ for different values of $\hat{\ell}\in(0,1)$ (rounded to $6$ d.p.):

\begin{table}[!htb]
\begin{center}
\caption{Fundamental QNM of the spin one field for $k=1$, obtained \textit{via} first-order WKB approximation.}\label{tab:1}
\hspace{-35pt}
\begin{tabular}{||c|c||}
\hline
\hline
\vphantom{\bigg|} $\hat{\ell}$ & WKB approx. for $\hat{\omega}$ \\
\hline
\hline
$0.0$ & $0.287050-0.091235i$ \\
\hline
$0.1$ & $0.293902-0.092012i$ \\
\hline
$0.2$ & $0.301291-0.092532i$ \\
\hline
$0.3$ & $0.309304-0.092708i$ \\
\hline
$0.4$ & $0.318051-0.092419i$ \\
\hline
$0.5$ & $0.327658-0.091486i$ \\
\hline
$0.6$ & $0.338285-0.089636i$ \\
\hline
$0.7$ & $0.350117-0.086433i$ \\
\hline
$0.8$ & $0.363377-0.081139i$ \\
\hline
$0.9$ & $0.378330-0.072338i$ \\
\hline
$1.0$ & $0.395289-0.056624i$ \\
\hline
\hline
\end{tabular}
\end{center}
\end{table}
\begin{figure}[htb!]
\begin{center}
\hspace{-2.5cm}
\includegraphics[scale=0.42]{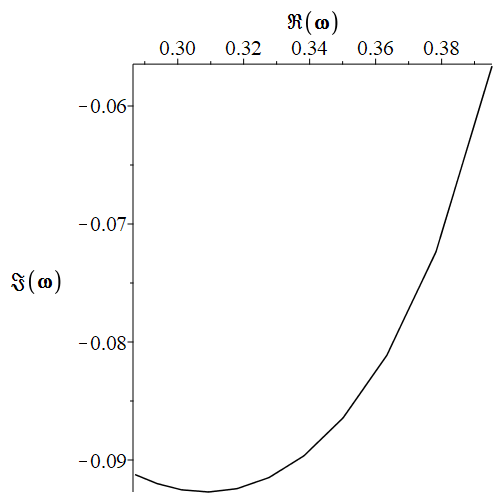}
\caption[Discrete plot with linear interpolation curve of the fundamental QNM of the spin one field with $k=1$ for aMc spacetime]{A plot of the points from Table~\ref{tab:1} with a linear interpolation curve. $\mbox{Re}(\hat{\omega})$ increases monotonically with $\hat{\ell}$, hence increasing $\hat{\ell}$ corresponds to one moving from left to right.}\label{F:4}
\end{center}
\end{figure}

Immediately there are the following qualitative observations:
\begin{itemize}
    \item As a sanity check, $\Im(\hat{\omega})<0$ for all values of $\hat{\ell}$, indicating that the propagation of electromagnetic fields in the background spacetime is stable --- an expected result;
    \item $\Re(\hat{\omega})$ increases monotonically with $\hat{\ell}$ --- this is the frequency of the corresponding QNMs;
    \item $\Im(\hat{\omega})$ decreases with $\hat{\ell}$ initially, up until $\hat{\ell}\approx 0.35$, and then it increases monotonically with $\hat{\ell}$ for the remainder of the domain --- this is the decay rate or damping rate of the QNMs;
    \item Given that throughout the historical literature, it is conventional to assert $\hat{\ell}\sim m_{p}$, it is of primary interest to examine the behaviour of this plot for small $\hat{\ell}$. In view of this, if one constrains the analysis of the qualitative behaviour for $\hat{\omega}$ \textit{prior} to the trough present in Fig.~\ref{F:4}, one would expect that the signals for electromagnetic radiation propagating in the presence of a regular black hole with asymptotically Minkowski core should have both a higher frequency as well as a faster decay rate than their Schwarzschild counterparts. This qualitative result \textit{may} translate to the spin two case, and speak directly to the LIGO/Virgo calculation. The fact that the signal is expected to be shorter-lived could present a heightened level of experimental difficulty when trying to delineate signals, though this may very well be offset by the fact that the signal also carries higher energy; further discussion on these points is left to both the numerical relativity and experimental communities.
\end{itemize}
\vspace*{20pt}


\subsection{Spin zero}
\vspace*{12pt}

For spin zero scalar fluctuations, specialising to the $s$-wave, similarly fix the fundamental mode $n=0$. First substitute this into Eq.~(\ref{Respin0}) and Eq.~(\ref{Imspin0}), which are the relevant equations to compute the real and imaginary approximations of $\hat{\omega}^2$ respectively (recall these have already specialised to the $s$-wave given $k$ was already fixed to be zero in the analysis of \S~\ref{S:QNMs}). Then, taking the appropriate square root yields the results from Table~\ref{tab:2} (to $6$ d.p.):
\clearpage

\vspace*{20pt}
\begin{table}[!htb]
\begin{center}
\caption{Fundamental QNM of the massless, minimally coupled spin zero scalar field for the $s$-wave ($k=0$), obtained \textit{via} first-order WKB approximation.}\label{tab:2}
\hspace{-35pt}
\begin{tabular}{||c|c||}
\hline
\hline
\vphantom{\bigg|} $\hat{\ell}$ & WKB approx. for $\hat{\omega}$ \\
\hline
\hline
$0.0$ & $0.187409-0.094054i$\\
\hline
$0.1$ & $0.189734-0.094530i$ \\
\hline
$0.2$ & $0.191948-0.094669i$ \\
\hline
$0.3$ & $0.194049-0.094425i$ \\
\hline
$0.4$ & $0.196027-0.093742i$ \\
\hline
$0.5$ & $0.197868-0.092557i$ \\
\hline
$0.6$ & $0.199552-0.090796i$ \\
\hline
$0.7$ & $0.201042-0.088385i$ \\
\hline
$0.8$ & $0.202285-0.085306i$ \\
\hline
$0.9$ & $0.203235-0.081735i$ \\
\hline
$1.0$ & $0.203894-0.078421i$ \\
\hline
\hline
\end{tabular}
\end{center}
\end{table}
\vspace*{25pt}
\begin{figure}[htb!]
\begin{center}
\hspace{-2.5cm}
\includegraphics[scale=0.45]{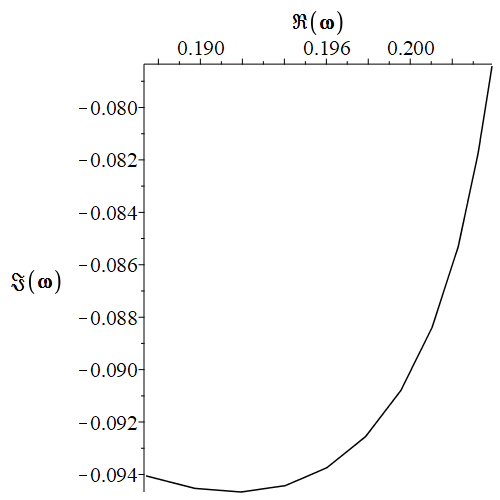}
\caption[Discrete plot with linear interpolation curve of the fundamental QNM of the spin zero field with $k=0$ for aMc spacetime]{A plot of the points from Table~\ref{tab:2} with a linear interpolation curve. $\mbox{Re}(\hat{\omega})$ increases monotonically with $\hat{\ell}$, hence increasing $\hat{\ell}$ corresponds to one moving from left to right.}\label{F:5}
\end{center}
\end{figure}
\clearpage

There are the following qualitative observations:
\begin{itemize}
    \item $\Re(\hat{\omega})$ once again increases monotonically with $\hat{\ell}$ --- higher $\hat{\ell}$-values correspond to higher frequency fundamental modes;
    \item $\Im(\hat{\omega})<0$ for all $\hat{\ell}$, indicating that the $s$-wave for minimally coupled massless scalar fields propagating in the background spacetime is stable;
    \item $\Im(\hat{\omega})$ decreases with $\hat{\ell}$ initially (down to a trough around $\hat{\ell}=0.25$), before monotonically increasing with $\hat{\ell}$ for the rest of the domain --- this is the decay/damping rate of the QNMs;
    \item Similarly as for the electromagnetic spin one case, when one examines the behaviour for small $\hat{\ell}$, the signals for the fundamental mode of spin zero scalar field perturbations in the presence of a regular black hole with asymptotically Minkowski core are expected to have a higher frequency and to be shorter-lived than for their Schwarzschild counterparts.
\end{itemize}

%
\subsection{Comparison with Bardeen and Hayward}

It is worth investigating whether these qualitative results are aligned with the analogous results for other well-known regular black hole geometries in GR. Analysis of the QNMs for both the Bardeen~\cite{Tbilisi:Bardeen:1968} and Hayward~\cite{FAEONBH:Hayward:2006} regular black holes has been performed in references~\cite{QMORBH:Flachi:2013, QMOBBH:Correa:2012, OQMFGPOBBH:Ulhoa:2014, QMOTFARBH:Abdujabbarov:2015}. The choices made in setting up a tractable numerical analysis make it very difficult to directly compare most of the findings, however in Appendix A of reference~\cite{QMORBH:Flachi:2013},
some analogous and comparable results are presented for the spin zero case for both the
Bardeen and Hayward models. The findings can be summarised as follows:
\begin{itemize}
    \item For the fundamental mode of the spin zero scalar $s$-wave for the Bardeen regular
    black hole, as the deviation from Schwarzschild increases, $\Re\left(\omega\right)$ increases and $\left\vert\Im\left(\omega\right)\right\vert$
    decreases. The signals are hence expected to have higher frequency but be longer-lived
    than for their Schwarzschild counterparts;
    \item For the fundamental mode of the spin zero scalar $s$-wave for the Hayward regular
    black hole, as deviation from Schwarzschild increases, both $\Re\left(\omega\right)$ and $\left\vert\Im\left(\omega\right)\right\vert$ decrease. The signals are hence expected to have lower frequency and be longer-lived than for their Schwarzschild counterparts.
\end{itemize}
These results suggest that for spin zero perturbations, the qualitative differences in the ringdown signal between the class of regular black hole models in static spherical symmetry and Schwarzschild is not uniform. As such, the ability to theoretically delineate between nonsingular candidate spacetimes in spherical symmetry based on the qualitative aspects of their QNM profiles should be possible. Whether this extends to the more astrophysically relevant domain of axisymmetry, or indeed to spin two axial and polar perturbations, is at this stage unclear. If so, if a measurement by LIGO/Virgo (or LISA) suggests that the object under observation has a ringdown signal qualitatively different from Kerr, and the margin of error on this measurement is sufficiently small, this \emph{should} eliminate a subset of the nonsingular candidate spacetimes from astrophysical consideration. However, given that the parameters which quantify the deviation from Schwarzschild/Kerr are often associated with quantum scales, it is reasonable to conjecture that the current margin of error present in the data from LIGO/Virgo is far too high to be able to form robust conclusions; this is left to the numerical and experimental community for further comment. Ongoing improvements to the accuracy of the numerical techniques used to generate the QNM profiles, and ongoing improvements to the technologies at humanity's disposal which enable astrophysical observations, will hopefully allow one to probe with the necessary level of accuracy.

\section{Perturbing the Regge--Wheeler potential}\label{S:RWperturb}

Suppose one perturbs the Regge--Wheeler potential itself, \textit{via} the replacement $\V(r) \to \V(r) + \delta\V(r)$. It is of interest to analyse what effect this has on the estimate for the QNMs. Classical perturbation of the potential to first-order in $\epsilon$ is performed, capturing any linear contributions from external agents that may disturb the propagating waveforms. First-order perturbation is well-motivated from the perspective of the historical literature, and ensures the analysis has the desired level of tractability. As such, one has the following: $\V(r) \to \V(r) + \delta\V(r) = \V(r) + \epsilon\,\delta\V_1(r) + \epsilon^2\,\delta\V_2(r) + \mathcal{O}(\epsilon^3) \approx \V(r) + \epsilon\,\delta\V_1(r)$. All terms of order $\epsilon^2$ or higher are therefore truncated. Consequently, for notational convenience it is advantageous to simply replace $\delta\V_1(r)$ with $\delta\V(r)$ in the discourse that follows, eliminating superfluous indices. Also for notational convenience, define $r_{max}=r_\sigma$ to be the generalised location of the peak of the potentials. One observes the following effects on the QNMs:
\begin{itemize}
\item 
Firstly, the \textit{position} of the peak shifts:
\begin{equation}
0 \approx [\V+\epsilon\,\delta\V]'(r)\Big|_{r=r_\sigma+\epsilon\,\delta r_\sigma} \ ,
\end{equation}
giving
\begin{equation}\label{perturb1}
\V'(r_\sigma+\epsilon\,\delta r_\sigma) + \epsilon\,[\delta\V]'(r_\sigma+\epsilon\,\delta r_\sigma) \approx 0 \ .
\end{equation}
Performing a first-order Taylor series expansion of the left-hand-side of Eq.~(\ref{perturb1}) about $\delta r_0=0$ then yields
\begin{equation}
    \V'(r_\sigma) + \epsilon\,[\delta\V]'(r_\sigma) + \epsilon\,\delta r_\sigma\left\lbrace \V''(r_\sigma) + \epsilon\,[\delta\V]''(r_\sigma)\right\rbrace \approx 0 \ ,
\end{equation}
and eliminating the term of order $\epsilon^2$, combined with the knowledge that $\V'(r_\sigma)=0$, gives
\begin{equation}\label{deltarsig}
    \delta r_\sigma \approx -\frac{[\delta\V]'(r_\sigma)}{\V''(r_\sigma)} \ .
\end{equation}
\item 
Secondly, the \textit{height} of the peak shifts:
\begin{equation}
[\V+\epsilon\,\delta\V](r_\sigma+\epsilon\,\delta r_\sigma)= \V(r_\sigma+\epsilon\,\delta r_\sigma) + \epsilon\,[\delta\V](r_\sigma+\epsilon\,\delta r_\sigma) \ ,
\end{equation}
and performing a first-order Taylor series expansion about $\delta r_0=0$ yields the following to first-order in $\epsilon$:
\begin{eqnarray}
[\V+\epsilon\,\delta\V](r_\sigma+\epsilon\,\delta r_\sigma) &\approx& \V(r_\sigma) + \epsilon\,\left\lbrace[\delta\V](r_\sigma)+\delta r_\sigma\V'(r_\sigma)\right\rbrace \nonumber \\
&=& \V(r_\sigma) + \epsilon\,[\delta\V](r_\sigma) \ .
\end{eqnarray}
\item
Third, the \textit{curvature} at the peak shifts
\begin{equation}
[\V+\epsilon\,\delta\V]''(r_\sigma+\epsilon\,\delta r_\sigma) = \V''(r_\sigma+\epsilon\,\delta r_\sigma) + \epsilon\,[\delta\V]''(r_\sigma+\epsilon\,\delta r_\sigma) \ ,
\end{equation}
and performing a first-order Taylor series expansion about $\delta r_\sigma=0$, to first-order in $\epsilon$, then gives
\begin{equation}
    [\V+\epsilon\,\delta\V]''(r_\sigma+\epsilon\,\delta r_\sigma) \approx \V''(r_\sigma) + \epsilon\,[\delta\V]''(r_\sigma) + \epsilon\,\delta r_\sigma\V'''(r_\sigma) \ ,
\end{equation}
which from Eq.~(\ref{deltarsig}) can then be approximated by the following:
\begin{equation}\label{Vprimeprime}
    [\V+\epsilon\,\delta\V]''(r_\sigma+\epsilon\,\delta r_\sigma) \approx \V''(r_\sigma) + \epsilon\,\left\lbrace[\delta\V]''(r_\sigma) - \frac{\V'''(r_\sigma)}{\V''(r_\sigma)}[\delta\V]'(r_\sigma)\right\rbrace \ .
\end{equation}
\end{itemize}
Given the following (one assumes the existence of some tortoise coordinate relation $\partial_{r_*}=T(r)\partial_{r}$):
\begin{eqnarray}
    \partial_{r_*}^2[\V+\epsilon\,\delta\V](r_\sigma+\epsilon\,\delta r_\sigma) &=& \eval{\partial_{r_*}\left\lbrace T(r)\partial_{r}[\V+\epsilon\,\delta\V](r)\right\rbrace}_{r=r_\sigma+\epsilon\,\delta r_\sigma} \nonumber \\[3pt]
    &=& \Bigg\lbrace T^2(r)[\V+\epsilon\,\delta\V]''(r) \nonumber \\
    && \qquad \quad +[\V+\epsilon\,\delta\V]'(r)T'(r)T(r)\Bigg\rbrace\Bigg\vert_{r=r_\sigma+\epsilon\,\delta r_\sigma} \ , \nonumber \\
    &&
\end{eqnarray}
and then performing Taylor series expansions about $\epsilon=0$ for the relevant terms, and truncating to first-order in $\epsilon$, one finds:
\clearpage
\begin{eqnarray}
    \partial_{r_*}^2[\V+\epsilon\,\delta\V](r_\sigma+\epsilon\,\delta r_\sigma) &=& [T(r_\sigma+\epsilon\,\delta r_\sigma)]^2[\V+\epsilon\,\delta\V]''(r_\sigma+\epsilon\,\delta r_\sigma) \nonumber \\[2pt]
    && \hspace{-0.7cm}+ [\V+\epsilon\,\delta\V]'(r_\sigma+\epsilon\,\delta r_\sigma)T'(r_\sigma+\epsilon\,\delta r_\sigma)T(r_\sigma+\epsilon\,\delta r_\sigma) \nonumber \\
    && \nonumber \\
    &\approx& \left[T(r_\sigma)+\epsilon\,\delta r_\sigma T'(r_\sigma)\right]^2\,[\V+\epsilon\,\delta\V]''(r_\sigma+\epsilon\,\delta r_\sigma) \nonumber \\[2pt]
    && \qquad \qquad \qquad \qquad + \epsilon\,\delta r_\sigma\V''(r_\sigma)T'(r_\sigma)T(r_\sigma) \nonumber \\
    && \nonumber \\
    &\approx& [T^2(r_\sigma) + 2\epsilon\,\delta r_\sigma T'(r_\sigma)T(r_\sigma)]\times \nonumber \\[2pt]
    && \hspace{-0.7cm}[\V+\epsilon\,\delta\V]''(r_\sigma+\epsilon\,\delta r_\sigma) - \epsilon\,T'(r_\sigma)T(r_\sigma)[\delta\V]'(r_\sigma) \ , \nonumber \\
    &&
\end{eqnarray}
and substituting the result from Eq.~(\ref{Vprimeprime}) then finally gives (all functions on the right-hand-side are evaluated at $r=r_\sigma$; notation suppressed for tractability):
\begin{eqnarray}
    \partial_{r_*}^2[\V+\epsilon\,\delta\V] &\approx& T^2\V'' - \epsilon\,\left\lbrace\left(3T'T+T^2\frac{\V'''}{\V''}\right)[\delta\V]'-T^2[\delta\V]''\right\rbrace \ . \nonumber \\
    &&
\end{eqnarray}
As such, for the square root in Eq.~(\ref{BLAH}), one has:
\begin{eqnarray}
    \sqrt{-2\partial_{r_*}^2[\V+\epsilon\,\delta\V](r_\sigma+\epsilon\,\delta r_\sigma)} &\approx& \Bigg\lbrace-2T^2\V'' \nonumber \\[2pt]
    && \hspace{-1.63cm} + 2\epsilon\,\left[\left(3T'T+T^2\frac{\V'''}{\V''}\right)[\delta\V]'-T^2[\delta\V]''\right]\Bigg\rbrace^{\frac{1}{2}} \ , \nonumber \\
    &&
\end{eqnarray}
and performing a first-order Taylor series expansion about $\epsilon=0$ gives the following (all functions on the right-hand-side are evaluated at $r=r_\sigma$, and $-\V''=\vert\V''\vert$ since one is at the \textit{peak}):
\begin{eqnarray}
    \sqrt{-2\partial_{r_*}^2[\V+\epsilon\,\delta\V]} &\approx& T\sqrt{-2\V''}+\frac{\epsilon}{\sqrt{2}}\Bigg\lbrace\left(\frac{3T'}{\vert\V''\vert^{\frac{1}{2}}}+\frac{T\V'''}{\vert\V''\vert^{\frac{3}{2}}}\right)[\delta\V]' \nonumber \\[2pt]
    && \qquad \qquad \qquad \qquad \qquad \qquad \qquad \quad -\frac{T}{\vert\V''\vert^{\frac{1}{2}}}[\delta\V]''\Bigg\rbrace \ . \nonumber \\
    &&
\end{eqnarray}
Assembling the pieces, to first-order in WKB and first-order in $\epsilon$ the approximate shift in the QNMs is given by:
\clearpage
\begin{eqnarray}\label{perturbativeform}
    \delta(\omega^2_n) &\approx& \epsilon\,[\delta\V] \nonumber \\
    && \hspace{-1cm}- i\left(n+\frac{1}{2}\right)\frac{\epsilon}{\sqrt{2}}\left\lbrace\left(\frac{3T'}{\vert\V''\vert^{\frac{1}{2}}}+\frac{T\V'''}{\vert\V''\vert^{\frac{3}{2}}}\right)[\delta\V]'-\frac{T}{\vert\V''\vert^{\frac{1}{2}}}[\delta\V]''\right\rbrace \ , \nonumber \\
    &&
\end{eqnarray}
where all expressions on the right-hand-side are evaluated at $r=r_\sigma$. This specific formula is general for all instances where the WKB approximation is appropriate. It is informative to now apply this to the most straightforward example of Schwarzschild spacetime.

\paragraph{Perturbing around Schwarzschild:}\

Specifying to spin one Schwarzschild, one sets $\ell\rightarrow0$, and has:
\begin{equation}
\V_{Sch.,1}(r) = \left(1-{2m\over r}\right){k(k+1)\over r^2} \ , \quad   T(r) = 1-\frac{2m}{r} \ , \quad r_\sigma = 3m \ . 
\end{equation}
Then the relevant quantities necessary to substitute into Eq.~(\ref{perturbativeform}) are:
\enlargethispage{20pt}
\begin{eqnarray}
    \V_{Sch.,1}''(r_\sigma) &=& -\frac{2k(k+1)}{81m^4} \ , \quad \V_{Sch.,1}'''(r_\sigma) = \frac{16k(k+1)}{243m^5} \ , \nonumber \\[3pt]
    T(r_\sigma) &=& \frac{1}{3} \ , \quad T'(r_\sigma) = \frac{2}{9m} \ ,
\end{eqnarray}
and the approximate shift in the QNMs is given by
\begin{eqnarray}
    \delta(\omega^2_n) &\approx& \epsilon\,[\delta\V](r_\sigma) \nonumber \\
    && \hspace{-0.75cm}-\epsilon\,i\left(n+\frac{1}{2}\right)\left\lbrace\frac{7m}{\sqrt{k(k+1)}}[\delta\V]'(r_\sigma)-\frac{3m^2}{2\sqrt{k(k+1)}}[\delta\V]''(r_\sigma)\right\rbrace\Bigg\vert_{r_\sigma=3m} \ . \nonumber \\
    &&
\end{eqnarray}
%
\subsubsection{Summary}\label{S:QNMs:discussion}

The spin-dependent Regge--Wheeler potentials for aMc spacetime have been extracted and their qualitative features thoroughly analysed. Subsequently, the spin one and spin zero fundamental QNM profiles have been examined \textit{via} first-order WKB approximation for the respective dominant multipole numbers. For small $\ell$, \textit{e.g.} $\ell\sim m_p$, each of the scalar spin zero and the spin one electromagnetic fluctuations propagating in an aMc spacetime background have both shorter-lived and higher-energy signals than their Schwarzschild counterparts. Analysis of perturbation of the Regge--Wheeler potential yields a general result explicating the associated shift in the QNM profiles under the perturbation $\V(r)\rightarrow\V(r)+\delta\V(r)$ to first-order in $\epsilon$. This general result can be easily applied to Schwarzschild spacetime.

Performing this analysis to higher-order in WKB, using the improved version of WKB with Pad\'{e} approximants, comparing the QNM profiles to those extracted using a different method to WKB (say, \textit{e.g.}, time domain integration), and numerical refinement of the approximations, are all prudent lines of inquiry for future research. Of the most importance is extending the discourse to the spin two axial mode. Also of interest is to explore the QNMs for when the candidate spacetime is modelling a compact massive object which is something other than a black hole; \textit{i.e.} when $\ell>2m/\e$. Given the knowledge and existence of aMcRN spacetime (see \S~\ref{S:amcRN}), one should also perform a standard QNM analysis for this candidate geometry.

%


%% file: 03-AMC/3-AMCEOS.tex
\chapter{The eye of the storm}\label{C:AMCEOS}

\def\tr{{\mathrm{tr}}}
\def\diag{{\mathrm{diag}}}
\def\cof{{\mathrm{cof}}}
\def\pdet{{\mathrm{pdet}}}
\def\d{{\mathrm{d}}}
\parindent0pt
\parskip7pt
\def\Kerr{{\scriptscriptstyle{\mathrm{Kerr}}}}
\def\eos{{\scriptscriptstyle{\mathrm{eos}}}}
As has already been discussed, exploration regarding the extraction of astrophysical observables for
nonsingular candidate geometries has been performed in a vast array of contexts~\cite{GLBARBH:Eiroa:2011, QMORBH:Flachi:2013, SORRBH:Abdujabbarov:2016, OTVORBH:Carballo-Rubio:2018, OTPBATCOBH:Carballo-Rubio:2020, GCBH:Carballo-Rubio:2020, OAW:Dai:2019, NWAGL:Cramer:1995, SSFW:Kavic:2021, PSISCOOSCO:Berry:2020, HOD:Carballo-Rubio:2021, GCBHILG:Carballo-Rubio:2022, GPOWSAIATSAQM:Bronnikov:2021, WWEM:Churilova:2021, AW:Bambi:2021, ROTRBHWAMC:Simpson:2021}. Appealing to the methodology of ``turning one knob at a time'' by working within the framework of ``GR$+$no curvature singularities'' (see \S~\ref{Intro:knob}), one can propose a list of further constraints on nonsingular geometries which render them as appropriate as possible for the experimental and observational communities, streamlining the discourse between theory and experiment. For instance, an ``idealised'' candidate spacetime should be asked to satisfy \textit{at least} the following constraints:
\begin{itemize}
    \item Astrophysical sources rotate --- impose axisymmetry.
    \item Impose asymptotic flatness at spacial infinity~\cite{PGLT1:Baines:2021}.
    \item
    The Hamilton--Jacobi equations should be separable ---  the geodesic equations should be at least numerically integrable to enable direct comparison with experimental data. A sufficient condition for this in axisymmetry is the existence of a nontrivial Killing two-tensor $K_{\mu\nu}$.
    \item 
    Impose separability of both Maxwell's equations and the equations governing the spin two polar and axial modes on the background spacetime --- this allows for the ``standard'' numerical techniques to be applied in analysing the quasi-normal modes of spin one electromagnetic and spin two gravitational perturbations, which are fundamental in analysing the ringdown phase of binary mergers. A good mathematical precursor for this constraint is the separability of the Klein--Gordon equation, allowing for standard analysis of spin zero scalar fluctuations. It should also be noted that for the spin two polar modes (even parity) the Zerilli potential~\cite{Zerilli:parity} takes the place of the usual Regge--Wheeler potential, and for this case the relativistic Cowling approximation is not particularly fit for purpose.
    \item 
    Impose a high degree of mathematical tractability. The complex process beginning with the inception of a candidate geometry, and finishing with a result able to be directly compared with experimental measurement involves many nontrivial steps --- candidate spacetimes amenable to highly tractable mathematical analysis will yield their astrophysical observables with \textit{far} more ease.
    \item 
    Constrain the amount of exotic matter and demand satisfaction of the relevant classical energy conditions outside horizons --- empirical evidence suggests any violation of the classical energy conditions should occur at a quantum scale, and (apart from the violations of the SEC due to a positive cosmological constant) exotic matter in an astrophysical context has not been observed~\cite{ECITEOGF:Visser:1997, GREC:Visser:1997, ECATCI:Barcelo:2000}. 
\end{itemize}
The above list serves only as a rough guideline. Naturally, there are many other constraints that are likely to be desirable --- forbid closed timelike curves, for instance, or impose separability of the Dirac equation, \textit{et cetera.} However the above list speaks \textit{directly} to the current observational and experimental community. Finding appropriate geometries which satisfy \textit{all} of these constraints is highly nontrivial, and generally speaking the best one can do is to use them as goalposts when constructing a candidate spacetime.

%
In Chapter~\ref{C:AMCEOS}, a rotating regular black hole with an asymptotically Minkow\-ski core is explored, dubbed the ``eye of the storm'' (eos). This geometry was in fact first proposed by Ghosh in reference~\cite{ANRBH:Ghosh:2015}; the current author discovered it independently by following a set of carefully chosen metric construction criteria which will be explored in \S~\ref{eosansatz}. Consequently, some results are repeated, though with rather different representations and emphasis. Numerous new and important results for this geometry are also presented. Set in stationary axisymmetry, this spacetime is a tightly controlled deviation from standard Kerr, is amenable to highly tractable mathematical analysis, and possesses the full ``Killing tower''~\cite{BHHSACI:Frolov:2017} of principal tensor, Killing--Yano tensor, and nontrivial Killing tensor. This induces an associated Carter constant~\cite{HASSSOEE:Carter:1968, GSOTKFOGF:Carter:1968}, giving a fourth constant of the motion and rendering the geodesic equations of motion for test particles in principle integrable (\textit{i.e.}, imposing separability of the Hamilton--Jacobi equation). Any energy-condition-violating physics is able to be pushed into the deep core, at a distance scale where GR is no longer empirically justified. Both the Klein--Gordon equation and Maxwell's equations are separable on the background spacetime, enabling quasi-normal modes analysis for spin zero and spin one perturbations \textit{via} the ``standard'' techniques.
\clearpage

With reference to the above list of proposed constraints, comparison with the existing literature on rotating regular black holes reveals that this geometry is \emph{very} close to ``idealised''.\footnote{It is also a good idea to bear in mind what \textit{cannot} be done: For instance the spacial slices of the Kerr spacetime cannot be put in conformally flat form~\cite{OTNOCFSITKAOSS:Kroon:2004}, nor can the three-metric  even be globally diagonalised~\cite{DDOTS3IKS:Baines:2021}.} The eos spacetime will be amenable to straightforward extraction of (potential) astrophysical observables in principle falsifiable/verifiable by the experimental and observational communities. Before segueing into the analysis of this specific geometry, it is worth exploring the choices made in constructing the metric.

%
\section{Metric ansatz}\label{eosansatz}

To set the stage, it is pertinent to first discuss static spherical symmetry for exposition, before migrating the discussion to the more astrophysically appropriate realm of stationary axisymmetry.

\subsection{Spherical symmetry}\label{EOSSS}

Recall the discussion in Chapter~\ref{C:AMCog}; one can always put any static, spherically symmetric line element into the standard form~\cite{G:Misner:2000, GR:Wald:2010, LW:Visser:1995}:
\begin{equation}
    \d s^2 = -\e^{-2\Phi(r)}\left(1-\frac{2m(r)}{r}\right)\d t^2 + \frac{\d r^2}{1-\frac{2m(r)}{r}} + r^2\d\Omega^2_2 \ .
\end{equation}
$\Phi(r)=0$ spacetimes then specialise to the one-function class of geometries characterised by~\cite{WIGTT:Jacobson:2007}:
\begin{equation}\label{Sch+1}
    \d s^2 = -\left(1-\frac{2m(r)}{r}\right)\d t^2 + \frac{\d r^2}{1-\frac{2m(r)}{r}} + r^2\d\Omega^2_2 \ .
\end{equation}
Specialising further to $m(r)=m$ yields the Schwarzschild solution in standard curvature coordinates. Consequently, one can think of Eq.~(\ref{Sch+1}) as the class of one-function modified Schwarzschild geometries. In the discourse that follows, it will be useful to think from this general perspective.

Within this one-function class of geometries, there are the following canonical regular black hole spacetimes:
\enlargethispage{40pt}

{\bf Bardeen~\cite{Tbilisi:Bardeen:1968}:}
\begin{equation}
m(r) = {\frac{mr^3}{(r^2+\ell^2)^{3/2}}} \ , \qquad 
\rho(r)={\frac{m \ell^2}{{\frac{4\pi}{3}} (r^2+\ell^2)^{5/2}}} \ .
\end{equation}
This implies an asymptotically de Sitter core with
\begin{equation}
\rho(0) ={\frac{m}{{\frac{4\pi}{3}} \ell^3}} \ .
\end{equation}
So the central density depends on asymptotic mass.

{\bf Hayward~\cite{FAEONBH:Hayward:2006}:}
\begin{equation}
m(r) = {\frac{m r^3}{r^3+ 2 m \ell^2}} \ , \qquad 
\rho(r)={\frac{m^2\ell^2}{{\frac{2\pi}{3}} (r^3+2m\ell^2)^{3}}} \ .
\end{equation}
This implies an asymptotically de Sitter core with
\begin{equation}
\rho(0) ={\frac{1}{{\frac{8\pi}{3}} \ell^2}} \ .
\end{equation}
So the central density is independent of asymptotic mass.

{\bf Asymptotically Minkowski core~\cite{RBHWAMC:Simpson:2019}:}
\begin{equation}
m(r) = m\,\e^{-\ell/r} \ , \qquad \rho(r) = \frac{\ell m\,\e^{-\ell/r}}{4\pi r^4} \ .
\end{equation}
This implies an asymptotically Minkowski core with
\begin{equation}
\rho(0) = 0 \ .
\end{equation}
So the central density is zero.

The inspiration for the construction of eos spacetime came directly from the standard aMc spacetime~\cite{RBHWAMC:Simpson:2019} of Chapters~\ref{C:AMCog} and~\ref{C:AMCQNM}. Recall that the aMc line element is given by:
\begin{equation}\label{AM}
    \d s^2 = -\left(1-\frac{2m\,\e^{-\ell/r}}{r}\right)\,\d t^2 + \frac{\d r^2}{1-\frac{2m\,\e^{-\ell/r}}{r}} + r^2\,\d\Omega^2_2 \ ,
\end{equation}
and that the $\ell\rightarrow0$ limit recovers Schwarzschild spacetime in the usual curvature coordinates; the ``supression parameter'' $\ell$ quantifies the deviation from Schwarzschild spacetime. When viewed as a modification of Schwarzschild, there is the ``regularising'' procedure:
\begin{itemize}
    \item Make the modification $m\rightarrow m(r)=m\,\e^{-\ell/r}$.
\end{itemize}
It should be emphasised that this is \textit{not} a coordinate transformation.

Recall that the severe mathematical discontinuity of the function $\e^{-\ell/r}$ at coordinate location $r=0$ implies that the line element Eq.~(\ref{AM}) is not $C^{\omega}$ at $r=0$ (it is in fact not even $C^0$). This implies that the region $r<0$ is grossly unphysical for this candidate spacetime. In and of itself, this is not a problem \textit{per se}, and physical analysis is valid for $r\geq0$. However, this raises the question: What are the most prudent mathematical choices one can make when attempting to ``regularise'' a candidate black hole \textit{via} exponential suppression? In the regime of static spherical symmetry, here are two other examples which are worth brief discussion.
\paragraph{Example:} Consider
\begin{equation}
\d s^2 = -\left(1-\frac{2m(r)}{r}\right)\,\d t^2 + \frac{\d r^2}{1-\frac{2m(r)}{r}} + r^2 \; \d \Omega^2_2 \ ,
\quad
m(r) = m\,\e^{-\frac{\ell^2}{r^2}} \ .
\end{equation}
For related ideas, see for instance~\cite{TSOGKBHWEM:Tinchev:2015}. See also reference~\cite{SATFOBHE:Xiang:2013}, where this mass function was employed directly.
Purely mathematically, $\exp\left(-{\ell^2}/{r^2}\right)$ is real analytic only for $r\neq0$, however it is $C^{+\infty}$ for all $r$. Superficially then, this example looks more general than that of Eq.~(\ref{AM}), due to the fact one can now extend the analysis to $r<0$. A deeper look reveals that this is not particularly useful. On each spacial slice $r=0$ is still a point, and in spacetime $r=0$ is a timelike curve. Consequently, one has two universes, corresponding to $r\geq0$ and $r\leq0$, with each being a copy of the geometry characterised by Eq.~(\ref{AM}), connected at the single point $r=0$. One may not traverse through this point, and the ``other'' universe is physically irrelevant. It is worth noting the resulting spacetime will still be curvature-regular; in fact \emph{all} spacetimes of the class $m(r)=m\,\e^{-\ell^n/r^n}$ will be curvature-regular.
\paragraph{Example:} Consider instead
\begin{equation}\label{Ex2}
\d s^2 = -\left(1-\frac{2m(r)}{r}\right)\,\d t^2 + \frac{\d r^2}{1-\frac{2m(r)}{r}} + (r^2+\ell^2) \; \d \Omega^2_2 \ ,
\quad
m(r) = m\,\e^{-\frac{\ell^2}{r^2}} \ .
\end{equation}
Because the angular part of the metric is modified, this is now intrinsically more general (from a physical perspective). Specifically, on any spacial slice $r=0$ now corresponds to a two-sphere of finite area $4\pi\ell^2$, and in spacetime $r=0$ is in fact a timelike hypertube (\textit{i.e.} a traversable wormhole throat). Now the two universes corresponding to $r\leq0$ and $r\geq0$ are connected at the traversable throat $r=0$, and a would-be timelike traveller may propagate between them. The qualitative causal structure has both an outer and an inner horizon, with a timelike traversable hypersurface in the deep core at $r=0$; this is qualitatively the same as for certain specialisations explored in Chapter~\ref{C:SVog}; see Fig.~\ref{fig:penrose2}.

\subsection{Kerr-like rotating spacetimes}
\enlargethispage{10pt}

Instead, herein one is interested in investigating a rotating version of aMc spacetime. This is better-motivated from an astrophysical standpoint, and hence more likely to speak to the relevant parties currently operating in observational and gravitational wave astronomy. Migrating the discourse to stationary axisymmetry, one begins with the Kerr spacetime in standard Boyer--Lindquist (BL) coordinates:
\begin{equation}\label{Kerr}
    \d s^2 = - \frac{\Delta_{\Kerr}}{\Sigma}(\d t-a\sin^2\theta\,\d\phi)^2 
    + \frac{\sin^2\theta}{\Sigma}[(r^2+a^2)\,\d\phi-a\,\d t]^2
    +\frac{\Sigma}{\Delta_{\Kerr}}\,\d r^2 + \Sigma\,\d\theta^2
     \ ,
\end{equation}
where as usual
\begin{equation}
    \Sigma = r^2+a^2\cos^2\theta \ , \qquad \Delta_{\Kerr}=r^2+a^2-2mr \ .
\end{equation}
The inverse metric can be written as
\begin{equation}
g_\Kerr^{\mu\nu} \partial_{\mu} \partial_{\nu} = \frac{-1}{\Sigma}\left\{ 
{\left[(r^2+a^2)\,\partial_t  +a\, \partial_\phi \right]^2\over\Delta_\Kerr}
 -{(\partial_\phi + a\sin^2\theta\, \partial_t)^2 \over\sin^2\theta}
-\Delta_\Kerr \partial_r^2 - \partial_\theta^2\right\}.
\end{equation}
In BL-coordinates, the ring singularity present in Kerr spacetime is located at $r=0$. Inspired by the aforementioned ``regularising'' procedure for aMc spacetime, in an attempt to regularise Kerr, leave the object $\d r$ in the metric undisturbed, and make a modification $m\rightarrow m(r)$.

Prosaically, this class of geometries can be viewed as a ``one-function off-shell'' extension of the Kerr geometry; this is a specialisation of Carter's ``two-function off-shell'' extension to Kerr~\cite{BHHSACI:Frolov:2017, HASSSOEE:Carter:1968, GSOTKFOGF:Carter:1968}. Due to results from reference~\cite{BHHSACI:Frolov:2017}, it is already known that geometries within this ``one-function off-shell'' extension to Kerr must possess a nontrivial Killing tensor $K_{\mu\nu}$. This implies the existence of an associated Carter constant, and hence separability of the Hamilton--Jacobi equations (and, in principle, integrable geodesics). This is yet another motivation for exploring this line of inquiry. In \S~{\ref{Killingtower}} it is explicitly verified that in general the ``one-function off-shell'' extension to Kerr always in fact possesses the full ``Kill\-ing tower'' of a Killing tensor, a Killing--Yano tensor, and a principal tensor~\cite{BHHSACI:Frolov:2017}.

With the ``one-function off-shell'' extension to Kerr in hand, another potential approach might be to instead make the modification $m\rightarrow m(r,\theta)$. One \textit{may} intuit from the fact that the slices of axisymmetry are $\theta$-dependent that any exponential ``supression mechanism'' also ought to have a $\theta$-depend\-ence; $\theta$-dependent modifications to certain mass functions in axisymmetry have been discussed in~\cite{IFOSRBHBOALP:Eichhorn:2021, FALFNPTIFORSBHWD:Eichhorn:2021}. However, in Kerr spacetime there are also geometric features of qualitative importance which are $\theta$-\textit{independent}, such as the horizon locations. Imposing this $\theta$-dependence also loses the guarantee that one can put the metric into the form of Carter's ``two-function off-shell'' extension to Kerr; one \textit{may} well lose the existence of a nontrivial Killing tensor $K_{\mu\nu}$ (and the associated ``Killing tower'')~\cite{BHHSACI:Frolov:2017}. Imposing this $\theta$-dependence also has severe implications on mathematical tractability.

For both approaches considered, fixing the most desirable $m(r)$ or $m(r,\theta)$ such that the candidate geometry is both \textit{regular} and \textit{tractable} is nontrivial, and all of the following examples are worth brief discussion.
\clearpage

\textbf{Example:}  Consider ``one-function off-shell'' Kerr (in BL-coordinates) with
\begin{equation}
m(r) = m\,\exp\left(-\frac{\ell}{r}\right) \ .
\end{equation}
Mathematically, one has the discontinuity at $r=0$, and hence the region $r<0$ is omitted from the analysis. In terms of Cartesian coordinates $r_{naive}^2 = x^2+y^2+z^2$ one has
\begin{equation}
r_{naive}^2 = r^2 + a^2 - {a^2 z^2\over r^2} \ , \qquad \cos\theta= z/r \ .
\end{equation}
Then
\begin{equation}
r_{naive}^2 = r^2 + a^2 \sin^2\theta \ ,
\end{equation}
while
\begin{equation}
\cos\theta_{naive}= {z\over r_{naive}} = {z\over r} \;{r\over r_{naive}} = 
\frac{\cos\theta}{\sqrt{ 1 + a^2 \sin^2\theta/r^2}}
=
\frac{r\; \cos\theta}{\sqrt{ r^2 + a^2 \sin^2\theta}} \ .
\end{equation}
So exponential suppression in the BL-coordinate $r$ (as $r\to 0^+$) suppresses the mass function $m(r)$ in the entire Cartesian disk $r_{naive} \leq a$, where $\cos\theta_{naive}=0$. This ought to render the spacetime curvature-regular (and indeed does so; this will be demonstrated shortly). This specific example is particularly useful in that the ``other universes'' in the maximal analytic extension of the usual version of Kerr are removed from the analysis due to the restriction $r\geq0$. Consequently the maximal analytic extension of this regularised spacetime will be trivial --- there will be no concerns arising from closed timelike curves in this candidate geometry.\footnote{It is perhaps worthwhile to note that even for standard Kerr spacetime, the closed timelike curves can arise only by dodging into the ``other'' universe ($r<0$). This is most obvious in Doran coordinates where, since $g^{tt}=-1$, the entire $r>0$ region is manifestly stably causal~\cite{ANFOTKS:Doran:2000, TKSABI:Visser:2007, TKSRBHIGR:Scott:2009, UVOTKS:Baines:2021}.}

\textbf{Example:} Consider instead ``one-function off-shell'' Kerr (in BL-coordinates) with
\begin{equation}
m(r) = m\,\exp\left(-\frac{\ell^2}{r^2}\right) \ .
\end{equation}
(See for instance~\cite{TSOGKBHWEM:Tinchev:2015, SATFOBHE:Xiang:2013}.)
Purely mathematically, one may now also consider $r<0$ given the metric is now $C^{+\infty}$ at $r=0$. In terms of Cartesian coordinates $r_{naive}^2 = x^2+y^2+z^2$ one still has
\begin{equation}
r_{naive}^2 = r^2 + a^2 \sin^2\theta \ ,
\end{equation}
\begin{equation}
\cos\theta_{naive}= \frac{r\; \cos\theta}{\sqrt{ r^2 + a^2 \sin^2\theta}} \ .
\end{equation}
\clearpage
So (quadratic) exponential suppression in the BL-coordinate $r$ (as $r\to 0^+$) suppresses the mass function $m(r)$ in the entire Cartesian disk $r_{naive} \leq a$, where $\cos\theta_{naive}=0$. However, physically there is now no point in continuing the $r$ coordinate to $r<0$. In the absence of the ring singularity at $r_{naive}=a$, there is nothing to generate an angle deficit or angle surfeit; the ring at $r_{naive}=a$ is utterly ordinary. Consequently, exploring $r\leq0$ is physically identical to exploring $r\geq0$. Notably, the curvature quantities and general analysis for this example are less tractable than for the example based on $\exp(-\ell/r)$.

%
\textbf{Example:} Consider modified Kerr (in BL-coordinates) with the somewhat messier $\theta$-dependent mass function
\begin{equation}
m(r,\theta) = m\,\exp\left(-\frac{\ell}{\sqrt{r^2+a^2\cos\theta^2}}\right) = m\,\exp\left(-\frac{\ell}{\sqrt{\Sigma}}\right) \ .
\end{equation}
Now $\sqrt{r^2+a^2\cos\theta^2}=0$ requires both $r=0$ and $\theta=\pi/2$. There will be the discontinuity at $r=0$ in the equatorial plane, where $m(r,\theta)\rightarrow m\,\exp\left(-\frac{\ell}{r}\right)$; one must omit $r<0$ from the analysis. Furthermore, \textit{via} the standard results
\begin{equation}
r_{naive}^2 = r^2 + a^2 \sin^2\theta \ ,
\end{equation}
and
\begin{equation}
\cos\theta_{naive}= \frac{r\; \cos\theta}{\sqrt{ r^2 + a^2 \sin^2\theta}} \ ,
\end{equation}
this implies that both 
$r_{naive}=a$ and $\theta_{naive}=\pi/2$. So exponential suppression in $\sqrt{r^2+a^2\cos^2\theta}$ suppresses the mass function $m(r,\theta)$ only at the \textit{edge} of the Cartesian disk $r_{naive} = a$, where $\cos\theta_{naive}=0$. The geometry is now not flat on the interior of the disk $r_{naive} < a$, with $\cos\theta_{naive}=0$. Supplementary to this, imposing the $\theta$-dependence in this specific manner has \textit{severe} implications on the tractability of the analysis.
%

\textbf{Example:} Consider instead modified Kerr (in BL-coordinates) with the messier $\theta$-dependent mass function
\begin{equation}
    m(r,\theta) = m\,\exp\left(-\frac{\ell^2}{r^2+a^2\cos^2\theta}\right) = m\,\exp\left(-\frac{\ell^2}{\Sigma}\right) \ .
\end{equation}
Note that one may now explore, purely mathematically, $r<0$. By the same logic as for the previous example, exponential suppression in $r^2+a^2\cos^2\theta$ suppresses the mass function $m(r,\theta)$ only at the \textit{edge} of the Cartesian disk $r_{naive}=a$, where $\cos\theta_{naive}=0$. The geometry is not flat on the interior of this disk. There is no ring singularity at $r_{naive}=a$, and so nothing to generate an angle deficit or angle surfeit; the ring at $r_{naive}=a$ is utterly ordinary. Consequently, even though one may mathematically explore $r<0$, there is no physical reason to do so, by the same logic as for previous examples. Furthermore, imposing the $\theta$-dependence in this manner \textit{severely} affects mathematical tractability.

Ultimately, deciding which candidate geometry is preferable for analysis is nontrivial. Exploring these examples leaves the following conclusions:
\begin{itemize}
    \item There is no physical point to forcing the analysis to be amenable to analytic extension to $r<0$ in this specific manner, and doing so has consequences concerning mathematical tractability;
    \item There may or may not be a physical point to forcing the suppression mechanism to have a $\theta$-dependence, however doing so has \textit{severe} implications on mathematical tractability, and also does not render the central disk Minkowski.
\end{itemize}
%

\subsection{The eye of the storm}

Consequently, it is intuitive to advocate for the most mathematically tract\-able of the aforementioned examples in axisymmetry; this is the example $m(r)=m \exp(-\ell/r)$. This results in the following specific and fully explicit metric, for now labelled the ``eye of the storm'' (eos) spacetime:
\begin{equation}\label{E:eos}
    \d s^2 = \frac{\Sigma}{\Delta_{\eos}}\,\d r^2 + \Sigma\,\d\theta^2 - \frac{\Delta_{\eos}}{\Sigma}(\d t-a\sin^2\theta\,\d\phi)^2 + \frac{\sin^2\theta}{\Sigma}[(r^2+a^2)\,\d\phi-a\,\d t]^2 \ ,
\end{equation}
where
\begin{equation}
    \Sigma = r^2+a^2\cos^2\theta \ , \qquad \Delta_{\eos}=r^2+a^2-2mr\,\e^{-\ell/r} \ .
\end{equation}
The inverse metric can be written as:
\begin{equation}
g_\eos^{\mu\nu} \partial_{\mu} \partial_{\nu} = \frac{1}{\Sigma}\left\{ 
-{\left[(r^2+a^2)\,\partial_t  +a\, \partial_\phi \right]^2\over\Delta_\eos}
 +{(\partial_\phi + a\sin^2\theta\, \partial_t)^2 \over\sin^2\theta}
+\Delta_\eos \partial_r^2 + \partial_\theta^2\right\} \ .
\end{equation}
Note again that this is the same geometry as presented by Ghosh in reference~\cite{ANRBH:Ghosh:2015}, now with a considerably more detailed physical justification as to why it is of interest. In the limit as $r\rightarrow+\infty$, asymptotic flatness is preserved. In the limit as $\ell\rightarrow0$, one returns the standard Kerr spacetime in BL coordinates. As such, enforce $\ell>0$ for nontrivial analysis, and the supression parameter $\ell$ can be viewed as quantifying the deviation from Kerr spacetime. In the limit as $a\rightarrow0$, one recovers Eq.~(\ref{AM}) precisely. In comparison with standard Kerr in BL coordinates, the domains for the temporal and angular coordinates are unaffected. However the discontinuity at $r=0$ restricts the domain for the $r$ coordinate to $r\geq0$. This removes concerns involving closed timelike curves which are present in the ``usual'' discourse surrounding maximally extended Kerr. Crucially, as shall be shortly observed, the ring singularity is excised; replaced instead by a region of spacetime which is asymptotically Minkowski. This renders the geometry globally nonsingular; this is a tractable model for a regular black hole with rotation.

\section{Geometric analysis}

From the form of the line element as in Eq.~(\ref{E:eos}), ordering the coordinates as $(t,r,\theta,\phi)$, it is straightforward to read off a convenient covariant tetrad (co-tetrad) which is a solution of $g_{\mu\nu}=\eta_{\hat{\mu}\hat{\nu}}\,e^{\hat{\mu}}{}_{\mu}\,e^{\hat{\nu}}{}_{\nu}$ (it should be noted this co-tetrad is not unique):
\begin{eqnarray}
    e^{\hat{t}}{}_{\mu} &=& \sqrt{\frac{\Delta_{\text{eos}}}{\Sigma}}\left(-1;0,0,a\sin^2\theta\right) \ , \qquad \qquad e^{\hat{r}}{}_{\mu} = \sqrt{\frac{\Sigma}{\Delta_{\text{eos}}}}\left(0;1,0,0\right) \ , \nonumber \\[3pt]
    e^{\hat{\theta}}{}_{\mu} &=& \sqrt{\Sigma}\left(0;0,1,0\right) \ , \qquad \qquad e^{\hat{\phi}}{}_{\mu} = \frac{\sin\theta}{\sqrt{\Sigma}}\left(-a;0,0,r^2+a^2\right) \ .
\end{eqnarray}
This co-tetrad uniquely defines a contra-tetrad (contravariant tetrad, or just tetrad) \textit{via} $e_{\hat{\mu}}{}^{\mu}=\eta_{\hat{\mu}\hat{\nu}}\; e^{\hat{\nu}}{}_{\nu}\; g^{\nu\mu}$. Explicitly:
\begin{eqnarray}
    e_{\hat{t}}{}^{\mu} &=& -\frac{1}{\sqrt{\Sigma\Delta_{\text{eos}}}}\left(r^2+a^2;0,0,a\right) \ , \qquad e_{\hat{r}}{}^{\mu} = \sqrt{\frac{\Delta_{\text{eos}}}{\Sigma}}\left(0;1,0,0\right) \ , \nonumber \\[3pt]
    e_{\hat{\theta}}{}^{\mu} &=& \frac{1}{\sqrt{\Sigma}}\left(0;0,1,0\right) \ , \qquad e_{\hat{\phi}}{}^{\mu} = \frac{1}{\sqrt{\Sigma\sin^2\theta}}\left(a\sin^2\theta;0,0,1\right) \ .
\end{eqnarray}
This tetrad will be employed to convert relevant tensor coordinate components into an orthonormal basis.

Now, it is useful to define the object
\begin{equation}
    \Xi = \frac{\ell\Sigma}{2r^3} \ .
\end{equation}
This will greatly simplify some of the following analysis. Where convenient for exposition, curvature quantities are displayed in the form:
\begin{equation}
\hbox{(something dimensionful)} \times \hbox{(something dimensionless)} \ .
\end{equation}
Note that $\Xi$ is dimensionless.

To confirm the assertion that the eye of the storm is curvature-regular, one must analyse the nonzero components of the Riemann curvature tensor with respect to this orthonormal basis. Fully explicitly, one has
\begin{eqnarray}
    R^{\hat{t}\hat{r}}{}_{\hat{t}\hat{r}} &=& \frac{2r^3m\,\e^{-\ell/r}}{\Sigma^3}\left[2\Xi^2-4\Xi+1-3\left(\frac{a}{r}\right)^2\cos^2\theta\right] \ , \nonumber \\[2pt]
    -\frac{1}{2}R^{\hat{t}\hat{r}}{}_{\hat{\theta}\hat{\phi}} = -R^{\hat{t}\hat{\theta}}{}_{\hat{r}\hat{\phi}} = R^{\hat{t}\hat{\phi}}{}_{\hat{r}\hat{\theta}} &=& \frac{a\cos\theta r^2m\,\e^{-\ell/r}}{\Sigma^3}\left[2\Xi+\left(\frac{a}{r}\right)^2\cos^2\theta-3\right] \ , \nonumber \\[2pt]
    R^{\hat{t}\hat{\theta}}{}_{\hat{t}\hat{\theta}} = R^{\hat{t}\hat{\phi}}{}_{\hat{t}\hat{\phi}} = R^{\hat{r}\hat{\theta}}{}_{\hat{r}\hat{\theta}} = R^{\hat{r}\hat{\phi}}{}_{\hat{r}\hat{\phi}} &=& \frac{r^3m\,\e^{-\ell/r}}{\Sigma^3}\left[2\Xi+3\left(\frac{a}{r}\right)^2\cos^2\theta-1\right] \ , \nonumber \\[2pt]
    R^{\hat{\theta}\hat{\phi}}{}_{\hat{\theta}\hat{\phi}} &=& \frac{2r^3m\,\e^{-\ell/r}}{\Sigma^3}\left[1-3\left(\frac{a}{r}\right)^2\cos^2\theta\right] \ .
\end{eqnarray}
All are of the general form
\begin{equation}
    R^{\hat{\alpha}\hat{\beta}}{}_{\hat{\mu}\hat{\nu}} = \frac{m\, \e^{-\ell/r}}{r^n\,\Sigma^3}X(r,\theta; a, \ell) \ ,
\end{equation}
where the object $X(r,\theta;a,\ell)$ is globally well-behaved. The only potentially dangerous behaviour comes from the $r^n\,\Sigma^3$ present in the denominators in the limit as $r\rightarrow0^{+}$. However, similarly as for aMc spacetime, the exponential dominates; $\lim_{r\rightarrow0^{+}}\e^{-\ell/r}/(r^n\,\Sigma^3)=0$ for all $\theta$. Consequently the ring singularity present at $r=0$ in BL coordinates for Kerr is replaced by a region of spacetime that is asymptotically Minkowski. This is already enough to conclude that the spacetime is globally regular in the sense of Bardeen~\cite{Tbilisi:Bardeen:1968}, and is consistent with the findings in reference~\cite{ANRBH:Ghosh:2015}.

Note that because the spacetime is now stationary rather than static, the Kretschmann scalar need no longer be positive definite~\cite{NBB:Lobo:2021}. 
It is now not sufficient to examine the Kretschmann scalar for regularity \textit{via} Theorem~\ref{Theorem:Kretsch}; one needs to inspect all the individual orthonormal Riemann components.

More generally, for the family of ``one-function off-shell'' Kerr geometries all nonzero orthonormal components of the Riemann tensor can be represented by
\begin{equation}
    R^{\hat{\alpha}\hat{\beta}}{}_{\hat{\mu}\hat{\nu}} = Z\left(r,\theta,m(r),m'(r),m''(r);m,a,\ell\right) \ ,
\end{equation}
for some function $Z$, and the condition for curvature regularity reduces to
\begin{equation}
    m(r) = \mathcal{O}(r^3) \ .
\end{equation}
The condition for an asymptotically Minkowski core reduces to
\begin{equation}
    m(r) = o(r^3) \ .
\end{equation}

\subsection{Other curvature quantities}

For the sake of rigour, it is prudent for one to examine the Riemann curvature invariants associated with the candidate geometry.

The Ricci scalar is given by
\begin{equation}
    R = \frac{2\ell^2m\,\e^{-\ell/r}}{\Sigma\,r^3} \ .
\end{equation}
The Ricci contraction $R_{\alpha\beta}R^{\alpha\beta}$ is given by
\begin{equation}
    R_{\alpha\beta}R^{\alpha\beta} = \frac{8\ell^2\left(m\,\e^{-\ell/r}\right)^2}{\Sigma^4}\left(\Xi^2-2\Xi+2\right) \ .
\end{equation}
Note that $\left(\Xi^2-2\Xi+2\right) = 1 + (\Xi-1)^2 \geq 1$ is manifestly positive.

The Kretschmann scalar ($K=R_{\alpha\beta\mu\nu}R^{\alpha\beta\mu\nu}$) is given by 
\begin{eqnarray}
    K &=& \frac{48r^6\left(m\,\e^{-\ell/r}\right)^2}{\Sigma^6}\Bigg\lbrace 1 -15\left(\frac{a}{r}\right)^2\cos^2\theta +15\left(\frac{a}{r}\right)^4\cos^4\theta-\left(\frac{a}{r}\right)^6\cos^6\theta \nonumber \\[2pt]
    && \qquad \qquad \qquad \qquad \qquad \qquad +\frac{4}{3}\Xi^4-\frac{16}{3}\Xi^3+8\Xi^2\left[1-\left(\frac{a}{r}\right)^2\cos^2\theta\right] \nonumber \\[2pt]
    && \qquad \qquad \qquad \qquad \qquad \quad -4\Xi\left[1-6\left(\frac{a}{r}\right)^2\cos^2\theta+\left(\frac{a}{r}\right)^4\cos^4\theta\right]\Bigg\rbrace \ . \nonumber \\
    &&
\end{eqnarray}
Note the presence of both positive definite and negative definite terms, with the negative definite terms depending on even powers of the spin parameter $a$, so that they switch off as the rotation is set to zero. Indeed as $a\to 0$ one has $\Xi \to {\ell\over 2r}$ and so
\begin{eqnarray}
    K_{a\to0} &\to& \frac{48\left(m\,\e^{-\ell/r}\right)^2}{r^6}
    \Bigg\lbrace \frac{4}{3}\Xi^4-\frac{16}{3}\Xi^3
    +8\Xi^2 -4\Xi  +1 \Bigg\rbrace \nonumber \\
    &\to &\frac{48\left(m\,\e^{-\ell/r}\right)^2}{r^6}
    \Bigg\lbrace 
     (1-\Xi)^4 +{\Xi^2\,(\Xi-2)^2\over 3} +{2\over3} \, \Xi^2
    \Bigg\rbrace \ .
\end{eqnarray}
This is now manifestly a positive definite sum of squares, as required.

To evaluate the Weyl contraction note that in this situation the (orthonormal) Weyl tensor has only two algebraically independent components
\begin{eqnarray}
C_{\hat t\hat\phi\hat t \hat\phi} &=& -{1\over2}C_{\hat t\hat r\hat t \hat r} =
C_{\hat t\hat\theta \hat t \hat\theta} = -C_{\hat\phi\hat r\hat \phi\hat r} 
= {1\over2} C_{\hat\phi\hat \theta\hat \phi\hat \theta} 
= -C_{\hat r \hat \theta\hat r \hat \theta} \ ,
\\[2pt]
C_{\hat t\hat \phi\hat r\hat\theta} &=&
 {1\over2} C_{\hat t\hat r\hat\phi\hat\theta} 
 = C_{\hat t\hat \theta\hat\phi\hat r} \ ,
\end{eqnarray}
where explicitly
\begin{eqnarray}
C_{\hat t\hat\phi\hat t \hat\phi} &=& \frac{r^3m\,\e^{-\ell/r}}{3\Sigma^3}\left\lbrace 2\Xi^2-6\Xi+3-9\left(\frac{a}{r}\right)^2\cos^2\theta\right\rbrace \ ,
\\[2pt]
C_{\hat t\hat r\hat\phi\hat\theta} &=& \frac{r^2m\,\e^{-\ell/r}a\cos\theta}{\Sigma^3}\left\lbrace2\Xi-3+\left(\frac{a}{r}\right)^2\cos^2\theta\right\rbrace \ .
\end{eqnarray}
The Weyl contraction ($C_{\alpha\beta\mu\nu}C^{\alpha\beta\mu\nu}$) 
is then given by 
\begin{equation}
    C_{\alpha\beta\mu\nu}C^{\alpha\beta\mu\nu} = 
    48 ([C_{\hat t\hat\phi\hat t \hat\phi}]^2 
    - [C_{\hat t\hat \phi\hat r\hat\theta}]^2) \ .
\end{equation}
Note the presence of both positive definite and negative definite terms, with the negative definite terms depending on even powers of the spin parameter $a$, so that they switch off as the rotation is set to zero.

It is then easy to check that $C_{\alpha\beta\mu\nu}C^{\alpha\beta\mu\nu}
= K + 2 R_{\mu\nu} R^{\mu\nu} - {1\over3} R^2$, as is also required.

All of these Riemann curvature invariants are globally well-behaved, remaining finite $\forall\,\,r\in[0,+\infty)$, cementing the fact that the eye of the storm is curvature-regular. In the limit as $\ell\rightarrow0$, the expected limiting behaviour when compared with standard Kerr spacetime is observed. In the limit as $a\rightarrow0$, all Riemann invariants tend to their counterparts for the spherically symmetric candidate geometry analysed in reference~\cite{RBHWAMC:Simpson:2019}.

\subsubsection{Ricci and Einstein tensors}

Both the Ricci and Einstein tensors are diagonal in the orthonormal basis. The Ricci tensor is given by
\begin{equation}
    R^{\hat{\mu}}{}_{\hat{\nu}} = \frac{2\ell m\,\e^{-\ell/r}}{\Sigma^2}\,\text{diag}\left(\Xi-1,\Xi-1,1,1\right) \ ,
\end{equation}
and the Einstein tensor is given by
\begin{eqnarray}
    G^{\hat{\mu}}{}_{\hat{\nu}} = -\frac{2\ell m\,\e^{-\ell/r}}{\Sigma^2}\,\text{diag}\left(1,1,\Xi-1,\Xi-1\right) \ .
\end{eqnarray}
These representations are \textit{highly} tractable when compared with the analogous results for other candidate rotating regular black holes in the literature~\cite{SORRBH:Abdujabbarov:2016, RRBH:Bambi:2013, MTKSPORBHFTS:Bambi:2014, GLBRW:Jusufi:2018, RRBHICMG:Jusufi:2020, QMQO:Jusufi:2021, KBHWPH:Herdeiro:2016, RRBHS:Abdujabbarov:2014, GRRBHIGRCTNE:Toshmatov:2017, NOTCOTGRCRBHIGRCNE:Toshmatov:2017, ANFORBHM:Franzin:2021, CBBS:Franzin:2021}.

\subsection{Horizons, surface gravity, and ergosurfaces}

Horizon locations are characterised by the roots of $\Delta_{\eos}(r)$, which are also the only coordinate singularities present in the line element Eq.~(\ref{E:eos}). Since $\Delta_{\eos}(r)$ is real,  while $\Delta_{\eos}(r=0) = a^2>0$ and $\Delta_{\eos}(r\to\infty) = \O(r^2)$, there are either two distinct roots, one double root, or zero roots. Since $\Delta_{\eos}(r) > \Delta_{\Kerr}(r)$ the location of the roots of $\Delta_{\eos}(r)$ is trivially bounded by the location of the roots of $\Delta_{\Kerr}(r)$. Specifically
\begin{equation}
m-\sqrt{m^2-a^2} < r_{H_{-}} \leq r_{H_{+}} < m+\sqrt{m^2-a^2} \ .
\end{equation}
In particular, if $m<a$ there certainly are no roots.

Analytically, one cannot explicitly solve for the roots of $\Delta_{\eos}(r)$.
However what one can do, assuming the existence of distinct roots $r_{H_{\pm}}$, is to ``reverse engineer'' by solving for 
$m(r_{H_{+}},r_{H_{-}})$ and $a^2(r_{H_{+}},r_{H_{-}})$. Note that by definition
\begin{equation}
   (r_{H_{+}})^2  -2 m \, r_{H_{+}} \exp( - \ell/r_{H_{+}}) + a^2 =0 \ ,
\end{equation}
\begin{equation}
   (r_{H_{-}})^2  -2 m \, r_{H_{-}} \exp( - \ell/r_{H_{-}}) + a^2 =0 \ .
\end{equation}
These are two simultaneous equations linear in $m$ and $a^2$. One finds
\begin{equation}
m(r_{H_{+}},r_{H_{-}}) =  {(r_{H_{+}})^2-(r_{H_{-}})^2 \over 
2 ( \e^{-\ell/r_{H_{+}}} \; r_{H_{+}} - \e^{-\ell/r_{H_{-}}} \; r_{H_{-}})} \ ,
\end{equation}
and
\begin{equation}
a^2 (r_{H_{+}},r_{H_{-}}) = {r_{H_{+}} r_{H_{-}}  (
\e^{-\ell/r_{H_{-}}} \; r_{H_{+}} - \e^{-\ell/r_{H_{+}}} \; r_{H_{-}})
\over  \e^{-\ell/r_{H_{+}}} \; r_{H_{+}} - \e^{-\ell/r_{H_{-}}} \; r_{H_{-}}} \ .
\end{equation}
In the degenerate extremal limit $r_{H_{+}}\to r_H \leftarrow r_{H_{-}}$, using the l'H\^opital rule,  this simplifies to
\begin{equation}
m(r_H) =  {(r_H)^2 \; e^{\ell/r_H} \over 
r_H + \ell} > r_H \ ,
\quad
\hbox{and}
\quad
a^2 (r_H) = {r_H^2  (r_H-\ell)\over r_H+\ell} < r_H^2 \ .
\end{equation}
For fixed $\ell$ and $r_H$, setting $a\to a(r_H)$, one has: (i) If $m>m(r_H)$ there will be two distinct roots, one above and one below $r_H$. (ii) If $m=m(r_H)$ there is one degenerate root exactly at $r_H$. (iii) If $m<m(r_H)$ there are no real roots.

%
Given this is the best one can say analytically, and that in this context the parameter $\ell$ is often associated with the Planck scale, one may Taylor series expand about $\ell=0$ for an approximation. 

First write
\begin{equation}
    r_{H} = m\,\e^{-\ell/r_{H_{\pm}}} + S_1 \sqrt{m^2\,\e^{-2\ell/r_{H}}-a^2} \ ,
\end{equation}
where $S_1=\pm 1$. For small $\ell$, expanding individual terms to second-order, one has:
\begin{equation}
    r_{H} = m - \ell - \mathcal{O}(\ell^2) + S_{1}\sqrt{m^2-a^2-2m\ell-\mathcal{O}(\ell^2)} \ .
\end{equation}
This has the correct limiting behaviour as $\ell\rightarrow0$. Investigating in more detail, instead of expanding about $\ell=0$ one can instead search for the approximate horizon locations by expanding about the Kerr horizon located at $r=r_{H,\Kerr}=m+S_{1}\sqrt{m^2-a^2}$. 

To second-order this gives
\begin{eqnarray}
    r_{H} &=& m + S_{1}\sqrt{m^2-a^2} \nonumber \\[3pt]
    && \ -S_{1}\frac{m\left[2m\sqrt{m^2-a^2}+S_{1}(2m^2-a^2)\right]}{(S_{1}m+\sqrt{m^2-a^2})\left[m\sqrt{m^2-a^2}+S_{1}(m^2-a^2)\right]}\ell + \mathcal{O}\left(\ell^2\right) \nonumber \\
    && \nonumber \\
    &=& r_{H,\Kerr} -\frac{2m\left(r_{H,\Kerr}\right)-a^2}{\left(r_{H,\Kerr}\right)\left[r_{H,\Kerr}-\frac{a^2}{m}\right]}\,\ell + \mathcal{O}(\ell^2) \ .
\end{eqnarray}
Notably, the surface area of each horizon is qualitatively unchanged from Kerr spacetime, given by
\begin{equation}
    A_{H} = 2\pi\int_{0}^{\pi}\sqrt{g_{\theta\theta}\,g_{\phi\phi}}\bigg\vert_{r_{H}}\,\d\theta = 4\pi(r_{H}^2+a^2) \ .
\end{equation}
The ergosurface is characterised by $g_{tt}=0$, \textit{implicitly} given by
\begin{equation}
    r_{\text{erg}}^2+a^2\cos^2\theta = 2m\,r_{\text{erg}}\:\e^{-\ell/r_{\text{erg}}} \ .
\end{equation}
Finding the implicit quadratic solutions for $r_{\text{erg}}$, taking the outer solution due to the constraint it must lie outside the \emph{outer} horizon, and expanding individual terms to second-order for small $\ell$ gives
\begin{equation}
    r_{\text{erg}} = m -\ell -\mathcal{O}(\ell^2)+ \sqrt{m^2-a^2\cos^2\theta-2m\ell-\mathcal{O}(\ell^2)} \ .
\end{equation}
This has the correct limiting behaviour as $\ell\rightarrow0$. Expanding instead around $r_{\text{erg,Kerr}}=m+\sqrt{m^2-a^2\cos^2\theta}$ yields
\begin{equation}
    r_{\text{erg}} = r_{\text{erg,Kerr}} - \frac{2m\left(r_{\text{erg,Kerr}}\right)-a^2\cos^2\theta}{\left(r_{\text{erg,Kerr}}\right)\left[r_{\text{erg,Kerr}}-\frac{a^2\cos^2\theta}{m}\right]}\,\ell + \mathcal{O}(\ell^2) \ .
\end{equation}
The surface gravity of the outer horizon in the $r>0$ universe is given by
\begin{equation}
    \kappa_{\text{out}} = \frac{1}{2}\,\frac{\d}{\d r}\left(\frac{\Delta_{\text{eos}}}{r^2+a^2}\right)\Bigg\vert_{r_{H}} = \frac{m\,\e^{-\ell/r_{H}}(r_{H}^3-\ell r_{H}^2-a^2r_{H}-a^2\ell)}{r_{H}(r_{H}^2+a^2)^2} \ .
\end{equation}
Imposing the extremality constraint $\kappa_{\text{out}}=0$ amounts to forcing inner and outer horizons to merge, and one recovers the condition $a^2\to a^2(r_H)$ discussed above.
Alternatively one could impose this constraint directly and find the extremal horizon location $r_H$ by solving a cubic, however this is not particularly informative as it gives the same qualitative information that has already been obtained.

\subsection{Killing tensor and Killing tower}\label{Killingtower}

It makes sense to first display the relevant results in full generality for the class of ``one-function off-shell'' Kerr geometries. When compared with the BL coordinate system of Eq.~(\ref{E:eos}), the generalised line element for ``one-function off-shell'' Kerr simply makes the replacement $\Delta_{\text{eos}}\rightarrow \Delta = r^2+a^2-2r\,m(r)$. The contravariant metric tensor can then be written in the following form
\begin{equation}\label{contramet}
    g^{\mu\nu} = -\frac{1}{\Sigma}\begin{bmatrix} \Sigma+\frac{2r(r^2+a^2)\,m(r)}{\Delta} & 0 & 0 & \frac{2ar\,m(r)}{\Delta}\\ 0 & -\Delta & 0 & 0 \\ 0 & 0 & -1 & 0 \\ \frac{2ar\,m(r)}{\Delta} & 0 & 0 & \frac{a^2}{\Delta}-\frac{1}{\sin^2\theta}\\\end{bmatrix}^{\mu\nu} \ .
\end{equation}
With the goal of finding a nontrivial  Killing two-tensor $K_{\mu\nu}$, satisfying $K_{(\mu\nu;\alpha)}=0$, one can apply the Papadopoulos--Kokkotas algorithm~\cite{PKSIDS:Kokkotas:2018, OKBHDAACC:Kokkotas:2021} for obtaining nontrivial Killing tensors on axisymmetric spacetimes. This algorithm is an extension of older results by Benenti and Francaviglia~\cite{ROCSSATATGR:Benenti:1979}, and the first step is to decompose the contravariant metric in BL coordinates into the following general form
\begin{equation}
    g^{\mu\nu} = \frac{1}{A_1(r)+B_1(\theta)}\begin{bmatrix} A_5(r)+B_5(\theta) & 0 & 0 & A_4(r)+B_4(\theta)\\ 0 & A_2(r) &
     0 & 0\\ 0 & 0 & B_2(\theta) & 0\\ A_4(r) + B_4(\theta) & 0 & 0 & A_3(r)+B_3(\theta)\\\end{bmatrix}^{\mu\nu} \ .
\end{equation}
Eq.~(\ref{contramet}) is readily interpreted to be of this form, with the explicit assignments
\begin{eqnarray}
    &A_1(r) = -r^2 \ , \quad &A_2(r) = -\Delta \ , \quad A_3(r) = \frac{a^2}{\Delta} \ , \nonumber \\[2pt]
    &  A_4(r) = \frac{2ar\,m(r)}{\Delta} \ , \quad &A_5(r) = r^2+\frac{2r(r^2+a^2)\,m(r)}{\Delta} \ ; \nonumber \\[3pt]
    & B_1(\theta) = -a^2\cos^2\theta \ , \quad &B_2(\theta) = -1 \ , \quad B_3(\theta) = -\frac{1}{\sin^2\theta} \ , \nonumber \\[2pt]
    &  B_4(\theta) = 0 \ , \quad &B_5(\theta) = a^2\cos^2\theta \ .
\end{eqnarray}
Given this decomposition, the Papadopoulos--Kokkotas algorithm~\cite{PKSIDS:Kokkotas:2018, OKBHDAACC:Kokkotas:2021} asserts that the following yields a nontrivial contravariant Killing tensor:
\begin{equation}
    K^{\mu\nu} = \frac{1}{A_1+B_1}\begin{bmatrix} B_1A_5-A_1B_5 & 0 & 0 & B_1A_4-A_1B_4\\ 0 & A_2B_1 & 0 & 0\\ 0 & 0 & -A_1B_2 & 0\\ B_1A_4-A_1B_4 & 0 & 0 & B_1A_3-A_1B_3\\\end{bmatrix}^{\mu\nu} \ .
\end{equation}
As such, one finds the following nontrivial rank two contravariant Killing tensor for ``one-function off-shell'' Kerr spacetimes:
\begin{equation}
    K^{\mu\nu} = \frac{a^2\cos^2\theta}{\Sigma}\begin{bmatrix} \frac{2r(r^2+a^2)\,m(r)}{\Delta} & 0 & 0 & \frac{2ar\,m(r)}{\Delta}\\ 0 & -\Delta & 0 & 0\\ 0 & 0 & \frac{r^2}{a^2\cos^2\theta} & 0\\ \frac{2ar\,m(r)}{\Delta} & 0 & 0 & \frac{a^2}{\Delta}+\left(\frac{r}{a\sin\theta\cos\theta}\right)^2\\\end{bmatrix}^{\mu\nu} \ .
\end{equation}
Lowering the indices, one finds covariant Killing two-tensor satisfying $K_{(\mu\nu;\alpha)}=0$ (easily verified using {\sf Maple}) in the BL coordinate basis.

Converting then to the orthonormal tetrad basis \textit{via} $K_{\hat{\mu}\hat{\nu}}=e_{\hat{\mu}}{}^{\mu}\,e_{\hat{\nu}}{}^{\nu}\,K_{\mu\nu}$, and raising the indices, gives the following
\begin{equation}
    K^{\hat{\mu}\hat{\nu}} = \begin{bmatrix}-a^2\cos^2\theta & 0 & 0 & 0\\ 0 & a^2\cos^2\theta & 0 & 0\\ 0 & 0 & -r^2 & 0\\ 0 & 0 & 0 & -r^2\\\end{bmatrix}^{\hat{\mu}\hat{\nu}} \ .
\end{equation}
Notice that in the tetrad basis this is identical to the nontrivial Killing two-tensor for Kerr spacetime. Specifically, notice that it is \textit{independent} of the  mass function $m(r)$. As such, the entire family of ``one-function off-shell'' Kerr geometries inherits the same Killing tensor as Kerr spacetime. Notably, both the Ricci tensor and Killing tensor are diagonal in this tetrad basis, and as such the commutator $[R,K]^{\hat{\mu}}{}_{\hat{\nu}}$ will vanish; by Theorem~\ref{Theorem:PGLT2} this is sufficient to conclude that the Klein--Gordon equation is separable on the background spacetime~\cite{PGLT2:Baines:2021}. As such, the eye of the storm is amenable to a standard spin zero quasi-normal modes analysis (invoke the relativistic Cowling effect~\cite{Cowling}, assume a separable wave form, and use a preferred numerical technique to approximate the ringdown signal). The same can be said for all candidate geometries in the class of ``one-function off-shell'' Kerr~\cite{BHHSACI:Frolov:2017, HASSSOEE:Carter:1968, GSOTKFOGF:Carter:1968}.

Furthermore, it is straightforward to verify that the following two-form square-root of the Killing tensor is a genuine Killing--Yano tensor, satisfying the Killing--Yano equation $f_{\hat{\mu}(\hat{\nu};\hat{\alpha})}=0$:
\begin{equation}
    f^{\hat{\mu}\hat{\nu}} = \begin{bmatrix} 0 & a\cos\theta & 0 & 0\\ -a\cos\theta & 0 & 0 & 0\\ 0 & 0 & 0 & -r\\ 0 & 0 & r & 0\\\end{bmatrix}^{\hat{\mu}\hat{\nu}} \ .
\end{equation}
It is straightforward to check that $K^{\hat{\mu}\hat{\nu}} = -f^{\hat{\mu}\hat{\alpha}}\;\eta_{\hat{\alpha}\hat{\beta}}\;f^{\hat{\beta}\hat{\nu}}$. The ``principal tensor''~\cite{BHHSACI:Frolov:2017} is then simply the Hodge dual of this two-form, and in full generality the family of ``one-function off-shell'' Kerr geometries possesses the full ``Killing tower''~\cite{BHHSACI:Frolov:2017} of Killing tensor, Killing--Yano tensor, and principal tensor. Separability of the Hamilton--Jacobi equations is guaranteed by the existence of $K_{\mu\nu}$ and the associated Carter constant, and in reference~\cite{PRSORRBHANS:Ghosh:2020l} the geodesics for the photon ring are computed. Notably, the eye of the storm is able to be delineated from Kerr, and results from reference~\cite{TRRMACFABH:Ghosh:2020} conclude that the data from the image of M87 provided by the EHT does not exclude the eye of the storm from being astrophysically viable. This is yet another highly desirable feature of the eye of the storm geometry.

The spacetime is also amenable to straightforward calculation of the black hole shadow~\cite{SORNBHS:Amir:2016, BHSIAASAAS:Tsukamoto:2018}. These calculations further demonstrate that the geometry falls within experimental constraints provided by the EHT. Furthermore, the fact that the eye of the storm falls within this class of ``one-function off-shell" Kerr geometries implies that Maxwell's equations also separate on the background spacetime; this is confirmed by the proof given in Appendix A of reference~\cite{SIKBH:Franzin:2021}. One conjectures that the equations governing the spin two polar and axial modes will also be separable on this geometry.

\section{Stress-energy and energy conditions}

Recall that the Einstein tensor in an orthonormal basis is given by
\begin{eqnarray}
    G^{\hat{\mu}}{}_{\hat{\nu}} = -\frac{2\ell m\,\e^{-\ell/r}}{\Sigma^2}\,\text{diag}\left(1,1,\Xi-1,\Xi-1\right) \ ,
\end{eqnarray}
where $\Xi=\frac{\ell\Sigma}{2r^3}$. Now fix the geometrodynamics by interpreting the spacetime through the lens of standard GR. As such, coupling the geometry to the Einstein equations, one has
\begin{equation}
    \frac{1}{8\pi}G^{\hat{\mu}}{}_{\hat{\nu}} = T^{\hat{\mu}}{}_{\hat{\nu}} = \text{diag}(-\rho,p_{r},p_{t},p_{t}) \ .
\end{equation}
Due to the fact that $-\rho=p_{r}$ this equation holds globally in the geometry, regardless of whether one is outside (inside) the outer (inner) horizons, or trapped in between them. The fact that the Einstein tensor is diagonal in an orthonormal basis implies that the stress-energy tensor is Hawking--Ellis type I~\cite{TLSSOS:Ellis:1973, GRCGSCATHC:Moruno:2017, ECOTHT:Moruno:2018, HCOS:Moruno:2021}. This leads to the following specific stress-energy components:
\begin{eqnarray}
    \rho &=& -p_{r} = \frac{\ell m\,\e^{-\ell/r}}{4\pi\Sigma^2} \ , \nonumber \\
    p_{t} &=& \frac{\ell m\,\e^{-\ell/r}}{4\pi\Sigma^2}\left(1-\Xi\right) \ .
\end{eqnarray}
An extremely desirable feature of the ``eye of the storm'' spacetime is its relationship with the classical energy conditions associated with GR. In view of $\ell>0$, one trivially globally satisfies $\rho>0$. The radial null energy condition (NEC) is trivially satisfied since $\rho+p_{r}=0$ globally. Analysing the transverse NEC:
\begin{equation}
    \rho + p_{t} = \frac{\ell m\,\e^{-\ell/r}}{4\pi\Sigma^2}\left(2-\Xi\right) \ .
\end{equation}
This changes sign when $\Xi=2$, or when $\frac{\Sigma}{r^3}=\frac{4}{\ell}$. On the equatorial plane this is when $r=\frac{\ell}{4}$. If $\frac{\Sigma}{r^3}>\frac{4}{\ell}$, the transverse NEC is violated, whilst if $\frac{\Sigma}{r^3}<\frac{4}{\ell}$, it is satisfied. For the equatorial plane, the violated region is when $r<\frac{\ell}{4}$.

Now analysing the strong energy condition (SEC), which implies $\rho+p_{r}+2p_{t}>0$:
\begin{equation}
    \rho + p_{r} + 2p_{t} = 2p_{t} = \frac{\ell m\,\e^{-\ell/r}}{2\pi\Sigma^2}\left(1-\Xi\right) \ .
\end{equation}
This changes sign when $\Xi=1$, or when $\frac{\Sigma}{r^3}=\frac{2}{\ell}$. If $\frac{\Sigma}{r^3}>\frac{2}{\ell}$, the SEC is violated, whilst if $\frac{\Sigma}{r^3}<\frac{2}{\ell}$, the SEC is satisfied. For the equatorial plane the violated region is whenever $r<\frac{\ell}{2}$.

Given the freedom to choose the suppression parameter $\ell$, this means that all of the energy-condition-violating physics can be forced into an arbitrarily small region in the deep core. One can conceive of three sensible categories of relativist in the present day:
\begin{itemize}
    \item Those who believe that GR holds \textit{everywhere}, other than at a distance scale where a mature and phenomenologically verifiable theory of quantum gravity must necessarily take over.
    \item Those who believe that GR can only be believed in regions external to any Cauchy horizon(s).
    \item Those who believe that GR only holds outside \textit{any} horizon full stop.
\end{itemize}
%

Regardless of one's personal subscription, the freedom to scale $\ell$ as required means all of the energy-condition-violating physics can be readily pushed into a region where GR is no longer an appropriate theory. Notably, no exotic matter is required in the exterior region of the spacetime. In the domain of outer communication, there is manifest satisfaction of all of the classical energy conditions. This is consistent with astrophysical observations, and is an extremely desirable feature of eye of the storm spacetime when compared with the remaining literature concerning rotating regular black holes; for instance in the spacetimes explored in references~\cite{ANFORBHM:Franzin:2021, CBBS:Franzin:2021} (see Chapters~\ref{C:SVog} and~\ref{C:SVCBB}) one has global violation of the NEC.

It should be noted that the modification to Kerr spacetime is in fact technically long
range, that is, with polynomial falloff at large distances. This is best seen by the fact that
as $r\rightarrow0$, $\rho\rightarrow0$, whilst for large $r$, $\rho=\mathcal{O}\left(r^{-4}\right)$; a rapid polynomial falloff. Classically, the effective matter distribution is that of an anisotropic fluid due to the fact that the stress-energy tensor is Hawking-Ellis type I and $p_{r}\neq p_{t}$. The density of this fluid has very rapid falloff with increasing $r$. In a minimally modified theory of gravity, where the presence of quantum fields alters the Einstein equations, it is conceivable that the unique stationary axisymmetric vacuum solution would in fact look like an anisotropic fluid when viewed through the lens of standard GR (this certainly holds in spherical symmetry). One would then expect
the effective $\rho$ to have rapid falloff for large $r$ due to the hypothetical quantum mechanism
only coming into effect at very small $r$. For an example of this line of inquiry in the literature,
see~\cite{OQDOTSS:Kazakov:1993}, where Kazakov and Solodukhin motivate a quantum deformation of the Schwarzschild solution. Providing a precise physical mechanism for a similarly motivated deformation of the Kerr solution is at this stage an avenue for further research.
\clearpage

Notably, the multipole moments of the Kerr solution are minimally affected when
compared with the eye of the storm. For large $r$ one has $g_{tt} = -1 + \frac{2m}{r} - \frac{2ma^2\cos^2\theta}{r^3}+m\ell\left(\frac{\ell}{r^3}-\frac{2}{r^2}\right)+\mathcal{O}\left(r^{-4}\right)$. The presence of the $\frac{1}{r^2}$ term in this expansion means that the
coordinate system is not ``mass-centered'' in the sense of Thorne~\cite{MEOGR:Thorne:1980}; consequently in order to perform a straightforward extraction of the quadrupole and higher multipole moments for the
eye of the storm one would need to find a coordinate transformation to de Donder gauge and
carefully perform the appropriate procedure outlined in~\cite{MEOGR:Thorne:1980}. However, one does have in general for large $r$ that $g_{\mu\nu} = g_{\mu\nu,\text{Kerr}}+m\ell\times\mathcal{O}\left(\frac{1}{r^2}\right)$. Outside horizons, one has $r>m$, hence the deviation from Kerr goes as $\mathcal{O}\left(\frac{\ell}{m}\right)$. Consequently, for small $\ell$ (i.e. $\ell\ll m$) the relevant series expansions are asymptotically close to those for Kerr, and hence the effect on the multipole
moments is arbitrarily small.

\section{Introducing the ``eosKN'' model}\label{eosKN}

Along very similar lines to the logic presented in \S~\ref{S:amcRN} for the construction of the aMcRN model, one can extend Kerr--Newman in BL coordinates to create the ``eosKN'' geometry. Begin with standard Kerr--Newman in BL coordinates:
\begin{eqnarray}
    \d s^2_\text{KN}  &=& -\frac{\Delta_\text{KN}}{\rho^2_\text{KN}}(a\sin^2\theta \d\phi - \d t)^2 
    + \frac{\sin^2\theta}{\rho^2_\text{KN}}\left[(r^2+a^2)\d\phi -a\d t\right]^2 \nonumber \\
    && \qquad \qquad \qquad \qquad \quad \qquad \qquad
    + \frac{\rho^2_\text{KN}}{\Delta_\text{KN}}\d r^2+\rho^2_\text{KN}\d\theta^2 \ ,
\end{eqnarray}
where
\begin{equation}
    \rho^2_\text{KN} = r^2+a^2\cos^2\theta \ , \qquad \Delta_\text{KN} = r^2+a^2-2mr+Q^2 \ .
\end{equation}
then simply perform the following ``regularising'' procedure:
\begin{itemize}
    \item Make the modification $m\to m(r)=m\,\e^{-\ell/r}$;
    \item Make the modification $Q^2\to Q^2(r)=Q^2\,\e^{-\ell/r}$.
\end{itemize}
The only significant change then is to replace
\begin{equation}
    \Delta_\text{KN}\to\Delta_\text{eosKN} = r^2+a^2-2mr\,\e^{-\ell/r}+Q^2\,\e^{-\ell/r} \ .
\end{equation}
The current author conjectures that the resulting spacetime will be globally regular, and model a charged regular black hole which acts as the analog to Kerr--Newman from the aMc family. This is the first presentation of this geometry in the literature, and at this stage verifying its regularity, and the existence of horizon locations, as well as analysis of the other salient features of the spacetime, is an avenue for further research.

\section{Conclusion of Part II}\label{S:discussion}

In Chapter~\ref{C:AMCEOS}, the class of ``one-function off-shell'' Kerr geometries has been defined, and the general existence of the full Killing tower demonstrated for the geometries within it. Within this class, the most desirable candidate spacetime, the eye of the storm, has been chosen according to a set of carefully chosen theoretically and experimentally motivated metric construction criteria. This spacetime models a rotating regular black hole with an asymptotically Minkowski core, is asymptotically Kerr for large $r$, manifestly satisfies all of the standard energy conditions of GR in both the region of theoretical validity and the region of experimental validity, has integrable geodesics in principle, and has the properties of separability of both the Klein--Gordon equation and Maxwell's equations. The eye of the storm is also the most mathematically tractable rotating regular black hole in the current literature, and is readily amenable to the extraction of (potential) astrophysical observables falsifiable/verifiable by the experimental community. It acts as the analog to Kerr from the family of regular black holes with asymptotically Minkowski cores.

The Schwarzschild analog, aMc spacetime, was thoroughly explored in Chapter~\ref{C:AMCog}. Similarly, it is well-behaved in the context of the point-wise energy conditions, and is highly mathematically tractable. This comes at the cost of important qualitative features being defined in terms of the Lambert $W$ function, and indeed in particular the analysis for the notable orbits of aMc spacetime (see \S~\ref{AMCISCOOSCO}) is far more intricate and subtle than one might expect. The complexity cost of the analysis is paid off \emph{via} the existence of several curious features, such as gravity turning ``repuslive'' at the coordinate location $r=\ell$, and the existence of two photon orbits for compact massive objects. Given the intricacy of performing this analysis for the static spherically symmetric aMc spacetime, one wonders what further challenges will be presented for attempting to extract important geodesics for the eye of the storm geometry. This is an important future research topic.
\enlargethispage{30pt}

In Chapter~\ref{C:AMCQNM} an approximation of the spin one and spin zero quasi-normal mode profiles for aMc spacetime was computed using a first-order WKB approximation. For sufficiently small $\ell$, the quasi-normal mode profiles of spin zero scalar and spin one electromagnetic perturbations on a background aMc spacetime exhibit shorter-lived and higher-energy signals than their Schwarzschild counterparts. A general analysis of perturbation of the Regge--Wheeler potential was also conducted in \S~\ref{S:RWperturb}, and the results applied to Schwarzschild spacetime. With respect to the eye of the storm, separability of both the Klein--Gordon equation and Maxwell's equations leads one to conjecture that the equations governing the spin two polar and axial modes will also separate on this spacetime. Verifying this is an important topic for future research. Performing a thorough quasi-normal modes analysis for this geometry is one of the more critical outstanding issues.
\clearpage

Notably, both aMc spacetime and eos spacetime are still ``models''. Elevating them to ``exact solution status'' by finding a combination of suitable source terms would be a fantastic result, and is the most important theoretical result still outstanding for these geometries. The community should try to attack this issue, along thematically similar lines to Bronnikov and Walia's work in reference~\cite{FSFSVS:Bronnikov:2022} (where SV and bbRN spacetime were promoted to ``exact solution status''), and see whether an associated action principle exists for aMc and eos in the Lagrangian formalism.
\vspace*{60pt}

\begin{center}
    This concludes Part~\ref{PartII}.
\end{center}


%% file: 02-Black-bounce/2-SVthinshell.tex
\part[Spacetime surgery: Exploring thin-shell \\\hspace*{1cm} constructions]{Spacetime surgery: Exploring thin-shell constructions}\label{PartIII}

\chapter{Thin-shell traversable wormholes}\label{C:SVthinshell}

%
%
Given a candidate spacetime, one can always call on the thin-shell formalism developed mathematically by Darmois~\cite{MDSMXXV:Darmois:1927}, and later W.~Israel~\cite{SHATSIGR:Israel:1966}, to perform ``surgery'' between two copies of the geometry, stitching them together at a hypersurface which is then interpreted as the timelike throat of a traversable wormhole~\cite{TWFSMSS:Visser:1989}. This procedure can be applied to a wide variety of candidate spacetimes, and is amenable to intriguing stability analyses when permitting dynamics of the wormhole throat. In Chapter~\ref{C:SVthinshell}, a dynamic thin-shell variant of Simpson--Visser (SV) spacetime is analysed, and a qualitatively similar analysis of aMc spacetime is briefly discussed. In Chapter~\ref{C:DMS}, a relativistic analysis of thin-shell Dyson mega-spheres is performed. One can, in principle, extend thin-shell analyses to ``thick-shell'' analyses by employing the anisotropic TOV equation; this line of inquiry is for now left for further research.

Focusing on the recently introduced SV spacetime~\cite{BTTW:Simpson:2019} as discussed in Chapter~\ref{C:SVog}, one can consider the construction of the related spherically symmetric thin-shell traversable wormholes within the context of standard GR. All of the really unusual physics is encoded in the ``bounce'' parameter $\ell$. The standard thin-shell traversable wormhole construction is modified, each bulk region now being a copy of SV spacetime, and with the physics of the thin-shell being (as much as possible) derivable from the Einstein equations. Furthermore, a dynamical analysis of the throat is applied by considering linearised radial perturbations around static solutions. It is demonstrated that the stability of the wormhole is equivalent to choosing suitable properties for the exotic material residing on the wormhole throat. The construction is sufficiently novel to be interesting, and sufficiently straightforward to be tractable.

First, consider the SV line element:
\begin{equation}
    \d s^{2}=-\left(1-\frac{2m}{\sqrt{u^{2}+\ell^{2}}}\right)\,\d t^{2}+
    \left(1-\frac{2m}{\sqrt{u^{2}+\ell^{2}}}\right)^{-1}\,\d u^2 +
    \left(u^{2}+\ell^{2}\right)\, \d\Omega^{2}_{2} \ .
\end{equation}
Here the $u$ and $t$ coordinates have the domains $u\in(-\infty,+\infty)$ and $t\in(-\infty,+\infty)$. In the original references~\cite{BTTW:Simpson:2019, VSBATW:Moruno:2019} and the previous Chapters, the $u$ coordinate was called $r$, however it is now prudent to use the symbol $r$ for other purposes. By considering the coordinate transformation $r^2=u^2 + \ell^2$, one can use the following completely equivalent line element:
\begin{equation}
    \d s^{2} = -\left(1-\frac{2m}{r}\right)\,\d t^{2}+
    \left(1-\frac{2m}{r}\right)^{-1}\left(1-\frac{\ell^2}{r^2}\right)^{-1}\,\d r^2    
    +r^{2}\, \d\Omega^{2}_{2} \ .
\end{equation}
Here the new $r$ coordinate is now a double-cover of the $u$ coordinate. It has the domain $r\in(\ell,+\infty)$, and can be interpreted as the Schwarzschild area coordinate --- all the two-spheres of constant $r$ have area $A(r)=4\pi r^2$. This choice has the additional advantage of making it easy to directly compare the current analysis with most other work in the literature. The $t$ coordinate has the usual domain $t\in(-\infty,+\infty)$.


\section{SV thin-shell formalism}\label{S:formalism}
\enlargethispage{20pt}

First one performs a general (and relatively straightforward) theoretical analysis, somewhat along the lines laid out in reference~\cite{GSSDTTWISGR:Garcia:2012}, but with appropriate specialisations, simplifications, and modifications. Subsequently, one can investigate a number of specific examples in the way of special cases and toy models. Bulk spacetimes are discussed in \S~\ref{SS:bulk}, the extrinsic curvature of the thin-shells is discussed in \S~\ref{SS:extrinsic}, before moving on to the Lanczos equations in \S~\ref{SS:Lanczos}. Then, the Gauss--Codazzi--Mainardi equations are discussed in \S~\ref{SS:Gauss}, before considering the equation of motion and its linearisation in \S~\ref{SS:eom} and \S~\ref{SS:linearized}. Finally, the master equations are developed in \S~\ref{SS:master}, before moving on to \S~\ref{S:applications}, where some specific examples and applications are presented.

\subsection{Bulk spacetimes}\label{SS:bulk}

Begin the discussion by considering two distinct ``bulk'' spacetime manifolds, ${\cal M_+}$ and ${\cal M_-}$, equipped with boundaries $\partial{\cal M_+}= \Sigma_+$  and $\partial{\cal M_-}= \Sigma_-$. As long as the boundaries are isometric, $ \Sigma_+\sim \Sigma_-$, then one can define a manifold ${\cal M} = {\cal M_+} \cup {\cal M_-}$, which is smooth except possibly for a thin-shell transition layer at $\Sigma_+\sim\Sigma_-$. In particular, consider two static spherically symmetric SV spacetimes given on ${\cal M_\pm}$ by the following two-parameter $(m,\ell)$ line elements $g_{\mu \nu}^+(x^{\mu}_+)$ and $g_{\mu \nu}^-(x^{\mu}_-)$:
\begin{equation}
    \d s^{2}=-\left(1-\frac{2m_{\pm}}{r_{\pm}}\right)\,\d t_{\pm}^{2} +
    \left(1-\frac{2m_{\pm}}{r_{\pm}}\right)^{-1}\left(1-\frac{\ell_{\pm}^2}{r_{\pm}^2}\right)^{-1}\,\d r_{\pm}^2    
    +r_{\pm}^{2}\, \d\Omega^{2}_{2\,\pm} \ .
    \label{generalmetric}
\end{equation}
The usual Einstein field equations, $G_{{\mu}{\nu}}=8\pi \,T_{{\mu}{\nu}}$ (with $c=G_{N}=1$), imply that the physically relevant 
orthonormal components of the stress-energy tensor are  (in the two bulk regions) specified by:
\begin{eqnarray}
\rho(r)&=&-\frac{1}{8\pi }\frac{\ell^2\left( r-4m \right)}{ r^5} \ ,\label{rho}\\
p_{r}(r)&=&-\frac{1}{8\pi }\frac{\ell^2}{ r^4} \ ,\label{pr}\\
p_{t}(r)&=&\frac{1}{8\pi }\frac{\ell^2\left( r-m \right)}{ r^5} \ .
\label{pt}
\end{eqnarray}
Here $\rho(r)$ is the energy density, $p_r(r)$ is the radial pressure, and $p_t(r)$ is the transverse pressure. Given the spherical symmetry, $p_t(r)$ is the pressure measured in the two directions orthogonal to the radial direction. The subscripts $\pm$ (on $m_\pm$ and $\ell_\pm$) have been (temporarily) suppressed for clarity.

The null energy condition (NEC) is satisfied provided, for any arbitrary null vector $k^\mu$, the stress-energy  $T_{\mu\nu}$ satisfies $T_{\mu\nu}\,k^\mu\,k^\nu\geq 0$. The radial null vector is $k^{\hat{\mu}}=(1,\pm 1,0,0)$ in the orthonormal frame where the stress energy is
$T_{\hat{\mu}\hat{\nu}}={\rm diag}[\rho(r),p_r(r),p_t(r),p_t(r)]$. Then
\begin{equation}
 (T_{\hat{\mu}\hat{\nu}}\,k^{\hat{\mu}}\,k^{\hat{\nu}})_\mathrm{radial}=\rho(r)+p_r(r)
 =-\frac{\ell^2(r-2m)}{4\pi r^{5}} < 0 \ .\label{generalNEC}
\end{equation}
To verify the negativity of this quantity, recall that the bulk spacetime models a regular black hole when $\ell\in(0, 2m)$, with horizons at $u_{H} = \pm\sqrt{(2m)^{2}-\ell^2}$, corresponding to $r_H= u_H^2+\ell^2 = 2m$. One therefore ``chops'' the spacetime outside any horizons that are present. Hence in both of the bulk regions there is the condition that $r>2m$. Thus the radial NEC will be manifestly violated in both of the bulk regions of the SV spacetime. In the context of static spherical symmetry, this is sufficient to conclude that all of the standard energy conditions associated with general relativistic analysis will be similarly violated.

In the transverse directions, one can choose the null vector to be $k^{\hat{\mu}}=(1, 0, \cos\zeta,\sin\zeta)$, and so
\begin{equation}
 (T_{\hat{\mu}\hat{\nu}}\,k^{\hat{\mu}}\,k^{\hat{\nu}})_\mathrm{transverse}
 =\rho(r)+p_t(r) =\frac{3 \ell^2m}{8\pi r^{5}}>0 \ .    \label{generalNEC2}
\end{equation}
While this is manifestly positive this is not enough to override the NEC violations coming from the radial direction.
\clearpage

\subsection{Normal four-vector and extrinsic curvature}\label{SS:extrinsic}

The two bulk manifolds, ${\cal M_+}$ and ${\cal M_-}$, are bounded by the hypersurfaces $\partial{\cal M_+}=\Sigma_+$ and ${\cal M_-}=\Sigma_-$. These two hypersurfaces possess induced three-metrics $g_{ij}^+$ and $g_{ij}^-$, respectively. The $\Sigma_{\pm}$ are chosen to be isometric. (In terms of the intrinsic coordinates, $g_{ij}^{\pm}(\xi)=g_{ij}(\xi)$, with $\xi^i=(\tau,\theta,\phi)$.) Thence a single 
manifold ${\cal M}$ is obtained by gluing together ${\cal M_+}$ and ${\cal M_-}$ at their boundaries, that is ${\cal M}={\cal M_+}\cup {\cal M_-}$, with the natural identification of the two boundaries $\Sigma_{\pm}=\Sigma$.

The boundary manifold $\Sigma$ possesses three tangent basis vectors ${\bf e}_{(i)}=\partial /\partial \xi^i$, with the holonomic components $e^{\mu}_{(i)}|_{\pm}=\partial x_{\pm}^{\mu}/\partial \xi^i$. This basis specifies the induced metric \textit{via} the scalar product $g_{ij}={\bf e}_{(i)}\cdot {\bf e}_{(j)}=g_{\mu \nu}e^{\mu}_{(i)}e^{\nu}_{(j)}|_{\pm}$. Finally, in explicit coordinates, the intrinsic metric to $\Sigma$ is given by
\begin{equation}
\d s^2_{\Sigma}= - \d\tau^2 + R^2(\tau) \,(\d\theta ^2+\sin
^2{\theta}\,\d\phi^2) \ .
\end{equation}
That is, the manifold ${\cal M}$ is obtained by gluing ${\cal M_+}$ and ${\cal M_-}$ at the three-surface $x^{\mu}(\tau,\theta,\phi)=(t(\tau),R(\tau),\theta,\phi)$. The respective four-velocities, tangent to the junction surface and orthogonal to the slices of spherical symmetry, are defined on the two sides of the junction surface. They are explicitly given by
\begin{eqnarray}
U^{\mu}_{\pm}=
\left(\frac{\sqrt{\left(1-\frac{2m_{\pm}}{R}\right)\left(1-\frac{\ell_{\pm}^2}{R^2}\right)+\dot{R}^{2}}}{\left(1-\frac{2m_{\pm}}{R}\right)\left(1-\frac{\ell_{\pm}^2}{R^2}\right)^{1/2}},\;
\dot{R},0,0 \right) \ .
\end{eqnarray}
Here $\tau$ is the proper time of an observer comoving with $\Sigma$, and the overdot denotes a derivative with respect to this proper time.
Furthermore, the timelike junction surface $\Sigma$ is given by the parametric equation $f(x^{\mu}(\xi^i))= r-R(\tau) = 0$, and the unit normal four-vector, $n^{\mu}$, is defined as
\begin{equation}\label{defnormal}
n_{\mu}=
{\nabla_\mu f\over ||\nabla f||} = 
\pm \,\left |g^{\alpha \beta}\,\frac{\partial f}{\partial
x ^{\alpha}} \, \frac{\partial f}{\partial x ^{\beta}}\right
|^{-1/2}\;\frac{\partial f}{\partial x^{\mu}} \ .
\end{equation}
Hence $n_{\mu}\,n^{\mu}=+1$ and $n_{\mu}e^{\mu}_{(i)}=0$. In the usual Israel formalism one chooses the normals to point from ${\cal M_-}$ to ${\cal M_+}$~\cite{SHATSIGR:Israel:1966}, so that the unit-normals to the junction surface are provided by the following expressions:
\begin{eqnarray}
n^{\mu}_{\pm} &=& \pm \left(
\frac{\dot{R}}{\left(1-\frac{2m_{\pm}}{R}\right)\left(1-\frac{\ell_{\pm}^2}{R^2}\right)^{1/2}},
\sqrt{\left(1-\frac{2m_{\pm}}{R}\right)\left(1-\frac{\ell_{\pm}^2}{R^2}\right)+\dot{R}^{2}},0,0
\right) \label{normal} \ . \nonumber \\
&&
\end{eqnarray}
Taking into account spherical symmetry one may also obtain the above expressions from consideration of the contractions $U^{\mu}n_{\mu}=0$ and $n^{\mu}n_{\mu}=+1$.

The extrinsic curvature, or second fundamental form, is typically defined as $K_{ij}=n_{{(}\mu;\nu{)}}e^{\mu}_{(i)}e^{\nu}_{(j)}$. 
Now, by differentiating $n_{\mu}e^{\mu}_{(i)}=0$ with respect to $\xi^j$,  one obtains the following useful relation
\begin{equation}
n_{\mu}\frac{\partial ^2 x^{\mu}}{\partial \xi^i \, \partial \xi^j}=
-n_{\mu,\nu}\, \frac{\partial x^{\mu}}{\partial \xi^i}\frac{\partial x^{\nu}}{\partial \xi^j} \ ,
\end{equation}
so that the extrinsic curvature $K_{ij}$ can therefore be represented in the form
\begin{eqnarray}
\label{extrinsiccurv}
K_{ij}^{\pm}=-n_{\mu} \left(\frac{\partial ^2 x^{\mu}}{\partial
\xi ^{i}\,\partial \xi ^{j}}+\Gamma ^{\mu \pm}_{\;\;\alpha
\beta}\;\frac{\partial x^{\alpha}}{\partial \xi ^{i}} \,
\frac{\partial x^{\beta}}{\partial \xi ^{j}} \right) \ .
\end{eqnarray}
Finally, using both spherical symmetry and Eq.~(\ref{extrinsiccurv}), the nontrivial components of the extrinsic curvature are given by:
\begin{eqnarray}
K^{\theta \;\pm }_{\;\;\theta}&=&\pm\frac{1}{R}\,\sqrt{\left(1-\frac{2m_{\pm}}{R}\right)\left(1-\frac{\ell_{\pm}^2}{R^2}\right)+\dot{R}^{2}} \ ,
\label{genKplustheta}
\\
K^{\tau\;\pm}_{\;\;\tau}&=&\pm\,
\left[\frac{\ddot R-
\frac{\ell_{\pm}^2}{R\left(R^2 - \ell_{\pm}^2\right)} \dot{R^2}+
\frac{m_{\pm}\left(R^2 - \ell_{\pm}^2\right)}{R^4}}{\sqrt{\left(1-\frac{2m_{\pm}}{R}\right)\left(1-\frac{\ell_{\pm}^2}{R^2}\right)+\dot{R}^{2}}} \right]  \ .
\label{genKminustautau}
\end{eqnarray}

\subsection{Lanczos equations and surface stress-energy}\label{SS:Lanczos}

For the case of a thin-shell, the extrinsic curvature need not be continuous across $\Sigma$. For notational clarity, denote the discontinuity in $K_{ij}$ as
$\kappa_{ij}=K_{ij}^{+}-K_{ij}^{-}$. The Einstein equations, when applied to the hypersurface joining the bulk spacetimes, now yield the Lanczos equations:
\begin{equation}
S^{i}_{\;j}=-\frac{1}{8\pi}\,(\kappa ^{i}_{\;j}-\delta
^{i}_{\;j}\; \kappa ^{k}_{\;k}) \ ,
\end{equation}
where $S^{i}_{\;j}$ is the surface stress-energy tensor on the junction interface $\Sigma$. Due to spherical symmetry $\kappa ^{i}_{\;j}={\rm diag} \left(\kappa ^{\tau}_{\;\tau},\kappa ^{\theta}_{\;\theta},\kappa^{\theta}_{\;\theta}\right)$, and the surface stress-energy tensor reduces to $S^{i}_{\;j}={\rm diag}(-\sigma,{\cal P},{\cal P})$, where $\sigma$ is the surface energy density, and ${\cal P}$ the surface pressure. The Lanczos equations imply
\begin{eqnarray}
\sigma &=&-\frac{1}{4\pi}\,\kappa ^{\theta}_{\;\theta} \ ;\label{sigma} \\
{\cal P} &=&\frac{1}{8\pi}(\kappa ^{\tau}_{\;\tau}+\kappa
^{\theta}_{\;\theta}) \ . \label{surfacepressure}
\end{eqnarray}
Using the computed extrinsic curvatures Eq's.~(\ref{genKplustheta}) and~(\ref{genKminustautau}), one may now evaluate the surface stresses:
\clearpage
\begin{eqnarray}
\sigma &=& -\frac{1}{4\pi R}\Bigg[\sqrt{\left(1-\frac{2m_{+}}{R}\right)\left(1-\frac{\ell_{+}^2}{R^2}\right)+\dot{R}^{2}} \nonumber \\[2pt]
&& \qquad \qquad \qquad \qquad \qquad \qquad +
\sqrt{\left(1-\frac{2m_{-}}{R}\right)\left(1-\frac{\ell_{-}^2}{R^2}\right)+\dot{R}^{2}}\Bigg] \ , \nonumber \\
&&
\label{gen-surfenergy2}
\\
{\cal P} &=& \frac{1}{8\pi R}\Bigg[
\frac{1+\dot{R}^2\left( \frac{R^2-2\ell_+^2}{R^2-\ell_+^2} \right)+R\ddot{R}-\frac{m_+ R^2 + \ell_+^2 \left(R-m_+\right)}{R^3}}{\sqrt{\left(1-\frac{2m_{+}}{R}\right)\left(1-\frac{\ell_{+}^2}{R^2}\right)+\dot{R}^{2}}} \nonumber \\[2pt]
&& \qquad \qquad \qquad \qquad +
\frac{1+\dot{R}^2\left( \frac{R^2-2\ell_-^2}{R^2-\ell_-^2} \right)+R\ddot{R}-\frac{m_- R^2 + \ell_-^2 \left(R-m_-\right)}{R^3}}{\sqrt{\left(1-\frac{2m_{-}}{R}\right)\left(1-\frac{\ell_{-}^2}{R^2}\right)+\dot{R}^{2}}}\Bigg] \ . \nonumber \\
&&
\label{gen-surfpressure2}
\end{eqnarray}
The surface mass of the thin-shell is defined by $m_s=4\pi R^2\sigma$, a result which shall be used extensively below. Furthermore the surface energy density $\sigma$ is always negative, implying energy condition violations in this thin-shell context. For the specific symmetric case, $m_+=m_-$, and for vanishing bounce parameters $\ell_{\pm}=0$, the analysis reduces to that of reference~\cite{TSW:Poisson:1995}.

\subsection{Gauss and Codazzi equations}\label{SS:Gauss}

The Gauss equation is sometimes called the first contracted Gauss--Codazzi equation. In standard GR it is more often referred to as the ``Hamiltonian constraint''. The Gauss equation is a purely mathematical statement relating bulk curvature to extrinsic and intrinsic curvature at the boundary:
\begin{eqnarray}
G_{\mu \nu}\;n^{\mu}\,n^{\nu}=\frac{1}{2}\,(K^2-K_{ij}K^{ij}-\,^3R) \ .
    \label{1Gauss}
\end{eqnarray}
Applying the Einstein equations, and evaluating the discontinuity across the junction surface, this becomes
\begin{equation}
8\pi \left[T_{\mu \nu}n^{\mu}n^{\nu}\right]^{+}_{-}
 =
 {1\over2} [K^2-K_{ij}K^{ij}]^{+}_{-} \ ,
\end{equation}
where the convention $\left[X \right]^+_-\equiv X^+|_{\Sigma}-X^-|_{\Sigma}$ is employed. Using $\overline{X} \equiv \frac{1}{2}
(X^+|_{\Sigma}+X^-|_{\Sigma})$ for notational simplicity, and applying the Lanczos equations, one deduces the constraint equation
\begin{eqnarray}
 \left[T_{\mu \nu}n^{\mu}n^{\nu} \right]^{+}_{-} = S^{ij}\;\overline{K}_{ij} \ .
\end{eqnarray}
In contrast the Codazzi equation (Codazzi--Mainardi equation), is often known as the second contracted Gauss--Codazzi equation. In standard GR more often referred to as the ``ADM constraint'' or ``momentum constraint''. The purely mathematical result is 
\begin{eqnarray}
G_{\mu \nu}\,e^{\mu}_{(i)}n^{\nu}=K^j_{i|j}-K,_{i} \ .
    \label{2Gauss}
\end{eqnarray}
Together with the Einstein and Lanczos equations, and considering the discontinuity across the thin-shell, this now yields the conservation identity:
\begin{eqnarray}\label{conservation}
\left[T_{\mu \nu}\; e^{\mu}_{\;(j)}n^{\nu}\right]^+_- = - S^{i}_{\;j|i} \ .
\end{eqnarray}
The left-hand-side of the conservation identity Eq.~(\ref{conservation}) can be interpreted in terms of momentum flux. Explicitly
\begin{eqnarray}
\left[T_{\mu\nu}\; e^{\mu}_{\;(\tau)}\,n^{\nu}\right]^+_-
 &=& \left[T_{\mu\nu}\; U^{\mu}\,n^{\nu}\right]^+_- \nonumber \\
&=&\left[\pm
\left(T_{\hat{t}\hat{t}}+T_{\hat{r}\hat{r}}\right)
\,\frac{\dot{R}\sqrt{\left(1-\frac{2m}{R}\right)\left( 1-\frac{\ell^2}{R^2} \right)+\dot{R}^{2}}}{\left(1-\frac{2m}{R}\right)\left( 1-\frac{\ell^2}{R^2} \right)} \;
\right]^+_- \nonumber \\[2pt]
&=&\left[\mp
\frac{\ell^2}{4\pi R^4}
\,\frac{\dot{R}\sqrt{\left(1-\frac{2m}{R}\right)\left( 1-\frac{\ell^2}{R^2} \right)+\dot{R}^{2}}}{\left( 1-\frac{\ell^2}{R^2} \right)} \;
\right]^+_- \ , 
\label{flux}
\end{eqnarray}
where $T_{\hat{t}\hat{t}}$ and $T_{\hat{r}\hat{r}}$ are the bulk stress-energy tensor components given in an orthonormal basis. Note that the flux term corresponds to the net discontinuity in the bulk momentum flux $F_\mu=T_{\mu\nu}\,U^\nu$ which impinges on the shell.
For notational simplicity, write
\begin{equation}
\left[T_{\mu\nu}\; e^{\mu}_{\;(\tau)}\,n^{\nu}\right]^+_-
=\dot R\, \Xi \ .
\end{equation}
Here, the useful quantity $\Xi$ is defined as
\begin{eqnarray}
\Xi
&=& - \frac{1}{4\pi R^2}
\Bigg[\frac{\ell_+^2}{\left( R^2-\ell_+^2 \right)} \sqrt{\left(1-\frac{2m_+}{R}\right)\left( 1-\frac{\ell_+^2}{R^2} \right)+\dot{R}^{2}} \nonumber \\[2pt]
&& \qquad \qquad \qquad \qquad + 
\frac{\ell_-^2}{\left( R^2-\ell_-^2 \right)} \sqrt{\left(1-\frac{2m_-}{R}\right)\left( 1-\frac{\ell_-^2}{R^2} \right)+\dot{R}^{2}} \;
\Bigg] \ . \nonumber \\
&&
\end{eqnarray}
Now $A=4\pi R^2$ is the surface area of the thin-shell. 
The conservation identity becomes
\begin{equation}
\frac{\d\sigma}{\d\tau}+(\sigma+{\cal P})\,\frac{1}{A} \frac{\d A}{\d\tau} = \Xi \, \dot{R} \ .
\label{E:conservation2}
\end{equation}
Equivalently
\begin{equation}
\frac{\d(\sigma A)}{\d\tau}+{\cal P}\,\frac{\d A}{\d\tau}=\Xi \,A \, \dot{R} \ .
\label{E:conservation3}
\end{equation}
The term on the right-hand side incorporates the flux term and encodes the work done by external forces, while the first term on the left-hand side is simply the variation of the internal energy of the shell, and the second term characterises the work done by the shell's internal forces. Provided that the equations of motion can be integrated to determine the surface energy density as a function of radius $R$, one can infer the existence of a suitable function $\sigma(R)$. Defining $ \sigma'= \d\sigma /\d R$, the conservation equation can then be written as
\begin{equation}
\sigma'=-\frac{2}{R}\,(\sigma +{\cal P})+\Xi \ .
\label{cons-equation}
\end{equation}

\subsection{Equation of motion}\label{SS:eom}

To analyse the stability of the wormhole, Eq.~(\ref{gen-surfenergy2}) can be rearranged to provide the thin-shell equation of motion, given by
\begin{equation}
\frac{1}{2} \dot{R}^2+V(R)=0  \ .
\end{equation}
The potential $V(R)$ is defined as
\begin{equation}
V(R)= \frac{1}{2}\left\{ 1-{\bar \Delta(R)\over R} -\left[\frac{m_{s}(R)}{2R}\right]^2-\left[\frac{\Delta(R)}{m_{s}
(R)}\right]^2\right\} \ ,
   \label{potential}
\end{equation}
where $m_s(R)=4\pi R^2\,\sigma(R)$ is the mass of the thin-shell, and the quantities $\bar \Delta(R)$ and $\Delta(R)$ are defined as
\begin{eqnarray}
\bar \Delta(R) &=& \left(m_{+}+m_{-}\right) + \frac{1}{2R} \left[ \ell_+^2 \left( 1-\frac{2m_+}{R} \right) + \ell_-^2 \left( 1-\frac{2m_-}{R} \right)  \right] \ , \nonumber \\
&& \\
\Delta(R) &=& \left(m_{+} - m_{-}\right) + \frac{1}{2R} \left[ \ell_+^2 \left( 1-\frac{2m_+}{R} \right) - \ell_-^2 \left( 1-\frac{2m_-}{R} \right)  \right] \ , \nonumber \\
&&
\end{eqnarray}
respectively.
Note that by differentiating with respect to {$\tau$}, the equation of motion implies $\ddot R = -V'({R}) $, which will be useful below.

As outlined in reference~\cite{GSSDTTWISGR:Garcia:2012}, one can reverse the logic flow and determine the surface mass as a function of the potential. More specifically, if one imposes a specific potential $V({R})$, this potential implicitly tells one how much surface mass one needs to distribute on the wormhole throat. This further places implicit demands on the equation of state of the exotic matter residing on the wormhole throat. This implies that, after imposing the equation of motion for the shell, one has:

\medskip
\noindent
\textbf{Surface energy density:}
\begin{eqnarray}
\sigma &=& -\frac{1}{4\pi R}\Bigg[\sqrt{\left(1-\frac{2m_{+}}{R}\right)\left(1-\frac{\ell_{+}^2}{R^2}\right) -2V(R)} \nonumber \\[2pt]
&& \qquad \qquad \qquad \qquad +
\sqrt{\left(1-\frac{2m_{-}}{R}\right)\left(1-\frac{\ell_{-}^2}{R^2}\right)-2V(R)} \;
\Bigg] \ . \nonumber \\
&&
\label{gen-surfenergy2b}
\end{eqnarray}
\textbf{Surface pressure:}
\begin{eqnarray}
{\cal P} &=& \frac{1}{8\pi R}\left[
\frac{1-2V(R)\left( \frac{R^2-2\ell_+^2}{R^2-\ell_+^2} \right)-RV'(R)-\frac{m_+ R^2 + \ell_+^2 \left(R-m_+\right)}{R^3}}{\sqrt{\left(1-\frac{2m_{+}}{R}\right)
\left(1-\frac{\ell_{+}^2}{R^2}\right){-2V(R)}}}
	\right.
	\nonumber \\
&& \qquad  \qquad \left.
+
\frac{1-2V(R)\left( \frac{R^2-2\ell_-^2}{R^2-\ell_-^2} \right)-RV'(R)-\frac{m_- R^2 + \ell_-^2 \left(R-m_-\right)}{R^3}}{\sqrt{\left(1-\frac{2m_{-}}{R}\right)
\left(1-\frac{\ell_{-}^2}{R^2}\right){-2V(R)}}}
\right] \ . \nonumber \\
&&
\label{gen-surfpressure2b}
\end{eqnarray}
\textbf{External energy flux:}
\begin{eqnarray}
\Xi
&=& - \frac{1}{4\pi R^2}
\Bigg[\frac{\ell_+^2}{\left( R^2-\ell_+^2 \right)} \sqrt{\left(1-\frac{2m_+}{R}\right)\left( 1-\frac{\ell_+^2}{R^2} \right)
-2V(R)} \nonumber \\
&& \nonumber \\
&& \qquad \qquad \qquad + 
\frac{\ell_-^2}{\left( R^2-\ell_-^2 \right)} \sqrt{\left(1-\frac{2m_-}{R}\right)\left( 1-\frac{\ell_-^2}{R^2} \right)
-2V(R)} \;
\Bigg] \ . \nonumber \\
&&
\end{eqnarray}
These three quantities, $\{\sigma,\,{\cal P},\,\Xi\}$, are inter-related by the differential conservation law, so at most two of them are functionally independent. One could equivalently work with the quantities $\{m_s,\,{\cal P},\,\Xi\}$.

\subsection{Linearised equation of motion}\label{SS:linearized}

Now consider the equation of motion $\frac{1}{2}\dot R^2 + V(R)=0$, which implies $\ddot R = -V'(R)$, and linearise around an assumed static solution at $R_0$. This implies that a second-order Taylor expansion of $V(R)$ around $R_0$ provides
\begin{equation}
V(R)=V(R_0)+V'(R_0)(R-R_0)+\frac{1}{2}V''(R_0)(R-R_0)^2+\mathcal{O}[(R-R_0)^3]
\ .\label{linear-potential0}
\end{equation}
Since one is expanding around a static solution, $\dot R_0=\ddot R_0 = 0$, both $V(R_0)=V'(R_0)=0$, so that Eq.~(\ref{linear-potential0}) reduces to
\begin{equation}
V(R)= \frac{1}{2}V''(R_0)(R-R_0)^2 + \mathcal{O}[(R-R_0)^3]
\,.   \label{linear-potential}
\end{equation}
The static solution at $R_0$ is stable if and only if $V(R)$ has a local minimum at $R_0$. This requires $V''(R_{0})>0$. This stability condition will be the fundamental tool in the subsequent analysis --- though reformulation in terms of more basic quantities will prove useful. For instance, it is useful to express the quantities $m_s'(R)$ and $m_s''(R)$ in terms of the potential and its derivatives ---  doing so allows one to develop a simple inequality on
$m_s''(R_0)$ by using the constraint $V''(R_0)>0$.
Similar formulae will hold for the pairs $\sigma'(R)$, $\sigma''(R)$, for ${\cal P}'(R)$, ${\cal P}''(R)$, and for $\Xi'(R)$, $\Xi''(R)$. In view of the multiple redundancies coming from the relations $m_s(R) = 4\pi\sigma(R) R^2$ and the differential conservation law, one can easily see that the only interesting quantities are  $\Xi'(R)$, $\Xi''(R)$.

In the applications analysed below, it is extremely useful to  consider the dimensionless quantity
\begin{eqnarray}
{m_s(R)\over R} = 4\pi \sigma(R) R &=& - \Bigg[\sqrt{\left(1-\frac{2m_{+}}{R}\right)\left(1-\frac{\ell_{+}^2}{R^2}\right) -2V(R)} \nonumber \\[2pt]
&& \qquad \qquad +
\sqrt{\left(1-\frac{2m_{-}}{R}\right)\left(1-\frac{\ell_{-}^2}{R^2}\right)-2V(R)} \;
\Bigg] \ . \nonumber \\
&&
\end{eqnarray}
Now express $[m_s(R)/R]'$ and $[m_s(R)/R]''$ in terms of the following quantities:
\begin{eqnarray}
\left[m_s(R)\over R\right]' &=& - \Bigg\{
{ \frac{m_+}{R^2}\left(1-\frac{\ell_{+}^2}{R^2}\right) + \frac{\ell_+^2}{R^3}\left(1-\frac{2m_{+}}{R}\right)- V'(R)
\over
\sqrt{\left(1-\frac{2m_{+}}{R}\right)\left(1-\frac{\ell_{+}^2}{R^2}\right)-2V(R)}} \nonumber \\[2pt]
&& \qquad \qquad \quad
+
{ \frac{m_-}{R^2}\left(1-\frac{\ell_{-}^2}{R^2}\right) + \frac{\ell_-^2}{R^3}\left(1-\frac{2m_{-}}{R}\right)- V'(R)
\over
\sqrt{\left(1-\frac{2m_{-}}{R}\right)\left(1-\frac{\ell_{-}^2}{R^2}\right)-2V(R)}}  \Bigg\} \ , \nonumber \\
&&
\end{eqnarray}
and
\begin{eqnarray}
\left[\frac{m_s(R)}{R}\right]'' &=&
{\left[ \frac{m_+}{R^2}\left(1-\frac{\ell_{+}^2}{R^2}\right) + \frac{\ell_+^2}{R^3}\left(1-\frac{2m_{+}}{R}\right)- V'(R) \right]^2
\over
\left[\left(1-\frac{2m_{+}}{R}\right)\left(1-\frac{\ell_{+}^2}{R^2}\right)-2V(R)\right]^{3/2}} \nonumber \\[2pt]
&& \quad +
{ \frac{2m_+}{R^3}\left(1-\frac{\ell_{+}^2}{R^2}\right) + \frac{3\ell_+^2}{R^4}\left(1-\frac{2m_{+}}{R}\right) - \frac{4m_+ \ell_+^2}{R^5} + V''(R)
\over
\sqrt{\left(1-\frac{2m_{+}}{R}\right)\left(1-\frac{\ell_{+}^2}{R^2}\right)-2V(R)}} \nonumber \\
&& \nonumber \\
&& \qquad +
{\left[ \frac{m_-}{R^2}\left(1-\frac{\ell_{-}^2}{R^2}\right) + \frac{\ell_-^2}{R^3}\left(1-\frac{2m_{-}}{R}\right)- V'(R) \right]^2
\over
\left[\left(1-\frac{2m_{-}}{R}\right)\left(1-\frac{\ell_{-}^2}{R^2}\right)-2V(R)\right]^{3/2}} \nonumber \\[2pt]
&& \quad \quad +
{ \frac{2m_-}{R^3}\left(1-\frac{\ell_{-}^2}{R^2}\right) + \frac{3\ell_-^2}{R^4}\left(1-\frac{2m_{-}}{R}\right) - \frac{4m_{-} \ell_{-}^2}{R^5} + V''(R)
\over
\sqrt{\left(1-\frac{2m_{-}}{R}\right)\left(1-\frac{\ell_{-}^2}{R^2}\right)-2V(R)}} \ . \nonumber \\
&&
\end{eqnarray}
Similarly, consider the useful dimensionless quantity
\begin{eqnarray}
4\pi R^2\, \Xi &=& - 
\Bigg[\frac{\ell_+^2}{\left( R^2-\ell_+^2 \right)} \sqrt{\left(1-\frac{2m_+}{R}\right)\left( 1-\frac{\ell_+^2}{R^2} \right)
-2V(R)} \nonumber \\[2pt]
&& \qquad \qquad + 
\frac{\ell_-^2}{\left( R^2-\ell_-^2 \right)} \sqrt{\left(1-\frac{2m_-}{R}\right)\left( 1-\frac{\ell_-^2}{R^2} \right)
-2V(R)} \;
\Bigg] \ . \nonumber \\
&&
\end{eqnarray}
This leads to the following relations:
\begin{eqnarray}
\left[4\pi R^2\, \Xi \right]'
&=& \frac{\ell_+^2}{\left( R^2-\ell_+^2 \right)}
\Bigg[\frac{2R}{\left( R^2-\ell_+^2 \right)} \sqrt{\left(1-\frac{2m_+}{R}\right)\left( 1-\frac{\ell_+^2}{R^2} \right)
-2V(R)} \nonumber \\[2pt]
&& \qquad \qquad \qquad \quad -
{ \frac{m_+}{R^2}\left(1-\frac{\ell_{+}^2}{R^2}\right) + \frac{\ell_+^2}{R^3}\left(1-\frac{2m_{+}}{R}\right)- V'(R)
\over
\sqrt{\left(1-\frac{2m_{+}}{R}\right)\left(1-\frac{\ell_{+}^2}{R^2}\right)-2V(R)}}  \Bigg]
\nonumber \\
&& \nonumber \\
&& +\frac{\ell_-^2}{\left( R^2-\ell_-^2 \right)}
\Bigg[\frac{2R}{\left( R^2-\ell_-^2 \right)} \sqrt{\left(1-\frac{2m_-}{R}\right)\left( 1-\frac{\ell_-^2}{R^2} \right)
-2V(R)} \nonumber \\[2pt]
&& \qquad \qquad \qquad \quad -
{ \frac{m_-}{R^2}\left(1-\frac{\ell_{-}^2}{R^2}\right) + \frac{\ell_-^2}{R^3}\left(1-\frac{2m_{-}}{R}\right)- V'(R)
\over
\sqrt{\left(1-\frac{2m_{-}}{R}\right)\left(1-\frac{\ell_{-}^2}{R^2}\right)-2V(R)}}  \Bigg] \ , \nonumber \\
&&
\end{eqnarray}
and
\clearpage
\begin{eqnarray}
\left[4\pi R^2\, \Xi \right]''
&=& -\frac{{2}\ell_{+}^2 ({3}R^2+\ell_{+}^2)}{\left( R^2-\ell_{+}^2 \right)^3} \sqrt{\left(1-\frac{2m_+}{R}\right)\left( 1-\frac{\ell_+^2}{R^2} \right)-2V(R)} \nonumber \\[2pt]
&& +
\frac{4\ell_{+}^2 R}{\left( R^2-\ell_{+}^2 \right)^2} 
{ \frac{m_+}{R^2}\left(1-\frac{\ell_{+}^2}{R^2}\right) + \frac{\ell_+^2}{R^3}\left(1-\frac{2m_{+}}{R}\right)- V'(R)
\over
\sqrt{\left(1-\frac{2m_{+}}{R}\right)\left(1-\frac{\ell_{+}^2}{R^2}\right)-2V(R)}} \nonumber \\[2pt]
&& +
\frac{\ell_{+}^2 }{\left( R^2-\ell_{+}^2 \right)} 
{\left[ \frac{m_+}{R^2}\left(1-\frac{\ell_{+}^2}{R^2}\right) + \frac{\ell_+^2}{R^3}\left(1-\frac{2m_{+}}{R}\right)- V'(R)\right]^2
\over
\left[\left(1-\frac{2m_{+}}{R}\right)\left(1-\frac{\ell_{+}^2}{R^2}\right)-2V(R)\right]^{3/2}} \nonumber \\[2pt]
%
&& \hspace{-1cm}+
\frac{\ell_{+}^2 }{\left( R^2-\ell_{+}^2 \right)} 
{ \frac{2m_+}{R^3}\left(1-\frac{\ell_{+}^2}{R^2}\right) + \frac{3\ell_+^2}{R^4}\left(1-\frac{2m_{+}}{R}\right) - \frac{4m_+ \ell_+^2}{R^5} + V''({R})
\over
\sqrt{\left(1-\frac{2m_{+}}{R}\right)\left(1-\frac{\ell_{+}^2}{R^2}\right)-2V(R)}} \nonumber \\
&& \nonumber \\	
&& -\frac{{2}\ell_{-}^2 ({3}R^2+\ell_{-}^2)}{\left( R^2-\ell_{-}^2 \right)^3} \sqrt{\left(1-\frac{2m_-}{R}\right)\left( 1-\frac{\ell_-^2}{R^2} \right)-2V(R)} \nonumber \\[2pt]
&& +
\frac{4\ell_{-}^2 R}{\left( R^2-\ell_{-}^2 \right)^2} 
{ \frac{m_-}{R^2}\left(1-\frac{\ell_{-}^2}{R^2}\right) + \frac{\ell_-^2}{R^3}\left(1-\frac{2m_{-}}{R}\right)- V'(R)
\over
\sqrt{\left(1-\frac{2m_{-}}{R}\right)\left(1-\frac{\ell_{-}^2}{R^2}\right)-2V(R)}} \nonumber \\[2pt]
&& +
\frac{\ell_{-}^2 }{\left( R^2-\ell_{-}^2 \right)} 
{\left[ \frac{m_-}{R^2}\left(1-\frac{\ell_{-}^2}{R^2}\right) + \frac{\ell_-^2}{R^3}\left(1-\frac{2m_{-}}{R}\right)- V'(R)\right]^2
\over
\left[\left(1-\frac{2m_{-}}{R}\right)\left(1-\frac{\ell_{-}^2}{R^2}\right)-2V(R)\right]^{3/2}} \nonumber \\[2pt]
&& \hspace{-1cm}+
\frac{\ell_{-}^2 }{\left( R^2-\ell_{-}^2 \right)} 
{ \frac{2m_-}{R^3}\left(1-\frac{\ell_{-}^2}{R^2}\right) + \frac{3\ell_-^2}{R^4}\left(1-\frac{2m_{-}}{R}\right) - \frac{4m_- \ell_-^2}{R^5} + V''(R)
\over
\sqrt{\left(1-\frac{2m_{-}}{R}\right)\left(1-\frac{\ell_{-}^2}{R^2}\right)-2V(R)}} \ . \nonumber \\
&&
\end{eqnarray}

\subsection{Master equation}\label{SS:master}
\enlargethispage{40pt}

Taking into account the extensive discussion above, one sees that in order to have a stable static solution at ${R}_0$, two equations and one inequality must be satisfied. Specifically:
\begin{eqnarray}
{m_s(R_0)\over R_0} = 4\pi \sigma(R_0) R_0 &=& - \Bigg[\sqrt{\left(1-\frac{2m_{+}}{R_0}\right)\left(1-\frac{\ell_{+}^2}{R_0^2}\right) } \nonumber \\[2pt]
&& \qquad \qquad +
\sqrt{\left(1-\frac{2m_{-}}{R_0}\right)\left(1-\frac{\ell_{-}^2}{R_0^2}\right)} \;
\Bigg] \ , \nonumber \\
&&
\end{eqnarray}
and
\begin{eqnarray}
\left[\frac{m_s(R_0)}{R_0}\right]' &=& -
{ \frac{m_+}{R_0^2}\left(1-\frac{\ell_{+}^2}{R_0^2}\right) + \frac{\ell_+^2}{R_0^3}\left(1-\frac{2m_{+}}{R_0}\right)
\over
\sqrt{\left(1-\frac{2m_{+}}{R_0}\right)\left(1-\frac{\ell_{+}^2}{R_0^2}\right)}} \nonumber \\[2pt]
&& \qquad \qquad -
{ \frac{m_-}{R_0^2}\left(1-\frac{\ell_{-}^2}{R_0^2}\right) + \frac{\ell_-^2}{R_0^3}\left(1-\frac{2m_{-}}{R_0}\right)
\over
\sqrt{\left(1-\frac{2m_{-}}{R_0}\right)\left(1-\frac{\ell_{-}^2}{R_0^2}\right)}} \ ,
\end{eqnarray}
and
\begin{eqnarray}
\left[\frac{m_s(R_0)}{R_0}\right]'' &\geq&
{\left[ \frac{m_+}{R_0^2}\left(1-\frac{\ell_{+}^2}{R_0^2}\right) + \frac{\ell_+^2}{R_0^3}\left(1-\frac{2m_{+}}{R_0}\right) \right]^2
\over
\left[\left(1-\frac{2m_{+}}{R_0}\right)\left(1-\frac{\ell_{+}^2}{R_0^2}\right)\right]^{3/2}} \nonumber \\[2pt]
&& +
{ \frac{2m_+}{R_0^3}\left(1-\frac{\ell_{+}^2}{R_0^2}\right) + \frac{3\ell_+^2}{R_0^4}\left(1-\frac{2m_{+}}{R_0}\right) - \frac{4m_+ \ell_+^2}{R_0^5} 
\over
\sqrt{\left(1-\frac{2m_{+}}{R_0}\right)\left(1-\frac{\ell_{+}^2}{R_0^2}\right)}} \nonumber \\[2pt]
&& +
{\left[ \frac{m_-}{R_0^2}\left(1-\frac{\ell_{-}^2}{R_0^2}\right) + \frac{\ell_-^2}{R_0^3}\left(1-\frac{2m_{-}}{R_0}\right) \right]^2
\over
\left[\left(1-\frac{2m_{-}}{R_0}\right)\left(1-\frac{\ell_{-}^2}{R_0^2}\right)\right]^{3/2}} \nonumber \\[2pt]
&& +
{ \frac{2m_-}{R_0^3}\left(1-\frac{\ell_{-}^2}{R_0^2}\right) + \frac{3\ell_-^2}{R_0^4}\left(1-\frac{2m_{-}}{R_0}\right) - \frac{4m_{-} \ell_{-}^2}{R_0^5} 
\over
\sqrt{\left(1-\frac{2m_{-}}{R_0}\right)\left(1-\frac{\ell_{-}^2}{R_0^2}\right)}} \ . \nonumber \\
&&
\label{stability_constraint1}
\end{eqnarray}
More specifically, Eq.~(\ref{stability_constraint1}) translates the stability condition $V''(R_0) \geq 0$ into an explicit inequality on $m_s''({R}_0)$, an inequality that can in particular cases be explicitly checked. In the absence of external forces this inequality is the only stability condition one requires. 
However, once one has external forces (that is, in the presence of fluxes $\Xi \neq 0$),  there is additional information:
\begin{eqnarray}
4\pi R_0^2\, \Xi_0
&=& - 
\Bigg[\frac{\ell_+^2}{\left( R_0^2-\ell_+^2 \right)} \sqrt{\left(1-\frac{2m_+}{R_0}\right)\left( 1-\frac{\ell_+^2}{R_0^2} \right)} \nonumber \\[2pt]
&& \qquad \qquad + 
\frac{\ell_-^2}{\left( R_0^2-\ell_-^2 \right)} \sqrt{\left(1-\frac{2m_-}{R_0}\right)\left( 1-\frac{\ell_-^2}{R_0^2} \right)} \;
\Bigg] \ . \nonumber \\
&&
\end{eqnarray}
%
This leads one to consider the quantity
\clearpage
\enlargethispage{40pt}
\begin{eqnarray}
\left[4\pi R_0^2\, \Xi_0 \right]'
&=& \frac{\ell_+^2}{\left( R_0^2-\ell_+^2 \right)}
\Bigg[\frac{2R_0}{\left( R_0^2-\ell_+^2 \right)} \sqrt{\left(1-\frac{2m_+}{R_0}\right)\left( 1-\frac{\ell_+^2}{R_0^2} \right)} \nonumber \\[2pt]
&& \qquad \qquad \quad \quad -
{ \frac{m_+}{R_0^2}\left(1-\frac{\ell_{+}^2}{R_0^2}\right) + \frac{\ell_+^2}{R_0^3}\left(1-\frac{2m_{+}}{R_0}\right)
\over
\sqrt{\left(1-\frac{2m_{+}}{R_0}\right)\left(1-\frac{\ell_{+}^2}{R_0^2}\right)}}  \Bigg] \nonumber \\[2pt]
&& + \frac{\ell_-^2}{\left( R_0^2-\ell_-^2 \right)}
\Bigg[\frac{2R_0}{\left( R_0^2-\ell_-^2 \right)} \sqrt{\left(1-\frac{2m_-}{R_0}\right)\left( 1-\frac{\ell_-^2}{R_0^2} \right)} \nonumber \\[2pt]
&& \qquad \qquad \quad \quad -
{ \frac{m_-}{R_0^2}\left(1-\frac{\ell_{-}^2}{R_0^2}\right) + \frac{\ell_-^2}{R_0^3}\left(1-\frac{2m_{-}}{R_0}\right)
\over
\sqrt{\left(1-\frac{2m_{-}}{R_0}\right)\left(1-\frac{\ell_{-}^2}{R_0^2}\right)}}  \Bigg] \ .
\end{eqnarray}
Furthermore, since $R_{0} > \ell_{\pm}$, the inequality on $\left[4\pi R_0^2\, \Xi_0 \right]''$ is given by
\begin{eqnarray}
\left[4\pi R_0^2\, \Xi_0 \right]''
&\geq& 
-\frac{3\ell_{+}^2 (2R_0^2+\ell_{+}^2)}{\left( R_0^2-\ell_{+}^2 \right)^3} \sqrt{\left(1-\frac{2m_+}{R_0}\right)\left( 1-\frac{\ell_+^2}{R_0^2} \right)} \nonumber \\
&& +
\frac{4\ell_{+}^2 R_0}{\left( R_0^2-\ell_{+}^2 \right)^2} 
{ \frac{m_+}{R_0^2}\left(1-\frac{\ell_{+}^2}{R_0^2}\right) + \frac{\ell_+^2}{R_0^3}\left(1-\frac{2m_{+}}{R_0}\right)
\over
\sqrt{\left(1-\frac{2m_{+}}{R_0}\right)\left(1-\frac{\ell_{+}^2}{R_0^2}\right)}} \nonumber \\
&& +
\frac{\ell_{+}^2 }{\left( R_0^2-\ell_{+}^2 \right)} 
{\left[ \frac{m_+}{R_0^2}\left(1-\frac{\ell_{+}^2}{R_0^2}\right) + \frac{\ell_+^2}{R_0^3}\left(1-\frac{2m_{+}}{R_0}\right)\right]^2
\over
\left[\left(1-\frac{2m_{+}}{R_0}\right)\left(1-\frac{\ell_{+}^2}{R_0^2}\right)\right]^{3/2}} \nonumber \\
&& \hspace{-1.1cm} +
\frac{\ell_{+}^2 }{\left( R_0^2-\ell_{+}^2 \right)} 
{ \frac{2m_+}{R_0^3}\left(1-\frac{\ell_{+}^2}{R_0^2}\right) + \frac{3\ell_+^2}{R_0^4}\left(1-\frac{2m_{+}}{R_0}\right) - \frac{4m_+ \ell_+^2}{R_0^5}
\over
\sqrt{\left(1-\frac{2m_{+}}{R_0}\right)\left(1-\frac{\ell_{+}^2}{R_0^2}\right)}}
\nonumber  \\
&& - \frac{3\ell_{-}^2 (2R_0^2+\ell_{-}^2)}{\left( R_0^2-\ell_{-}^2 \right)^3} \sqrt{\left(1-\frac{2m_-}{R_0}\right)\left( 1-\frac{\ell_-^2}{R_0^2} \right)} \nonumber \\
&& +
\frac{4\ell_{-}^2 R_0}{\left( R_0^2-\ell_{-}^2 \right)^2} 
{ \frac{m_-}{R_0^2}\left(1-\frac{\ell_{-}^2}{R_0^2}\right) + \frac{\ell_-^2}{R_0^3}\left(1-\frac{2m_{-}}{R_0}\right)
\over
\sqrt{\left(1-\frac{2m_{-}}{R_0}\right)\left(1-\frac{\ell_{-}^2}{R_0^2}\right)}} \nonumber \\
&& +
\frac{\ell_{-}^2 }{\left( R_0^2-\ell_{-}^2 \right)} 
{\left[ \frac{m_-}{R_0^2}\left(1-\frac{\ell_{-}^2}{R_0^2}\right) + \frac{\ell_-^2}{R_0^3}\left(1-\frac{2m_{-}}{R_0}\right)\right]^2
\over
\left[\left(1-\frac{2m_{-}}{R_0}\right)\left(1-\frac{\ell_{-}^2}{R_0^2}\right)\right]^{3/2}} \nonumber \\
&& \hspace{-1.1cm} +
\frac{\ell_{-}^2 }{\left( R_0^2-\ell_{-}^2 \right)} 
{ \frac{2m_-}{R_0^3}\left(1-\frac{\ell_{-}^2}{R_0^2}\right) + \frac{3\ell_-^2}{R_0^4}\left(1-\frac{2m_{-}}{R_0}\right) - \frac{4m_- \ell_-^2}{R_0^5} 
\over
\sqrt{\left(1-\frac{2m_{-}}{R_0}\right)\left(1-\frac{\ell_{-}^2}{R_0^2}\right)}} \ .
	\label{stability_constraint2}
\end{eqnarray}
In summary, the inequalities Eq's.~(\ref{stability_constraint1}) and~(\ref{stability_constraint2}) dictate the stability regions of the wormhole solutions under consideration, and in the following section specific applications and examples are examined.

\section{Applications and Examples}\label{S:applications}

In this section, the general formalism described above shall be applied to some specific examples. Several of these special cases are particularly important in order to emphasise some of the specific features of SV spacetime. Some examples are essential to assess the simplifications due to symmetry between the two asymptotic regions, while other cases are useful to understand the asymmetry between the two universes used in traversable wormhole construction. In the following analysis, specific cases are considered by tuning the parameters of the bulk spacetimes, namely, the two bounce parameters $\ell_{\pm}$, and the two masses $m_{\pm}$.

\subsection{Vanishing flux term: $\ell_{\pm}=0$}\label{SS1a:specific}

\begin{figure}[!h]
\includegraphics[scale=0.38]{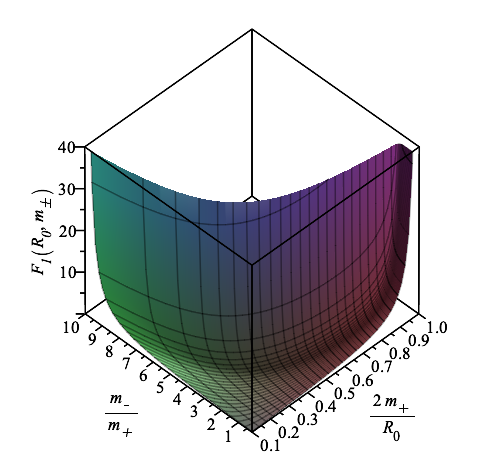}
\includegraphics[scale=0.38]{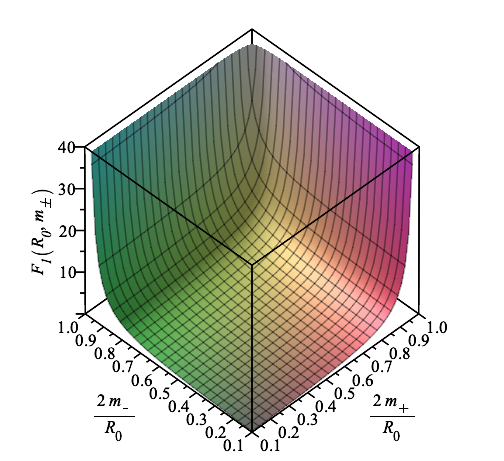}
\caption[Stability analysis for thin-shell Schwarzschild traversable wormholes]{Stability analysis for thin-shell Schwarzschild traversable wormholes, taking into account that $\ell_{\pm}=0$ and $m_{-} \neq  m_{+}$. The surfaces are given by the dimensionless quantity $F_{1}(R_0,m_\pm)$, defined by the right-hand side of inequality Eq.~(\ref{stabilitySchw}). The stability regions lie above the surfaces depicted in the plots. The range $0<x=2m_+/R_0 <1$ is considered, and $0< y_1=m_-/m_+ < 1/x$ in the left plot, and  $0<y_2=2m_-/R_0 <1$ in the right plot, respectively. Note that large stability regions exist for low values of $x=2m_+/R_{0}$ and of $y_{1,2}$. For regions close to the event horizon, $x \rightarrow 1$, the stability region decreases in size and only increases significantly for low values of $y_1$. See the text for details.}
\label{fig:stable1}
\end{figure}

Here, consider the case of a vanishing flux term, that is $\Xi =0$, which is induced by imposing $\ell_{\pm}=0$. Thus, the only stability constraint arises from inequality Eq.~(\ref{stability_constraint1}). Note that this case corresponds to the thin-shell Schwarzschild traversable wormholes analysed in references~\cite{TSW:Poisson:1995, GSSDTTWISGR:Garcia:2012}.
For the specific case of $\ell_{\pm}=0$, and considering an asymmetry in the masses $m_{-} \neq  m_{+}$, inequality Eq.~(\ref{stability_constraint1}) reduces to
\begin{equation}
R_0^2 \left[\frac{m_s(R_0)}{R_0}\right]''    \geq  F_{1}(R_0,m_{\pm}) = \frac{\frac{2m_{+}}{R_0} \left( 1-\frac{3m_{+}}{2R_0} \right)}{\left( 1- \frac{2m_{+}}{R_0} \right)^{3/2}} +
\frac{\frac{2m_{-}}{R_0} \left( 1-\frac{3m_{-}}{2R_0} \right)}{\left( 1- \frac{2m_{-}}{R_0} \right)^{3/2}} \ .
\label{stabilitySchw}
\end{equation}
In order to plot the stability regions, it is prudent to define the following dimensionless form of the constraint as $F_{1}(R_0,m_{\pm})= R_0^2 \left[m_s(R_0)/R_0\right]''  $, which is depicted as the surfaces given in the plots of Fig.~\ref{fig:stable1}. The stability regions lie above these surfaces. In order to visualise the whole range of the parameters, so as to bring infinite $R_0$ in to a finite region of the plot, the definition $x=2m_{+}/R_0$ is imposed for convenience. For instance, the limit $R_0 \rightarrow +\infty$ corresponds to $x\rightarrow 0$, and $R_0 = 2m_{+}$ is equivalent to $x=1$. Thus, the range $0<x <1$ is considered. 

In the left-hand plot of Fig.~\ref{fig:stable1}, the parameter $y_1=m_-/m_+$ is considered, which lies within the range $0< y_1 < 1/x$. This parameter provides information on the relative variation of the masses. However, one may also consider a more symmetrical form of the stability analysis, by considering the definition  $y_2=2m_{-}/R_0$, which possesses the range $0<y_2 <1$, and the stability region is depicted in the right-hand plot of Fig.~\ref{fig:stable1}. These two plots provide complementary information.

Regarding the stability of the solution, from Fig.~\ref{fig:stable1} it is verified that large stability regions exist for low values of $x=2m_+/R_0$ and of $y_1=m_-/m_+$ (and of $y_2=2m_-/R_0$). For regions close to the event horizon, $x \rightarrow 1$, the stability region decreases in size and only exists for low values of $y_{1,2}$. The specific case of $y_1=m_-/m_+=1$, corresponds to the thin-shell Schwarzschild wormholes analysed in reference~\cite{GSSDTTWISGR:Garcia:2012}, and one verifies that the size of the stability regions increases as the junction interface of the thin-shell increases. Namely, as $x=2m_+/R_0 \rightarrow 0$, as is transparent from Fig.~\ref{fig:stable1}.

\subsection{Vanishing mass: $\ell_{+} \neq \ell_{-}$ and $m_{\pm}=0$}\label{SS2a:symmetric}

Consider now the case of vanishing mass terms $m_{\pm}=0$, with an asymmetry of the bounce parameters $\ell_{+} \neq \ell_{-}$.

\begin{figure}[!h]
\begin{center}
\includegraphics[scale=0.50]{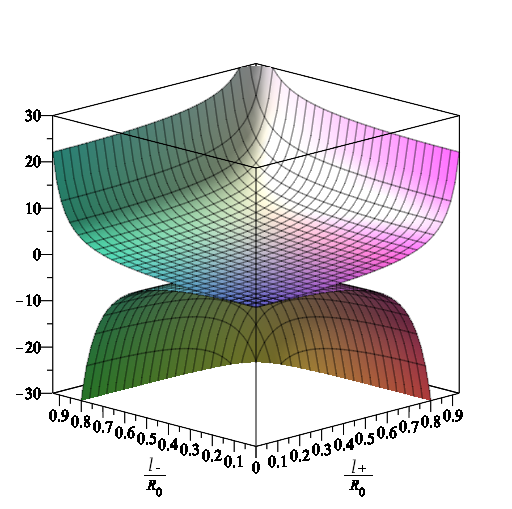}
\caption[Stability region for the case of vanishing masses and nonzero $\ell_+\neq\ell_-$]{The upper surface depicts the quantity $F_{2}(R_0,\ell_{\pm})= R_0^2 \left[m_s(R_0)/R_0\right]''  $, and the stable region lies above the surface of that curve. On the other hand, the function $G_{2}(R_0,\ell_{\pm})=R_0^2 \left[4\pi R_0^2\, \Xi_0 \right]''$ is depicted by the lower surface, and the stable region also lies above the surface of that curve. Thus the final stability region of the solution lies above the upper surface. 
See the text for more details.}
\label{fig:zeromass}
\end{center}
\end{figure}

For this case, inequality Eq.~(\ref{stability_constraint1}) reduces to
\begin{equation}
R_0^2 \left[\frac{m_s(R_0)}{R_0}\right]''    \geq  F_{2}(R_0,\ell_{\pm}) = 2 \left[ \frac{\frac{\ell_{+}^2}{R_0^2} \left( \frac{3}{2} -\frac{\ell_{+}^2}{R_0^2} \right)}{\left( 1- \frac{\ell_{+}^2}{R_0^2} \right)^{3/2}} +
\frac{\frac{\ell_{-}^2}{R_0^2} \left( \frac{3}{2} -\frac{\ell_{-}^2}{R_0^2} \right)}{\left( 1- \frac{\ell_{-}^2}{R_0^2} \right)^{3/2}}  \right] \ ,
\label{stabilityzeromass1}
\end{equation}
and inequality Eq.~(\ref{stability_constraint2}) takes the following form
\begin{eqnarray}
R_0^2 \left[4\pi R_0^2\, \Xi(R_0)\right]''    &\geq&  G_{2}(R_0,\ell_{\pm}) \nonumber \\[2pt]
&=& - \frac{\frac{\ell_{+}^2}{R_0^2} \left( 6 -4\frac{\ell_{+}^2}{R_0^2} + 2 \frac{\ell_{+}^4}{R_0^4} \right)}{\left( 1- \frac{\ell_{+}^2}{R_0^2} \right)^{5/2}} -
\frac{\frac{\ell_{-}^2}{R_0^2} \left( 6 - 4\frac{\ell_{-}^2}{R_0^2} + 2 \frac{\ell_{-}^4}{R_0^4} \right)}{\left( 1- \frac{\ell_{-}^2}{R_0^2} \right)^{5/2}} \ . \nonumber \\
&&
\label{stabilityzeromass2}
\end{eqnarray}
Now, consider the definition of the parameters $x=\ell_+/R_0$ and $y=\ell_-/R_0$ for convenience, so as to bring infinite $R_0$ within a finite region of the plot. That is, $R_0 \rightarrow +\infty$ is represented as $x\rightarrow 0$; and $R_0 = \ell_{+}, \ell_{-}$ is equivalent to $x, y=1$. Thus, the parameters $x$ and $y$ are restricted to the ranges $0<x<1$ and $0<y<1$.
Inequality Eq.~(\ref{stabilityzeromass1}) is depicted as the upper surface in Fig.~\ref{fig:zeromass}, and inequality Eq.~(\ref{stabilityzeromass2}) is depicted as the lower surface, and the stability regions are given above the respective surfaces. Thus, the final stability region lies above the upper surface.

\subsection{Asymmetric vanishment: $\ell_{+}=0$ and $m_{-}=0$}\label{SS2:asymmetric}

\begin{figure}[!h]
\begin{center}
\includegraphics[scale=0.50]{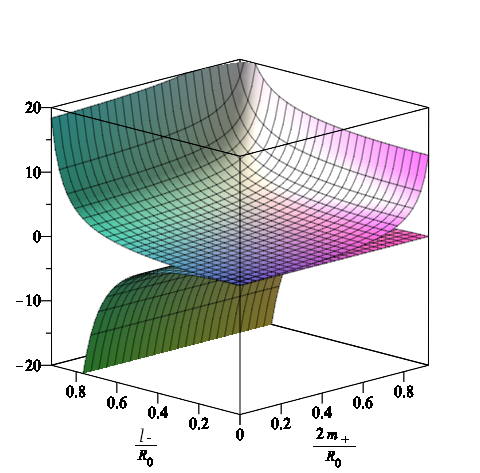}
\caption[Stability region for the case of asymmetric vanishing of parameters: $\ell_+=0$ and $m_-=0$]{The upper surface depicts the quantity $F_{3}(R_0,m_{+},\ell_{-})= R_0^2 \left[m_s(R_0)/R_0\right]''$, and the stable region lies above that surface. On the other hand, the function $G_{3}(R_0,\ell_{-})=R_0^2 \left[4\pi R_0^2\, \Xi_0 \right]''$ is depicted by the lower surface, and the stable region also lies above the surface of that plot. Thus, the final stability region of the solution lies above the upper surface. See the text for more details.}
\label{fig_zerom_zeroA.png}
\end{center}
\end{figure}

Consider now the case of vanishing interior mass $m_{-}=0$, and vanishing exterior parameter $\ell_{+}=0$. For this case, the inequality Eq.~(\ref{stability_constraint1}) reduces to
\begin{equation}
R_0^2 \left[\frac{m_s(R_0)}{R_0}\right]''    \geq  F_{3}(R_0,m_{+},\ell_{-}) =  \left[ \frac{\frac{2m_{+}}{R_0} \left( 1 -\frac{3m_{+}}{2R_0} \right)}{\left( 1- \frac{2m_{+}}{R_0} \right)^{3/2}} 
+
\frac{\frac{3\ell_{-}^2}{R_0^2} \left( 1 -\frac{2\ell_{-}^2}{3R_0^2} \right)}{\left( 1- \frac{\ell_{-}^2}{R_0^2} \right)^{3/2}}  \right] \ ,
\label{stabilityzerom_zeroA1}
\end{equation}
and inequality Eq.~(\ref{stability_constraint2}) is given by
\begin{equation}
R_0^2 \left[4\pi R_0^2\, \Xi(R_0)\right]''    \geq  G_{3}(R_0,\ell_{-}) = -
\frac{2\frac{\ell_{-}^2}{R_0^2} \left( 3 - 2\frac{\ell_{-}^2}{R_0^2} +  \frac{\ell_{-}^4}{R_0^4} \right)}{\left( 1- \frac{\ell_{-}^2}{R_0^2} \right)^{5/2}}   \ .
\label{stabilityzerom_zeroA2}
\end{equation}
These are depicted as the upper and lower surfaces, respectively, in Fig.~\ref{fig_zerom_zeroA.png}. As in the previous example, the final stability region of the solution lies above the upper surface.

\clearpage

\subsection{Mirror symmetry: $\ell_{\pm}=\ell$ and $m_{\pm}=m$}\label{SS3a:symmetric}

Consider, for simplicity, the symmetric case, \textit{i.e.}, $\ell_{\pm}=\ell$ and $m_{\pm}=m$, so that the stability conditions reduce to:
\begin{eqnarray}
\frac{1}{2}\,R_0^2 \left[\frac{m_s(R_0)}{R_0}\right]'' 
&\geq&
{\left[ \frac{m}{R_0}\left(1-\frac{\ell^2}{R_0^2}\right) + \frac{\ell^2}{R_0^2}\left(1-\frac{2m}{R_0}\right) \right]^2
\over
\left[\left(1-\frac{2m}{R_0}\right)\left(1-\frac{\ell^2}{R_0^2}\right)\right]^{3/2}} \nonumber \\[2pt]
&& \qquad \qquad +
{ \frac{2m}{R_0}\left(1-\frac{\ell^2}{R_0^2}\right) + \frac{3\ell^2}{R_0^2}\left(1-\frac{2m}{R_0}\right) - \frac{4m \ell^2}{R_0^3} 
\over
\sqrt{\left(1-\frac{2m}{R_0}\right)\left(1-\frac{\ell^2}{R_0^2}\right)}} \ , \nonumber \\
&&
   \label{stablesymmass}
\end{eqnarray}
and
\begin{eqnarray}
R_0^2 \left[4\pi R_0^2\, \Xi_0 \right]''
&\geq& 
\frac{2\ell^2 }{\left( R_0^2-\ell^2 \right)} 
\Bigg\{
-\frac{3 R_0^2(2R_0^2+\ell^2)}{\left( R_0^2-\ell^2 \right)^2} \sqrt{\left(1-\frac{2m}{R_0}\right)\left( 1-\frac{\ell^2}{R_0^2} \right)} \nonumber \\[2pt]
&& \qquad \qquad \qquad +
\frac{4 R_0^2}{\left( R_0^2-\ell^2 \right)} 
{ \frac{m}{R_0}\left(1-\frac{\ell^2}{R_0^2}\right) + \frac{\ell^2}{R_0^2}\left(1-\frac{2m}{R_0}\right)
\over
\sqrt{\left(1-\frac{2m}{R_0}\right)\left(1-\frac{\ell^2}{R_0^2}\right)}} \nonumber \\[2pt]
&& \qquad \qquad \qquad \quad +
{ \frac{2m}{R_0}\left(1-\frac{\ell^2}{R_0^2}\right) + \frac{3\ell^2}{R_0^2}\left(1-\frac{2m}{R_0}\right) - \frac{4m \ell^2}{R_0^3}
\over
\sqrt{\left(1-\frac{2m}{R_0}\right)\left(1-\frac{\ell^2}{R_0^2}\right)}} \nonumber \\[2pt]
&& \qquad \qquad \qquad \qquad +
{\left[ \frac{m}{R_0}\left(1-\frac{\ell^2}{R_0^2}\right) + \frac{\ell^2}{R_0^2}\left(1-\frac{2m}{R_0}\right)\right]^2
\over
\left[\left(1-\frac{2m}{R_0}\right)\left(1-\frac{\ell^2}{R_0^2}\right)\right]^{3/2}} \Bigg\} \ , \nonumber \\
&&
\label{stablesymXi}
\end{eqnarray}
respectively.

\begin{figure}[!h]
\begin{center}
\includegraphics[scale=0.50]{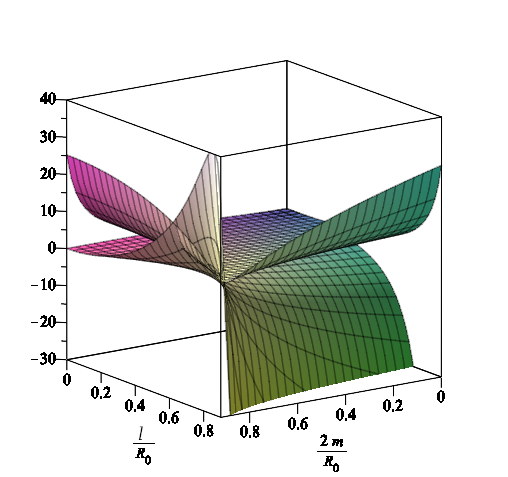}
\caption[Stability region for the case of mirror symmetry: $\ell_{\pm}=\ell$ and $m_{\pm}=m$]{The upper surface depicts the quantity $F_{4}(R_0,m,\ell)= R_0^2 \left[m_s(R_0)/R_0\right]''$, and the stable region lies above that surface. On the other hand, the function $G_{4}(R_0,m,\ell)=R_0^2 \left[4\pi R_0^2\, \Xi_0 \right]''$ is depicted by the lower surface, and the stable region also lies above that surface. The final stability region of the solution lies above the intersection of both surfaces. Note that inequality Eq.~(\ref{stablesymXi}) serves to decrease the stability region for high values of $x$ and $y$. See the text for more details.}
\label{fig:stable2}
\end{center}
\end{figure}

It is useful to express Eq.~(\ref{stablesymmass}) as the dimensionless $F_{4}(R_0,m,\ell) =\newline R_0^2 \left[m_s(R_0)/R_0\right]''$, and Eq.~(\ref{stablesymXi}) as $G_{4}(R_0,m,\ell)= R_0^2 \left[4\pi R_0^2\, \Xi_0 \right]''$. Both surfaces are depicted in Fig.~\ref{fig:stable2}, and the final stability region is situated above the intersection of the surfaces.

Note the definition $x=2m/R_0$ for convenience, so as to bring infinite $R_0$ within a finite region of the plot. That is, $R_0 \rightarrow +\infty$ is represented as $x\rightarrow 0$; and $R_0 = 2m$ is equivalent to $x=1$. Thus, the parameter $x$ is restricted to the range $0<x<1$. Furthermore, defining the parameter $y=\ell/R_0$, this also lies in the range $0<y<1$. It is interesting to note that the inequality Eq.~(\ref{stablesymXi}) serves to decrease the stability region for high values of $x$ and $y$, as is transparent from Fig.~\ref{fig:stable2}.

\vspace*{22pt}

\subsection{Two specific asymmetric cases}\label{SS3b:asymmetric}

\subsubsection{$\ell_{+} \neq \ell_-$ and $m_{+}=m_-$}\label{SS3b:asymmetric1}

\begin{figure}[!h]
\includegraphics[scale=0.4]{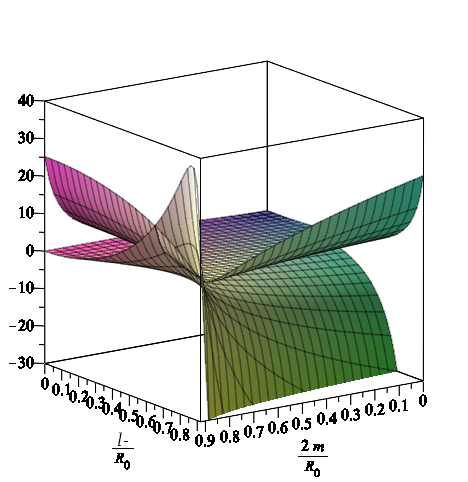}
\includegraphics[scale=0.4]{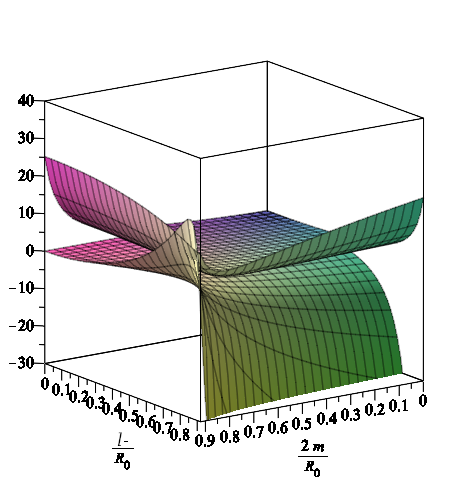}
\caption[Stability regions for the asymmetric case $\ell_+ = \alpha\ell_-$ and $m_+=m_-$, for $\alpha<1$]{Specific case of $\ell_{+} = \alpha\, \ell_-$ and $m_{+}=m_-$ for $\alpha<1$: The upper surface depicts the quantity $F_{5}(R_0,m,\ell_{\pm})= R_0^2 \left[m_s(R_0)/R_0\right]''$, and the function $G_{5}(R_0,m,\ell_{\pm})=R_0^2 \left[4\pi R_0^2\, \Xi_0 \right]''$ is depicted by the lower surface. The stable regions are given above the surfaces. $\alpha=0.9$ is considered in the left plot, and $\alpha=0.4$ is considered in the right plot. Note that decreasing the value of $\alpha$ serves to decrease the lower surface representing inequality Eq.~(\ref{stability_constraint2}), and thus increase the final stability region. See the text for more details.}
\label{fig:stable5a}
\end{figure}

\begin{figure}[!h]
\includegraphics[scale=0.35]{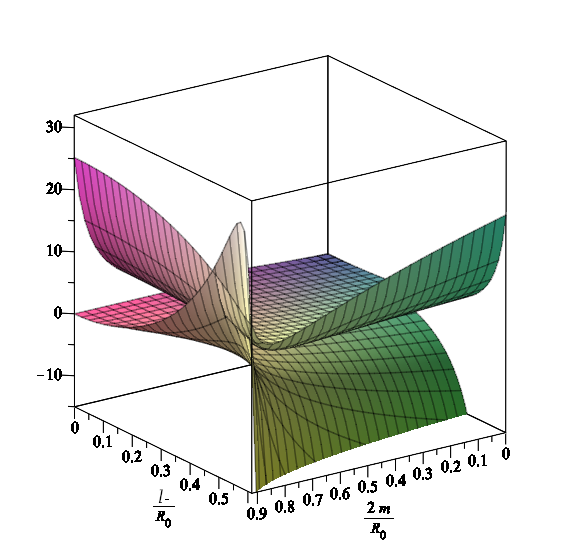}
\includegraphics[scale=0.35]{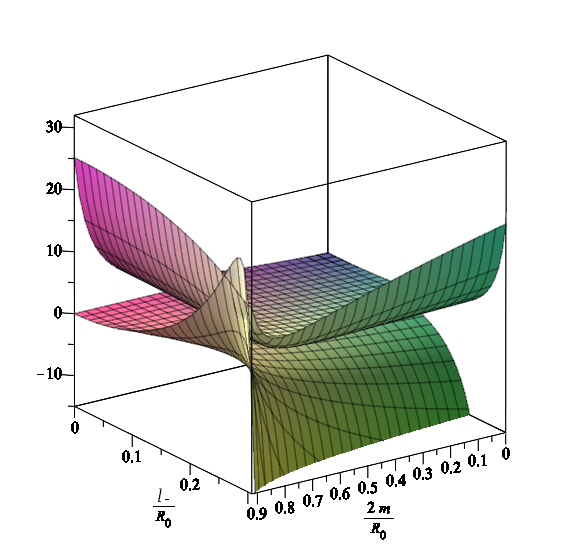}
\caption[Stability regions for the asymmetric case $\ell_+ = \alpha\ell_-$ and $m_+=m_-$, for $\alpha>1$]{Specific case of $\ell_{+} = \alpha\, \ell_-$ and $m_{+}=m_-$ for $\alpha>1$: The upper surface and lower surfaces depict the quantities $F_{5}(R_0,m,\ell_{\pm})= R_0^2 \left[m_s(R_0)/R_0\right]''$ and $G_{5}(R_0,m,\ell_{\pm})=R_0^2 \left[4\pi R_0^2\, \Xi_0 \right]''$, respectively. The stable regions are given above the surfaces. We have considered $\alpha=1.5$ in the left plot and $\alpha=3$ in the right plot. Note that increasing the value of $\alpha$, serves to decrease the lower surface representing inequality Eq.~(\ref{stablesymXi}), and thus increase the final stability region. However, the range for $y$ decreases for increasing values of $\alpha$ (for $\alpha=1.5$, the range is $0<y<2/3$ and for $\alpha=3$, it is $0<y<1/3$). See the text for more details.}
\label{fig:stable5b}
\end{figure}

Consider the case of symmetric masses $m_{\pm}=m$, but with asymmetric bounce parameters $\ell_{+} \neq \ell_-$. In order to analyse the stability regions, one can define $\ell_{+} = \alpha\, \ell_-$, where $\alpha \in \mathbb{R}^{+}$. Rather than write down the explicit form of the inequalities which are rather lengthy and messy, the dimensionless form of inequalities Eq.~(\ref{stability_constraint1}) and Eq.~(\ref{stability_constraint2}) are presented as the upper and lower surfaces in the plots of Fig.~\ref{fig:stable5a} and Fig.~\ref{fig:stable5b}. Note that the specific case of $\alpha=1$ reduces to the analysis of the mirror symmetry considered in the previous example \S~\ref{SS3a:symmetric}. The dimensionless parameters $x=2m/R_0$ and $y=\ell_{-}/R_0$ are employed in order to analyse the stability regions. In the following the two cases $\alpha<1$ and $\alpha>1$ are separated.
\clearpage
\enlargethispage{30pt}

\begin{itemize}
\item  
{\bf Specific case of $\alpha<1$}: The upper surface and lower surfaces in Fig.~\ref{fig:stable5a} depict the functions $F_{5}(R_0,m,\ell_{\pm})= R_0^2 \left[m_s(R_0)/R_0\right]''$ and $G_{5}(R_0,m,\ell_{\pm})=R_0^2 \left[4\pi R_0^2\, \Xi_0 \right]''$, respectively, and the stability regions representing the inequalities Eq.~(\ref{stability_constraint1}) and Eq.~(\ref{stability_constraint2}) are the regions above these surfaces. The range of the dimensionless parameters is $0< x,y <1$. $\alpha=0.9$ is considered in the left plot, and $\alpha=0.4$ is considered in the right plot of Fig.~\ref{fig:stable5a}. As the final stability region of the solution lies above the intersection of both surfaces, decreasing the value of $\alpha$ qualitatively serves to decrease the lower surface representing inequality Eq.~(\ref{stability_constraint2}), and thus increase the final stability region. This is transparent for high values of $x$ and $y$.
\item 
{\bf Specific case of $\alpha>1$}: The analysis is analogous to the above case and is depicted in Fig.~\ref{fig:stable5b}, however, here the left plot is given by $\alpha=1.5$ and the right plot by $\alpha=3$. The range of the dimensionless parameters is given by $0< x <1$ and $0< y <1/\alpha$. As in the previous example, the final stability region of the solution lies above the intersection of both surfaces depicted in Fig.~\ref{fig:stable5b}. Note that increasing the value of $\alpha$, serves to decrease the lower surface representing inequality Eq.~(\ref{stability_constraint2}), and thus increase the final stability region. However, the range for $y$ decreases for increasing values of $\alpha$ (for $\alpha=1.5$, the range is $0<y<2/3$, and for $\alpha=3$, it is $0<y<1/3$). This analysis is transparent for high values of $x$ and $y$ in Fig.~\ref{fig:stable5b}.
\end{itemize}

\clearpage

\enlargethispage{10pt}

\subsubsection{$\ell_{+} = \ell_-$ and $m_{+} \neq m_-$}\label{SS3b:asymmetric2}

\begin{figure}[!h]
\includegraphics[scale=0.34]{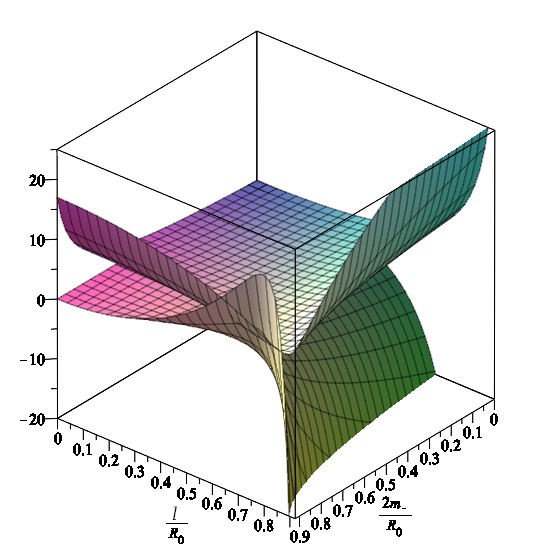}
\includegraphics[scale=0.34]{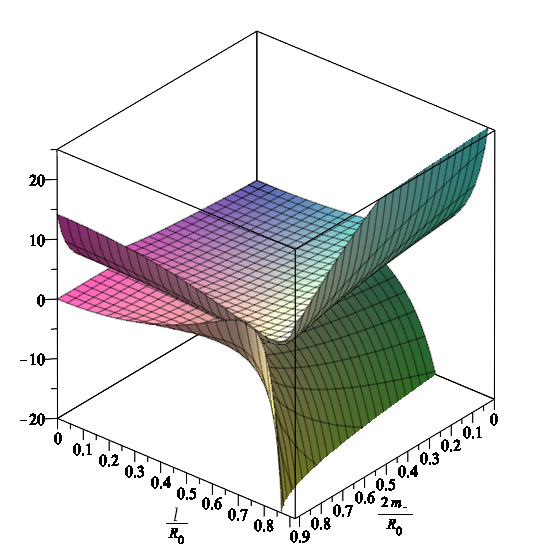}
\caption[Stability regions for the asymmetric case $\ell_+ = \ell_-$ and $m_+ = \alpha m_-$, for $\alpha<1$]{Specific case of $\ell_{+} = \ell_-$ and $m_{+} = \alpha m_-$ for $\alpha<1$. The dimensionless parameters $x=2m_{-}/R_0$, and $y=\ell/R_0$ are considered. $\alpha=0.9$ is considered in the left plot, and $\alpha=0.7$ in the right plot. Note that decreasing the value of $\alpha$, serves to decrease the lower surface representing inequality Eq.~(\ref{stability_constraint2}), and thus increase the final stability region.}
\label{fig:stable7a}
\end{figure}

\begin{figure}[!h]
\includegraphics[scale=0.34]{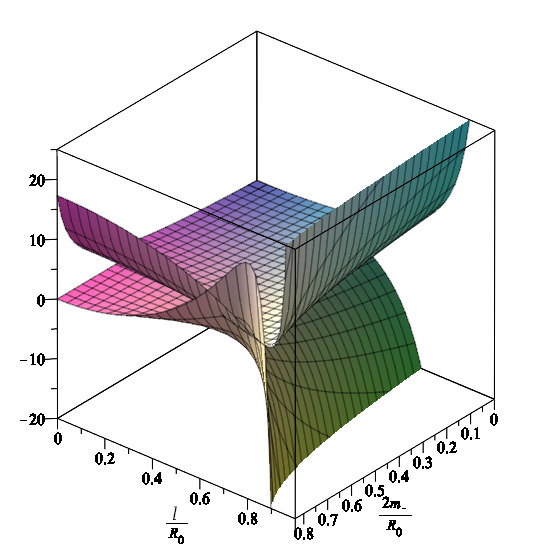}
\includegraphics[scale=0.34]{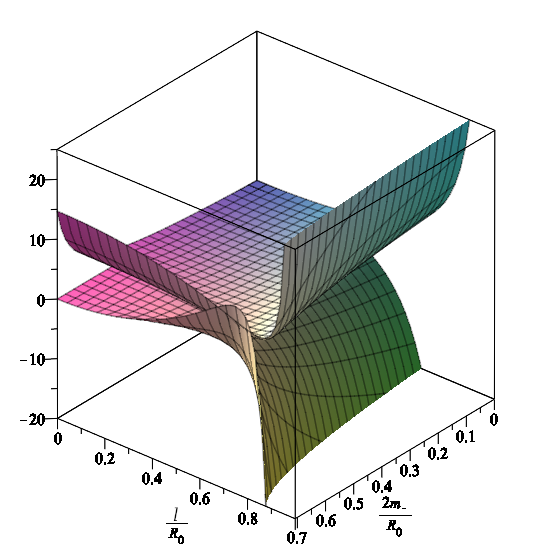}
\caption[Stability regions for the asymmetric case $\ell_+ = \ell_-$ and $m_+ = \alpha m_-$, for $\alpha>1$]{Specific case of $\ell_{+} = \ell_-$ and $m_{+} = \alpha m_-$ for $\alpha>1$. The dimensionless parameters $x=2m{-}/R_0$, and $y=\ell/R_0$ are considered. $\alpha=1.1$ is considered in the left plot, and $\alpha=1.3$ in the right plot. Note that increasing the value of $\alpha$, serves to decrease the lower surface representing inequality Eq.~(\ref{stability_constraint2}), and thus increase the final stability region.}
\label{fig:stable7b}
\end{figure}

Here the case of asymmetric masses $m_{\pm} \neq m$ is considered, but with symmetric bounce parameters $\ell_{\pm} = \ell$, and for simplicity define $m_{+} = \alpha\, m_-$, where $\alpha \in \mathbb{R}^{+}$. As in the specific case previously given above, the dimensionless form of inequalities Eq.~(\ref{stability_constraint1}) and Eq.~(\ref{stability_constraint2}) are presented as the upper and lower surfaces in the plots of Fig.~\ref{fig:stable7a} and Fig.~\ref{fig:stable7b}. The dimensionless parameters $x=2m_{-}/R_0$ and $y=\ell/R_0$ are employed in order to analyse the stability regions, and as in the previous example the cases $\alpha<1$ and $\alpha>1$ are separated.

\begin{itemize}
\item  {\bf Specific case of $\alpha<1$}:\newline The functions $F_{6}(R_0,m,\ell_{\pm})= R_0^2 \left[m_s(R_0)/R_0\right]''$ and $G_{6}(R_0,m,\ell_{\pm})=R_0^2 \left[4\pi R_0^2\, \Xi_0 \right]''$ are depicted by the upper and lower surfaces in Fig.~\ref{fig:stable7a}, respectively, and the stability regions representing the inequalities Eq.~(\ref{stability_constraint1}) and Eq.~(\ref{stability_constraint2}) are given above these surfaces. The range of the dimensionless parameters is $0< x,y <1$. $\alpha=0.9$ is considered in the left plot, and $\alpha=0.7$ is considered in the right plot of Fig.~\ref{fig:stable7a}. The final stability region of the solution lies above the intersection of both surfaces. Note that decreasing the value of $\alpha$, serves to decrease the lower surface representing inequality Eq.~(\ref{stability_constraint2}), and thus increase the final stability region.
\item {\bf Specific case of $\alpha>1$}: Analogously to the above case, the stability regions are depicted in Fig.~\ref{fig:stable7b}. The left plot is given by $\alpha=1.1$ and the right plot by $\alpha=1.3$. The range of the dimensionless parameters is given by $0< x <1$ and $0< y <1/\alpha$. As in the previous example, the final stability region of the solution lies above the intersection of both surfaces depicted in Fig.~\ref{fig:stable7b}. Note that increasing the value of $\alpha$, serves to decrease the lower surface representing inequality Eq.~(\ref{stability_constraint2}), and thus increase the final stability region. However, the range for $y$ decreases for increasing values of $\alpha$.
\end{itemize}
\enlargethispage{60pt}

\vfill
\clearpage

\subsubsection{Summary of thin-shell SV construction}\label{S:conclusion}

Thin-shell wormholes based on the recently introduced SV spacetime have been considered. Specifically, by matching two copies of SV spacetime using the cut-and-paste procedure, the stability and evolution of dynamic thin-shell SV wormholes has been rigorously analysed. The parameter space of various models depending on the bulk masses $m_\pm$, and the values of the bulk bounce parameters $\ell_\pm$, has been explored, investigating the internal dynamics of the thin-shell connecting the two bulk spacetimes, and demonstrating the existence of suitable stability regions in the parameter space. Several of these models are particularly useful in order to emphasise the specific features of SV spacetime.
Indeed, the interesting physics is encoded in the parameter $\ell_{\pm}$ which characterises the scale of the bounce, and emphasises the spacetime features. In fact, the presence of the parameter $\ell_{\pm}$ induces the flux term in the conservation identity, which is responsible for the net discontinuity in the bulk momentum flux which impinges on the shell. Thus, due to this flux term, the bounce parameter places an additional constraint on the stability analysis of the spacetime geometry.

From the linearised stability analysis one may assess and understand the (a)symmetry between the two universes in the traversable wormhole construction. For instance, consider the simple case of vanishing mass $m_{\pm}=0$ and $\ell_{\pm} \neq 0$ analysed in \S~\ref{SS2a:symmetric}, where large stability regions exist for $\ell_{\pm} \ll R_0$. However, the lower stability surface for the specific case of $\ell_+ \sim R_0$ decreases significantly for the asymmetric case of decreasing the value of $\ell_-$. Another interesting example is the asymmetric case analysed in \S~\ref{SS2:asymmetric}, where $\ell_+=0$ and $m_-=0$ were considered. Here, the lower stability surfaces for $\ell_{-} \sim R_0$ and $m_{+} \sim R_0$ may be significantly decreased by decreasing the asymmetric parameters, until large stability regions are present for $\ell_- \ll R_0$ and $m_+ \ll R_0$. 
The mirror symmetric case of $\ell_{\pm}=\ell$ and $m_{\pm}=m$, considered in \S~\ref{SS3a:symmetric}, is of particular interest. Despite the fact that large stability regions exist for $\ell \ll R_0$ and $m \ll R_0$, the region at $\ell \approx R_0$ and $m \approx R_0$ is notable. Here the flux term constraint kicks in and decreases the size of the stability region significantly. In fact, analysing this specific region for the asymmetric case is particularly interesting, as one may explore the asymmetry between the universes in the wormhole construction by considering the case of $\ell_+ \neq \ell_-$ in \S~\ref{SS3b:asymmetric}. More specifically, by varying the relative values of $\ell_+$ and $\ell_-$, the analysis showed that one could increase or decrease the size of the stability regions, and one may assess and understand the (a)symmetry between the two universes in the traversable wormhole construction. The constructions considered are sufficiently novel to be interesting, and sufficiently straightforward to be tractable.
\vfill
\clearpage

\section{aMc thin-shell construction}\label{aMcthinshell}

A qualitatively similar construction of a thin-shell traversable wormhole, using the aMc spacetime of Chapter~\ref{C:AMCog} in the bulk manifolds as opposed to SV spacetime, was analysed rigorously in reference\cite{TTWCFARBHWAMC:Berry:2020}. The wormhole construction was analysed similarly \emph{via} the thin-shell formalism, allowing the
four-velocity of the wormhole throat to be calculated along with the junction surface unit normal vectors, the extrinsic curvature, and the junction surface stress-energy. A key difference with the aMc thin-shell construction is that the flux term, $\Xi$, is manifestly zero. This leads to the following set of master equations constraining the stability of the model under linearised radial perturbations of the wormhole throat:
\begin{eqnarray}\label{stabilityineq0}
    \frac{m_{s}(R_{0})}{R_{0}} &=& -\left[\sqrt{1-\frac{2m_{+}\e^{-\ell_{+}/R_{0}}}{R_{0}}} + \sqrt{1-\frac{2m_{-}\e^{-\ell_{-}/R_{0}}}{R_{0}}}\right] \,,
    \end{eqnarray}
    \begin{eqnarray}
    \left[\frac{m_{s}(R_{0})}{R_{0}}\right]' &=& -\left[\frac{m_{+}\e^{-\ell_{+}/R_{0}}(R_{0}-\ell_{+})}{R_{0}^{3}\sqrt{1-\frac{2m_{+}\e^{-\ell_{+}/R_{0}}}{R_{0}}}} + \frac{m_{-}\e^{-\ell_{-}/R_{0}}(R_{0}-\ell_{-})}{R_{0}^{3}\sqrt{1-\frac{2m_{-}\e^{-\ell_{-}/R_{0}}}{R_{0}}}}\right] \,, \nonumber \\
    &&
    \end{eqnarray}
     \begin{eqnarray}
    \left[\frac{m_{s}(R_{0})}{R_{0}}\right]'' &\geq& \frac{\left[\frac{m_{+}\e^{-\ell_{+}/R_{0}}}{R_{0}^{2}}\left(1-\frac{\ell_{+}}{R_{0}}\right)\right]^{2}}{\left[1-\frac{2m_{+}\e^{-\ell_{+}/R_{0}}}{R_{0}}\right]^{\frac{3}{2}}} - \frac{\frac{m_{+}\ell_{+}\e^{-\ell_{+}/R_{0}}}{R_{0}^{4}}\left(4-\frac{\ell_{+}}{R_{0}}\right)}{\sqrt{1-\frac{2m_{+}\e^{-\ell_{+}/R_{0}}}{R_{0}}}} \nonumber \\[2pt]
    && + \frac{\left[\frac{m_{-}\e^{-\ell_{-}/R_{0}}}{R_{0}^{2}}\left(1-\frac{\ell_{-}}{R_{0}}\right)\right]^{2}}{\left[1-\frac{2m_{-}\e^{-\ell_{-}/R_{0}}}{R_{0}}\right]^{\frac{3}{2}}} - \frac{\frac{m_{-}\ell_{-}\e^{-\ell_{-}/R_{0}}}{R_{0}^{4}}\left(4-\frac{\ell_{-}}{R_{0}}\right)}{\sqrt{1-\frac{2m_{-}\e^{-\ell_{-}/R_{0}}}{R_{0}}}} \ , \nonumber \\
    &&
    \label{stabilityineq}
\end{eqnarray}
with the final inequality giving the stability regions for various cases of the parameters $m_{\pm}$ and $\ell_{\pm}$.

For suitable choices of the parameters, the wormhole under consideration violates the NEC in the bulk spacetime, however the SEC can be globally satisfied. The surface energy at the wormhole throat is negative; much like other traversable wormholes, exotic matter is required to keep the wormhole throat open. This class of wormholes permits a clean and quite general stability
analysis, with wide swathes of stable behaviour. Furthermore the stability plateau exhibits a ``pit'' of enhanced stability when the wormhole throat is close to where a near-extremal horizon would have existed in the bulk spacetime. Finally, the quantity of exotic matter required to support the wormhole throat can be minimised (and in some cases made arbitrarily small), through suitable choices of the parameters. For more details, please see reference~\cite{TTWCFARBHWAMC:Berry:2020}.


%% file: 04-Dyson.tex
\chapter{\hbox{Thin-shell Dyson mega-spheres}}\label{C:DMS}

%
%
%
%
Loosely inspired by the somewhat fanciful notion of detecting an arbitrarily advanced alien civilization, in Chapter~\ref{C:DMS} a general-relativistic thin-shell Dyson mega-sphere completely enclosing a central star-like object is considered, \emph{via} a full general-relativistic treatment using the Israel--Lanczos--Sen junction conditions~\cite{SHATSIGR:Israel:1966, TSIGRAC:Barrabes:1991, FVDMIDEG:Lanczos:1924, Unpublished:Lanczos:1922, UDGDSAU:Sen:1924}. Attention is placed on the surface mass density, the surface stress, the classical energy conditions, and the forces between hemispheres. It is demonstrated that in the physically acceptable region, the NEC, WEC, and SEC are always satisfied, while the DEC can be violated if the Dyson mega-sphere is sufficiently close to forming a black hole. 
It is also demonstrated that the original version of the maximum force conjecture, 
$F \leq {1\over 4} F_{Stoney}= {1\over 4} F_{Planck}$, can easily be violated if the Dyson mega-sphere is sufficiently compact, that is, sufficiently close to forming a black hole. Interestingly there is a finite region of parameter space where one can violate the original version of the maximum force conjecture \textit{without} violating the DEC. Finally, the possibility of nested thin-shell mega-spheres (Matrioshka configurations), and thick-shell 
Dyson mega-spheres, are briefly discussed.

%
Somewhat strangely, the vast majority of currently available analyses along these lines use Newtonian  gravity, however herein the analysis uses the thin-shell formalism of Chapter~\ref{C:SVthinshell}.\footnote{Much more rarely one may sometimes encounter analyses using nonstandard theories of modified  gravity~\cite{GHODSVCAHB:Kaloper:2011}.} For now, in the interests of tractability, attention is restricted to investigating systems with spherical symmetry.
%
%
For calculations, unless explicitly stated, geometrodynamic units are adopted where $G_N\to1$ and $c\to 1$; see for instance~\cite{G:Misner:2000, GR:Wald:2010}. Sometimes key results are presented in physical SI units; this is made explicit as and when required.

\section{Spacetime metric ansatz}\label{ansatz}
%

To set the stage, let the central star have mass $m$, and take the mega-sphere to have a radius $a$ larger than the radius of the star; in fact for almost all purposes one may safely idealise the radius of the central star to be zero. Let the total mass of the (star)+(mega-sphere) system be $M$. Then, outside the mega-sphere, the spacetime geometry can be described by
\begin{equation}
\d s^2 = -\left(1-{2M\over r}\right) \d t^2 + {\d r^2\over 1-2M/r} + r^2 \;\d\Omega^2_2 \ ; 
\qquad (r>a) \ ,
\end{equation}
while inside the mega-sphere
\begin{equation}
\d s^2 = -\xi^2 \left(1-{2m\over r}\right) \d t^2 + {\d r^2\over 1-2m/r} + r^2 \;\d\Omega^2_2 \ ;
\qquad (r<a) \ .
\end{equation}
%
Here, to maintain continuity of the $t$ coordinate as one crosses the mega-sphere at $r=a$, it is useful to set
\begin{equation}
\xi^2 = {1-2M/a\over 1-2m/a} \ .
\end{equation}
Physically one must demand $ 0 \leq m \leq M < a/2$. First one demands $m < a/2$ and $M < a/2$ to prevent horizon formation from enclosing the mega-sphere. Then one demands $M\geq m$ so that the total mass is not smaller than the mass of the central object. Finally $m\geq 0$ so that the central object does not have negative mass. Putting this all together, one has $ 0 \leq m \leq M < a/2$, so that $0\leq \xi^2 \leq 1$.

It is furthermore useful to define a proper distance radial coordinate, normalised so that $\ell=0$ on the shell at $r=a$:
\begin{equation}
\ell(r) = \left\{ \begin{array}{l l}
\displaystyle{+\int_a^r {\d r \over \sqrt{1-2M/r}} }& \qquad (r>a) \ ;\\[20pt]
\displaystyle{-\int_r^a {\d r \over \sqrt{1-2m/r}}} & \qquad (r<a) \ .
\end{array}
\right.
\end{equation}
Then $n_\mu = \nabla_\mu \ell$ is by construction a unit outward-pointing spacelike vector, which can be used to define the projection operator
\begin{equation}
h_{\mu\nu} = g_{\mu\nu} - n_\mu n_\nu \ .
\end{equation}
%
Finally, formally inverting $\ell(r)$ to implicitly extract $r(\ell)$, one can re-express the spacetime geometry as
\begin{equation}
\d s^2 = -\left(1-{2M\over r(\ell)}\right) \d t^2 + {\d\ell^2} 
+ r(\ell)^2 \;\d\Omega^2_2 \ ;
\qquad (r>a) \ ,
\end{equation}
\begin{equation}
\d s^2 = -\xi^2 \left(1-{2m\over r(\ell)}\right) \d t^2 + {\d\ell^2} 
+ r(\ell)^2 \;\d\Omega^2_2 \ ;
\qquad (r<a) \ .
\end{equation}

\section{Israel--Lanczos--Sen junction conditions}

It is now straightforward to calculate the extrinsic curvature tensors:
\begin{equation}
(K^\pm) _{\mu\nu} = 
{1\over2} \left. {\partial g_{\mu\nu}\over\partial \ell} 
\right|_{\ell= 0^\pm}
=
{1\over2} \left. {\partial r \over\partial\ell} \; {\partial g_{\mu\nu}\over\partial r } 
\right|_{\ell= 0^\pm}
=
{1\over2} \left. {1 \over\sqrt{g_{rr}}} \; {\partial g_{\mu\nu}\over\partial r }
\right|_{\ell= 0^\pm} \ ,
\end{equation}
 just above and below the thin shell at $r=a$.

Then in the coordinate basis the extrinsic curvatures are:
\begin{equation}
(K^+)_{tt} = -\sqrt{1-2M/a} \; {M\over a^2} \ , \qquad 
(K^+)_{\theta \theta} = {(K^+)_{\phi \phi} \over \sin^2\theta} 
= \sqrt{1-2M/a} \; a \ ,
\end{equation}
and
\begin{equation}
(K^-)_{tt} =  -\xi^2 \sqrt{1-2m/a}\; {m\over a^2} \ , \qquad 
(K^-)_{\theta \theta} = {(K^-)_{\phi\phi}\over \sin^2\theta} 
= \sqrt{1-2m/a}\;  a \ .
\end{equation}
Rewriting this in the natural orthonormal basis one has:
\begin{equation}
(K^+)_{\hat t \hat t} = - {M/a^2\over\sqrt{1-2M/a}} \ , \qquad 
(K^+)_{\hat \theta \hat \theta} = (K^+)_{\hat \phi \hat \phi} 
= {\sqrt{1-2M/a}\over a} \ ,
\end{equation}
and
\begin{equation}
(K^-)_{\hat t \hat t} = - {m/a^2\over\sqrt{1-2m/a}} \ , \qquad 
(K^-)_{\hat \theta \hat \theta} = (K^-)_{\hat \phi \hat \phi} 
= {\sqrt{1-2m/a}\over a} \ .
\end{equation}
Furthermore, the (distributional) stress-energy on the shell is then easily obtained in terms of the discontinuity $[K]_{\mu\nu}$ of the extrinsic curvatures ~\cite{SHATSIGR:Israel:1966, TSIGRAC:Barrabes:1991, FVDMIDEG:Lanczos:1924, Unpublished:Lanczos:1922, UDGDSAU:Sen:1924, LW:Visser:1995, TWSSE:Visser:1989, TWFSMSS:Visser:1989, TSW:Poisson:1995, ANATTWAA:Lobo:2015}:
\begin{equation}
[K]_{\mu\nu} = (K_+)_{\mu\nu} - (K_-)_{\mu\nu} \ .
\end{equation}
Explicitly, working in the natural orthonormal basis, $[K]_{\hat \mu\hat \nu}$ is given by the following matrix:
\begin{equation}
\left[ \begin{array}{cc|cc}
 -{M/a^2\over\sqrt{1-2M/a}} +  {m/a^2\over\sqrt{1-2m/a}} &0\;&0&0\\[10pt]
0&0&0&0\\[1pt]
\hline
0&0&{\vphantom{\Bigg|}{\sqrt{1-2M/a}\over a}}-{\sqrt{1-2m/a}\over a}&0\\
0&0&0&{\sqrt{1-2M/a}\over a}-{\sqrt{1-2m/a}\over a}
\end{array}
\right] \ .
\end{equation}
%
For the discontinuity in the trace of the extrinsic curvature, $K= g^{\mu\nu} K_{\mu\nu}=
 g^{\hat \mu\hat \nu} K_{\hat \mu\hat \nu}\,$, one has
\begin{equation}
[K] = + {M/a^2\over\sqrt{1-2M/a}} -  {m/a^2\over\sqrt{1-2m/a}} +
2\left( {\sqrt{1-2M/a}\over a}-{\sqrt{1-2m/a}\over a} \right) \ ,
\end{equation}
which can be re-written as
\begin{equation}
[K] = +{1\over a} \left\{ {2-3M/a\over \sqrt{1-2M/a}} -  {2-3m/a\over \sqrt{1-2m/a}} \right\} \ .
\end{equation}
%
By invoking the 
Israel--Lanczos--Sen junction conditions~\cite{SHATSIGR:Israel:1966,TSIGRAC:Barrabes:1991,FVDMIDEG:Lanczos:1924,Unpublished:Lanczos:1922, UDGDSAU:Sen:1924, LW:Visser:1995, TWSSE:Visser:1989, TWFSMSS:Visser:1989, TSW:Poisson:1995, ANATTWAA:Lobo:2015}, the distributional stress-energy tensor is easily found:
\begin{equation}
T_{\mu\nu} = -{1\over 8\pi} \left( [K]_{\mu\nu}  -  [K] h_{\mu\nu} \right) \, \delta(\ell) 
= S_{\mu\nu}\, \delta(\eta) \ ,
\end{equation}
or better yet, its orthonormal form:
\begin{equation}
T_{\hat \mu\hat \nu} = -{1\over8\pi} \left( [K]_{\hat \mu\hat \nu}  -  [K] h_{\hat \mu\hat \nu} \right) \, \delta(\ell) = S_{\hat \mu\hat \nu}\, \delta(\eta) \ .
\end{equation}
Here the surface energy density $\sigma$, and surface stress $\Pi$ (this is the \textit{negative} of what is usually called the surface tension $\vartheta$, that is, $\Pi=-\vartheta$), are defined by
\begin{equation}
S_{\hat \mu\hat \nu}  = \left[ \begin{array}{cc|cc}
 \sigma &0\;&0&0\\
0&0&0&0\\
\hline
0&0&\Pi&0\\
0&0&0&\Pi
\end{array}
\right] \ .
\end{equation}
Thence
\begin{equation}
\sigma = {1\over 4\pi \, a } \; \left( \sqrt{1-2m/a}-\sqrt{1-2M/a}\right) \ ,
\end{equation}
and
\begin{equation}
\Pi = {1\over 8\pi \, a } \left( {1-M/a\over\sqrt{1-2M/a}} -
{1-m/a\over\sqrt{1-2m/a}} 
\right) \ .
\end{equation}
Now, as long as $M \geq m$, it is immediate that the surface energy density is positive; $\sigma \geq 0$.
To bound the surface stress $\Pi$, note that
\begin{equation}
{\partial\Pi\over\partial M} = {M\over 8\pi \, a^3 (1-2M/a)^{3/2}} >0 \ .
\end{equation}
Thence, as long as one satisfies the physically motivated constraint that $M \geq m$, (the total mass of the system is greater than the mass of the central star), it is immediate that $\Pi \geq 0$. So in the physically relevant regime, where $a > 2M \geq 2m\geq 0$, all the surface stress-energy components are guaranteed non-negative.

Somewhat similar calculations are relevant to thin-shell gravastar models~\cite{SG:Visser:2004, GTG:Garcia:2012, NSAOTG:Garcia:2018, LSAOGTS:Garcia:2015}, though in that context many of the technical details, and all of the physics goals, are very different.

\section{Classical energy conditions}
\subsubsection{Definitions}

Recall that for any type I stress energy tensor~\cite{TLSSOS:Ellis:1973} (which is certainly and explicitly the case in the current spherically symmetric situation), the standard classical energy conditions are based on considering various linear combinations of the energy density $\rho$ and the principal pressures $p_i$.
%
\begin{description}
\item[NEC:]  $\rho+p_i\geq 0$.
\item[WEC:] $\rho\geq 0$ and $\rho+p_i\geq 0$.
\item[SEC:]  $\rho+ \sum_i p_i \geq 0$ and $\rho+p_i\geq 0$.
\item[DEC:] $|p_i| \leq \rho$.
\end{description}
Thence, in this specific thin-shell configuration, with principal stresses of the form $\mathrm{diag}\{\sigma,0,\Pi,\Pi\}$, these standard classical energy conditions will specialise to:
\begin{description}
\item[NEC:]  $\sigma\geq 0$ and $\sigma+\Pi \geq 0$.
\item[WEC:] Equivalent to NEC.
\item[SEC:] NEC and $\sigma+2\Pi \geq 0$.
\item[DEC:] NEC and $\sigma-\Pi \geq 0$. (That is, $|\Pi| \leq \sigma$.) 
\end{description}
It is pertinent to analyse these energy conditions in detail.

\subsection{Null, weak, and strong energy conditions}

In the physically relevant regime, where $M \geq m$,  both $\sigma$ and $\Pi$  are guaranteed non-negative, so all three of the NEC, WEC, and SEC are automatically satisfied. 

\subsection{Dominant energy condition}

To investigate the DEC it is useful to define the dimensionless quantity
\begin{equation}
w = {\Pi\over\sigma} \ ,
\end{equation}
and ask whether or not $w\leq1$. (It is already known that $w\geq 0$.)

Explicitly, to investigate the DEC one needs to consider the dimensionless parameter
\begin{equation}
w = { 1\over 2}{ 
\left( {1-M/a\over\sqrt{1-2M/a}} -{1-m/a\over\sqrt{1-2m/a}} \right)
\over
\left( \sqrt{1-2m/a}-\sqrt{1-2M/a}\right)
} \ .
\end{equation}
%
This quantity is more easily analysed in terms of two dimensionless ``compactness'' parameters $\Chi = 2M/a$ and $\chi=2m/a$, where now the physically acceptable range corresponds to $0 \leq \chi \leq \Chi \leq 1$. Then,
\begin{equation}
w = { 1\over 2}{ 
\left( {1-\Chi/2\over\sqrt{1-\Chi}} -{1-\chi/2\over\sqrt{1-\chi}} \right)
\over
\left( \sqrt{1-\chi}-\sqrt{1-\Chi}\right)
}
= {1\over4} \left\{{1\over\sqrt{1-\chi}\sqrt{1-\Chi} } -1 \right\} \ .
\end{equation}
Saturation of the DEC, the condition $w=1$, then easily translates to
\begin{equation}
(1-\Chi)(1-\chi) = {1\over25} \ .
\end{equation}
Less symmetrically, but perhaps more usefully, this can be rewritten as
\begin{equation}
\chi = 1 - {1\over25(1-\Chi)} \ , \quad\quad
\Chi = 1 - {1\over25(1-\chi)} \ .
\end{equation}
For a thin-shell Dyson mega-sphere on the verge of violating the DEC one has
\begin{equation}
\sigma = \Pi = {1\over 4\pi \, a } \; 
\left( \sqrt{1-\chi}-\sqrt{1-\Chi}\right)=
{1\over 4\pi \, a } \; 
\left( {1\over5\sqrt{1-\Chi}} -\sqrt{1-\Chi}\right) \ .
\end{equation}
To enforce $\sigma\geq 0$ while saturating the DEC one needs $\Chi\geq 4/5$, (and $\Chi<1$). 

Furthermore, the DEC-violating region is easily seen to be
\begin{equation}
1 - {1\over25(1-\Chi)} < \chi<\Chi \leq 1 \ .
\end{equation}
That is,
\begin{equation}
{{24\over25}-\Chi\over1-\Chi} < \chi<\Chi \leq 1 \ .
\end{equation}
Thus violations of the DEC are relatively easy to arrange (at least theoretically; the implied construction materials would be somewhat \textit{outr\'e}).

It should be re-emphasised that violations of the classical energy conditions are not absolute prohibitions on ``interesting physics'',  see~\cite{TFTEC:Barcelo:2002, APOEC:Curiel:2017, ECIGRAQFT:Kontou:2020}, and specifically~\cite{GWDVTNEC:Santiago:2022, TBPBASBIGR:Santiago:2021, TBPBASBWTCOGR:Visser:2021}, but they certainly are invitations to think very carefully about the underlying physics.

\begin{figure}[!htbp]
\begin{center}
\includegraphics[scale=0.5]{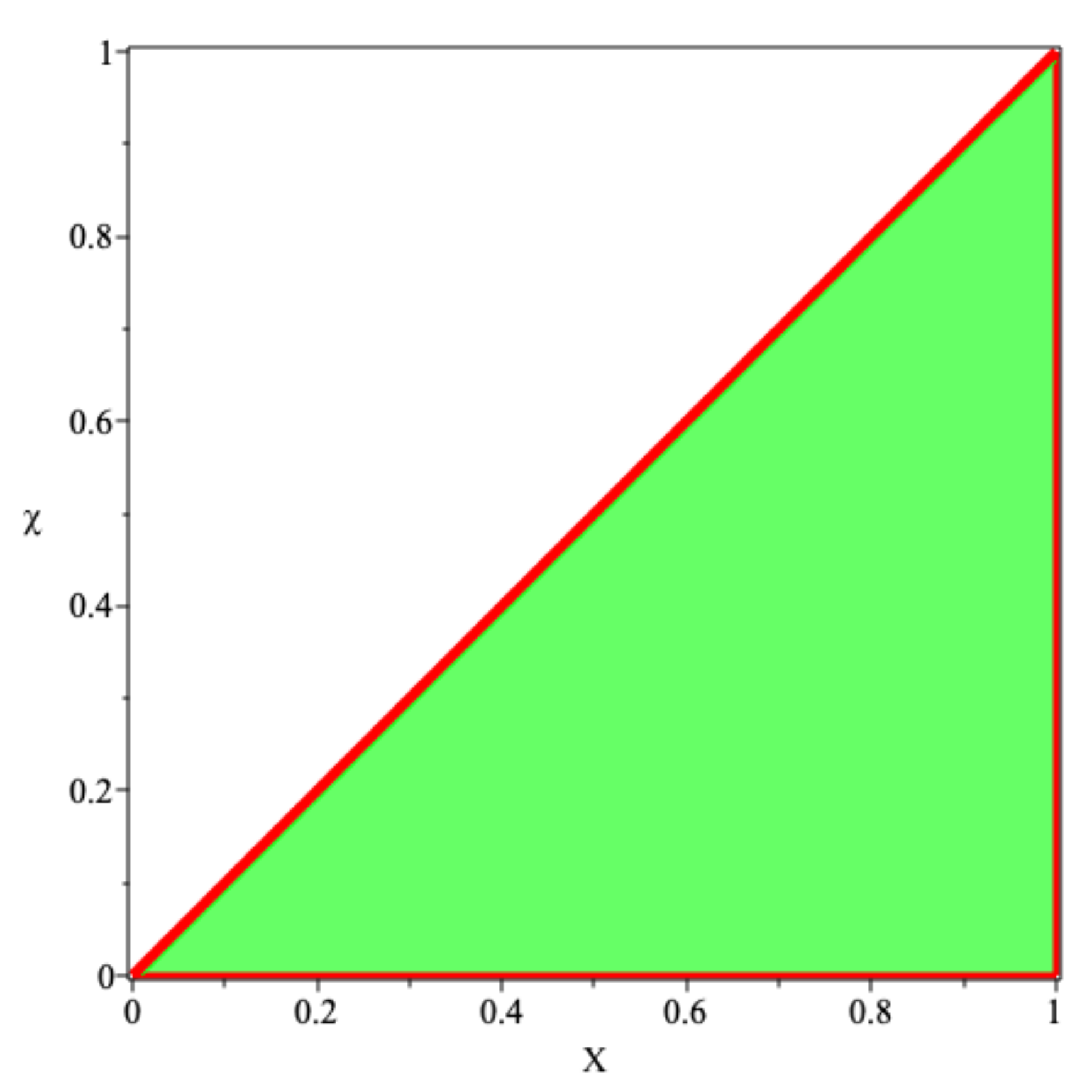}
\caption{Boundary and interior of the physically acceptable region $0\leq\chi\leq\Chi<1$.}
\label{F-DEC-0}
\end{center}
\end{figure}

\begin{figure}[htbp]
\begin{center}
\includegraphics[scale=0.5]{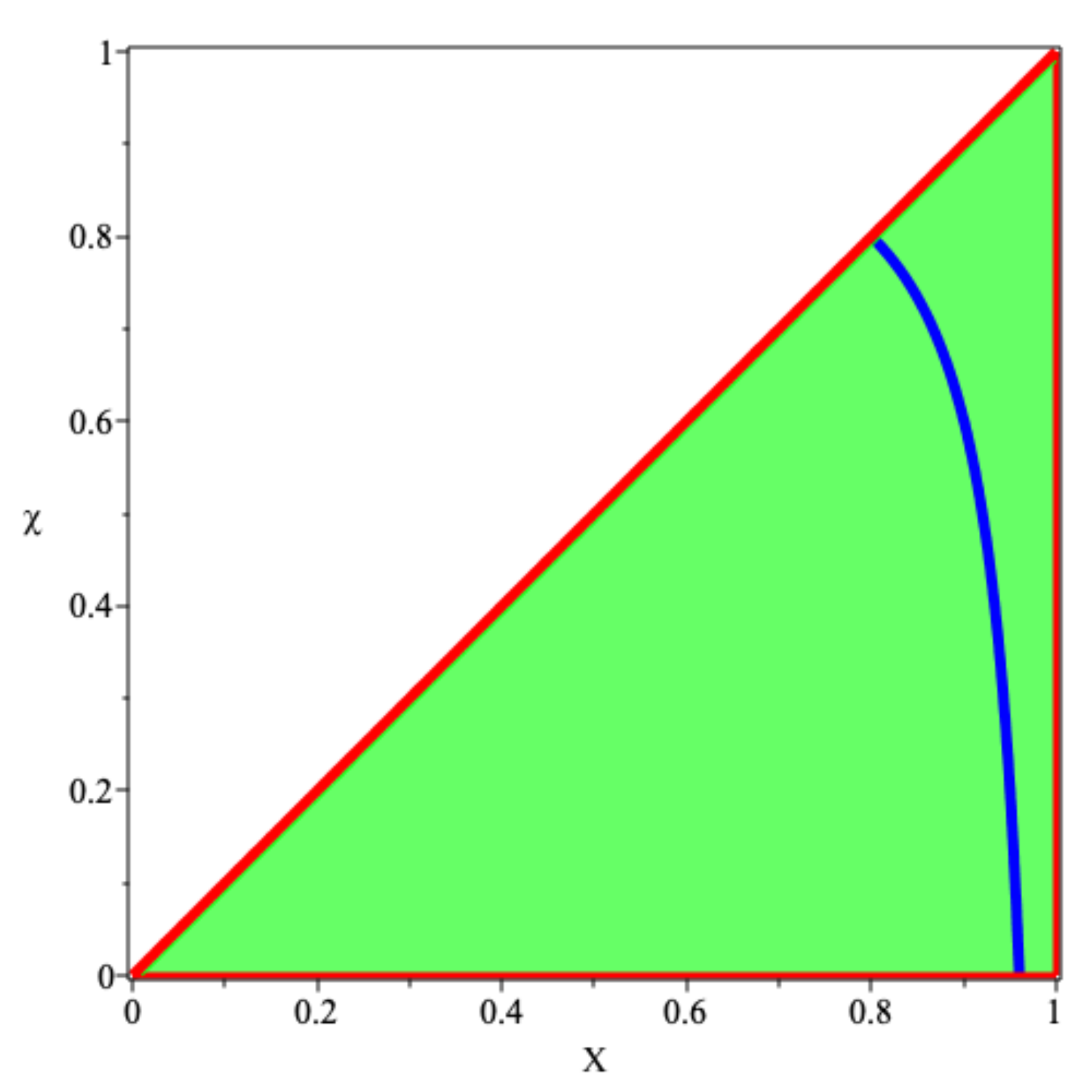}
\caption[DEC violations within the physically acceptable region]{DEC violations --- The green region is the physically acceptable region; that is $0\leq\chi\leq\Chi<1$. The blue curve corresponds to $w=1$, the boundary of the DEC violating region. To the left of the blue curve $w<1$ and the DEC is satisfied. To the right of the blue curve $w>1$ and the DEC is violated.
}
\label{F-DEC-1}
\end{center}
\end{figure}

\begin{figure}[htbp]
\begin{center}
\includegraphics[scale=0.7]{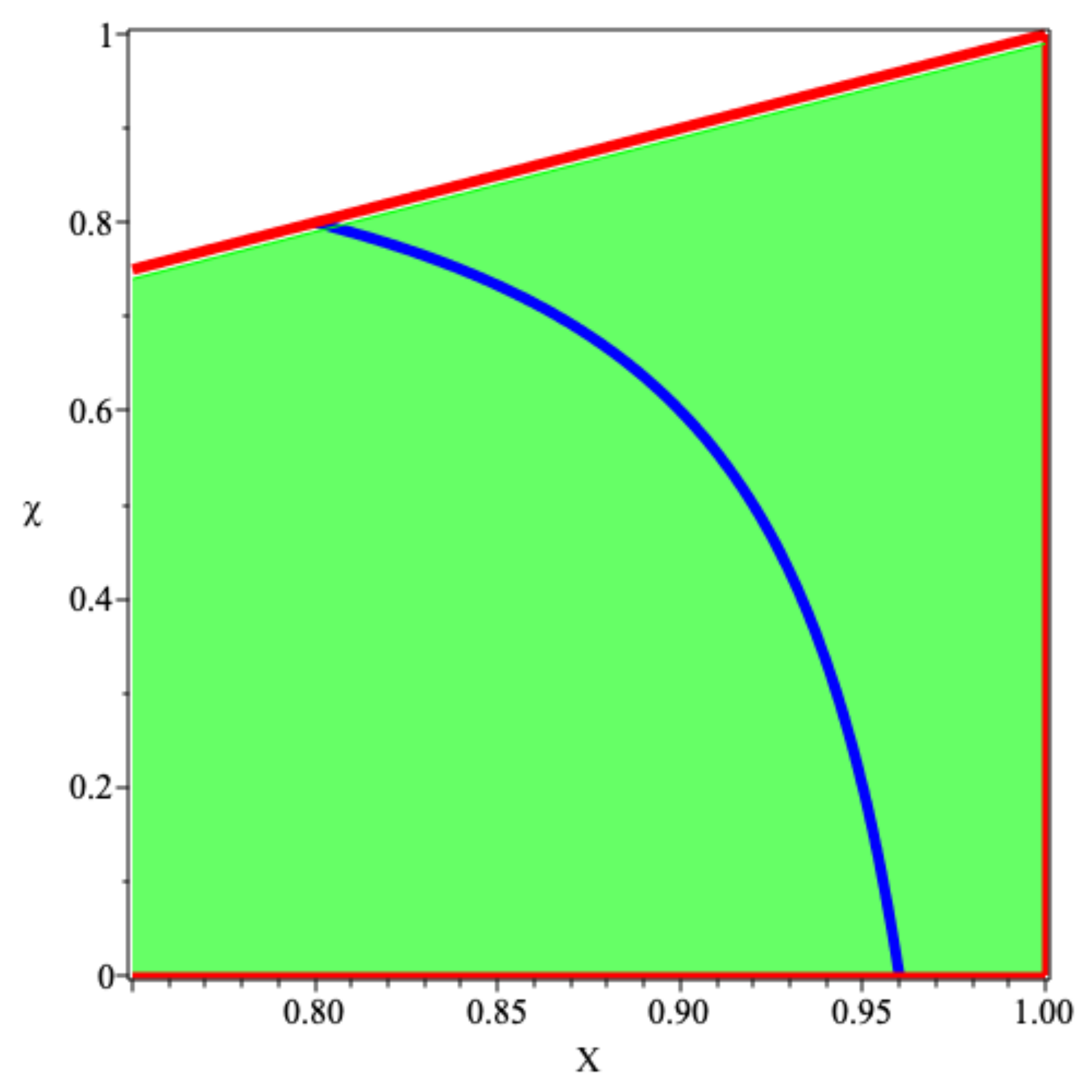}
\caption[Zoomed in plot of DEC violations within the physically acceptable region]{DEC violations, now focusing on the region above  $\Chi = {3\over 4}$. The green region is the physically acceptable region, that is $0\leq\chi\leq\Chi<1$. The blue curve corresponds to $w=1$, the boundary of the DEC violating region. To the left of the blue curve $w<1$ and the DEC is satisfied. To the right of the blue curve $w>1$ and the DEC is violated.
}
\label{F-DEC-2}
\end{center}
\end{figure}

The three regions where $w<1$, $w=1$, and $w>1$ in the physically acceptable part of the $(\chi,\Chi)$-plane, $0\leq\chi\leq\Chi\leq1$, are plotted in Figs.~\ref{F-DEC-0}--\ref{F-DEC-2}.

\clearpage
%
\section{Force between two hemispheres}
%

The force between any two hemispheres separated by a great circle on the Dyson mega-sphere is simply given (in the usual geometrodynamic units) by the dimensionless quantity
\begin{equation}
F = \Pi \times (2\pi a) = {1\over 4 } \left( {1-M/a\over\sqrt{1-2M/a}} -
{1-m/a\over\sqrt{1-2m/a}} 
\right) \ .
\end{equation}
First, rewrite this in terms of the two dimensionless compactness parameters, $\Chi = 2M/a$ and $\chi=2m/a$, as
\begin{equation}
F = {1\over 4 } \left( {1-\Chi/2\over\sqrt{1-\Chi}} -
{1-\chi/2\over\sqrt{1-\chi}} 
\right) \ .
\end{equation}
This can further be recast as
\begin{equation}
F = {1\over 8 } \left( {1\over\sqrt{1-\Chi}}+ \sqrt{1-\Chi} -
{1\over\sqrt{1-\chi}} - \sqrt{1-\chi} 
\right) \ .
\end{equation}
The maximum force conjecture, in its strong form, would require $F\leq 1/4$.
But this is easily seen to be generically violated; see discussion below.

\subsubsection{Physical units}

To rephrase this discussion in physical units, one can consider the Stoney
force~\cite{OTPUON:Stoney:1881}\footnote{ See ``Stoney units'';~\href{https://en.wikipedia.org/wiki/Stoney_units}{StoneyUnits-Wikipedia}.}, (also known as the Planck force, though the analysis by Stoney predates that of Planck by some 20 years).~\footnote{See ``Planck units''; ~\href{https://en.wikipedia.org/wiki/Planck_units}{PlanckUnits-Wikipedia}.} This is defined by
\begin{equation}
 F_{Stoney} = F_{Planck} = {c^4 \over G_N} \approx 1.2\times10^{44} \hbox{ N} \ ,
 \label{stoney-planck}
\end{equation}
in terms of which, re-inserting the factors of $c$ and $G_N$ as in Eq.~\eqref{stoney-planck}, one has
\begin{equation}
F_{physical} = {1\over 4 } \left( {1-\Chi/2\over\sqrt{1-\Chi}} -
{1-\chi/2\over\sqrt{1-\chi}} 
\right) F_{Stoney} \ .
\end{equation}
Alternatively,
\begin{equation}
F_{physical} = {1\over 8 } \left( {1\over\sqrt{1-\Chi}}+ \sqrt{1-\Chi} -
{1\over\sqrt{1-\chi}} - \sqrt{1-\chi} 
\right) F_{Stoney} \ .
\end{equation}

\subsubsection{Special case $m=0$}

First, note that
\begin{equation}
{\partial F\over\partial M} = {M\over 4 a^2 (1-2M/a)^{3/2} } >0 \ ,
\end{equation}
and
\begin{equation}
{\partial F\over\partial m} = -{m\over 4 a^2 (1-2m/a)^{3/2} } <0 \ .
\end{equation}
So to maximise the force between any two hemispheres one should maximise $M$, (without forming a black hole, so one requires $M<a/2$), and minimise $m$, (for instance by setting $m\to 0$, so that the Dyson mega-sphere is empty; there is no central star).

It is then easy to check that for the special case $m=0$ one has
\begin{equation}
F_{m=0} = {1\over 4 } \left( {1-M/a\over\sqrt{1-2M/a}} - 1 \right)
= {1\over 4 } \left( {1-\Chi/2\over\sqrt{1-\Chi}} - 1 \right) \ .
\end{equation}
Therefore, when forcing $m=0$, certainly $F_{m=0} > 1/4$ over the range:
\begin{equation}
\Chi  \in \Big( \left\{4\sqrt3-6\right\}  , 1\Big) = \Big(0.928203232..., 1\Big) \ .
\end{equation}
Thence (sufficiently close to black hole formation) the original form of the maximum force conjecture is manifestly false, and if one desires to rescue the maximum force conjecture one needs to carefully (and somewhat artificially) restrict the domain of discourse to only include certain forces. See related discussion in~\cite{CTTMFC:Jowsey:2021,RML:Jowsey:2021, MFACC:Faraoni:2021}, with countervailing opinions in~\cite{COMFACC:Schiller:2021,TFMFAMP:Schiller:2021}.

\subsubsection{General case}
\enlargethispage{30pt}

\begin{figure}[!htbp]
\begin{center}
\includegraphics[scale=0.4]{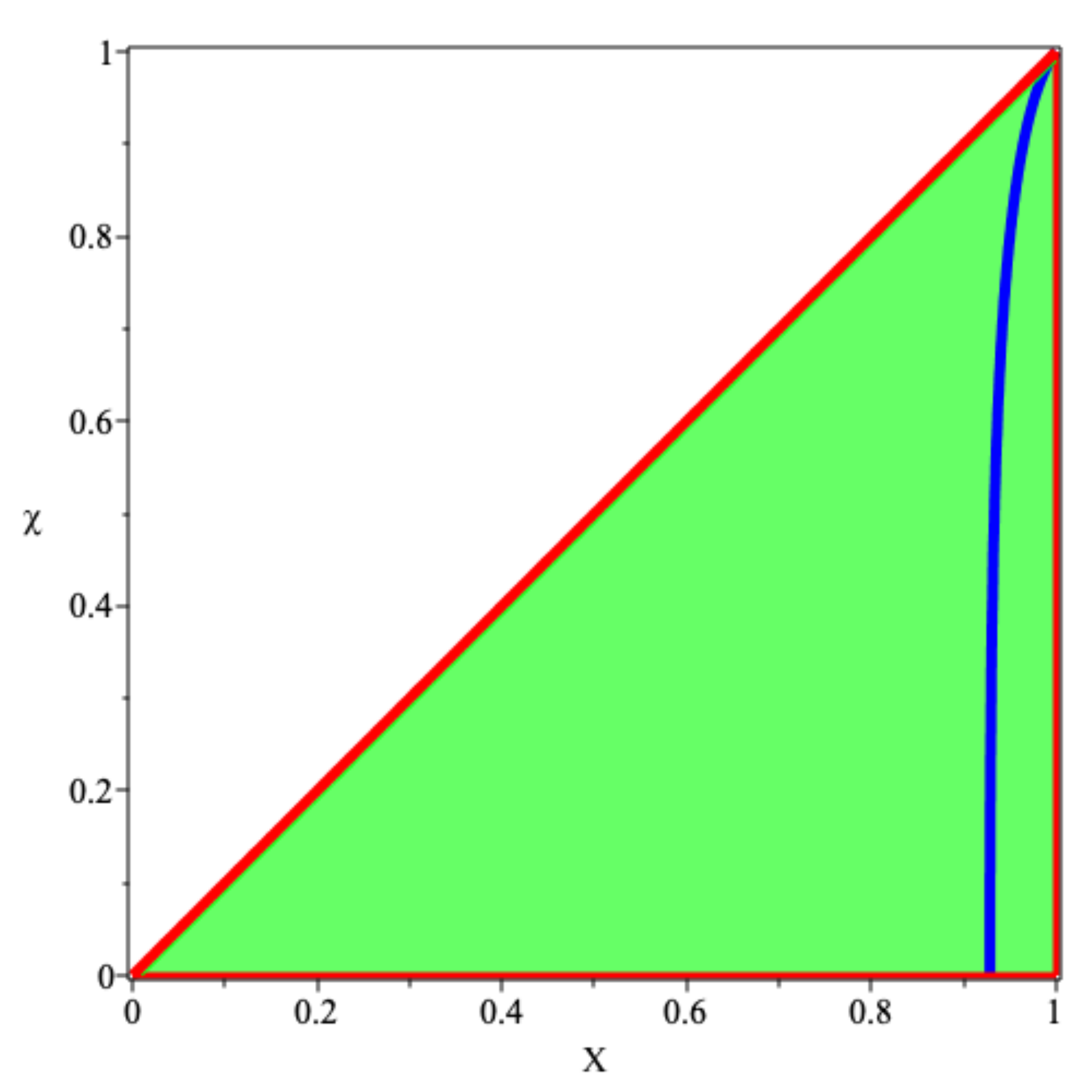}
\caption[Violations of the maximum force conjecture]{
Violations of the (naive) maximum force conjecture. The green region is the physically acceptable region;  $0\leq\chi\leq\Chi<1$. The blue curve corresponds to $F=1/4$, the boundary of the region violating the (naive) maximum force conjecture. This curve intersects the $\Chi$-axis at $\Chi = 4\sqrt{3}-6 = 0.928203232...$. To the left of the blue curve $F<1/4$ and the (naive) maximum force conjecture is satisfied. To the right of the blue curve $F>1/4$ and the (naive) maximum force conjecture is violated.
}
\label{F-force-1}
\end{center}
\end{figure}

\begin{figure}[!htbp]
\begin{center}
\includegraphics[scale=0.4]{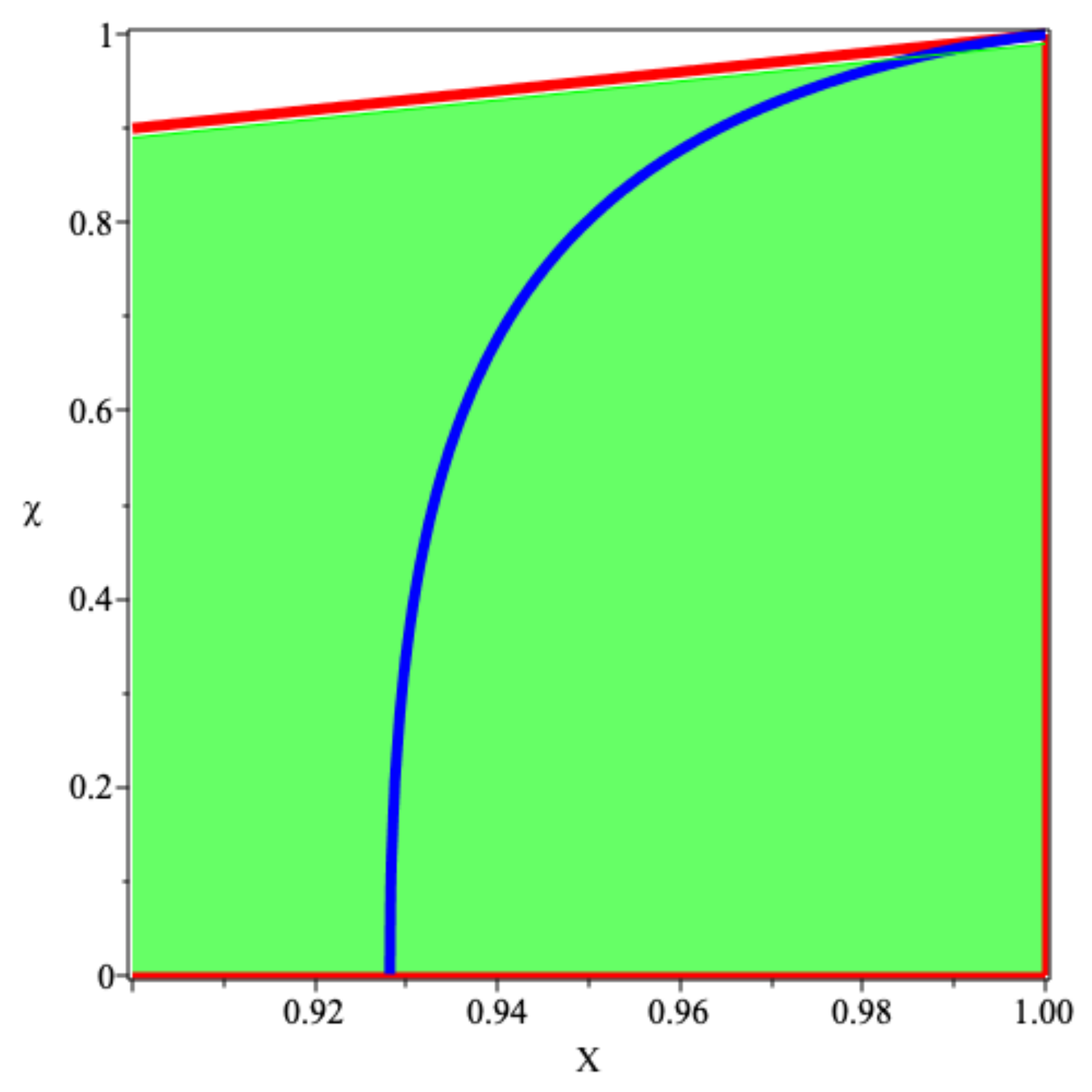}
\caption[Zoomed in plot of the violations of the maximum force conjecture]{
Violations of the (naive) maximum force conjecture, now focusing on the region $\Chi > {9\over10}$. The green region is the physically acceptable region; $0\leq\chi\leq\Chi<1$. The blue curve corresponds to $F=1/4$, the boundary of the region violating the (naive) maximum force conjecture. This curve intersects the $\Chi$-axis at $\Chi = 4\sqrt{3}-6 = 0.928203232...$. To the left of the blue curve $F<1/4$ and the (naive) maximum force conjecture is satisfied. To the right of the blue curve $F>1/4$ and the (naive) maximum force conjecture is violated.
}
\label{F-force-2}
\end{center}
\end{figure}

More generally in terms of the two compactness parameters $\Chi$ and $\chi$, it has already been seen that:
\begin{eqnarray}
F &=&  {1\over 4 } \left( {1-\Chi/2\over\sqrt{1-\Chi}} -
{1-\chi/2\over\sqrt{1-\chi}} \right)
\\
&=& {1\over 8 } \left( {1\over\sqrt{1-\Chi}}+ \sqrt{1-\Chi} -
{1\over\sqrt{1-\chi}} - \sqrt{1-\chi} 
\right) \ .
\end{eqnarray}
For any fixed value of $\chi$, this force $F$ becomes arbitrarily large as $\Chi\to 1$. Specifically, the boundary of the region satisfying both $F <1/4$ and the physicality constraints $0 \leq \chi \leq \Chi \leq 1$ is plotted in figures \ref{F-force-1} and \ref{F-force-2}.~\footnote{To obtain the curved line in figures \ref{F-force-1} and \ref{F-force-2} one sets $F=1/4$ and solves resulting quadratic for $\chi$ as a function of $\Chi$. This quadratic yields two somewhat messy roots, only one of which is physical. Even then, one has to be careful to restrict attention to the physically relevant region.} Again, the original form of the maximum force conjecture is manifestly false. For this maximum force violation to occur one must be ``close'' to black hole formation: $\chi\lesssim\Chi\lesssim1$.


\clearpage
\section{Violating the maximum force conjecture without violating the DEC}

\begin{figure}[!htbp]
\begin{center}
\includegraphics[scale=0.55]{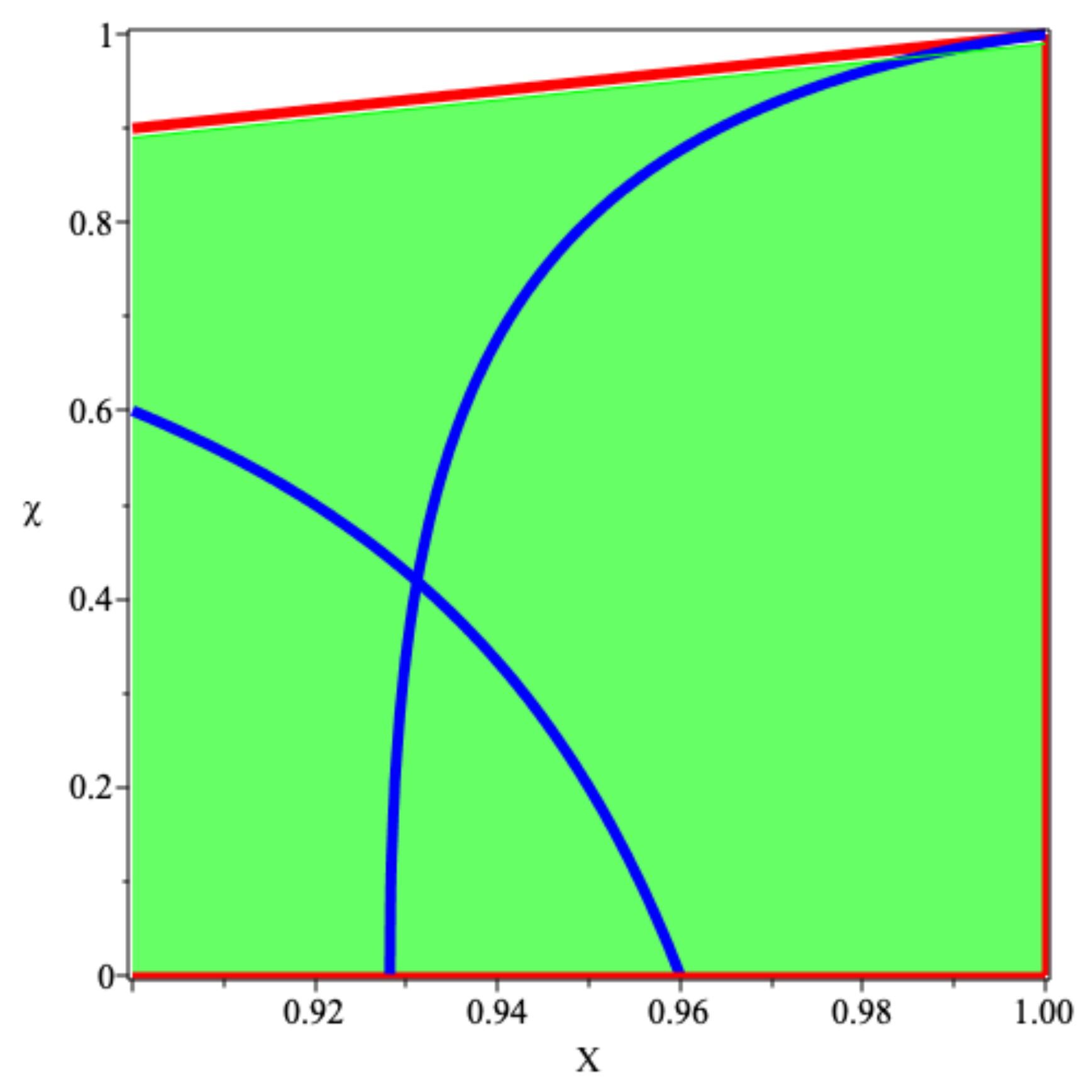}
\caption[Superposed plot of both the violations of the DEC and the violations of the maximum force conjecture]{
Subset of the physical region $0 \leq \chi \leq \Chi \leq 1$ where both $w<1$ and $F>1/4$. The condition $w<1$ (the DEC) is satisfied below the monotone decreasing curved line. The condition $F>1/4$ (violating the strong maximum force conjecture) is satisfied below the monotone increasing curved line.
There is a region between $\Chi = 4\sqrt{3}-6 = 0.928203232...$ and $\Chi = {24\over25} = 0.96$ where for a finite range of $\chi$ it is possible to satisfy the DEC while violating the strong maximum force conjecture. By considering the region above the two curved lines, it is also possible to violate the DEC while satisfying  the (strong) maximum force conjecture. Similarly to the left of the two curved lines both DEC and maximum force conjectures are satisfied, while to the right of the two curved lines both DEC and maximum force conjectures are violated.
}
\label{F-superimpose}
\end{center}
\end{figure}

By superposing the parts of the physical regime $0 \leq \chi \leq \Chi \leq 1$ where both $w<1$ and $F>1/4$ one easily sees that it is possible to violate the (strong) maximum force conjecture without violating the DEC. See figure~\ref{F-superimpose}. It is also possible to violate the DEC while satisfying the (strong) maximum force conjecture. For this to occur one must be ``close'' to black hole formation; $\chi\lesssim\Chi\lesssim1$.

\section{Weak-gravity regime}\label{S:weak-gravity}

In the weak-gravity regime, (where both $\Chi = 2M/a\ll 1$ and $\chi=2m/a\ll1$), one has
\begin{equation}
\sigma \approx {1\over 4\pi } {M-m\over a^2}  \geq 0 \ ,
\end{equation}
and
\begin{equation} 
\Pi \approx {1\over 16\pi } {M^2-m^2\over a^3}  \geq 0 \ . 
\end{equation}
%
So in the weak-gravity regime one has the physically plausible result that
\begin{equation}
M\approx m+ 4\pi \sigma \,a^2 \ ,
\end{equation}
while the surface stress is
\begin{equation}
\Pi \approx  {1\over 4} \; {(M+m)\over a} \; \sigma \;\ll\;  \sigma \ . 
\end{equation}
That is
\begin{equation}
w = {\Pi\over\sigma} \approx {1\over 4} \; {(M+m)\over a}
={\Chi+\chi\over 8}
 \; \;\ll\; 1 \ . 
\end{equation}
So all of the classical energy conditions, including the DEC, are automatically satisfied in the weak-gravity regime. 

Finally note that in the weak-gravity regime the force between two hemispheres is
\begin{equation}
F \approx {1\over 8 } \; {M^2-m^2\over a^2} 
= {1\over2} (\Chi^2-\chi^2)  \ll 1 \ . 
\end{equation}
However, as seen above there are much more general and precise results available, fully applicable to the highly relativistic strong-gravity regime.

\section{Low-mass thin-shell configurations}\label{S:low-mass}
\newcommand*{\wasyfamily}{\fontencoding{U}\fontfamily{wasy}\selectfont}
\newcommand*{\astrosun}{{\odot}}
\newcommand*{\mercury}{{\text{\wasyfamily\char39}}}
\newcommand*{\venus}{{\text{\wasyfamily\char25}}}
\newcommand*{\earth}{{\oplus}}
\newcommand*{\mars}{{\text{\wasyfamily\char26}}}
\newcommand*{\jupiter}{{\text{\wasyfamily\char88}}}
\newcommand*{\saturn}{{\text{\wasyfamily\char89}}}
\newcommand*{\uranus}{{\text{\wasyfamily\char90}}}
\newcommand*{\neptune}{{\text{\wasyfamily\char91}}}
\newcommand*{\pluto}{{\text{\wasyfamily\char92}}}

Another interesting approximation is that of a low-mass thin-shell, where
\begin{equation}
M - m \ll m \ .
\end{equation}
One can then approximate
\begin{equation}
\sigma(m,M;a) = {M-m\over 4\pi a^2 \sqrt{1-2M/a} }+\O([M-m]^2) \ ,
\end{equation}
while
\begin{equation}
\Pi(m,M;a) ={m(M-m)\over  8\pi a^3 \sqrt{1-2M/a}} +\O([M-m]^2) \ .
\end{equation}
%
Partially eliminating $M-m$ between these two equations gives
\begin{equation}
\Pi(m,M;a) = {1\over 4}\;  {2m\over a} \; \sigma(m,M;a)  + \O([M-m]^2) \ .
\end{equation}
Trivially,
\begin{equation}
w(m,M;a) ={m/a\over  2(1-2m/a)} + \O([M-m]) \ ,
\end{equation}
and
\begin{equation}
F(m,M;a) ={m(M-m)\over  a^2 \sqrt{1-2M/a}} +\O([M-m]^2) \ ,
\end{equation}
while with a little more work one can obtain
\begin{equation}
w(m,M;a) ={m/a\over  2(1-2m/a)} 
+{(M-m)\over 4 a(1-2m/a)^2} +
\O([M-m]^2) \ .
\end{equation}
Thus even for a low mass shell, $M - m \ll m$, by driving $\chi \to 1$, one can still easily arrange violations of the DEC and the original form of the maximum force conjecture.

To calibrate the approximate size of these quantities, note that in geometrodynamic units the mass of the Sun is $M_\astrosun \approx 1.5 \;\hbox{km}$, while the mass of Jupiter is $M_\jupiter\approx 1.1 \;\hbox{m}$. The astronomical unit is $1 \;\hbox{AU} \approx 1.5\times 10^8 \;\hbox{km}$. Then disassembling Jupiter to build a mega-sphere with radius 1 AU leads to the dimensionless estimates
\begin{equation}
w \approx 5 \times 10^{-9} \ , \qquad 
F \approx 7.333 \times 10^{-20} \ .
\end{equation}
That is, solar system scale Dyson mega-spheres would be well into the Newtonian regime. For general relativistic effects to be appreciable one must be ``close'' to black hole formation; $\chi\lesssim\Chi\lesssim1$.

Furthermore in physical units $M_\jupiter\approx 1.898 \times 10^{27}\;\hbox{kg}$, so that  for the surface energy density one has
\begin{equation}
\sigma\approx {M_\jupiter\over (4\pi \,1 \, {\hbox{AU}}^2)} 
\approx 7003 \;\hbox{kg/m}^2 \ . 
\end{equation}
Locally, this is actually human scale --- since the density of iron is approximately $7800 \;\hbox{kg/m}^3$, this would correspond to a Dyson mega-sphere a little under $1$ metre thick. Globally speaking however, this is well beyond human technology, and would at the very least require the intervention of a Kardashev type II civilization~\cite{TOIBEC:Kardashev:1964}.

\section{Possible generalizations}\label{S:geeralize}

\subsubsection{Matrioshka configurations}\label{S:Matrioshka}

It would in principle be possible to extend the current  discussion to deal with the more general Matrioshka configurations (multiply-nested thin-shell Dyson mega-spheres). Such configurations have been mooted in the context of Matrioshka brains.\footnote{See ``Matrioshka brain'' (nested Dyson mega-spheres); ~\href{https://en.wikipedia.org/wiki/Matrioshka_brain}{MatrioshkaBrain-Wikipedia}.} No really new matters of principle would be involved, but the algebra would become considerably more complex. Such issues are put aside for now.

\subsubsection{Thick-shell Dyson mega-spheres}\label{S:thick}

In contrast, the potential consideration of thick-shell Dyson mega-spheres would require a completely different set of techniques; one would need to study the \textit{anisotropic} Tolman--Oppenheimer--Volkoff (TOV) equation, paying careful attention to suitable boundary conditions and equations of state. Once again, such issues are put aside for now.

\subsubsection{Adding rotation}\label{S:rotation}

Adding rotation to these Dyson mega-sphere configurations would likely be extremely tedious, if not outright impractical. Even if the angular momentum of the central object and Dyson mega-sphere are parallel, the required calculations would be daunting. One could either try to work with the full Kerr geometry~\cite{GFOASMAAEOASM:Kerr:1963, GCAR:Kerr:1965, MOARCM:Couch:1965, TKSABI:Visser:2007, TKSRBHIGR:Scott:2009, TGOKBH:ONeill:2014}, or with the Lense--Thirring slow rotation approximation~\cite{UDEDEDZADBDPUMNDEG:Lense:1918, OTHOTSLE:Pfister:2007, PGLT1:Baines:2021, PGLT2:Baines:2021, PGLT3:Baines:2022, PGLT4:Baines:2022}. This is a topic for future research.
\vspace*{15pt}

\section{Conclusion of Part III}\label{S:discussion}

Arbitrarily advanced alien civilizations are the \textit{sine qua non} of much speculative physics --- developed with a view to testing the ultimate limitations of putative technological advancement. Notable examples of speculative physics are Dyson mega-structures, (specifically the Dyson mega-spheres of Chapter~\ref{C:DMS}), Lorentzian wormholes, warp drives, and tractor/pressor/\newline stressor beams. A common theme arising in such speculative physics is the violation of some or all of the classical energy conditions, and/or violations of the original strong form of the maximum force conjecture. Such violations are not the absolute prohibition they were once thought to be, but they are certainly invitations to think deeply about the underlying physics. The Dyson mega-structures in particular, are in principle accessible to astronomical observational tests. In Chapter~\ref{C:DMS} the simple spherically symmetric thin-shell Dyson mega-sphere has been analysed, in a fully general relativistic manner using the Israel--Lanczos--Sen thin-shell formalism. There are regions in parameter-space where either, both, or neither of the DEC and strong maximum force conjecture are violated.
\clearpage

In Chapter~\ref{C:SVthinshell}, the SV thin-shell traversable wormhole has been thoroughly discussed, and the aMc equivalent briefly presented. These are both further examples of nonsingular black hole mimickers, and are amenable to a straightforward stability analysis when permitting linearised radial perturbations of the wormhole throat. Several specific exmaples of the SV construction are considered, fixing the parameters to allow for nontrivial analysis. Both thin-shell wormholes require a violation of the NEC; an expected result. Overall, the constructions are highly tractable, and are exemplars of the types of spacetimes able to be generated \emph{via} the ``cut-and-paste'' procedure applied to nonsingular black hole spacetimes.

Significant future work involving thin-shell constructions would be to apply the procedure to axisymmetric candidate spacetimes, \emph{e.g.} the bbKN geometry of Chapter~\ref{C:SVCBB}, or the eye of the storm geometry from Chapter~\ref{C:AMCEOS}. If such constructions are amenable to the extraction of (potential) astrophysical observables, this would give a new class of objects to the astronomers for consideration.
\vspace*{60pt}

\begin{center}
    This concludes Part~\ref{PartIII}.
\end{center}

%
%

%% file: 05-Conclusion.tex
\chapter{Conclusions}\label{C:con}

%
%
In this thesis, two families of regular spacetimes have been thoroughly explored through the lens of ``GR$+$no curvature singularities'', modelling either regular black holes or traversable wormholes, and the associated key results codified. These are the family of black-bounce spacetimes (see Part~\ref{PartI}), and the family of regular black holes with asymptotically Minkowski cores (see Part~\ref{PartII}), respectively. Furthermore, related thin-shell traversable wormhole constructions have been discussed and analysed, as well as a brief examination of thin-shell Dyson mega-spheres (see Part~\ref{PartIII}). All spacetimes considered are examples of ``black hole mimickers'' amenable to the extraction of (potential) astrophysical observables in principle falsifiable/verifiable by the observational and experimental communities.

When excising curvature singularities from standard GR, it has been discussed that the following constraints ought to be placed on the candidate geometries to make them as appropriate as possible for the experimental and observational communities:
\begin{itemize}
    \item Impose axisymmetry.
    \item Impose asymptotic flatness at spacial infinity.
    \item
    The Hamilton--Jacobi equations should be separable to give in principle integrable geodesics; the geometry should possess a nontrivial Killing two-tensor $K_{\mu\nu}$.
    \item 
    Impose separability of the Klein--Gordon equation and Maxwell's equations on the background spacetime.
    \item
    Ensure the spacetime is amenable to straightforward Regge--Wheeler and Zerilli analyses; the spin two axial and polar modes must be able to be extracted.
    \item 
    Impose a high degree of mathematical tractability.
    \item 
    Constrain the amount of exotic matter and demand satisfaction of the relevant classical energy conditions outside horizons (except possibly allowing for violation of the SEC due to the presence of a positive cosmological constant).
    \item Forbid closed timelike curves.
    \item Impose separability of the Dirac equation.
\end{itemize}
Finding appropriate geometries which satisfy \textit{all} of these constraints is highly nontrivial. The bbKN spacetime of Chapter~\ref{C:SVCBB} satisfies some, however one of the most notable results in this thesis is the analysis of the eye of the storm geometry from Chapter~\ref{C:AMCEOS}. This spacetime is \emph{very} close to an idealised candidate geometry. It models a rotating regular black hole with an asymptotically Minkowski core, is asymptotically Kerr for large $r$, possesses the full ``Killing tower'' of nontrivial Killing tensor, Killing--Yano tensor, and principal tensor, has integrable geodesics in principle, has the properties of separability of both the Klein--Gordon equation and Maxwell's equations, and the only violation of any of the classical energy conditions occurs at an arbitrarily small distance scale in the deep core where GR is no longer appropriate. The eye of the storm is also the most mathematically tractable rotating regular black hole in the current literature. Further examination of this geometry, and the charged ``eosKN'' geometry presented in \S~\ref{eosKN}, is an important topic for future research. The analyses of all candidate geometries discussed in this thesis are extremely clean and tractable; this is certainly one of the major reasons they have become popular in the literature.

More abstractly, the cleanliness and tractability of the constructions analysed herein can be seen as a direct byproduct of the fact that exploration of classical black hole mimickers falls within the methodology of ``turning one knob at a time'' when attempting to progress theoretical physics (see \S~\ref{Intro:knob}). When compared to proposed resolutions of singularities \emph{via} more heavily modified theories of gravity, or candidate theories of quantum gravity, these discussions are highly cogent, more tractable, and do not run the risk of reformulating known results. By exploring the full space permitted by the framework of ``GR$+$no curvature singularities'', at the very least the community should be able to gain an understanding of the correct direction in which to head to progress quantum gravity, if not discover highly significant results.

With a look to the future, there are two ``directions'' for this framework that are worth further discussion. Firstly, establishing a generalised (insofar as is possible) method for extracting the appropriate Lagrangian for a given geometry in axisymmetry would be an incredible result. The work by Bronnikov and Walia in reference~\cite{FSFSVS:Bronnikov:2022} is the usual approach, and reveals that there is still an element of ``guess and check'' involved when trying to elevate candidate geometries to exact solutions by finding suitable source terms. Were the community able to expedite this process, then the full list of nonsingular Lagrangians associated to desirable candidate geometries could be compiled and directly compared to hypothesised candidate particles arising from quantum field theory.

The other ``direction'' of importance is to streamline the discourse between theory and experiment. While cataloguing Lagrangians is a prudent task, in parallel there must be constant checking against observationally or experimentally obtained data to determine whether certain candidate geometries are forbidden by measurement. In order to do this, all of the theoretical, numerical, and instrumental/experimental communities must communicate frequently and work together. An example of the complexities involved: Chapter~\ref{C:AMCQNM} extracts a spin one and spin zero first-order WKB approximation of the quasinormal modes for aMc spacetime. In order to speak to the LIGO/Virgo (or LISA) calculations, one must furthermore perform this task for the spin two polar and axial modes, as well as ensure that the results have an appropriate margin of error. This is extremely nontrivial. There are many different numerical techniques to extract these results, all with their sets of (dis)advantages, and knowing what order to perform the calculations to, or what assumptions are permitted versus those which critically affect the usability of the results, needs to be made more transparent in the literature. While theoreticians have an obligation to try and communicate with the observational/experimental communities, instrumentalists also have an obligation to inform the precision of their instruments as improvements are made on the LIGO/Virgo (or LISA) devices; these constraints should directly inform the types of calculations and models which are being explored. Clarifying this epistemological gap will enable the communities involved to forge an extremely exciting path towards the so-called ``theory of everything''. The newly available technology is precisely the type of ``step-function-esque'' leap which could enable another Renaissance in physics.

%
\part{Appendices}\label{PartIV}

\appendix
\chapter{Formulation of GR}\label{A:GR}
%
%
%

Spacetime can be intuitively thought of as the ``stage'' on which the ``play'' of the universe is set. Given this intuition, one must impose mathematically rigorous constraints on this ``stage'' such that it may have a sensible physical interpretation in the framework of GR. Firstly one constructs spacetime by imagining a four-dimensional backdrop consisting of three dimensions of space and one dimension of time. Coupled with this is the idea that in the presence of an object equipped with mass and/or momentum, the four-dimensional spacetime exhibits curvature. John A. Wheeler puts it excellently~\cite{G:Misner:2000}:

\textit{``Space acts on matter, telling it how to move. In turn, matter reacts back on space, telling it how to curve.''
}

Specifically, Frederic P. Schuller provides a compact definition of spacetime as follows~\cite{Schuller}:

\textit{
``Spacetime is a four-dimensional topological manifold with a smooth atlas carrying a torsion-free connection compatible with a Lorentzian metric and a time orientation satisfying the Einstein equations.''
}


\section{Four-dimensional topological manifold}

A topological manifold $\mathcal{M}$ of dimension $d$ is a mathematical object characterised by the following~\cite{ITT:Adams:2008}:
\begin{itemize}
    \item It is a topological space $(\mathcal{E}, \mathcal{T})$, consisting of a set $\mathcal{E}$ together with a topology of open sets $\mathcal{T}$, which is locally Euclidean:
    \begin{eqnarray}\label{Euclideanspace}
        && \forall \ x\in\mathcal{E}, \ \exists \ \mathcal{O}\in\mathcal{T} \ \mbox{and} \ n\in\mathbb{Z}^{+}\,: \nonumber \\
        && x\in\mathcal{O} \ \mbox{and} \ \exists \ \mathcal{X}\subset\mathbb{R}^{n} \ \mbox{and} \ \exists \ \mbox{homeomorphism} \ f:\mathcal{O}\leftrightarrow\mathcal{X} \ . \nonumber \\
        &&
    \end{eqnarray}
    \item The dimensionality of the space, $d$, is the same everywhere. With regard to Eq.~(\ref{Euclideanspace}), $n=d\, \ \forall \ \mbox{such} \ \mathcal{X}\subset\mathbb{R}^{n}$; the ``chunks'' of Euclidean space all come from the same Euclidean space. Terminologically such a space is a ``$d$-manifold''.
    \item The manifold is Hausdorff:
    \begin{equation}
        \forall \ x_{1}, x_{2}\in\mathcal{E}, \ \exists \ \mathcal{O}_{1}, \mathcal{O}_{2}\in\mathcal{T}: x_{1}\in \mathcal{O}_{1}, x_{2}\in \mathcal{O}_{2} \ \mbox{and} \ \mathcal{O}_{1}\cap \mathcal{O}_{2}=\emptyset \ .
    \end{equation}
    \item The manifold has at least one countable atlas~\cite{MATH465, MOM:Gauld:2009} --- a set is ``countable'' if all its elements can be put in injective correspondence with the natural numbers, $\mathbb{N}$.\footnote{For further elaboration on these very basic concepts from elementary topology please see reference~\cite{ITT:Adams:2008}; \emph{e.g.} rigorous definition of ``homeomorphism'', \emph{etc.}}
\end{itemize}
\vspace*{20pt}


\section{Smooth atlas}
\vspace*{15pt}

First define an atlas:
\begin{itemize}
\item An atlas is a collection of charts which covers the entire locally Euclidean space $\mathcal{E}$.
\item A chart $(\mathcal{O}, f, U)$ on a member of the topology $\mathcal{O}\in\mathcal{T}$ is a subset $U\subseteq\mathbb{R}^{d}$ together with a homeomorphism $f:\mathcal{O}\leftrightarrow U=f(\mathcal{O})$.
\item A countable atlas is hence a set of charts\\ $\mathcal{A}=\left\lbrace\left(\mathcal{O}_{i}, f_{i}, U_{i}\right)\right\rbrace: \ f_{i}:\mathcal{O}_{i}\leftrightarrow U_{i}=f_{i}\left(\mathcal{O}_{i}\right)\subseteq\mathbb{R}^{d}$, with $i\in\mathcal{I}$ (some countable indexing set $\mathcal{I}$), such that $\bigcup_{i\in\mathcal{I}} \ {\mathcal{O}_{i}}=\mathcal{E}$.
\end{itemize}
Now define smooth:
\begin{itemize}
    \item An atlas is ``smooth'' if all transition maps in the atlas are smooth maps.
    \item A transition map is defined as: Let $(\mathcal{O}_{1}, f_{1}, U_{1})$ and $(\mathcal{O}_{2}, f_{2}, U_{2})$ be charts in $\mathcal{M}$ such that $\mathcal{O}_{1}\cap\mathcal{O}_{2}$ is nonempty. Then the map defined by $g:=f_{2}\circ f_{1}^{-1}$ is the transition map $g:f_{1}(\mathcal{O}_{1}\cap \mathcal{O}_{2})\rightarrow f_{2}(\mathcal{O}_{1}\cap \mathcal{O}_{2})$.
    \item A smooth map is a map for which derivatives of all orders are defined everywhere in its domain. Hence a smooth manifold is a $C^{\infty}$-manifold.
\end{itemize}
\clearpage


\section{Torsion-free connection}

The manifold must now be equipped with a geometry --- one must have a means of making accurate statements concerning distance, angles, and curvature. Let there exist some arbitrary smooth $d$-manifold $\mathcal{M}$. The first step towards establishing a geometry on $\mathcal{M}$ is to impose a coordinate system such that each point in $\mathcal{M}$ can be uniquely identified. Given a countable atlas for $\mathcal{M}$, some $\mathcal{A}=\left\lbrace\left(\mathcal{O}_{i}, f_{i}, U_{i}\right)\right\rbrace: \ \bigcup_{i\in\mathcal{I}}\mathcal{O}_{i}=\mathcal{M}$, observe:
\begin{eqnarray}\label{coordinatesystem}
    && \forall \ \textbf{p}\in\mathcal{M}, \exists \ i\in\mathcal{I}, \ \mbox{and} \ \left(\mathcal{O}_{i}, f_{i}, U_{i}\right)\in\mathcal{A}: \ \textbf{p}\in\mathcal{O}_{i} \nonumber \\
    \Longrightarrow \ && \exists \ \textbf{x}=\left(x_{1},\cdots,x_{d}\right)\in U_{i}\subseteq\mathbb{R}^{d}: \ \textbf{x}=f_{i}(\textbf{p}) \nonumber \\
    \Longrightarrow \ && \textbf{p}=f_{i}^{-1}\left(\textbf{x}\right) \ .
\end{eqnarray}
So the point $\textbf{p}\in\mathcal{M}$ has coordinate location $\left(x_{1},\cdots,x_{d}\right)$ with respect to the chosen coordinate patch $\left(\mathcal{O}_{i}, f_{i}, U_{i}\right)$. This is a generalised process for establishing a coordinate system on a given manifold.\footnote{Restricting the $x_{1},\cdots,x_{d}$ to the reals is not strictly necessary, but is standard practice for most candidate spacetimes. Bold font to indicate the multi-dimensionality of objects is henceforth dropped; dimensionality is obvious from context.}

Imposing some coordinate system on $\mathcal{M}$, let there exist $x, y \in \mathcal{M}: \ x\neq y$. This construction ultimately implies the existence of the respective tangent spaces, $T_{x}$ and $T_{y}$, allowing one to discuss the behaviour of tangent vectors on $\mathcal{M}$. One must now define ``parallel transport'' of these vectors (this terminology dates back to Knebelman, 1951~\cite{SORP:Knebelman:1951}; see also the detail provided in reference~\cite{MATH465}).

Parallel transport is the process by which a vector transports along smooth curves in $\mathcal{M}$. Let there exist a smooth curve parameterised by an arbitrary scalar parameter, $\gamma$, connecting $x,y\in\mathcal{M}$. Then define the ``transport'' function, $T_{\left[x\rightarrow y;\gamma\right]}: T_{x}\rightarrow T_{y}$ (this is a $T^{1}_{1}$ bi-tensor; see reference~\cite{tensors} for details). This function must possess several properties:

\begin{itemize}
    \item The null path $\gamma_{0}$ must define the identity of the function, $T_{\left[x\rightarrow x;\gamma_{0}\right]}=I:T_{x}\rightarrow T_{x}$.
    \item $T_{\left[x\rightarrow y;\gamma\right]}$ should be an everywhere-invertible mapping.
    \item Reversing a path should correspond to the inverse of the transport function, $T_{\left[x\rightarrow y;\gamma\right]}=\left(T_{\left[y\rightarrow x;\tilde{\gamma}\right]}\right)^{-1}$.\footnote{Note that $\tilde{\gamma}$ is just some arbitrary scalar-valued reparameterisation of the reverse path; there is a degree of freedom in choosing these parameters.}
    \item The transport operator ought to be a linear operator between vector spaces. If two vectors $V_{1}, V_{2} \in T_{x}$ are \emph{both} propagated along a smooth curve from $x$ to $y$ in $\mathcal{M}$, then\\ $T_{\left[x\rightarrow y;\gamma\right]}\left(V_{1}+V_{2}\right)=T_{\left[x\rightarrow y;\gamma\right]}\left(V_{1}\right)+T_{\left[x\rightarrow y;\gamma\right]}\left(V_{2}\right)$.
\end{itemize}
\clearpage
Note that when a chosen coordinate basis is ``curved'' with respect to $\mathcal{M}$, or \textit{vice versa}, parallel transport of the basis vectors along smooth curves in $\mathcal{M}$ alters the direction of the basis vectors. Quantifying this alteration, one must define a rate of change, or a notion of derivative, of the transport function $T$. There are multiple methods of tensorial differentiation; the most commonly used is the covariant derivative, defined \textit{via} an object called ``the connection''.

Given some orthonormal basis $\lbrace \hat{e}_{i}\rbrace\,; i\in\mathcal{I}$, the connection is the triple-indexed object $\Gamma^{i}{}_{jk}$ which possesses the following properties~\cite{G:Misner:2000}:
\begin{itemize}
    \item $i, j, k \in \mathcal{I}$ .
    \item The index $i$ indicates the basis vector being stretched/multiplied.
    \item The index $j$ indicates the basis vector being transported.
    \item The index $k$ indicates the basis vector defining the direction of the motion.
\end{itemize}
A \emph{torsion-free} connection is a connection which obeys the condition~\cite{G:Misner:2000}
\begin{equation}
    \Gamma^{i}{}_{jk} = \Gamma^{i}{}_{kj}
\end{equation}
in a coordinate basis. Setting torsion to zero is a choice, and by doing so one specialises the notion of the connection to the ``Christoffel connection''. In order to explicitly define the Christoffel connection, one must first understand the metric tensor $g_{\mu\nu}$.
\vspace*{12pt}


\section{Lorentzian metric}\label{metric}
\vspace*{7pt}

Given an arbitrary coordinate system on some manifold $\mathcal{M}=\left(\mathcal{E},\mathcal{T}\right)$, the distance between any two points is defined by an object called the ``metric''. The form that the metric takes is context-dependent. Let $x, y, z \in \mathcal{E}$ be points expressed with respect to some chosen coordinate system; the purely mathematical notion of a metric on a set $\mathcal{E}$ is defined as a function\\ $g:\mathcal{E} \ \mbox{x} \ \mathcal{E} \rightarrow \mathbb{R}$ possessing the following properties~\cite{ITT:Adams:2008}:
\begin{eqnarray}\label{metricproperties}
    &\bullet& \ g(x, y)\geq 0 \ \forall \ x, y \in \mathcal{E}, \ \mbox{and} \ g(x, y) = 0 \ \mbox{iff} \ x = y \ . \nonumber \\
    &\bullet& \ g(x, y) = g(y, x) \ \forall \ x, y \in \mathcal{E} \ . \nonumber \\
    &\bullet& \ g(x, y) + g(y, z)\geq g(x, z) \ \forall \ x, y, z \in \mathcal{E} \ .
\end{eqnarray}
$g(x, y)$ is the distance between the points $x$ and $y$, and the pair $(\mathcal{E}, g)$ is called a ``metric space''.
\clearpage

Migrating from the purely mathematical definition of the metric to the realm of physics, one relaxes the three properties of Eq.~(\ref{metricproperties}) to attribute physical meaning to the notion of distance. This difference in definition is a direct result of the important role that time plays in the universe; the physical motivation behind it is to be able to clearly separate between events which are timelike, null, or spacelike-separated:
\begin{itemize}
    \item ``Timelike-separated'' events in a spacetime are causally connected --- events that can be reached from each other by traveling within either the future or past-directed light cones on a spacetime diagram.
    \item ``Null-separated'' events in a spacetime are marginally causally connected --- events that can only be reached from  one another by particles traveling at the speed of light $c$, \emph{i.e.} along the boundaries of the future or past-directed light cones in a spacetime diagram.
    \item ``Space-like separated'' events in a spacetime are not causally connected --- events that lie outside of both the future and past-directed light cones on a spacetime diagram.
\end{itemize}
To separate these classes of event, one must introduce metric signature. The signature of a metric is determined by the number of positive or negative eigenvalues that arise from its matrix representation~\cite{G:Misner:2000}. A metric with Riemannian signature has exclusively positive eigenvalues; this yields a positive definite metric tensor and the notion of distance is simply the length of the \emph{shortest} possible curve between two points in the manifold. Alternatively, a ``Lorentzian metric'' is a metric of Lorentz-signature --- the matrix representation possesses an odd number of negative eigenvalues whilst the remainder are positive. In this case the notion of distance corresponds to the principle of least action in the manifold; this is an ``extremal distance'' defined by the principles of variational calculus.

In the context of GR, the distinction between timelike, null, and spacelike-separated events is achieved by imposing a Lorentz-signature metric on the spacetime four-manifold. There is a choice in how to do this; the most common metric signatures are either $(+,-,-,-)$, or $(-,+,+,+)$ (where the sign indicates whether that basis component is inheriting a positive or negative eigenvalue in the matrix representation). For detailed discussion, see reference~\cite{G:Misner:2000}. The generalised definition for the metric in GR can then be given as follows:

The metric is a symmetric, nondegenerate, and position-dependent matrix that is a rank-two tensor, denoted canonically by $g_{\mu\nu}$, which has constant Lorentzian signature. Its form is governed by the ``line element'' --- the distance in the manifold $\mathcal{M}$ between points which are infinitesimally displaced with respect to the chosen coordinate basis. The generalised line element is given by
\begin{equation}
    \d s^2 = g_{\mu\nu}\,\d x^{\mu}\,\d x^{\nu} \ ,
\end{equation}
where $\d x^{\mu}$ are the infinitesimal displacements of each of the coordinates. The generalised notion of distance between two points is found by extremising the action of the following integral:
\begin{equation}\label{arclengthgen}
    L_{\mathcal{M}} = \int_{\gamma_{0}}^{\gamma_{1}}\sqrt{g_{\mu\nu}\frac{\d x^{\mu}}{\d\gamma}\frac{\d x^{\nu}}{\d\gamma}} \ \d\gamma \ ,
\end{equation}
where the curve connecting the points is parameterised by some scalar parameter $\gamma$. The metric obeys the metricity condition: $\nabla_{\alpha}g_{\mu\nu}=0$.

One may now explicitly define the Christoffel connection in terms of the metric. For some arbitrary metric tensor $g_{\mu\nu}$, the Christoffel symbols are defined by
\begin{equation}\label{Christoffel}
    \Gamma^{\mu}{}_{\alpha\beta} = \frac{1}{2}g^{\mu\nu}\left(\partial_{\alpha}g_{\nu\beta}+\partial_{\beta}g_{\nu\alpha}-\partial_{\nu}g_{\alpha\beta}\right) \ ,
\end{equation}
where the index $\nu$ is contracted over in accordance with the Einstein summation convention.


\section{Time orientation}

Suppose one has a construction $\left(\mathcal{M}, g_{\mu\nu}\right)$; a four-manifold equipped with a Lorentzian metric, with a corresponding Christoffel connection as defined by Eq.~(\ref{Christoffel}). Imposing a time orientation on $\left(\mathcal{M}, g_{\mu\nu}\right)$ allows one to extend the local causal structure defined by the light cones to a  global causal structure. Locally, the causal structure is akin to that of special relativity --- this ensues due to the fact that one may always approximate local flatness in GR. Globally however, topology may not be so trivial, and objects such as manifold singularities or self-identified ``twists'' within $\mathcal{M}$ may result in a confused notion of causality. This is something to be avoided for standard GR. One may make the following mathematical statement~\cite{GPOPILS:Rajan:2016, GRAECS:Hawking:1979}: A spacetime $\left(\mathcal{M}, g_{\mu\nu}\right)$ is time orientable if and only if there exists a globally defined timelike vector field on $\mathcal{M}$. All candidate spacetimes examined in this thesis are time-orientable.


\section{The axioms of GR}\label{einsteineqs}


%
%
\subsection{Einstein equivalence principle}

Gravity is encoded by the Christoffel connection $\Gamma^{\mu}{}_{\alpha\beta}$ on a topological four-manifold equipped with an associated metric tensor $g_{\mu\nu}$ of Lorentzian signature such that:
\begin{itemize}
    \item The universality of free fall is defined by the geodesic equations of motion of the Christoffel connection with respect to the chosen coordinate system. This means that a curve in the manifold parameterised by some arbitrary scalar parameter $\gamma$ is a geodesic if and only if the tangent vectors to the curve, given by $t^{\mu}=\frac{dX^{\mu}}{d\gamma}$, satisfy the following differential equation (known as the geodesic equation):
    \begin{equation}\label{geodesic1}
        \frac{\d^2X^{\mu}}{\d\gamma^2}+\Gamma^{\mu}{}_{\alpha\beta}\frac{\d X^{\alpha}}{\d\gamma}\frac{\d X^{\beta}}{\d\gamma} = f(\gamma)\frac{\d X^{\mu}}{\d\gamma} \ .
    \end{equation}
    The parameter $\gamma$ is deemed to be ``affine'' if $f(\gamma)=0$ in Eq.~(\ref{geodesic1}), reducing the geodesic equation to
    \begin{equation}\label{affinegeodesic}
        \frac{\d^2X^{\mu}}{\d\gamma^2}+\Gamma^{\mu}{}_{\alpha\beta}\frac{\d X^{\alpha}}{\d\gamma}\frac{\d X^{\beta}}{\d\gamma} = 0 \ .
    \end{equation}
    \item The flat space limit of spacetime recovers the theory of special relativity, where the metric tensor becomes the Minkowski metric:\\ $g_{\mu\nu} = \eta_{\mu\nu} = \ \mbox{diag}\left(-1, 1, 1, 1\right)$.
\end{itemize}
This provides the framework for how objects propagate through gravitational ``fields'', \emph{i.e.} curved spacetime. Still required is to fix the geometrodynamics; a codifying set of field equations which relate the curvature of the manifold to the source of that curvature --- in this case the specific distribution of mass and energy within the manifold itself.
\vspace*{12pt}

\subsection{Einstein field equations}\label{energyconditions}

The final step to establishing the framework of GR is to examine the Einstein field equations. They are canonically presented as~\cite{G:Misner:2000}:\footnote{An alternative form includes a term containing the cosmological constant $\Lambda$; approximating $\Lambda\approx 0$ is common practice since $\Lambda$ has been observed to be extremely small.}
\begin{equation}\label{EEQ}
    R_{\mu\nu} - \frac{1}{2}R \, g_{\mu\nu} = G_{\mu\nu} = \frac{8\pi G_{N}}{c^4} T_{\mu\nu} \ .
\end{equation}
The left-hand side of the equations describes the ``fields'' themselves, \emph{i.e} the curvature of spacetime. The right-hand side describes the source of the curvature; this is the distribution of stress-energy, energy density and momentum throughout the spacetime.

\subsubsection{The quasi-local curvature of spacetime}

This is encoded in the expression $R_{\mu\nu}-\frac{1}{2}Rg_{\mu\nu}$.\footnote{This is only the Ricci curvature. To fully express global curvature one must also factor into account the Weyl curvature. Qualitatively, Ricci curvature contains information pertaining to how volumes of objects are distorted in the presence of tidal forces, while Weyl curvature encodes the changes in shape. For more information on the nature of Weyl and Ricci curvature, please see~\cite{Weylcurv}.} To understand what these symbols mean, observe the following:
\begin{itemize}
    \item $g_{\mu\nu}$ --- This is the metric tensor (see \S~\ref{metric}). The desired physics means one enforces Lorentzian signature.
    \item $R_{\mu\nu}$ --- This is the Ricci curvature tensor, constructed \textit{via} the following steps: First the connection, specialised to the Christoffel symbols of the second kind, given by
    \begin{equation}
        \Gamma^{\mu}{}_{\alpha\beta} = \frac{1}{2}g^{\mu\nu}\left(\partial_{\alpha}g_{\nu\beta}+\partial_{\beta}g_{\nu\alpha}-\partial_{\nu}g_{\alpha\beta}\right) \ ,
    \end{equation}
    then the four-indexed Riemann curvature tensor defined as~\cite{G:Misner:2000}:
    \begin{equation}
        R^{\mu}{}_{\nu\alpha\beta} = \partial_{\alpha}\Gamma^{\mu}{}_{\nu\beta}-\partial_{\beta}\Gamma^{\mu}{}_{\nu\alpha}+\Gamma^{\mu}{}_{\sigma\alpha}\Gamma^{\sigma}{}_{\nu\beta}-\Gamma^{\mu}{}_{\sigma\beta}\Gamma^{\sigma}{}_{\nu\alpha} \ ,
    \end{equation}
    and finally the Ricci curvature tensor $R_{\mu\nu}$ is obtained \textit{via} contraction of the Riemann tensor on the first and third indices~\cite{G:Misner:2000}:
    \begin{equation}
        R_{\mu\nu} = R^{\sigma}{}_{\mu\sigma\nu} \ .
    \end{equation}
    \item The Ricci scalar $R$. This is obtained by contracting the inverse metric with the Ricci tensor~\cite{G:Misner:2000};
    \begin{equation}
        R = g^{\mu\nu}R_{\mu\nu} \ .
    \end{equation}
\end{itemize}
Gathering all of these terms defines the Einstein curvature tensor, $G_{\mu\nu}=R_{\mu\nu}-\frac{1}{2}R g_{\mu\nu}$. Note that this implies that there are in total ten equations in the Einstein field equations at each point in spacetime; this comes from the fact that both $R_{\mu\nu}$ and $g_{\mu\nu}$ have a four-by-four matrix representation which is strictly symmetric. The left-hand side of the Einstein field equations is now fully understood; it is important to note that all information encoded in the curvature tensors and scalar invariants ultimately stems directly from the metric tensor, $g_{\mu\nu}$. It therefore follows that defining a line element (and hence a metric environment) is all one requires in order to characterise all the relevant information pertaining to the curvature of a particular spacetime. All analyses of candidate spacetimes are hence conducted beginning solely with the metric as the starting point. It is worth noting that there are numerous other curvature tensors and curvature invariants which provide geometric information about the four-manifold but are not explicitly part of the Einstein field equations; specifically the tensors and invariants which will form additional parts of the analyses in this thesis include the Weyl tensor $C_{\mu\nu\alpha\beta}$, the Ricci contraction $R_{\mu\nu}R^{\mu\nu}$, the Kretschmann scalar $R_{\mu\nu\alpha\beta}R^{\mu\nu\alpha\beta}$, and the Weyl contraction $C_{\mu\nu\alpha\beta}C^{\mu\nu\alpha\beta}$ (for details on these mathematical objects see reference~\cite{G:Misner:2000}).
\clearpage

\subsubsection{The stress-energy-momentum tensor}

The right-hand side of the Einstein field equations encodes the source of the curvature in spacetime. Before proceeding note the following --- until now all expressions unless otherwise stated have assumed the international system of units (SI-units); henceforth geometrodynamic units shall be used to simplify calculations as is conventional in physics. As such, the speed of light in a vacuum $c=1$, and Newton's gravitational constant $G_{N}=1$. Multiplication of combinations of $c$ and $G_{N}$ where appropriate will return SI-units if desired. Hence the rephrased Einstein field equations are:
\begin{equation}\label{einsteineqsfinal}
    G_{\mu\nu} = R_{\mu\nu}-\frac{1}{2}Rg_{\mu\nu} = 8\pi T_{\mu\nu} \ .
\end{equation}
The general form of $T_{\mu\nu}$ can be given as follows~\cite{GAITEGR:Hartle:2003}:
\begin{equation}
    T_{\mu\nu} = \begin{bmatrix}
    \rho & F_{i} \\
    F_{j} & \pi_{ij}
    \end{bmatrix} \ ,
\end{equation}
where the latin indices index the three spacial dimensions;\footnote{US English uses ``spatial'' as opposed to the British English ``spacial''; this is a matter of taste.} $i,j\in\left\lbrace 1, 2, 3\right\rbrace$. Defining these objects:
\begin{itemize}
    \item $\rho$ is the energy density of the relativistic masses present in the spacetime.
    \item $F_{i}$ and $F_{j}$ represent the directional energy flux of relativistic mass across each $x_{i}$ surface (these are analogous to the Poynting vectors from electromagnetism), encoding momentum density.
    \item $\pi_{ij}$ represents the spacial shear stress tensor, with the diagonal entries (those independent of direction) defining radial and transverse pressure.
\end{itemize}
Eq.~(\ref{EEQ}) can now be fully understood from first principles, completing the elementary formulation of GR.


\chapter{GR: Standard definitions}\label{B:GRbasics}
%

 It is worth presenting and briefly analysing several key concepts which are referred to frequently when conducting standard GR analysis of candidate spacetimes.
 

\section{Killing symmetries}\label{Killing}

%
\subsubsection{Killing vector fields}

A Killing vector field in the context of spacetime is defined as a vector field which preserves the metric~\cite{G:Misner:2000}. This means that $\xi^{\mu}$ is a Killing vector if and only if any displacement of a set of points by some $\xi^{\mu}\d x_{\mu}$ leaves all distance relationships are unchanged. It follows that displacements along Killing vector fields are isometries; bijective maps $f:\mathbb{R}^{n}\rightarrow\mathbb{R}^{n}$ such that $g_{\mu\nu}\left(f(x), f(y)\right)=g_{\mu\nu}\left(x, y\right)$. A physical corollary of this displacement being an isometric map is that worldlines of particles displaced infinitesimally in an arbitrary direction along the Killing vector field are congruent. From this definition, one can derive Killing's equation: $\mathcal{L}_{\xi} \ g_{\mu\nu} = 0$; that the Lie derivative of the metric tensor with respect to a given Killing vector field $\xi$ is manifestly zero.\footnote{For details on the Lie derivative and its relationship with the Killing vector, please see reference~\cite{G:Misner:2000}.}

Killing's equation is often written as $\nabla_{\mu}\xi_{\nu} + \nabla_{\nu}\xi_{\mu} = 0$. Using the definition of the Lie derivative in terms of partial derivatives, the metricity condition, and the lack of torsion, it is straightfroward to prove this:
\begin{eqnarray}
    \mathcal{L}_{\xi}\,g_{\mu\nu} &=& \xi^{\alpha}\partial_{\alpha}g_{\mu\nu} + g_{\mu\alpha}\partial_{\nu}\xi^{\alpha} + g_{\alpha\nu}\partial_{\mu}\xi^{\alpha} \nonumber \\
    &=& \xi^{\alpha}\nabla_{\alpha}g_{\mu\nu} + g_{\mu\alpha}\nabla_{\nu}\xi^{\alpha} + g_{\alpha\nu}\nabla_{\mu}\xi^{\alpha} \nonumber \\
    &=& \nabla_{\nu}\xi_{\mu} + \nabla_{\mu}\xi_{\nu} = 0 \ .
\end{eqnarray}

\subsubsection{Conserved physical quantities}

%
%
%
Given an affinely parameterised geodesic, parameterised by some affine scalar parameter $\gamma$, the following result ensues as a corollary from the Einstein equivalence principle (for the affinely parameterised tangent vector to the curve, $X^{\nu}$): $X^{\mu}\nabla_{\mu}X^{\nu}=0$. \textit{i.e.} the four-acceleration for geodesic motion is zero. Combine this with Killing's equation, and the metricity condition, and one can easily prove the following assertion: for any Killing vector field $\xi^{\mu}$ on a spacetime $\left(\mathcal{M}, g_{\mu\nu}\right)$, and some affinely parameterised tangent vector field $X^{\nu}$ to an affinely parameterised geodesic, $\xi^{\mu}g_{\mu\nu}X^{\nu}=K$, $K\in\mathbb{R}$.

\textit{Proof:}
\begin{eqnarray}
\frac{\d}{\d\gamma}\left(\xi^{\mu}g_{\mu\nu}X^{\nu}\right) &=& \left(X^{\sigma}\nabla_{\sigma}\xi^{\mu}\right)g_{\mu\nu}X^{\nu}
+ \xi^{\mu}\left(X^{\sigma}\nabla_{\sigma}g_{\mu\nu}\right)X^{\nu}
+ \xi^{\mu}g_{\mu\nu}\left(X^{\sigma}\nabla_{\sigma}X^{\nu}\right) \nonumber \\[3pt]
&=& X^{\sigma}\nabla_{\sigma}\xi_{\nu}X^{\nu}
= X^{\sigma}\nabla_{(\sigma}\xi_{\nu)}X^{\nu} = 0 \ . \qquad \mbox{\textit{QED.}}
\end{eqnarray}
It follows that given some arbitrary spacetime, where there are symmetries in the spacetime geometry with associated Killing vector fields, there will be conserved physical quantities in accordance with the same conservation laws that arise from classical analytic mechanics~\cite{G:Misner:2000}. To see that these symmetries exist, one may use the example of a stationary metric expressed with respect to some coordinate patch $\left(t, r, \theta, \phi\right)$, where $t$ is the temporal coordinate (this argument can be generalised to arbitrary coordinate patches in various domains with ease). Stationary implies time-independent; it follows that $\partial g_{\mu\nu}/{\partial t}=0$. The geometric interpretation of this relation is that any curve in the manifold can be shifted by some $\Delta t=\epsilon, \ \epsilon\in\mathbb{R}$, to form a congruent curve; \emph{i.e.} the transformation $t\rightarrow t+\epsilon$ preserves the metric. One may conclude that $\xi^{\mu}=\partial_{t}=\left(1, 0, 0, 0\right)=\delta^{\mu}{}_{t}$ is a Killing vector.

Given a timelike test particle's worldline, the affinely parameterised tangent vector field to the worldline will be the four-momentum of the particle: $P^{\nu} = m_0V^{\nu}$, where $V^{\nu}$ is the four-velocity. For the specific stationary example where $\xi^{\mu}=\partial_{t}$ is a Killing vector, in suitable coordinates, one has the following:
\begin{eqnarray}\label{Energy}
    \xi^{\mu}g_{\mu\nu}P^{\nu} &=& \delta^{\mu}{}_{t}g_{\mu\nu}P^{\nu}
    = g_{t\nu}P^{\nu}
    = E \ .
\end{eqnarray}
It follows that the symmetry of the spacetime in the $t$-coordinate ultimately yields the conservation of energy, $E$, along the worldlines of test particles. This specific conserved quantity is relied upon heavily in the subsequent analyses of candidate spacetimes; also utilised (where appropriate) is the metric independence of the azimuthal $\phi$-coordinate, which implies the conservation of the quantity $g_{\phi\phi}P^{\phi}$. This quantity in turn implies the conservation of the azimuthal angular momentum $L_\phi$, along similar lines as Eq.~(\ref{Energy}). See reference~\cite{G:Misner:2000} for more details on these conserved quantities in GR.


\section{Geometric constraints}\label{sphericalsym}

All candidate spacetimes analysed in this thesis possess either a spherically symmetric or axisymmetric matter distribution. Furthermore, the majority of geometries are either stationary or static (the exception to this is the Vaidya black-bounce geometry briefly discussed in \S~\ref{SVVaidya}).

A spacetime is static if it admits a global hypersurface-orthogonal timelike Killing vector field~\cite{TLSSOS:Ellis:1973}. Physically this means that the geometry is time-independent and also irrotational (note that spherical symmetry implies lack of rotation). Mathematically one may always diagonalise the matrix representation of a static metric. This is a special case of the more general stationary spacetime --- one which is time-independent but permits rotation (hence mathematically only requires that it admits a global timelike Killing vector field).

Axisymmetric spacetimes possess at least one spacelike Killing vector field along a line of manifest symmetry; typically this is denoted as a symmetry in one of the chosen angular coordinates, \textit{e.g.} $\xi^{\phi}$. Spherical symmetry has even more structure, with manifest symmetry in both angular directions; \textit{i.e.} $\xi^{\theta}$ is a global spacelike Killing vector field as well (in the suitably chosen coordinate basis).

A corollary of spherical symmetry is that one may fix either angular coordinate arbitrarily when discussing the worldlines of particles and simplify any physical problems to a reduced equatorial state --- for example when calculating innermost stable circular orbits (ISCOs) and photon spheres\footnote{Discussion and definition of these objects is presented in \S~\ref{ISCOintro}.} this greatly reduces the complexity of calculations.

Also, in general for spherically symmetric matter distributions one has the following form of the stress-energy tensor~\cite{TLSSOS:Ellis:1973}:
\begin{equation}\label{stress}
    T_{\mu\nu} = \begin{bmatrix}
    \rho & 0 & 0 & 0 \\
    0 & p_r & 0 & 0 \\
    0 & 0 & p_t & 0 \\
    0 & 0 & 0 & p_t
    \end{bmatrix} \ .
\end{equation}
To give two examples, the Kerr geometry is stationary axisymmetric, whilst Schwarz\-schild is static spherically symmetric.
\clearpage

%

\section{Classical energy conditions}

As well as the form of the components of $T_{\mu\nu}$ being dictated by the Einstein field equations, there are numerous (at least seven) different energy conditions in the context of classical GR which imply mathematical constraints on $T_{\mu\nu}$~\cite{LW:Visser:1995}. The most fundamental of these conditions is the null energy condition (NEC), the satisfaction of which is mathematically represented by~\cite{LW:Visser:1995}
\begin{equation}
    \mbox{NEC} \quad \Longleftrightarrow \quad T_{\mu\nu}t^{\mu}t^{\nu}\geq 0 \ ,
\end{equation}
where $t^{\mu}$ is any arbitrary null vector, and the inequality must hold globally. Given a stress-energy tensor of the form in Eq.~(\ref{stress}), \textit{i.e.} for any static spherically symmetric spacetime in a suitably chosen coordinate system, this assertion simplifies to the following in terms of the pressures within the matter distribution:
\begin{equation}
    \mbox{NEC} \quad \Longleftrightarrow \quad \rho+p_r \geq 0, \ \mbox{and} \ \rho+p_t\geq 0 \ .
\end{equation}
The primary energy conditions of interest are the null, weak, strong, and dominant energy conditions. In static spherical symmetry one has the specialised result that if the NEC is violated it implies direct violation of the weak, strong, and dominant energy conditions also; this fact is used in the analyses of the various candidate spacetimes in this thesis (for details pertaining to the corollaries of the violation of the NEC, please see reference~\cite{LW:Visser:1995}). Canonically, traversable wormhole geometries (see \S~\ref{BHMint}) have required a violation of the NEC~\cite{WISATUFIT:Morris:1988}, and as such represent what are known as ``exotic'' solutions to the Einstein equations.

Regular black hole geometries (see \S~\ref{BHMint}) typically require a violation of \emph{some} of the energy conditions, however the NEC has a chance to be satisfied. This is due to the fact that in the wormhole context the violation of the radial NEC is directly related to the ``flare-out'' condition pertaining to wormhole throats~\cite{LW:Visser:1995}. For regular black hole spacetimes, one instead expects a violation of the strong energy condition (SEC). This is a consequence of the fact that the lack of curvature singularities implies geodesic completeness on the manifold, and geodesic completeness implies the SEC will not be satisfied (for details on the corollaries of geodesic completeness, please see reference~\cite{TLSSOS:Ellis:1973}). In terms of the principal pressures for the case presented in Eq.~(\ref{stress}), satisfaction of the SEC amounts to:
\begin{equation}
    \mbox{SEC} \quad \Longleftrightarrow \quad \rho+p_r+2p_t \geq 0 \ .
\end{equation}
This condition will be utilised occasionally as part of the analyses in this thesis.


\section{Horizons}\label{Horizon}

Horizons in the context of GR are subtle physical objects, and there are various different types of horizon corresponding to disparate technical definitions. Fundamentally, all classes of horizon are characterised as a physical surface within the four-manifold permitting the passage of massive and massless particles in one direction only, and whose location is such that the notion of time for all external observers comes to a stop at the horizon~\cite{G:Misner:2000, GR:Wald:2010, TLSSOS:Ellis:1973, LW:Visser:1995}.

The most commonly encountered class of horizon is the event horizon, or absolute horizon --- this is primarily due to the popularity of the event horizon in science fiction, although some physicists and applied mathematicians also advocate using the definition of the event horizon as it enables a more straightforward environment in which to prove mathematical theorems~\cite{POOH:Visser:2014}. One has the following definition of the event horizon~\cite{LW:Visser:1995}: For each asymptotically flat region the associated future/past event horizon is defined as the boundary of the region from which causal curves (that is, null or timelike curves) can reach asymptotic future/past null infinity.\footnote{Future/past null infinity is defined as~\cite{GAITEGR:Hartle:2003}: the spacial surface corresponding to the set of all coordinate locations which outgoing/ingoing null curves (\emph{e.g.} light rays) are able to asymptotically approach as $\vert t\vert\rightarrow+\infty$. This boundary is only defined in an asymptotically flat region of spacetime, and is one of the ``conformal infinities'' used in constructing the Carter--Penrose diagrams which diagrammatically represent the global causal structure of specific spacetimes (the others being future/past timelike infinity and spacelike infinity). For details on conformal infinities, timelike and spacelike infinity, and their uses in general relativity, please see reference~\cite{conformal}.} It turns out that the event horizon may not in fact be the most well-informed category of horizon when describing physical reality --- one runs into many technical issues, one of which concerns black hole evaporation across large timescales and the potential recovery of information deemed to be strictly ``lost'' by the definition of these event/absolute horizons (for details see reference~\cite{POOH:Visser:2014}).

Another type of horizon is the apparent horizon~\cite{LW:Visser:1995}, defined locally in terms of trapped surfaces. Pick any closed, spacelike, two-dimensional surface (a two-surface). At any point on the two-surface there are two null geodesics that are orthogonal to the surface. They can be used to define inward and outward propagating wavefronts. If the area of both inward and outward propagating wavefronts decreases as a function of time, then the original two-surface is a trapped surface and one is inside the apparent horizon. More precisely, if the expansion of both sets of orthogonal null geodesics is negative, then the two-surface is a trapped surface. So the apparent horizon is the boundary between these trapped and untrapped surfaces, characterised by conditions on the focusing/defocusing of null geodesics. It should be noted that in a spherically symmetric, time-independent metric environment these two horizon definitions are equivalent. Given the fact the geometry remains unchanged with time, any apparent horizon is also an event horizon, however this is not the case with a dynamical metric environment, and it is important to distinguish between them. When specialised to spherical symmetry, the apparent horizon simply corresponds to the locus of coordinate locations such that radially propagating light rays have zero coordinate velocity (in most coordinate patches this will be when $\d\theta=0, \d\phi=0, \ \mbox{and} \ \frac{\d r}{\d t}=0$).\footnote{In the special case of both a diagonal metric environment \emph{and} spherical symmetry, this definition implies that the location of the horizon is simply defined by the surface which forces $g_{tt}=0$; this simplification is utilised where appropriate.}

%
Crucially, one typically defines a black hole region within a spacetime as the region of the geometry which lies strictly within an event horizon. The outer horizons in this thesis, unless otherwise stated, are event horizons.

For the purposes of this thesis, the final type of horizon discussed is the Cauchy horizon (often referred to as an ``inner'' horizon). For some spacelike surface $\Sigma$ within a spacetime, the associated future Cauchy horizon is defined as the boundary of the region from which all past-directed causal curves intersect $\Sigma$~\cite{LW:Visser:1995}. The inner horizons present in this thesis, unless otherwise stated, are Cauchy horizons.


\section{Singularity}\label{singularity}

When discussing the presence/location of a singularity within a specific geometry, it is important to differentiate between a gravitational singularity and a coordinate singularity. Both can be mathematically characterised by coordinate locations which correspond to poles of a coefficient function or functions in the metric; the qualitative difference is that a gravitational singularity is representative of a physical source of infinite curvature in the spacetime (a tear in the topological manifold), whilst coordinate singularities represent nothing physically special at all and may always be removed through an alternative choice of coordinate patch. One therefore needs a method of mathematically separating the two so as to draw meaningful physical conclusions. There are multiple ways of doing this, for example \textit{via} geodesic incompleteness or \textit{via} analysis of the Riemann curvature tensor with respect to an orthonormal tetrad~\cite{TLSSOS:Ellis:1973}. For the purposes of this thesis, a gravitational singularity will generally be defined as a coordinate location which forces one or more of the nonzero Riemann curvature tensor components $R^{\hat{\alpha}\hat{\beta}}{}_{\hat{\mu}\hat{\nu}}$ to have infinite magnitude with respect to an orthonormal basis.
\clearpage


\section{ISCO and photon sphere}\label{ISCOintro}
\enlargethispage{20pt}

The innermost stable circular orbit (ISCO) and the photon sphere correspond to specific locations within a given spacetime which are of significant observational interest. The ISCO is defined as the innermost stable circular orbit which a massive particle is able to maintain around some massive object~\cite{G:Misner:2000}, and physically corresponds to the innermost edge of the accretion disc; an important astrophysical object for empirical observation. The photon sphere corresponds to the locus of coordinate locations sufficiently near the centralised massive object such that photons (or any massless particle) are forced to propagate in circular geodesic orbits (which may be stable or unstable)~\cite{G:Misner:2000}. Given an appropriate test particle for each case, both of these mathematical concepts are characterised by coordinate locations corresponding to stationary points of the effective energy potentials of the test particles --- this is a corollary of the desired physics inherent in orbital mechanics, that a test particle orbiting a massive body be in mechanical equilibrium (in the classical sense; see reference~\cite{POM:Campbell:1943} for details). To find the coordinate location of the ISCO one reparameterises the line element using tangent vectors to a timelike worldline, whilst for the photon sphere one reparameterises the line element with respect to the tangent vectors of a null worldline.

Given the fact that both the ISCO and the photon sphere are strictly \emph{circular} orbits, they are typically only discussed in the context of spherically symmetrical geometries. There are much messier generalisations, ISCOs and photon rings, for the equatorial plane of axisymmetric geometries such as Kerr spacetime. Without loss of generality one may always (through an appropriate choice of coordinate patch) define the effective energy potentials of arbitrary test particles as some function $V(r, \theta)$, the form of which is found \textit{via} analysis of the conserved quantities implied by the Killing symmetries of each candidate spacetime. In spherical symmetry, $V(r,\theta)\rightarrow V(r)$, and the coordinate locations of the circular orbits can be found at the $r$-values which satisfy $V^{'}(r)=0$. The stability of each circular orbit is then determined by the sign of $V^{''}(r)$ in the following manner~\cite{POM:Campbell:1943}:
\begin{eqnarray}
V^{''}(r)<0 \qquad &\Longrightarrow& \qquad \mbox{unstable orbit} \ . \nonumber \\
V^{''}(r)=0 \qquad &\Longrightarrow& \qquad \mbox{marginally stable orbit} \ . \nonumber \\
V^{''}(r)>0 \qquad &\Longrightarrow& \qquad \mbox{stable orbit} \ .
\end{eqnarray}
Generalisations to the axisymmetric effective potentials $V(r,\theta)$ are discussed in context on a case-by-case basis.

The notion of stability is defined to be whether small peturbations orthogonal to the geodesic orbit on either side cause the particle to remain in the circular orbit (stable), or cause it to follow some altogether qualitatively different worldline (unstable). This raises a notable point: terminologically ``ISCO'' is standard, but one would assume that use of the word ``stable'' in the term ISCO implies that the corresponding orbit must always be stable. In general, this is not the case. Quite often the ISCO is at best one-sided stable, and in the specific case for the Morris--Thorne traversable wormhole~\cite{WISATUFIT:Morris:1988}, the ISCO is in fact two-sided \emph{unstable}. This is merely an oddity in use of language; there are no significant ramifications pertaining to whether the ISCO is stable/unstable in the context of the desired astronomical analysis.

The motivation for examining these objects is to compare given specific solutions to the Einstein equations to observational data provided by astronomers. Astronomers will also be informed as to where to point their telescopes (with respect to a specified coordinate patch) in order to gain as much pertinent information regarding the behaviour of both massless and massive particles as they near the region of spacetime with the highest curvature. Due to the fact that the ISCO corresponds to the innermost edge of the accretion disc, it is inherently crucial for the astrophysical imaging of black hole regions. The photon sphere is also of significance for astronomers; for instance these objects are of particular importance to the EHT~\cite{FM87EHTR:Akiyama:2019, FM87EHTR1:Akiyama:2019, FM87EHTR2:Akiyama:2019, FM87EHTR3:Akiyama:2019, FM87EHTR4:Akiyama:2019, FM87EHTR5:Akiyama:2019}, and the James Webb Space Telescope~\cite{JamesWebb1, JamesWebb2, JamesWebb3, JamesWebb4, JamesWebb5, JamesWebb6}.

%
It should be noted that in Newtonian gravity, there is no concept of an ISCO, as one may easily stabilise the orbits of test particles which are arbitrarily close to a mass source. It follows that ISCO locations are intrinsically general relativistic.


\section{Regge-Wheeler equation}
\enlargethispage{20pt}

In order to use the Regge--Wheeler equation to conduct tractable analysis, one first rewrites the metric in terms of a naturally defined tortoise coordinate. In a spherically symmetric geometry with a diagonal metric environment, the tortoise coordinate $r_{*}$ is constructed such that it must satisfy the first-order differential equation~\cite{RELSAGFFDBH:Boonserm:2013}
\begin{eqnarray}
    \d r_{*} &=& \sqrt{-\frac{g_{rr}}{g_{tt}}} \, \d r \ ; \nonumber \\[4pt]
    \Longrightarrow \quad r_{*} &=& \bigintssss \sqrt{-\frac{g_{rr}}{g_{tt}}} \, \d r \ .
\end{eqnarray}
For the specialised case where $g_{rr}g_{tt}=-1$, one has the simplified tortoise coordinate given by
\begin{equation}
    \d r_{*} = g_{rr} \, \d r \ ; \qquad \Longrightarrow \qquad r_{*} = \int g_{rr} \, \d r \ .
\end{equation}
In a spherically symmetric \emph{and} static environment, without loss of generality, this enables one to rewrite the metric as follows:
\begin{equation}
    \d s^{2} = A(r)\left\lbrace -\d t^{2} + \d r_{*}^{2}\right\rbrace+g_{\theta\theta}\,\d\Omega^{2}_{2} \ .
\end{equation}
The use of the tortoise coordinate normalises the relation between $\d t^{2}$ and $\d r^{2}$, such that radially propagating test particles (\emph{i.e.} $\d\Omega^{2}_{2}=0$) have worldlines which correspond to $\pi/4$-radian lines on a spacetime diagram. This is a useful feature, however it comes at the cost of some hitherto unknown conformal factor in the form of $A(r)$ (for radial null propagation use of the tortoise coordinate is particularly nice as one may simply ignore the effect of the conformal factor in view of the fact that $\d s^{2}=0$).

Having rewritten the metric in terms of $r_{*}$, the Regge--Wheeler equation enables one to draw conclusions pertaining to the energy potentials of the following objects, subject to linear peturbations induced by greybody factors~\cite{RELSAGFFDBH:Boonserm:2013}:
\begin{itemize}
    \item The spin zero massive or massless scalar field minimally coupled to gravity.
    \item The spin one Maxwell vector field.
    \item The spin two axial peturbation mode (odd parity).
\end{itemize}
It follows that the Regge--Wheeler equation is crucial in analysing the ringdown of the associated quasi-normal modes for candidate spacetimes, and indeed it is specifically used in Chapter~\ref{C:AMCQNM} for approximation of the spin one and spin zero quasi-normal modes. The general form of the Regge--Wheeler equation is given by~\cite{RELSAGFFDBH:Boonserm:2013}
\begin{equation}
    \partial_{r_{*}}^{2}\hat{\phi}+\lbrace \omega^2-\mathcal{V}\rbrace\hat\phi = 0 \ ,
\end{equation}
where $\hat\phi$ is the scalar or vector field of interest, $\mathcal{V}$ is the spin-dependent Regge--Wheeler potential for the test particle, and $\omega$ is a temporal frequency component in the Fourier domain.


\section{Surface gravity and Hawking temperature}
\enlargethispage{20pt}

Surface gravity in GR requires the existence of a stationary Killing horizon, defined to be a null hypersurface at a coordinate location where the norm of the Killing vector field goes to zero~\cite{G:Misner:2000, GR:Wald:2010}. Given a stationary Killing horizon in a spacetime, and some suitably normalised Killing vector field $\xi^{\mu}$, the surface gravity $\kappa$ is then calculated by evaluating the following equation at the coordinate location of the Killing horizon~\cite{G:Misner:2000, GR:Wald:2010}:
\begin{equation}
    \xi^{\mu}\nabla_{\nu}\xi_{\mu} = -\kappa \; \xi_{\nu} \ ,
\end{equation}
and \textit{via} Killing's equation and appropriate contraction with the metric tensor, one may rewrite this as
\begin{equation}
    \xi^{\mu}\nabla_{\mu}\xi_{\nu} = \kappa \; \xi_{\nu} \ .
\end{equation}
Surface gravity $\kappa$ is then directly related to the Hawking temperature $T_{H}$ as a consequence of Hawking evaporation by the following formula~\cite{TLSSOS:Ellis:1973}:
\begin{equation}
    T_{H} = \frac{\hbar\kappa}{2\pi k_{B}} \ .
\end{equation}

%
%
%

\chapter{List of publications}\label{C:Publications}


\section{Publications}

Presented is a full list of the current author's relevant works to-date, beginning in late October 2018. This consists of 22 published regular articles (one solo-author), one article published as part of conference proceedings (solo-author), and one MSc thesis. Works with an asterisk (*) at the beginning of their entry in the list are those from which significant content has been reproduced in this thesis. All works are also available on the arXiv.
\vspace*{7pt}

\textbf{Regular Articles:}
\begin{enumerate}
    \item Boonserm, Petarpa and Ngampitipan, Tritos and Simpson, Alex and Visser, Matt, \textit{``Exponential metric represents a traversable wormhole''},\newline Phys. Rev. D \textbf{98}, 8, 084048 (2018). \doi{10.1103/PhysRevD.98.084048}.
    \item *Simpson, Alex and Visser, Matt, \textit{``Black-bounce to traversable wormhole''}, JCAP \textbf{02}, 042 (2019). \doi{10.1088/1475-7516/2019/02/042}.
    \item *Simpson, Alex and Mart\'{i}n-Moruno, Prado and Visser, Matt, \textit{``Vaidya spacetimes, black-bounces, and traversable wormholes''}, Class. Quantum Grav. \textbf{36}, 14, 145007 (2019). \doi{10.1088/1361-6382/ab28a5}.
    \item *Simpson, Alex and Visser, Matt, \textit{``Regular Black Holes with Asymptotically Minkowski Cores''}, Universe \textbf{2020}, 6(1), 8 (2019).\newline \doi{10.3390/universe6010008}.
    \item Boonserm, Petarpa and Ngampitipan, Tritos and Simpson, Alex and Visser, Matt, \textit{``Decomposition of the total stress energy for the generalized Kiselev black hole''}, Phys. Rev. D \textbf{101}, 2, 024022 (2020).\newline \doi{10.1103/PhysRevD.101.024022}.
    \clearpage
    \item Boonserm, Petarpa and Ngampitipan, Tritos and Simpson, Alex and Visser, Matt, \textit{``Innermost and outermost stable circular orbits in the presence of a positive cosmological constant''}, Phys. Rev. D \textbf{101}, 2, 024050 (2020). \doi{10.1103/PhysRevD.101.024050}.
    \item *Lobo, Francisco S. N. and Simpson, Alex and Visser, Matt, \textit{``Dynamic thin-shell black-bounce traversable wormholes''}, Phys. Rev. D \textbf{101}, 12, 124035 (2020). \doi{10.1103/PhysRevD.101.124035}.
    \item *Berry, Thomas and Lobo, Francisco S. N. and Simpson, Alex and Visser, Matt, \textit{``Thin-shell traversable wormhole crafted from a regular black hole with asymptotically Minkowski core''}, Phys. Rev. D \textbf{102}, 6, 064054 (2020). \doi{10.1103/PhysRevD.102.064054}.
    \item *Berry, Thomas and Simpson, Alex and Visser, Matt, \textit{``Photon Spheres, ISCOs, and OSCOs: Astrophysical Observables for Regular Black Holes with Asymptotically Minkowski Cores''}, Universe \textbf{2021}, 7(1), 2 (2020). \doi{10.3390/universe7010002}.
    \item Baines, Joshua and Berry, Thomas and Simpson, Alex and Visser, Matt, \textit{``Painlev\'e\textendash{}Gullstrand Form of the Lense\textendash{}Thirring Spacetime''}, Universe \textbf{2021}, 7(4), 105 (2021). \doi{10.3390/universe7040105}.
    \item Baines, Joshua and Berry, Thomas and Simpson, Alex and Visser, Matt, \textit{``Unit-lapse versions of the Kerr spacetime''}, Class. Quantum Grav. \textbf{38}, 5, 055001 (2021). \doi{10.1088/1361-6382/abd071}.
    \item Baines, Joshua and Berry, Thomas and Simpson, Alex and Visser, Matt, \textit{``Darboux diagonalization of the spatial 3-metric in Kerr spacetime''}, Gen. Rel. Grav. \textbf{53}, 3 (2021). \doi{10.1007/s10714-020-02765-0}.
    \item *Lobo, Francisco S. N. and Rodrigues, Manuel E. and Silva, Marcos V. de S. and Simpson, Alex and Visser, Matt, \textit{``Novel black-bounce spacetimes: Wormholes, regularity, energy conditions, and causal structure''}, Phys. Rev. D \textbf{103}, 8, 084052 (2021). \doi{10.1103/PhysRevD.103.084052}.
    \item Berry, Thomas and Simpson, Alex and Visser, Matt, \textit{``Regularity of a General Class of ``Quantum Deformed'' Black Holes''}, Universe \textbf{2021}, 7(6), 165 (2021). \doi{10.3390/universe7060165}.
    \item *Franzin, Edgardo and Liberati, Stefano and Mazza, Jacopo and Simpson, Alex and Visser, Matt, \textit{``Charged black-bounce spacetimes''}, JCAP \textbf{07}, 036 (2021). \doi{10.1088/1475-7516/2021/07/036}.
    \item *Simpson, Alexander Marcus, \textit{``Ringing of the Regular Black Hole with Asymptotically Minkowski Core''}, Universe \textbf{2021}, 7(11), 418 (2021).\newline \doi{10.3390/universe7110418}.
    \clearpage
    \item Baines, Joshua and Berry, Thomas and Simpson, Alex and Visser, Matt, \textit{``Killing Tensor and Carter Constant for Painlev\'{e}--Gullstrand Form of Lense--Thirring Spacetime''}, Universe \textbf{2021}, 7(12), 473 (2021).\newline \doi{10.3390/universe7120473}.
    \item Baines, Joshua and Berry, Thomas and Simpson, Alex and Visser, Matt, \textit{``Geodesics for the Painlev\'{e}--Gullstrand form of Lense--Thirring Spacetime''}, Universe \textbf{2022}, 8(2), 115 (2021). \doi{10.3390/universe8020115}.
    \item *Simpson, Alex and Visser, Matt, \textit{``The eye of the storm: A regular Kerr black hole''}, JCAP \textbf{03}, 011 (2022). \doi{10.1088/1475-7516/2022/03/011}.
    \item *Simpson, Alex and Visser, Matt, \textit{``Astrophysically viable Kerr-like spacetime -- into the eye of the storm''}, Phys. Rev. D \textbf{105}, 6, 064065 (2022). \doi{10.1103/PhysRevD.105.064065}.
    \item Baines, Joshua and Berry, Thomas and Simpson, Alex and Visser, Matt, \textit{``Non-equatorial circular geodesics for the Painlev\'{e}--Gullstrand form of Lense--Thirring spacetime''}, Gen. Rel. Grav. \textbf{54}, 79 (2022).\newline \doi{10.1007/s10714-022-02963-y}.
    \item *Berry, Thomas and Simpson, Alex and Visser, Matt, \textit{``General-relati\-vistic thin-shell Dyson mega-spheres''}, Phys. Rev. D. \textbf{106}, 8, 084001 (2022).
    \doi{10.1103/PhysRevD.106.084001}.
\end{enumerate}
\vspace*{15pt}

\textbf{Conference Proceedings Article:}
\begin{itemize}
    \item *Simpson, Alex, \textit{``From black-bounce to traversable wormhole, and beyond''}, 16th Marcel Grossmann Meeting on~Recent Developments in Theoretical and Experimental General Relativity, Astrophysics and Relativistic Field Theories (2021). arXiv:\href{https://arxiv.org/abs/2110.05657}{2110.05657}.
\end{itemize}
\vspace*{15pt}

\textbf{MSc Thesis:}
\begin{itemize}
    \item *Simpson, Alex, \textit{``Traversable Wormholes, Regular Black Holes, and Black-Bounces''}, Victoria University of Wellington Te Herenga Waka (2019). arXiv:\href{https://arxiv.org/abs/2104.14055}{2104.14055}.
\end{itemize}
\clearpage


\section{Declaration}
In an effort to keep this thesis somewhat self-contained, some results are quoted from, and some content has, (in suitably modified form, that is, modulo prudent changes in presentation and updates and extensions in scientific content), been extracted from both the current author's MSc thesis~\cite{Masters:Simpson:2019} and Mr Thomas Berry's MSc thesis~\cite{Berrythesis}.

Work updated and modified from the current author's MSc thesis:
\begin{itemize}
    \item Portions of Appendices~\ref{A:GR} and~\ref{B:GRbasics};
    \item Summaries of relevant content presented in articles [2--4] from the above list; that is, references~\cite{BTTW:Simpson:2019, VSBATW:Moruno:2019, RBHWAMC:Simpson:2019}. These summaries form some of the content in Chapters~\ref{C:SVog} and~\ref{C:AMCog}.
\end{itemize}
Work updated and modified from Mr Thomas Berry's MSc thesis:
\begin{itemize}
    \item Summaries of relevant content presented in articles [8--9] from the above list; that is, references~\cite{PSISCOOSCO:Berry:2020, TTWCFARBHWAMC:Berry:2020}. These summaries form some of the content in Chapters~\ref{C:AMCog} and~\ref{C:SVthinshell}.
\end{itemize}


\section{Bibliometrics}

As at March 25th, 2023, data obtained from~\href{https://inspirehep.net/authors/1720242}{AlexSimpson-InspireHEP}:
\begin{itemize}
    \item Total citations: \textbf{736}
    \item H-index: \textbf{15} (15 papers with at least 15 citations each)
    \item i10-index: \textbf{18} (18 papers with at least 10 citations each)
    \item G-index: \textbf{6} (6 papers with at least $6^2=36$ citations each)
    \item M-index: \textbf{3.5} (h-index/number of years since first publication)
\end{itemize}
\clearpage

%
%